\documentclass[aps,a4paper, preprint, superscriptaddress,preprintnumbers,floatfix,nofootinbib,amsmath,amssymb]{revtex4}
\usepackage{graphicx}
\usepackage{epsfig}
\usepackage{amsmath}
\usepackage{amsfonts}
\usepackage{amssymb}
\usepackage{url}
\usepackage{hyperref}
\usepackage{subfigure}
\usepackage{hhline}
\usepackage{color}
\usepackage{bm}
\usepackage{array,multirow}
\usepackage{cancel}
\usepackage{cleveref}
\newcommand{\beqa}{\begin{eqnarray}}
\newcommand{\eeqa}{\end{eqnarray}}
\newcommand{\beq}{\begin{equation}}
\newcommand{\eeq}{\end{equation}}

\newcommand{\bmt}{\begin{pmatrix}}
\newcommand{\emt}{\end{pmatrix}}
\usepackage[toc,page]{appendix}
\usepackage{comment}
\newcommand{\be}{\begin{equation}}
\newcommand{\ee}{\end{equation}}
\newcommand{\bea}{\begin{eqnarray}}
\newcommand{\eea}{\end{eqnarray}}
\newcommand{\nn}{\nonumber}

\begin{document}
\title{Investigating the  role of new physics in $b \to c \tau \bar \nu_\tau$ transitions}

\author{ Suchismita Sahoo}
\email{suchismita8792@gmail.com}
\affiliation{\, Theoretical Physics Division, Physical Research Laboratory, Ahmedabad-380009, India}                                         

\author{Rukmani Mohanta }
\email{rukmani98@gmail.com}
\affiliation{\,School of Physics, University of Hyderabad,
              Hyderabad - 500046, India }              
              

\begin{abstract}
In recent times, the charged-current mediated semileptonic  $b \to c \tau \bar \nu_\tau$ processes have attracted a lot of attention after the observation of lepton non-universality ratios, $R_{D^{(*)}}$, $R_{J/\psi}$ and the measurements on  $D^*$ and $\tau$ longitudinal polarization fractions in $\bar B\to D^* \tau \bar \nu_\tau$ processes.  We present a model-independent analysis of $ \bar B\to D^{(*)} \tau \bar \nu_\tau$, $ B_s\to D_s^{(*)} \tau \bar \nu_\tau$, $ B_c^+ \to (\eta_c, J/\psi) \tau^+   \nu_\tau$,  $\Lambda_b \to \Lambda_c \tau \bar \nu_\tau$ and $\bar B \to D^{**} \tau \bar \nu_\tau$   (where $D^{**} = \{D^*_0, D_1^*, D_1, D_2^*\}$ are the four lightest excited charm mesons)  processes  involving  $b \to c \tau \bar \nu$ quark level transitions  by considering the most general effective Lagrangian in the presence of new physics. We perform a global fit to  various set of new coefficients, including the measurements on $R_{D^{(*)}}$, $R_{J/\psi}$ and the upper limit on Br($B_c^+ \to \tau^+ \bar \nu_\tau$). We then show the implications of constrained new couplings on the  branching fractions,  lepton non-universality ratios and various angular observables of these decay modes in four different bins of $q^2$.
\end{abstract}
\maketitle

\section{Introduction}
Although,  we have not seen any unambiguous signal  of new physics (NP) at the LHC experiment so far, the
observation of lepton universality violating (LUV) ratios in $b \to s ll$ ($R_{K^{(*)}}$) \cite{Aaij:2014ora, Aaij:2019wad,Aaij:2017vbb, Abdesselam:2019wac} and $b \to c \tau \bar \nu_\tau$ ($R_{D^{(*)}}$, $R_{J/\psi}$) \cite{Huschle:2015rga,Abdesselam:2016cgx, Abdesselam:2016xqt, Hirose:2017dxl, Hirose:2016wfn, Aaij:2017tyk, Aaij:2015yra, Aaij:2017deq, Aaij:2017uff, Lees:2012xj, Lees:2013uzd, HFLAV} decay modes have provided an indirect hint for the   existence of NP beyond the Standard Model (SM).  The measurements on  lepton non-universality (LNU) observables, $R_K \equiv  \Gamma(B^+ \to K^+ \,\mu^+\,\mu^-)/\Gamma(B^+ \to K^+\,e^+\,e^-)$ along with $R_{K^*} \equiv \Gamma (B^0 \to K^{*0} \mu^+\mu^-)/\Gamma(B^0 \to K^{*0} e^+ e^-)$ by LHCb Collaborations  disagree with their SM predictions  at $\sim 2.5\sigma$ level \cite{Aaij:2014ora, Aaij:2019wad, Bobeth:2007dw, Aaij:2017vbb, Abdesselam:2019wac, Capdevila:2017bsm}.   
In the $b \to c \tau \bar \nu_\tau$ sector, the LHCb \cite{Aaij:2017tyk, Aaij:2015yra, Aaij:2017deq, Aaij:2017uff} as well as Belle \cite {Huschle:2015rga,Abdesselam:2016cgx, Abdesselam:2016xqt, Hirose:2017dxl, Hirose:2016wfn}  and BaBar \cite {Lees:2012xj, Lees:2013uzd} have measured the  LNU ratios, $R_{D^{(*)}}\equiv \Gamma(\bar B \to D^{(*)}\, \tau \bar{\nu}_\tau)/\Gamma(\bar B \to D^{(*)}\, l \bar{\nu}_l)$ and $R_{J/\psi}\equiv \Gamma(B_c^+ \to J/\psi\, \tau^+ \nu_\tau)/\Gamma(B_c^+ \to J/\psi\, l^+ \nu_l)$, where $l=e,\mu$. Combining the $R_{D^{(*)}}$ data of all the experiments, the world average values by HFLAV Collaboration \cite{HFLAV} are
\bea 
R_D^{\rm Expt}=0.340\pm 0.027 \pm 0.013\,,~~~~R_{D^*}^{\rm Expt}=0.295\pm 0.011\pm 0.008\,,\label{RD-Expt}
\eea 
which disagree with their SM predictions \cite{Na:2015kha, Fajfer:2012vx, Fajfer:2012jt}
\bea 
R_D^{\rm SM}=0.299 \pm 0.003\,,~~~~R_{D^*}^{\rm SM}=0.258 \pm 0.005\,, \label{RD-SM}
\eea 
by $3.08\sigma$. The measured value of $R_{J/\Psi}$ by LHCb   \cite{Aaij:2017tyk}
\bea 
R_{J/\psi}^{\rm Expt}= 0.71 \pm 0.17 \pm 0.18\,,\label{Rjpsi-Expt}
 \eea
 also shows a $1.7\sigma$ deviation from the corresponding SM prediction \cite{Wen-Fei:2013uea, Ivanov:2005fd, Dutta:2017xmj}
 \bea 
R_{J/\psi}^{\rm SM}= 0.289\pm 0.01\,.\label{Rjpsi-SM}
 \eea
 The uncertainties from the CKM matrix elements  and the  form factors are canceled out to a large extent in  all these LNU ratios associated with $b \to sll $ and $b \to c \tau \bar \nu_\tau$ and hence, these observed anomalies indirectly point  towards the possible interplay of new physics. Besides these LNU ratios, the Belle Collaboration has also measured  $D^*$ and $\tau$ longitudinal polarization fractions in $\bar B \to D^* \tau \bar \nu_\tau$ channel. The measured values of  $P_\tau^{D^*}$  by Belle \cite{Hirose:2016wfn}
\bea
&&P_\tau^{D^*}|^{\rm Expt}=-0.38\pm 0.51^{+0.21}_{-0.16}\,,  \label{Ptau-Expt}
\eea
is almost consistent with its SM prediction, $P_\tau^{D^*}|^{\rm SM}=-0.497\pm 0.013$ \cite{Tanaka:2012nw}.
However, disagreement of $1.6\sigma$ is found between the Belle measurement \cite{Adamczyk:2019wyt, Abdesselam:2019wbt}
\bea
&&F_L^{D^*}|^{\rm Expt}=0.60\pm 0.08\pm 0.04\,, \label{FL-Expt}
\eea
and its SM prediction, $F_L^{D^*}|^{\rm SM}=0.46\pm 0.04$ \cite{Alok:2016qyh}. 

The investigation of  charge-current mediated semileptonic decays like $\bar B \to D^{(*)} \tau \bar \nu_\tau$, $B_s \to D_s^{(*)} \tau \bar \nu_\tau$, $B_c^+ \to (\eta_c, J/\psi) \tau^+ \nu_\tau$,  $\Lambda_b \to \Lambda_c \tau \bar \nu_\tau$  and $\bar B \to D^{**} \tau \bar \nu_\tau$  driven by  $b \to c \tau \bar \nu_\tau$ transition are  utterly interesting, as the  NP contributions would have to be significantly large  enough to  provide visible impact  in these   tree level decay channels in the SM,  which in turn would requires new particles to be either rather light or strongly coupled to the SM particles.  Both the decay processes  $\bar B \to D^{(*)}$ and $B_s \to D_s^{(*)}$  are related by $SU(3)$ flavor symmetry, i.e., differ only in their spectator quark,  mediated by   $b\to c$ transition, involving the same  CKM matrix element $V_{cb}$.  Therefore, a detailed comprehensive  analysis of these channels will help in  determining the value of $|V_{cb}|$.    The  semileptonic decays of $B_{(s)}$ and $B_c$ mesons have been studied intensively in the literature \cite{Bhol:2014jta, Li:2009wq, Li:2010bb, Atoui:2013mqa, Atoui:2013zza, Bailey:2012rr, Monahan:2016qxu, Na:2012kp, Monahan:2018lzv, Chen:2011ut, Fan:2013kqa,  Monahan:2017uby, Dutta:2018jxz}. The heavy-heavy   baryonic decay modes of $b$-flavored baryons can also serve as an additional source for the determination of the CKM matrix element $V_{cb}$ \cite{Aaij:2015bfa, Fiore:2015cmx, Patrignani:2016xqp, Hsiao:2017umx}, which are investigated by many authors in the SM as well as in the presence of NP \cite{Woloshyn:2014hka, Wu:2015yqa, Shivashankara:2015cta, Gutsche:2015rrt, Gutsche:2015mxa, Detmold:2015aaa, Dutta:2015ueb, Pervin:2005ve, Faustov:2016yza, Datta:2017aue, Li:2016pdv, DiSalvo:2018ngq, Bernlochner:2018kxh, Ray:2018hrx}. In  Ref. \cite{Bernlochner:2016bci, Bernlochner:2017jxt}, the rare semileptonic decays of $B$ meson to higher excited charmed mesons $(D^{**})$ with a lepton and a neutrino in the final state i.e., $\bar B \to D^{**} l\bar \nu_l$, where $D^{**}\in {D_0^*, D_1^*, D_1, D_2}$ are scrutinized in the SM and in the model independent approach. In this work, we would like to study  all the above discussed  decay processes involving  $b \to c \tau \bar \nu_\tau$ quark level transition in a  model independent way by extending the operator structure of Lagrangian beyond standard model. In this approach, we  find additional  Wilson coefficients contributions to the SM coefficients.  We constrain the new parameters from the $\chi^2$ fit of $R_{D^{(*)}}$, $R_{J/\psi}$ and upper limit on Br($B_c^+ \to \tau^+ \nu_\tau$). The bin-wise branching ratio, forward-backward asymmetry, LNU ratios, $\tau$ and $V(=D_{(s)}^*, J/\psi)$ polarization asymmetry of $b \to c \tau \bar \nu_\tau$ decay modes for both real and complex  new coefficients are estimated in this analysis. 

The plan of the paper is as follows. In section II, we give the most general interaction Lagrangian and the theoretical framework for the analysis  of $b \to c \tau \bar \nu_\tau$ transition. Our methodology to constrain the new coefficients is presented in section III. Section IV describes the bin-wise numerical analysis of branching ratios and various angular observables of  $\bar B \to D^{(*)} \tau \bar \nu_\tau$, $B_s \to D_s^{(*)} \tau \bar \nu_\tau$, $B_c^+ \to (\eta_c, J/\psi) \tau^+ \nu_\tau$ decay processes. The $\Lambda_b \to \Lambda_c \tau \bar \nu_\tau$  and $\bar B \to D^{**} \tau \bar \nu_\tau$ processes are discussed in section V and VI respectively. Section VII summarizes our results.

\section{ Theoretical Framework }

 The most general effective Lagrangian of  $b \to c \tau \bar{\nu}_l$ process can be written as   \cite{Sakaki:2013bfa}
\bea \label{ham-bc}
\mathcal{H}_{\rm eff}=\frac{4G_F}{\sqrt{2}} V_{cb} \Big [ \left(\delta_{l\tau} + V_L \right) \mathcal{O}_{V_L}^l + V_R \mathcal{O}_{V_R}^l +  S_L \mathcal{O}_{S_L}^l +  S_R \mathcal{O}_{S_R}^l + T \mathcal{O}_T^l \Big ],
\eea
where $G_F$ is the Fermi constant, $V_{cb}$ is the CKM matrix element, $\mathcal{O}_X$'s ($X=V_{L,R}$, $S_{L,R}$, $T$) are the six-dimensional operators 
\bea
\mathcal{O}_{V_L}^l &=& \left(\bar{c}_L \gamma^\mu b_L \right) \left(\bar{\tau}_L \gamma_\mu \nu_{lL} \right), ~~~~~
\mathcal{O}_{V_R}^l = \left(\bar{c}_R \gamma^\mu b_R \right) \left(\bar{\tau}_L \gamma_\mu \nu_{lL} \right), \nn \\
\mathcal{O}_{S_L}^l &=& \left(\bar{c}_L  b_R \right) \left(\bar{\tau}_R \nu_{l L} \right),  ~~~~~
\mathcal{O}_{S_R}^l = \left(\bar{c}_R b_L \right) \left(\bar{\tau}_R \nu_{l L} \right), \nn \\
\mathcal{O}_{T}^l &=& \left(\bar {c}_R \sigma^{\mu \nu}  b_L \right) \left(\bar{\tau}_R \sigma_{\mu \nu} \nu_{lL} \right),
\eea
 and the corresponding   Wilson coefficients $(X)$ are zero in the SM, which  can only be generated in NP models. Here $q_{L(R)} = L(R)q$ are the chiral quark fields with $L(R)=(1\mp \gamma_5)/2$ as the projection operators.
 
Including all the new physics operators, the  differential decay rate of $\bar B \to  P l \nu_l$ processes, where $P=D_{(s)}, \eta_c$ are the pseudo-scalar mesons,  with respect to $q^2$ is given by \cite{Sakaki:2013bfa}
\bea
\frac{d\Gamma(\bar B \to  P l \bar \nu_l)}{dq^2} &=& {G_F^2 |V_{cb}|^2 \over 192\pi^3 M_B^3} q^2 \sqrt{\lambda_P(q^2)} \left( 1 - {m_l^2 \over q^2} \right)^2  \nn \\   && \times \Bigg \lbrace \Big | 1 + V_L + V_R \Big |^2 \left[ \left( 1 + {m_l^2 \over 2q^2} \right) H_{0}^{2} + {3 \over 2}{m_l^2 \over q^2}  H_{t}^{2} \right] \nn \\ && + {3 \over 2} \left |S_L + S_R \right |^2 \, H_S^{2} + 8 \left |T \right |^2 \left(1+ \frac{2m_l^2}{q^2} \right) H_T^2 \nn \\ && +3{\rm Re}\left[ ( 1 + V_L + V_R ) (S_L^* + S_R^* ) \right] {m_l \over \sqrt{q^2}} \, H_S H_{t}  \nn \\ && -12{\rm Re}\left[ \left( 1 + V_L + V_R \right) T^* \right] \frac{m_l}{\sqrt{q^2}} H_T H_0  \Bigg \rbrace,  \label{br-exp}
\eea
where 
\bea
\lambda_P =\lambda (M_B^2, M_P^2, q^2), ~~ ~~ {\rm with}~~~~~\lambda(a,b,c)=a^2+b^2+c^2-2(ab+bc+ca)\,.
\eea
$M_B~(M_P)$ is the mass of the $B~(P)$ meson, $m_l$ is the charged lepton mass and $H_{0,t,S,T}$ are the  helicity amplitudes which include the form factors $(F_{0,1,T})$ \cite{Sakaki:2013bfa}. 

The differential decay distribution of $\bar B \to  V l \bar \nu_l$ processes where $V$ denotes the vector mesons ($V=D_{(s)}^*, J/\psi$),    in terms of helicity amplitudes ($H_{i,\pm},~ H_{i,0}, H_t$, where $i=V,T$)   with respect to $q^2$ is given by  \cite{Sakaki:2013bfa}
\bea
 {d\Gamma(\bar B \to  V l \bar \nu_l) \over dq^2} &=& {G_F^2 |V_{cb}|^2 \over 192\pi^3 M_B^3} q^2 \sqrt{\lambda_{V} (q^2)} \left( 1 - {m_l^2 \over q^2} \right)^2 \times \nn  \\ && \bigg \{  \left( \left|1 + V_L \right|^2 + \left| V_R\right|^2 \right)  \left[ \left( 1 + {m_l ^2 \over 2q^2} \right) \left( H_{V, +}^2 + H_{V,-}^2 + H_{V,0}^2 \right) + {3 \over 2}{m_l^2 \over q^2} \, H_{V,t}^2 \right]  \nn \\ && - 2{\rm Re}\left[\left(1+ V_L \right) V_R^* \right] \left[ \left( 1 + {m_l^2 \over 2q^2} \right) \left( H_{V,0}^2 + 2 H_{V,+} H_{V,-} \right) + {3 \over 2}{m_l^2 \over q^2} \, H_{V,t}^2 \right] \nn \\ && +  {3 \over 2} |S_L - S_R|^2 \, H_S^2 + 8 |T|^2 \left(1+\frac{2m_l^2}{q^2} \right) \left(H_{T, +}^2 + H_{T, -}^2 + H_{T, 0}^2 \right) \nn \\ &&  + 3{\rm Re}\left [ \left ( 1 + V_L - V_R \right) \left (S_L^* - S_R^* \right) \right ] {m_l \over \sqrt{q^2}} \, H_S H_{V,t} \nn \\ && -12{\rm Re}\left[ \left(1+V_L^* \right)T^* \right] \frac{m_l}{\sqrt{q^2}} \left(H_{T,0}H_{V,0}+H_{T,+}H_{V, +} - H_{T,-} H_{V, -} \right) \nn \\ && 
 +12{\rm Re}\left[ V_R^* T^* \right ] \frac{m_l}{\sqrt{q^2}}  \left(H_{T,0}H_{V,0}+H_{T,+}H_{V, -} - H_{T,-} H_{V, +} \right) \bigg \}, 
\eea
where  $\lambda_{V}= \lambda (M_B^2, M_{V}^2, q^2)$.    
Alongside the decay rate, we also consider the following  angular observables to probe NP in semileptonic  $B$ decays. 
\begin{description}
\item[Forward-backward asymmetry]:
\bea
  A_{FB}\left(q^2\right)  =  \left[ \int_{-1}^0 d\cos\theta_l \frac{d^2\Gamma}{dq^2 d\cos\theta_l}
 - \int_{0}^1 d\cos\theta_l \frac{d^2\Gamma}{dq^2 d\cos\theta_l}\right] \;.
   \eea
\item[Lepton non-universality]:
\bea
&&R_P=\frac{{\rm Br}(\bar B \to  P \tau \bar \nu)}{{\rm Br}(\bar B \to  P l \bar \nu)}\,,\nn \\
&&R_V=\frac{{\rm Br}(\bar B \to  V \tau \bar \nu)}{{\rm Br}(\bar B \to  V l \bar \nu)}\,, ~~l=e,\mu\,.
\eea
\item[Tau polarization parameter]:
\bea
P_\tau (q^2) = \frac{ d\Gamma (\lambda_\tau = 1/2)/dq^2 - d\Gamma (\lambda_\tau = -1/2)/dq^2}{d\Gamma (\lambda_\tau = 1/2)/dq^2 + d\Gamma (\lambda_\tau = -1/2)/dq^2}\,.
\eea
The detailed expressions for the decay distributions $d\Gamma(\lambda=\pm 1/2)/dq^2$ can be found in the Ref. \cite{Sakaki:2013bfa}.
\item[ Polarization of $\boldsymbol{V}$]:
The longitudinal $(L)$ and transverse $(T)$  polarization components of daughter vector meson $(V)$ are given by \cite{Biancofiore:2013ki}
\bea
F_{L, T}^{D^*}(q^2) = \frac{d\Gamma_{L, T} \left(\bar B \to D^* \tau \bar{\nu} \right)/ dq^2}{d\Gamma \left(\bar B \to D^* \tau \bar{\nu} \right) / dq^2}\,.
\eea
\end{description}
\section{Constraints on new coefficients}

In this section, we perform the $\chi^2$ fitting to obtain the values of the new coefficients, from the observables  $R_{D^{(*)}}$, $R_{J/\psi}$ and Br$(B_c^+ \to \tau^+ \nu_\tau)$, where  $\chi^2$  is defined as 
\bea
\chi^2(X)=\sum_i \frac{(\mathcal{O}_i^{\rm th}(X)-\mathcal{O}_i^{\rm Expt})^2}{(\Delta \mathcal{O}_i^{\rm Expt})^2+(\Delta \mathcal{O}_i^{\rm SM})^2}\,.
\eea
Here $\mathcal{O}_i^{\rm th}(X)$ are the total  theoretical predictions for the observables with  $X(=V_{L,R},~S_{L,R}, ~T)$ as the new Wilson coefficients and $\mathcal{O}_i^{\rm Expt}$ represent  the corresponding  measured central values.   $\Delta \mathcal{O}_i^{\rm Expt}$  and $\Delta \mathcal{O}_i^{\rm SM}$  are respectively the experimental and SM uncertainties of the observables. The complete expression for  $\bar B (B_{c}^+) \to D^{(*)}(J/\psi) l \bar \nu_l$ decay rates, required to compute the $R_{D^{(*)}}$, $R_{J/\psi}$ LNU  ratios are provided in the previous section. We use the $\bar B \to D$ hadronic form factors from \cite{Lattice:2015rga} that calculated using lattice QCD techniques and for  the  $\bar B \to D^*$ decays, we use the  heavy quark effective theory (HQET) form factors from \cite{Caprini:1997mu, Bailey:2014tva, Amhis:2014hma}. For the  $B_c^+ \to J/\psi$ form factors, we consider the perturbative QCD (PQCD) calculation from  \cite{Kurimoto:2002sb, Watanabe:2017mip}.  Including the NP contributions, the branching ratios of  $B_c^+ \to \tau^+ \nu_\tau$ processes is  given by \cite{Biancofiore:2013ki}
\bea \label{BR-Bclnu}
{\rm BR} (B_{c}^+ \to \tau^+  \nu_\tau)&=&\frac{G_F^2 M_{B_c} m_\tau^2}{8 \pi} \Big( 1-\frac{m_\tau^2}{M_{B_c}^2}\Big)^2 f_{B_c}^2 \left | V_{cb} \right |^2 \tau_{B_c^+} \nn \\ && \times \Big | \left(1 + V_L-V_R \right)-\frac{M_{B_c}^2}{m_\tau (m_b +m_c)}  \left(S_L - S_R \right) \Big |^2.
\eea
Using the decay constant, $f_{B_c}=489\pm 4\pm 3$ MeV from \cite{Aoki:2013ldr, Chiu:2007km} and the CKM matrix elements, particle masses and  life time of $B_c$ meson from \cite{Tanabashi:2018oca}, the predicted   branching fraction  in the SM   is given as 
\bea
{\rm BR} (B_c^+ \to \tau^+ \nu_\tau)|^{\rm SM}= (3.6 \pm 0.14) \times 10^{-2}\,, \label{BR-Bctaunu-SM}
\eea
and its current  experimental upper limit is \cite{Akeroyd:2017mhr}
\bea
{\rm BR} (B_c^+ \to \tau^+ \nu_\tau)|^{\rm Expt} < 30\%.
\label{BR-Bctaunu-Expt}
\eea
In this analysis, we fit all possible cases of new coefficients which are classified  as
\begin{itemize}
\item[Case A:] Presence of only one  new real  coefficient at a time.
\item[Case B:] Presence of only one new complex  coefficient at a time.
\item[Case C:] Presence of various combinations of two  new real coefficients at a time.
\end{itemize}

Case A contains individual $5$ real new coefficients (vector, scalar and tensor types), whose best-fit values are presented in Table  \ref{Tab:Best-fit}\,. In this Table,  we have also presented the $\chi^2_{\rm min, ~SM+NP}/{\rm d.o.f}$ as well as the pull values,  defined as: pull=$\sqrt{\chi^2_{\rm SM}-\chi^2_{\rm SM+NP}}$\,. The degrees of freedom in this case is $3$, since we consider four observables with one additional new parameter. We find $\chi^2_{\rm SM}=11.193$ for the SM. The $\chi^2_{\rm min, ~SM+V_L}/{\rm d.o.f}=0.767$ and pull=$2.982$ for the additional $V_L$ coefficient, which implies the NP contribution due to  $V_L$ coupling, fit the measurement very well. Whereas the presence of $V_R$, scalar and tensor type couplings give poor fit to the $R_{D^{(*)}}$, $R_{J/\psi}$ and Br($B_c^+ \to \tau^+ \nu_\tau$) data. 
\begin{figure}[htb]
\includegraphics[width=0.31\textwidth]{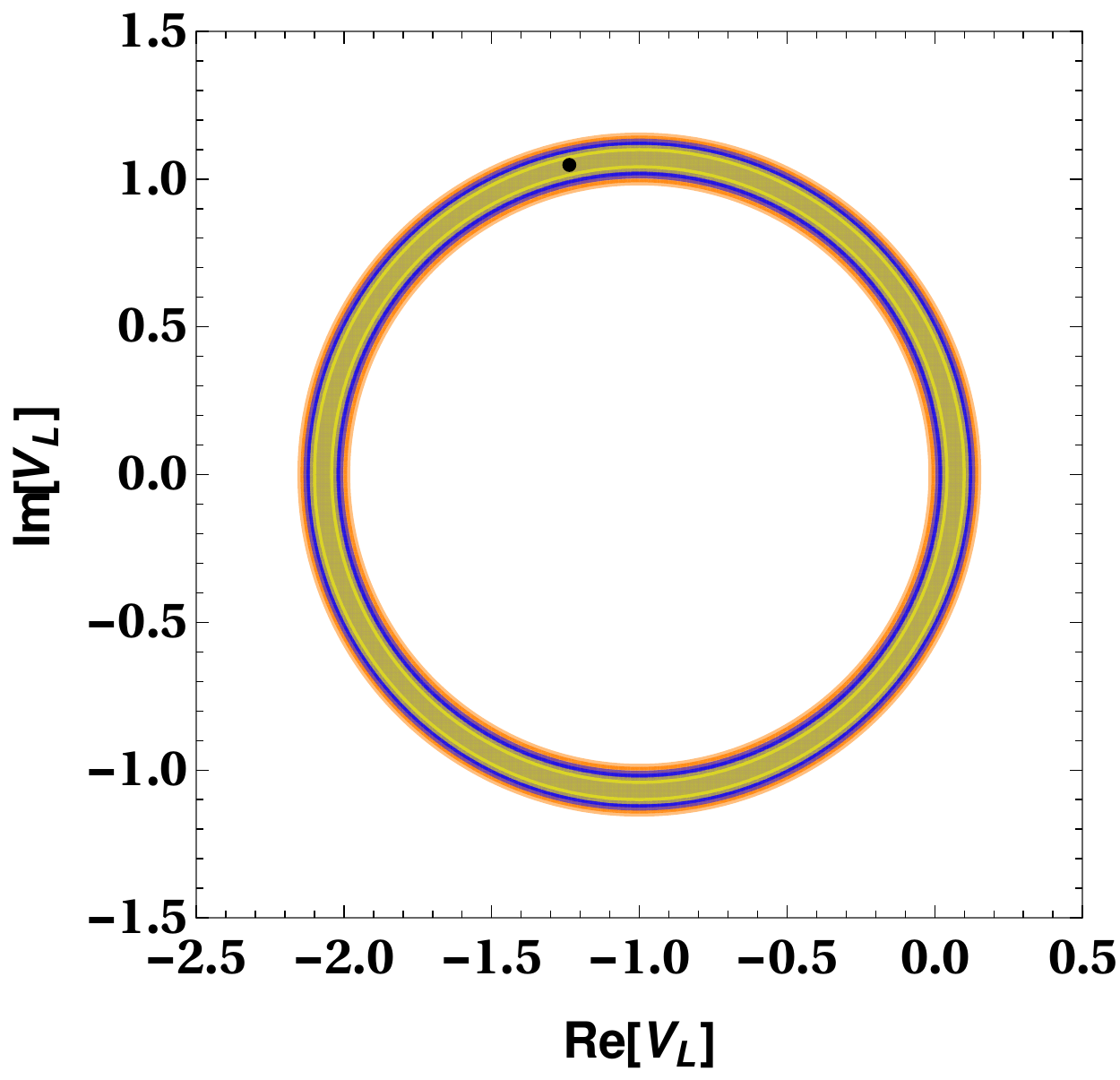}
\quad
\includegraphics[width=0.31\textwidth]{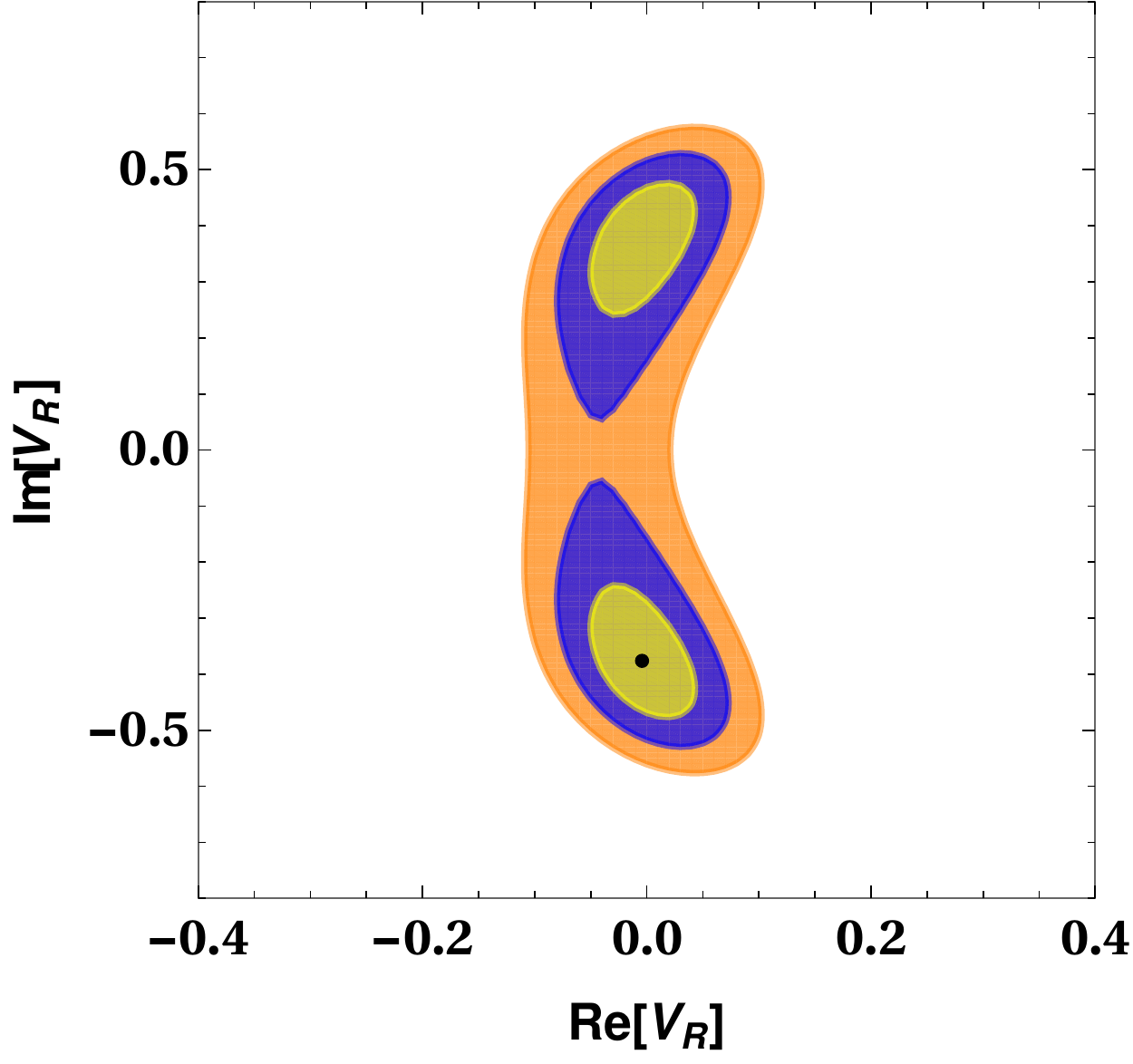}
\quad
\includegraphics[width=0.31\textwidth]{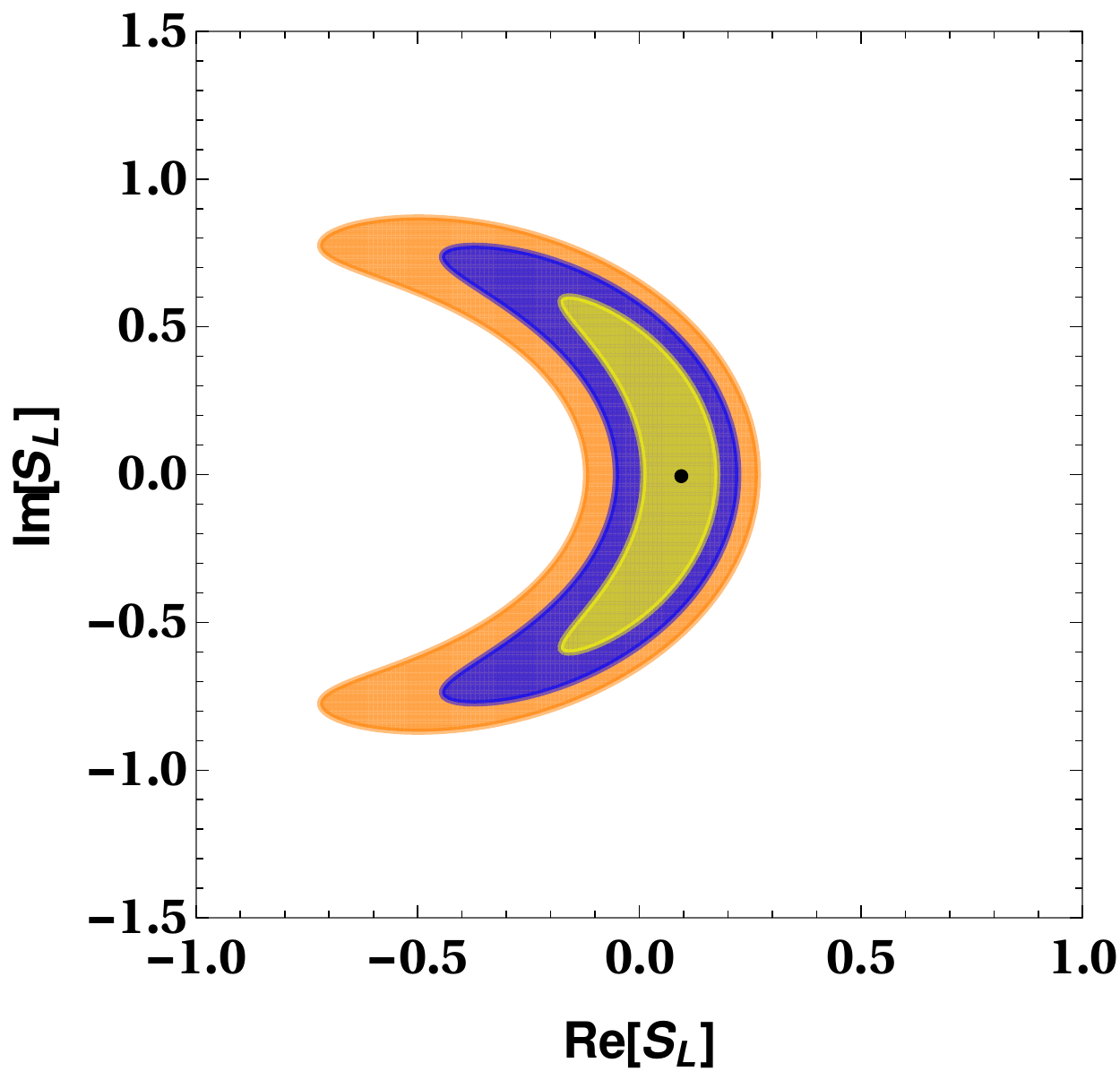}
\quad
\includegraphics[width=0.31\textwidth]{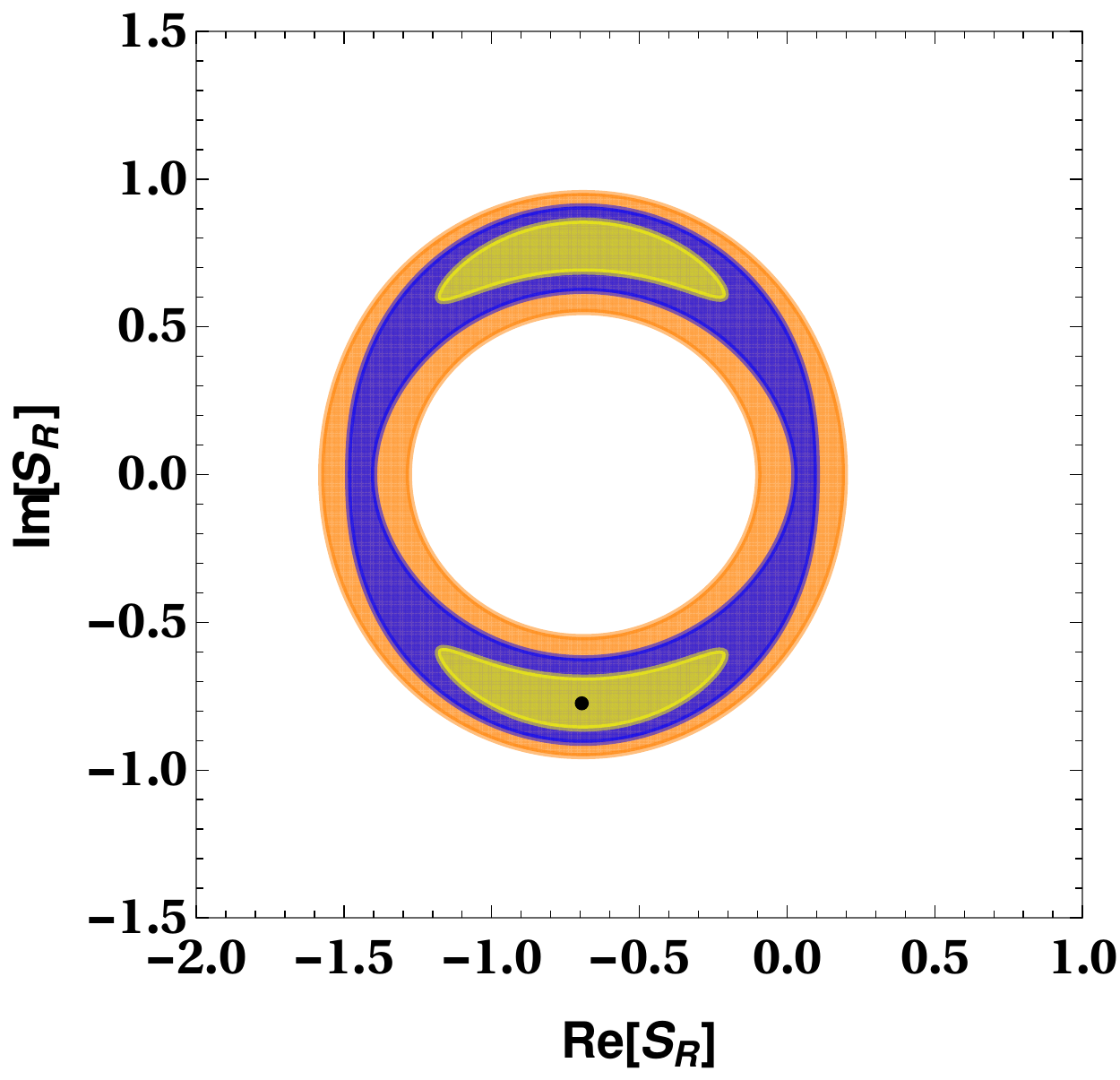}
\quad
\includegraphics[width=0.31\textwidth]{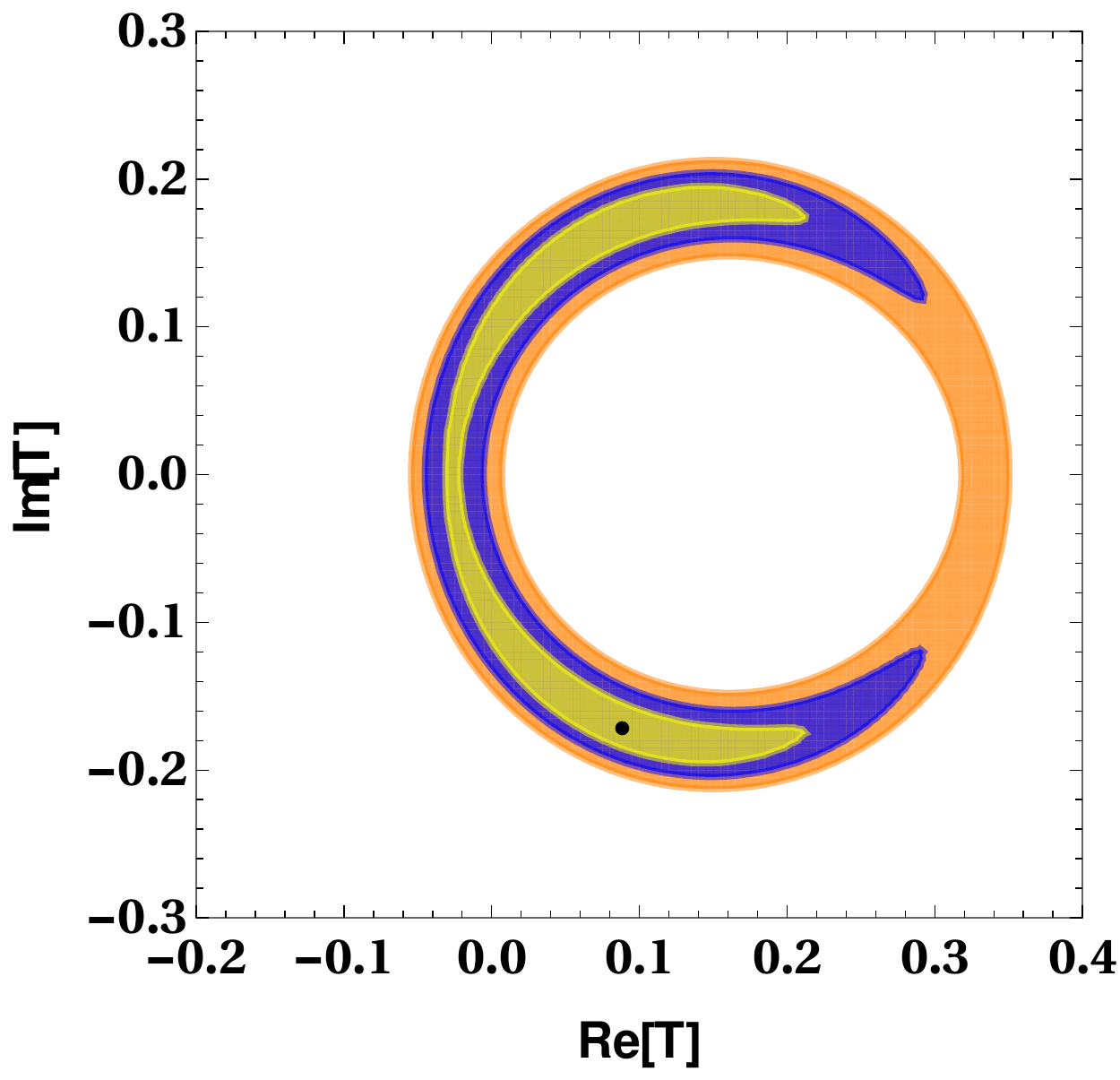}
\caption{Constraints on real and imaginary part of the new coefficients obtained from $\chi^2$ fit to $R_{D^{(*)}}$, $R_{J/\psi}$ and  Br($B_c^+ \to \tau^+ \nu_\tau$) observables. Here the black dots represent the best-fit values. }\label{Fig:Case-B}
\end{figure}

Case B includes the presence of  individual  complex vector, scalar and  tensor type Wilson coefficients. The constrained plots on real and imaginary parts of individual complex Wilson coefficients associated with  $b \to c \tau \bar \nu_\tau$ transitions, obtained from fit to $R_{D^{(*)}}$, $R_{J/\psi}$ and the upper limit on branching ratio of $B_c^+ \to \tau^+ \nu_\tau$ are depicted in Fig. \ref{Fig:Case-B}\,. Here  yellow, blue and orange colors represent $1\sigma$, $2\sigma$ and $3\sigma$ contours respectively and black dots are the respective best-fit values. The pull for only complex $V_L$ coefficient remain same as previous case and fits the measured data well. We found $\chi^2_{\rm min,~ SM+V_R}=2.29$ for the complex $V_R$ coefficient and the pull value has increased from $1.57$ to $2.984$. The $S_L$ coefficient fits very poorly with data and the  pull remains same as case A. Though their is increment in the pull value of  $S_R$ and $T$ coupling as compared to previous case, till they provide poor fit.  
 \begin{figure}[htb]
\includegraphics[width=0.31\textwidth]{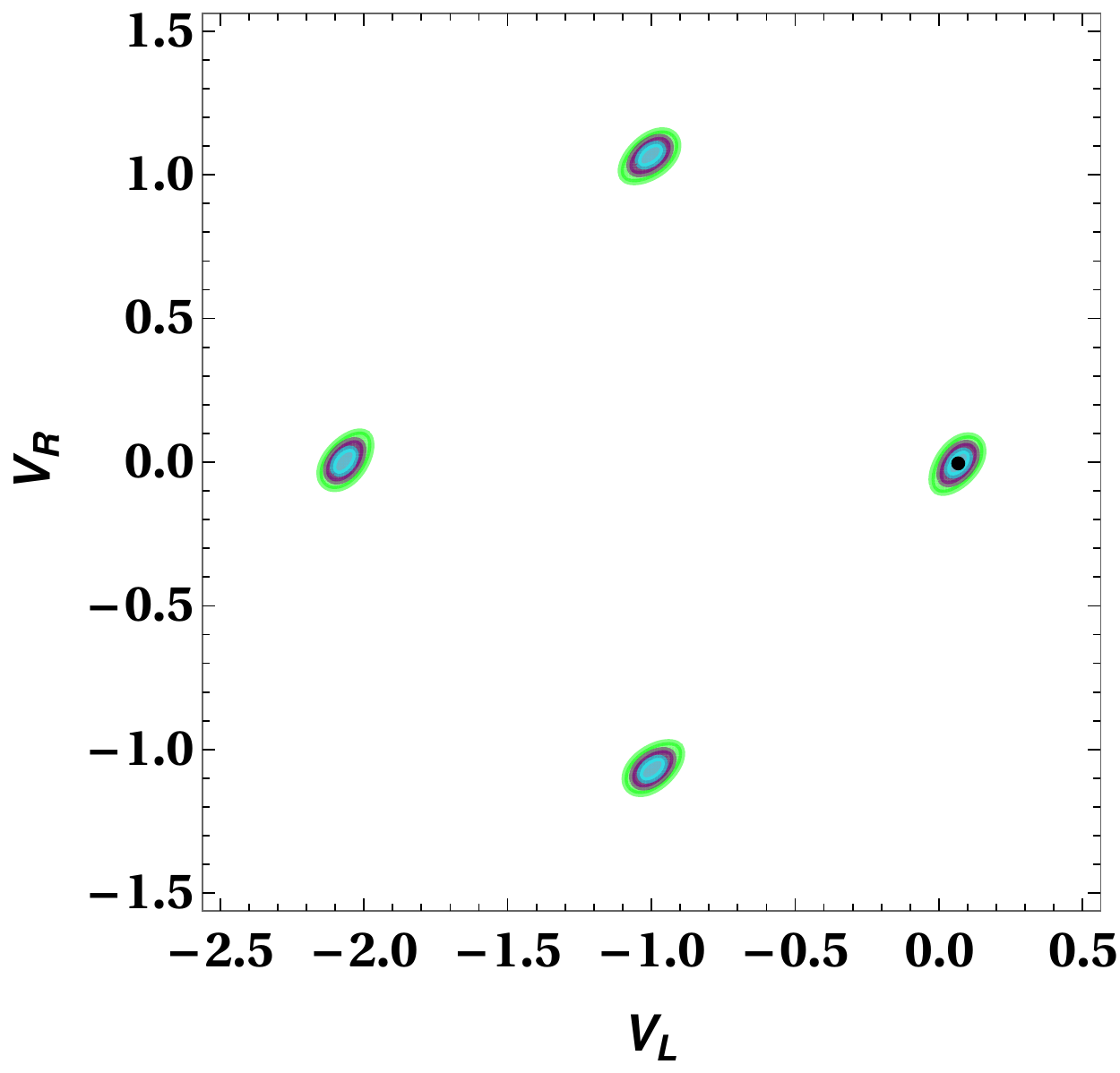}
\quad
\includegraphics[width=0.31\textwidth]{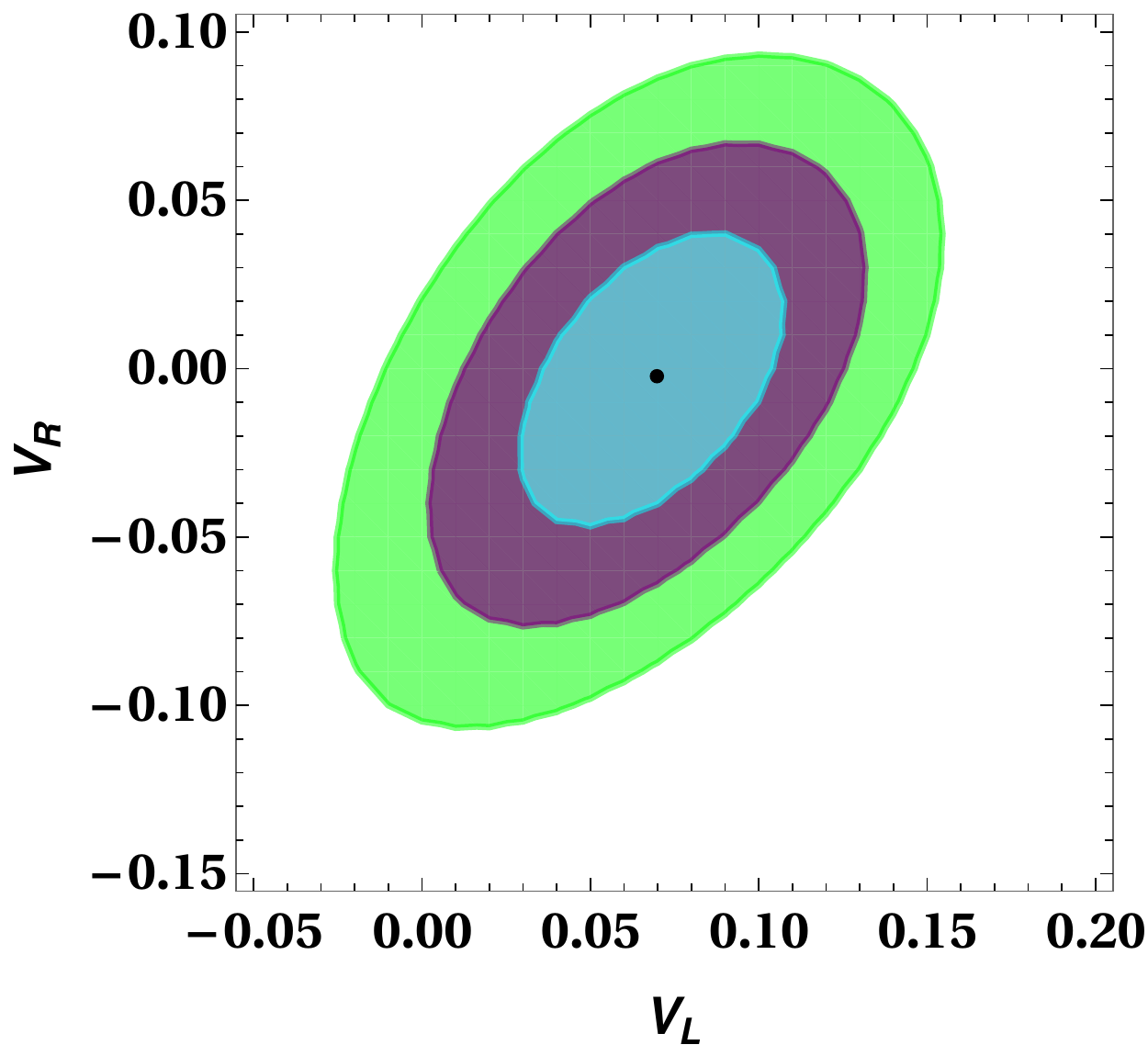}
\quad
\includegraphics[width=0.31\textwidth]{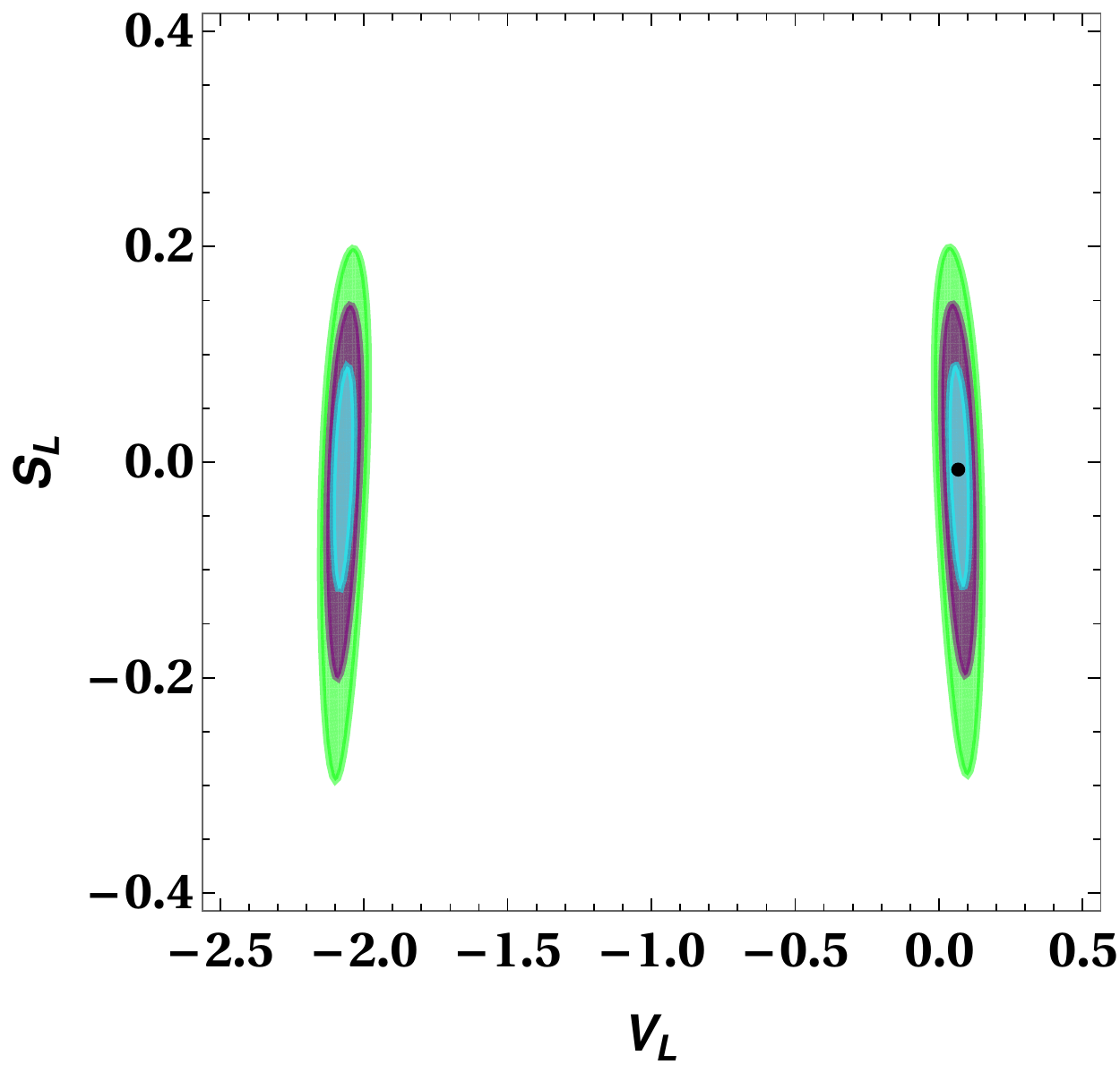}
\quad
\includegraphics[width=0.31\textwidth]{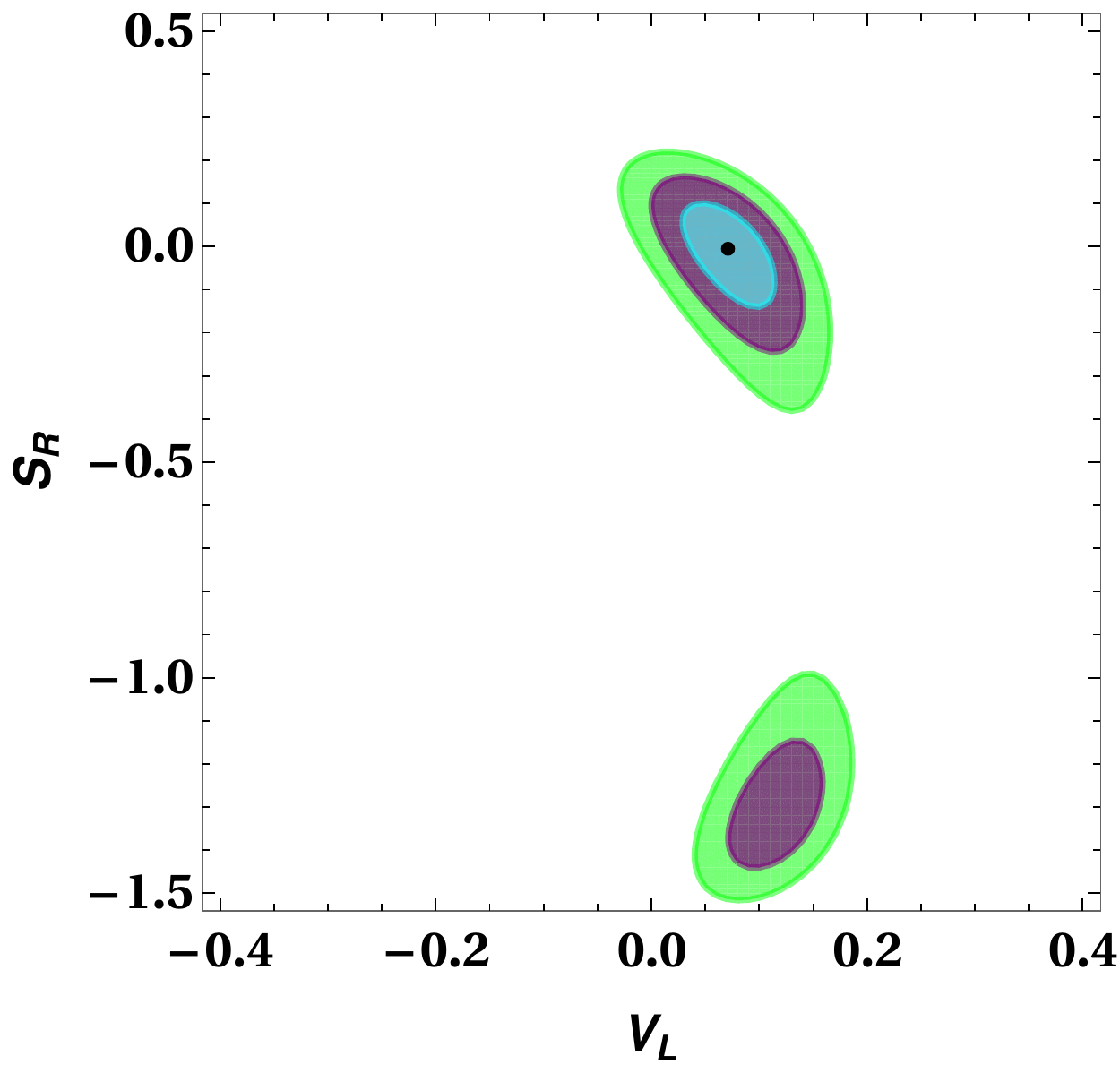}
\quad
\includegraphics[width=0.31\textwidth]{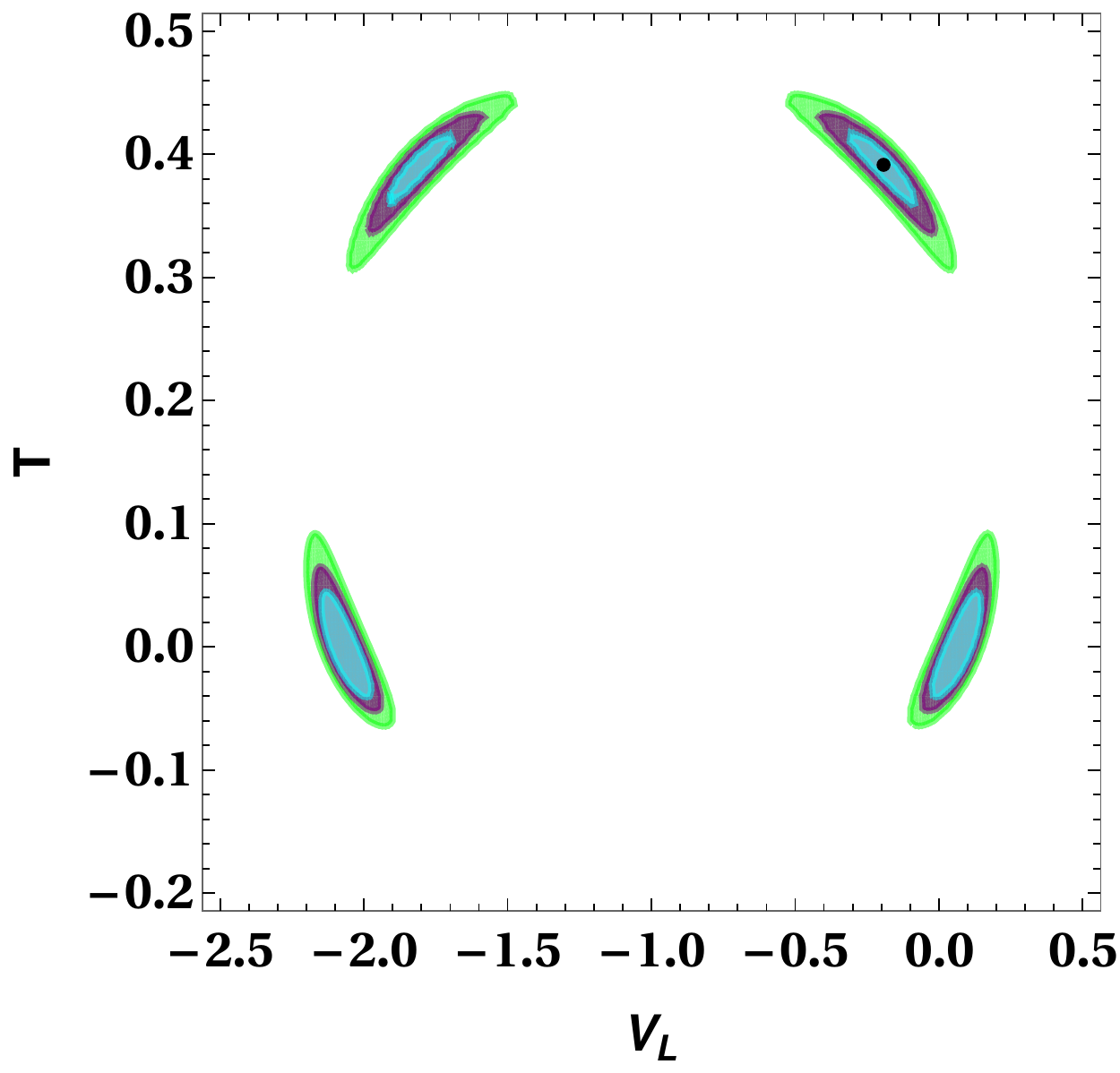}
\quad
\includegraphics[width=0.31\textwidth]{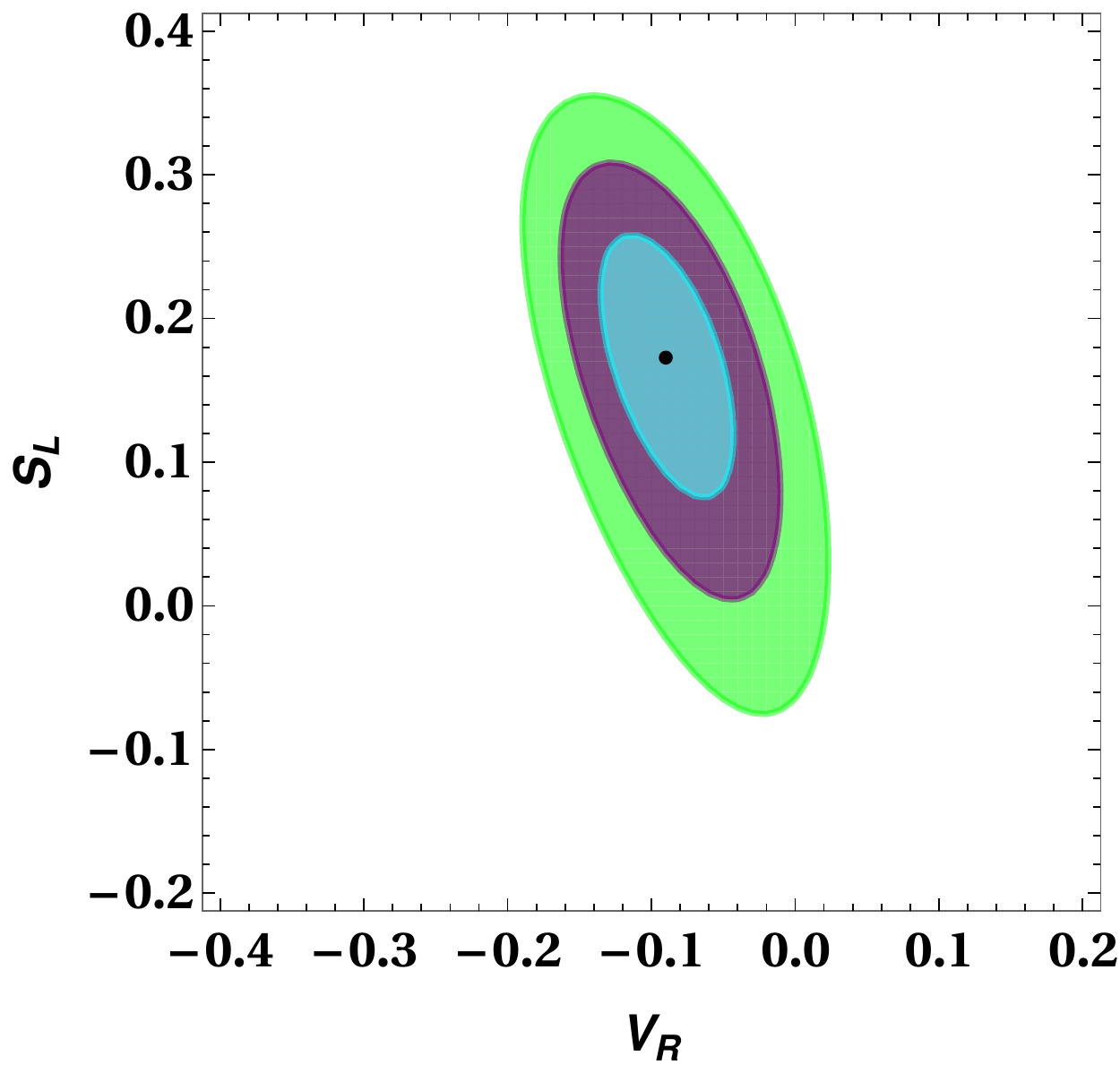}
\quad
\includegraphics[width=0.31\textwidth]{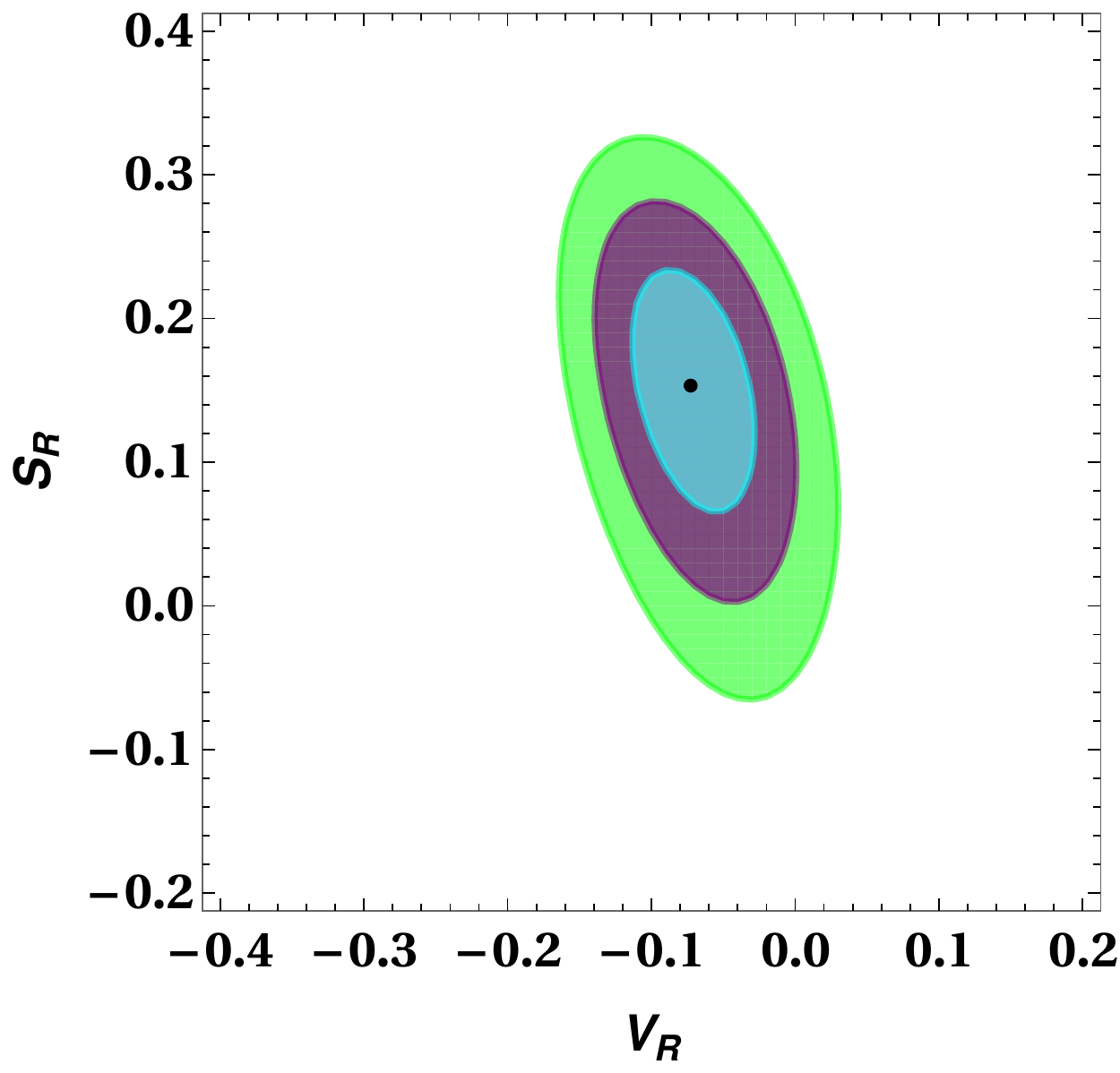}
\quad
\includegraphics[width=0.31\textwidth]{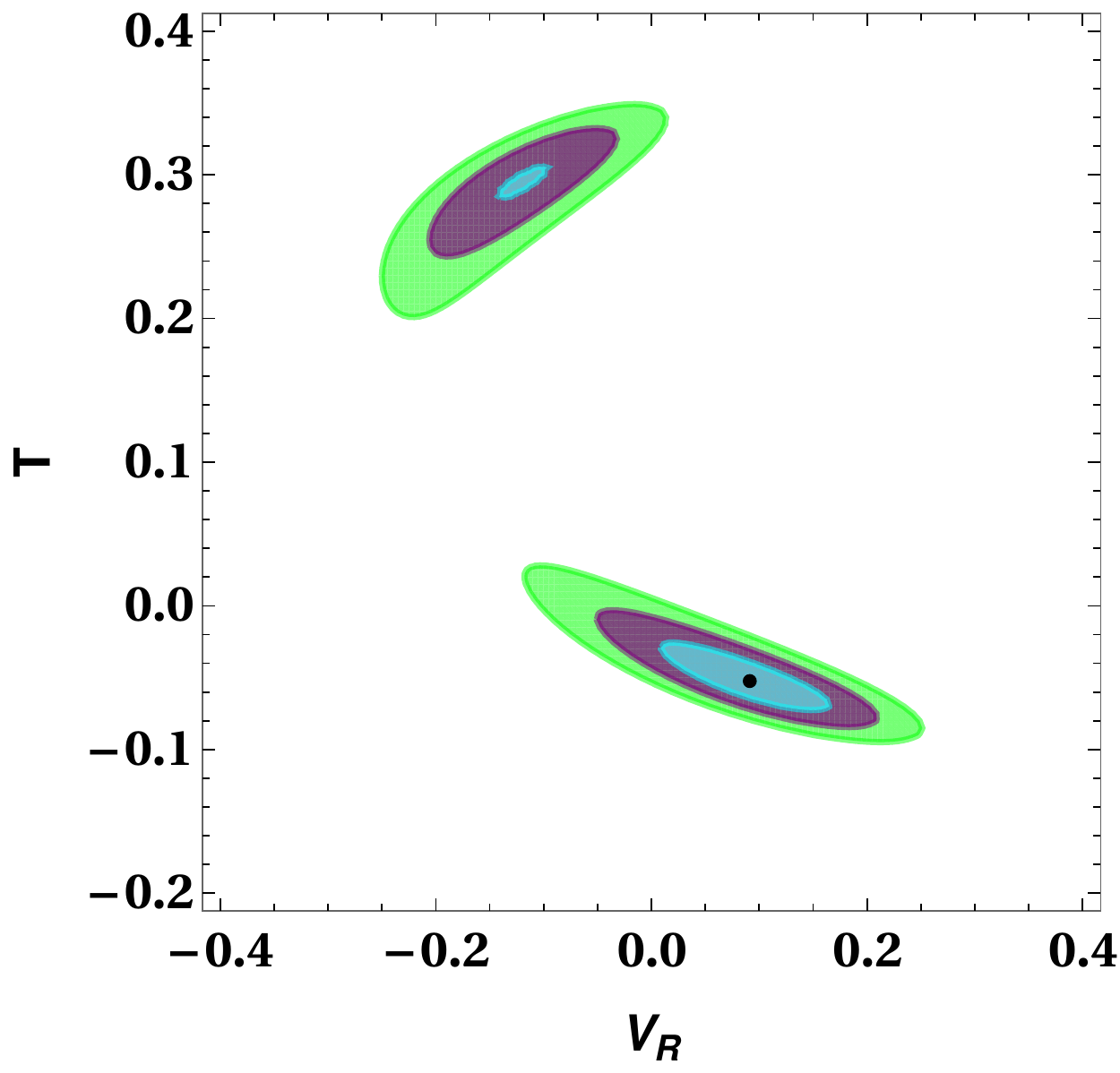}
\quad
\includegraphics[width=0.31\textwidth]{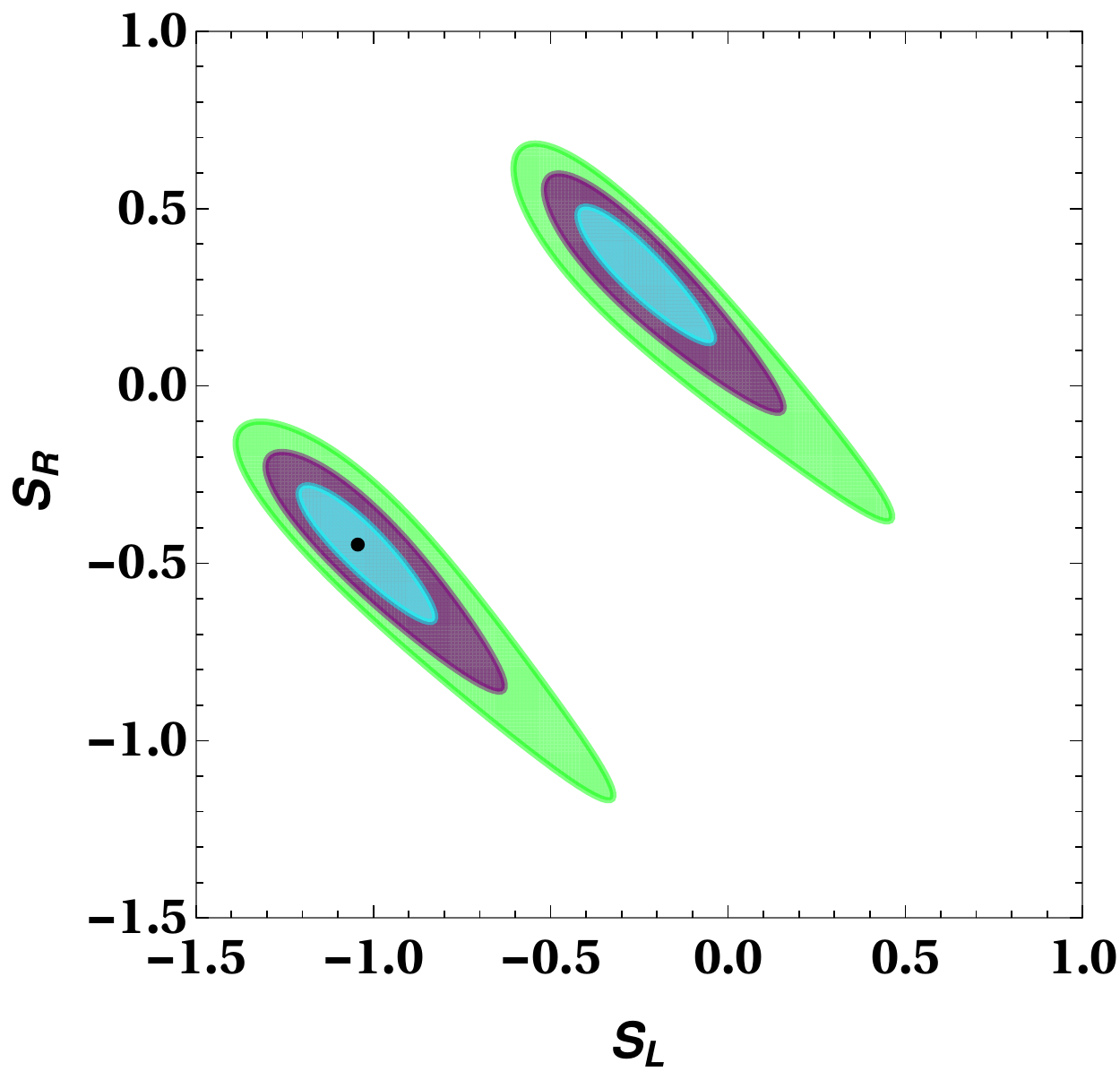}
\quad
\includegraphics[width=0.31\textwidth]{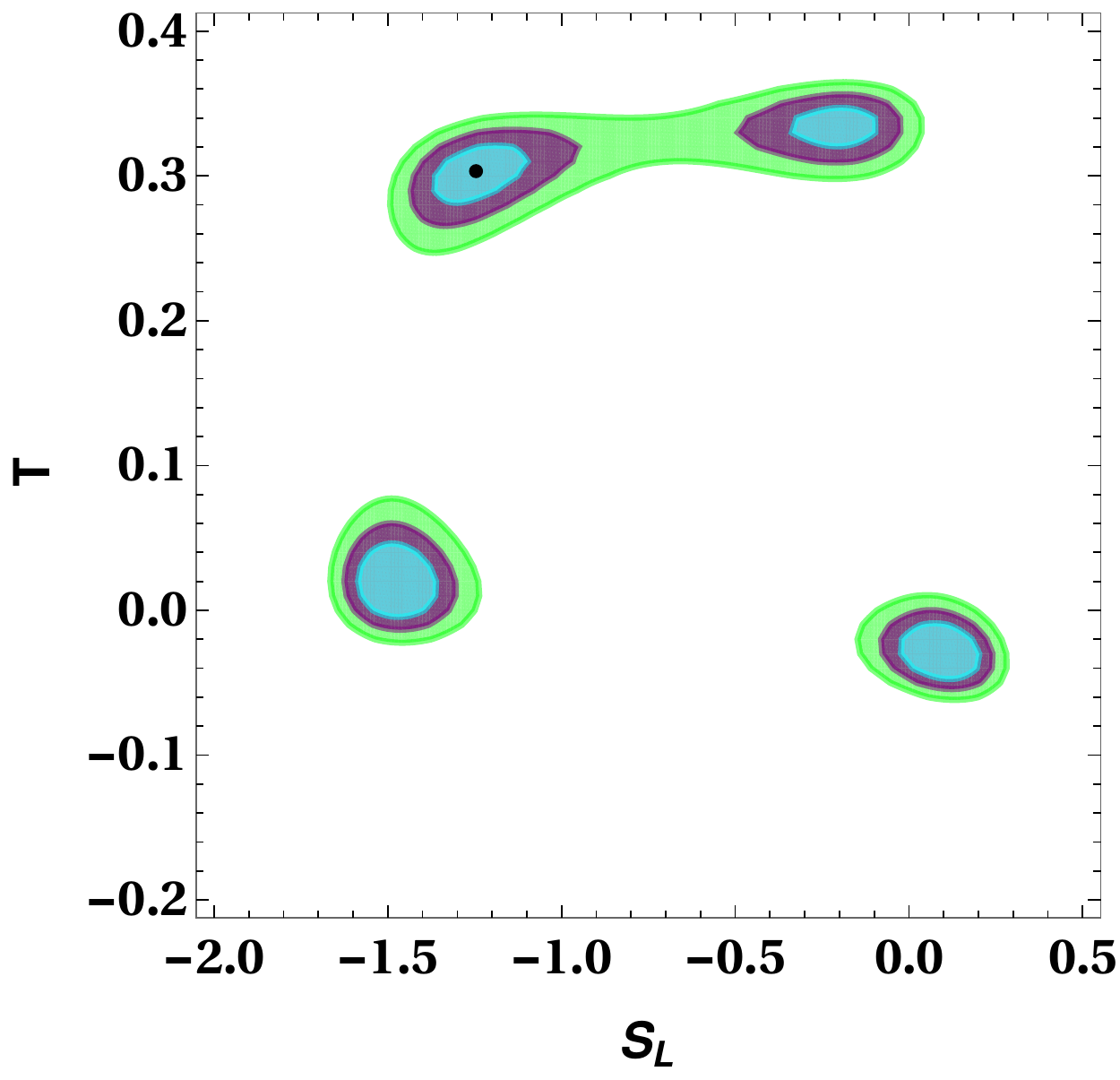}
\quad
\includegraphics[width=0.31\textwidth]{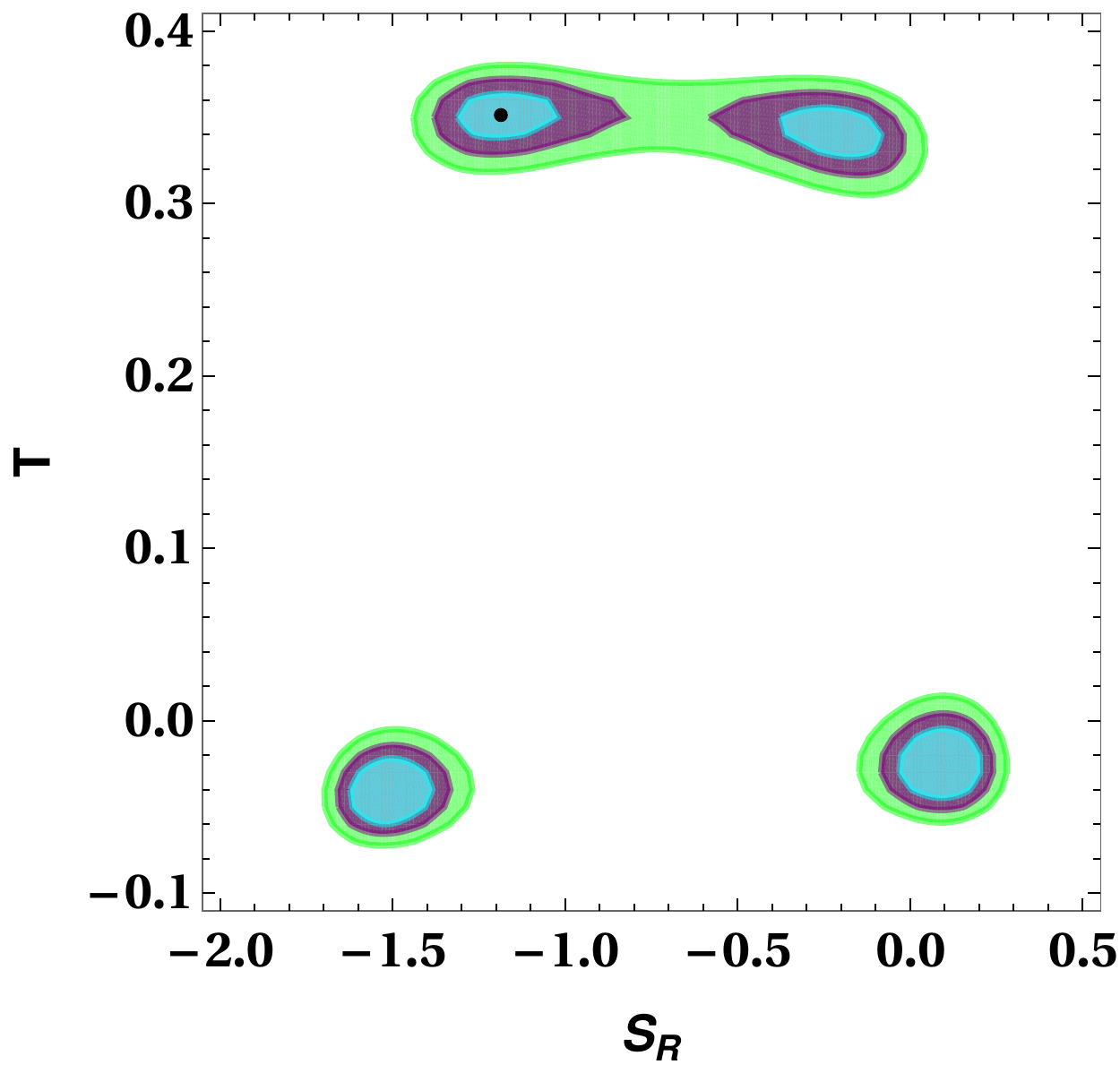}
\caption{Constraints on various combination of real new coefficients obtained from $\chi^2$ fit to $R_{D^{(*)}}$, $R_{J/\psi}$ and  Br($B_c^+ \to \tau^+ \nu_\tau$) observables for case C of our analysis. We show the zoom plot for $V_L-V_R$ plane in the top-middle panel. Here the black dots represent the best-fit values.}\label{Fig:Case-C}
\end{figure} 

Now coming to case C, which includes various possible combination of two real Wilson coefficients at a time. The constrained plots for $10$ possible sets of new real coefficients are shown in Fig. \ref{Fig:Case-C}\,. We present the zoomed plot for $V_L-V_R$ coefficients in the top-middle panel of this figure.  We find that, the NP contribution arising due to the presence of $V_L-V_R$, $V_L-S_L$, $V_L-S_R$, $V_R-S_L$ and $V_R-S_R$  sets of new coefficients provide an acceptable fit with pull values $\sim 2.96$. The $V_R-T$ combination  fit the experimental data  quite  effectively  with the highest pull value $3.02$.   The fit for remaining possible sets of real Wilson coefficients, such as $V_L\&T,~S_L\& S_R,~S_L~\&~ T,~S_R\& T$ are not robust. 
 \begin{table}
\begin{center}
\caption{Predicted best-fit values of new Wilson coefficients for all possible cases. We also provide the $\chi^2_{\rm min}/{\rm d.o.f}$ and pull values.}\label{Tab:Best-fit}
\begin{tabular}{|c| c | c | c |c|}
\hline
Cases~&~New Wilson coefficients~& ~Best-fit values~&~ $\chi^2_{\rm min}/{\rm d.o.f}$~&~Pull~\\
\hline 
\hline
Case A~&~$V_L$~&~$-2.07$~&~$0.767$~&~$2.982$~\\
&~$V_R$~&~$-0.0434$~&~$2.91$~&~$1.57$~\\
&~$S_L$~&~$0.097$~&~$2.81$~&~$1.663$~\\
&~$S_R$~&~$-1.443$~&~$3.319$~&~$1.112$~\\
&~$T$~&~$-0.0263$~&~$1.6$~&~$2.527$~\\
\hline
 Case B~&~$({\rm Re}[V_L],{\rm Im}[V_L])$~&~$(-1.233,1.045)$~&~$1.151$~&~$2.982$~\\
&~$({\rm Re}[V_R],{\rm Im}[V_R])$~&~$(-0.0034,-0.3783)$~&~$1.145$~&~$2.984$~\\
&~$({\rm Re}[S_L],{\rm Im}[S_L])$~&~$(0.97,0)$~&~$4.213$~&~$1.663$~\\
&~$({\rm Re}[S_R],{\rm Im}[S_R])$~&~$(-0.695,-0.777)$~&~$2.175$~&~$2.616$~\\
&~$({\rm Re}[T],{\rm Im}[T])$~&~$(0.0886,-0.17)$~&~$1.416$~&~$2.892$~\\
\hline
 Case C~&~$(V_L,V_R)$~&~$(0.0694,-0.0026)$~&~$1.147$~&~$2.983$~\\
 &~$(V_L,S_L)$~&~$(0.0714,-0.0063)$~&~$1.147$~&~$2.983$~\\
&~$(V_L,S_R)$~&~$(0.0724,-0.0086)$~&~$1.145$~&~$2.984$~\\
&~$(V_L,T)$~&~$(-0.194,0.3913)$~&~$2.42$~&~$2.52$~\\

&~$(V_R,S_L)$~&~$(-0.09,0.1726)$~&~$1.167$~&~$2.976$~\\
&~$(V_R,S_R)$~&~$(-0.072,0.154)$~&~$1.15$~&~$2.96$~\\
&~$(V_R,T)$~&~$(0.091,-0.0519)$~&~$1.02$~&~$3.02$~\\

&~$(S_L,S_R)$~&~$(-1.04,-0.449)$~&~$2.72$~&~$2.4$~\\
&~$(S_L,T)$~&~$(-1.25,0.303)$~&~$1.989$~&~$2.686$~\\

&~$(S_R,T)$~&~$(-1.1875,0.352)$~&~$2.23$~&~$2.596$~\\
\hline

\end{tabular}
\end{center}
\end{table}
 
\section{ Numerical analysis for $\bar B \to D^{(*)} \tau^- \bar \nu_\tau$, $\bar B_{s} \to D^{(*)}_{s}  \tau^- \bar \nu_\tau$ and $B_c^+ \to (\eta_c, J/\psi) \tau^+  \nu_\tau$  processes }

After gathering all the information about the  constraints on various  new coefficients, we now proceed to  analyze the impact of  NP, for all the possible combinations as  discussed  above,  on branching ratios and various angular observables of $\bar B \to D^{(*)} \tau^- \bar \nu_\tau$, $B_c^+ \to (\eta_c, J/\psi) \tau^+  \nu_\tau$ and $B_s \to D_s^{(*)} \tau^- \bar \nu_\tau$ decay modes. We study the angular observables of all these decay processes in four $q^2$ (in GeV$^2$) bins: $m_\tau^2\to 5$, $5\to 7$, $7\to 9$ and $9 \to (M_B-M_{P(V)})^2$.   All the input parameters required for numerical estimation are taken from \cite{Tanabashi:2018oca}\,.  The values of the form factors used in this analysis for various processes are as: 
(a) lattice QCD results \cite{Lattice:2015rga} for  $\bar B\to D$ form factors (b) HQET \cite{Caprini:1997mu, Bailey:2014tva, Amhis:2014hma} results on $\bar B \to D^*$ form factors (c) the form factors for  $B_c^+ \to \eta_c $ ($B_c^+ \to J/\psi$)
 transitions are from   light-front quark model (perturbative QCD) \cite{Huang:2018nnq} (\cite{Kurimoto:2002sb, Watanabe:2017mip}) (d)  lattice QCD results on $B_s \to D_s$ form factors \cite{Monahan:2017uby} and for $B_s \to D_s^*$ form factors  from perturbative QCD approach \cite{Fan:2013kqa}. As the  tensor form factors for the processes  $B_c^+ \to (\eta_c, J/\psi)$ and $B_s \to D_s^{(*)}$  are not available,  we relate them with the respective (axial)vector form factors through the  equations of motion. In the following subsections, we discuss our predicted results on all these $b \to c \tau \bar \nu_\tau$ decay modes for each case. 
\subsection{Case A}
\begin{figure}[htb]
\includegraphics[scale=0.5]{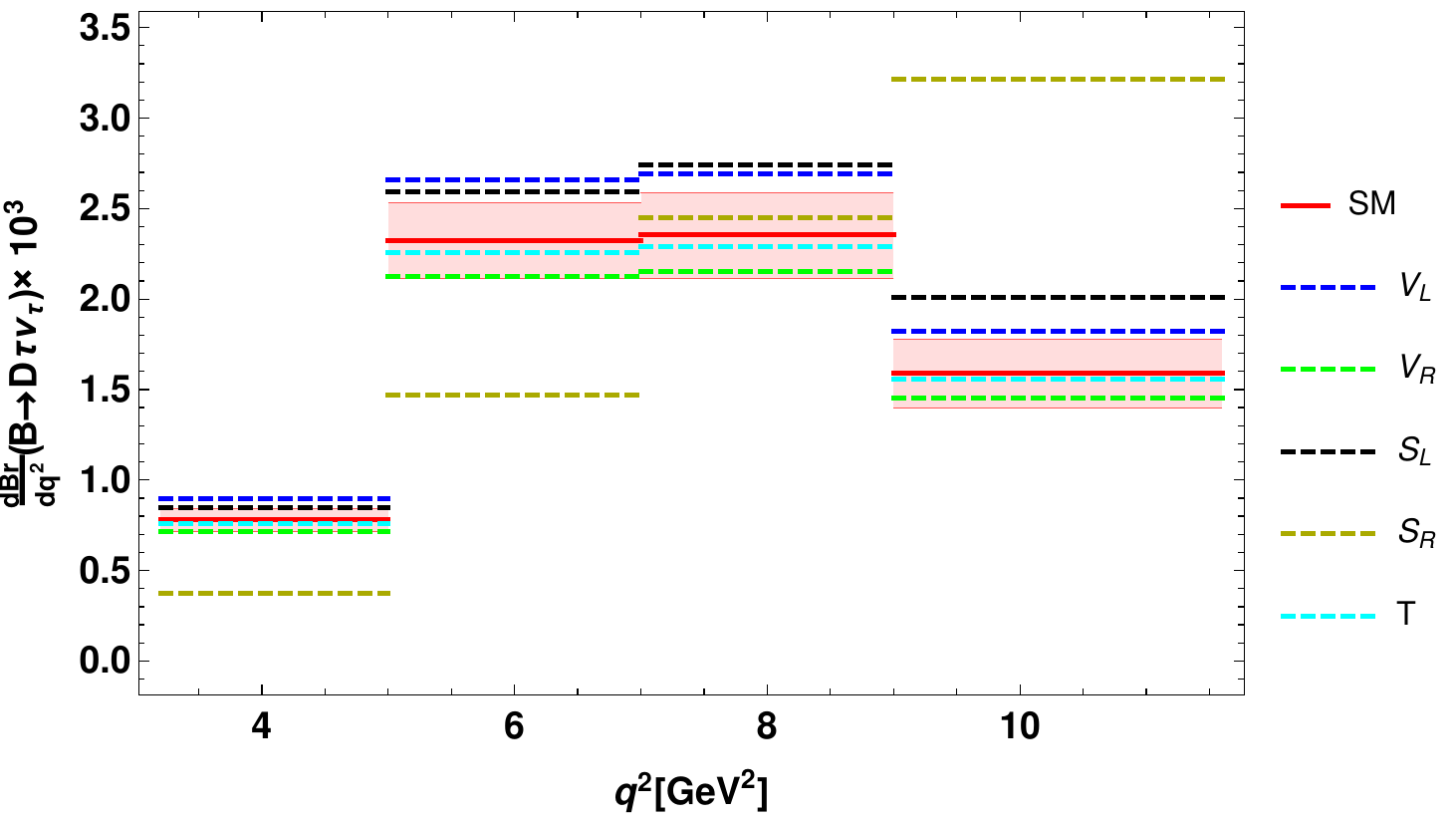}
\quad
\includegraphics[scale=0.5]{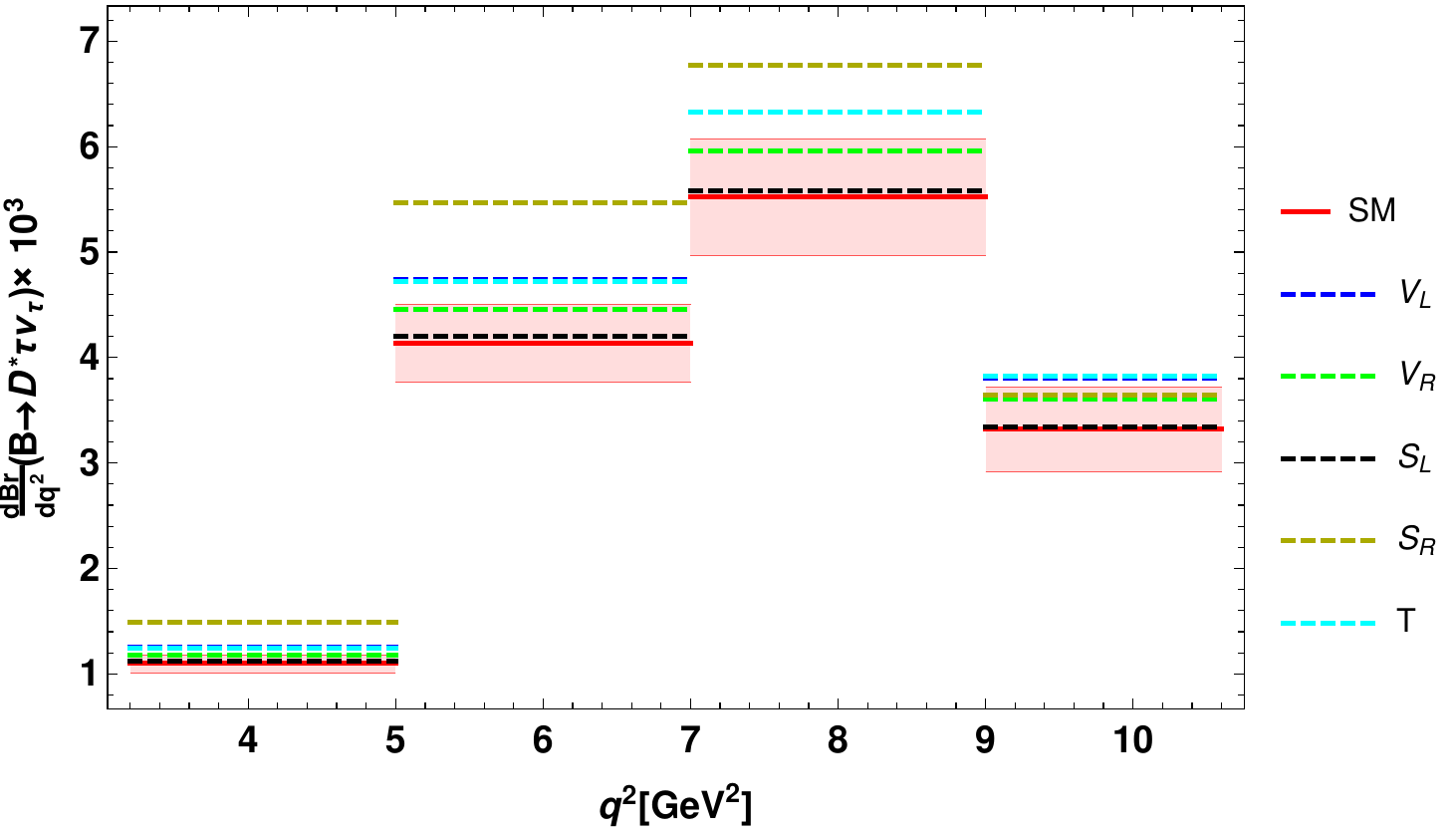}
\quad
\includegraphics[scale=0.5]{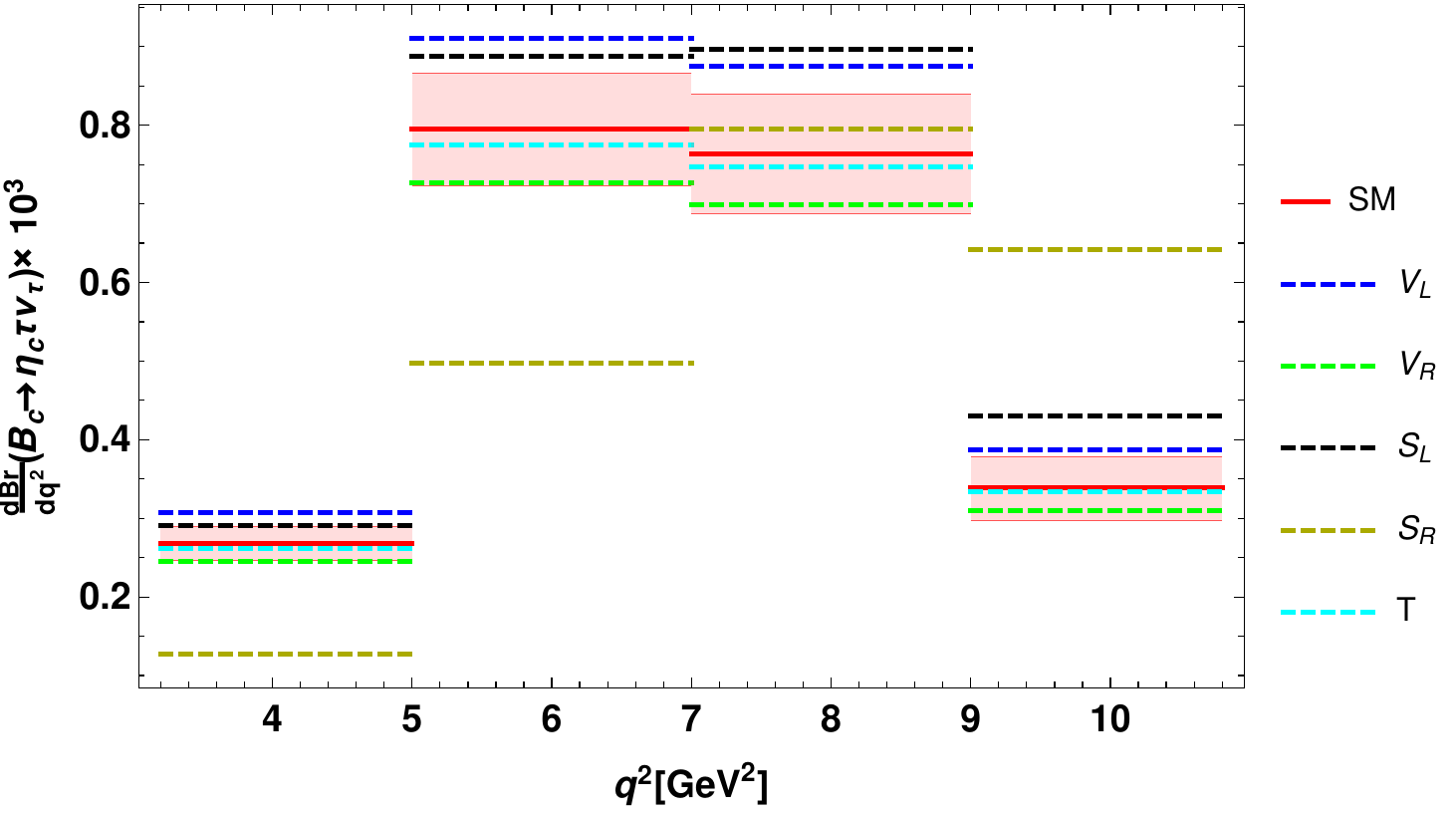}
\quad
\includegraphics[scale=0.5]{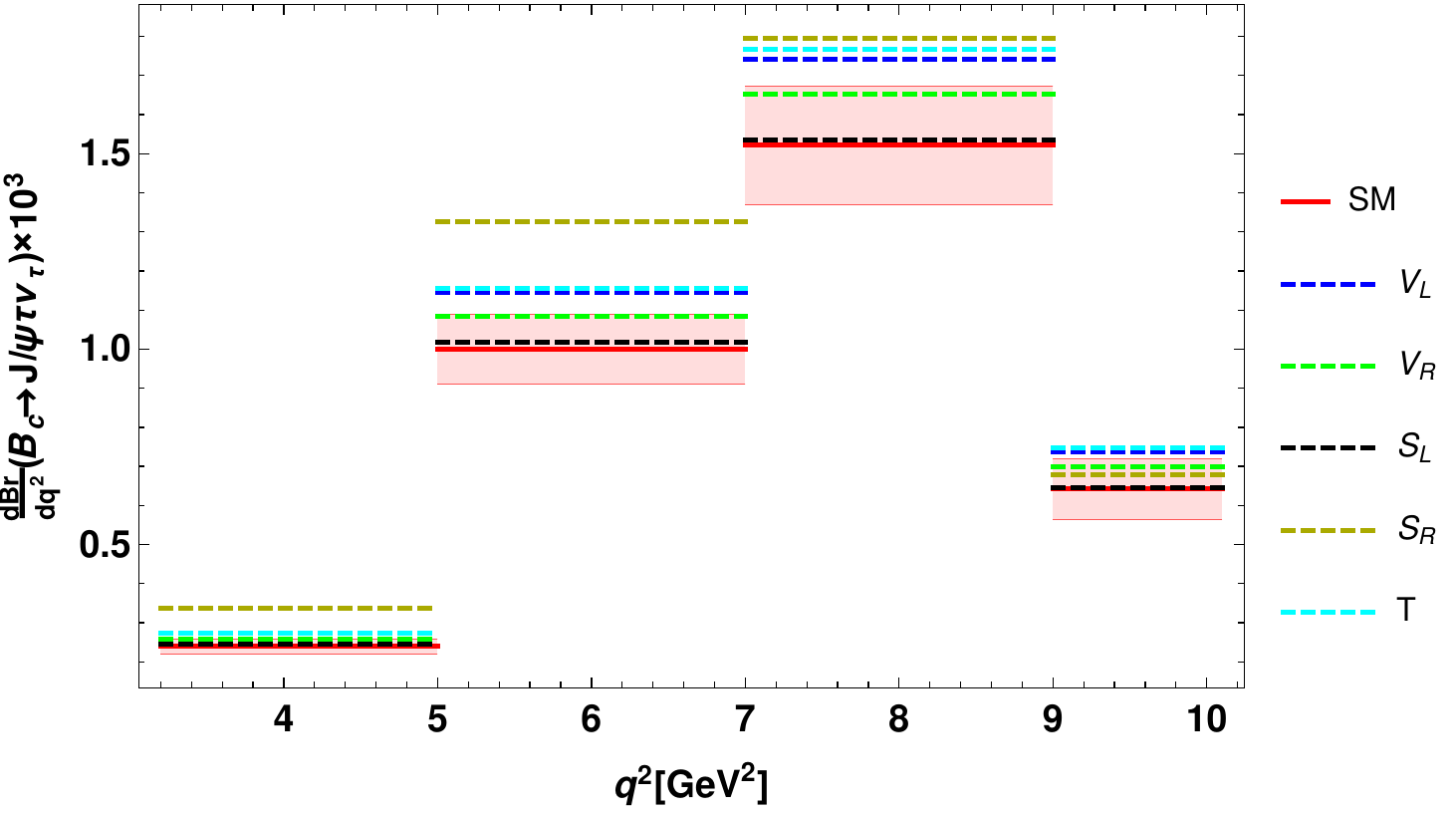}
\quad
\includegraphics[scale=0.5]{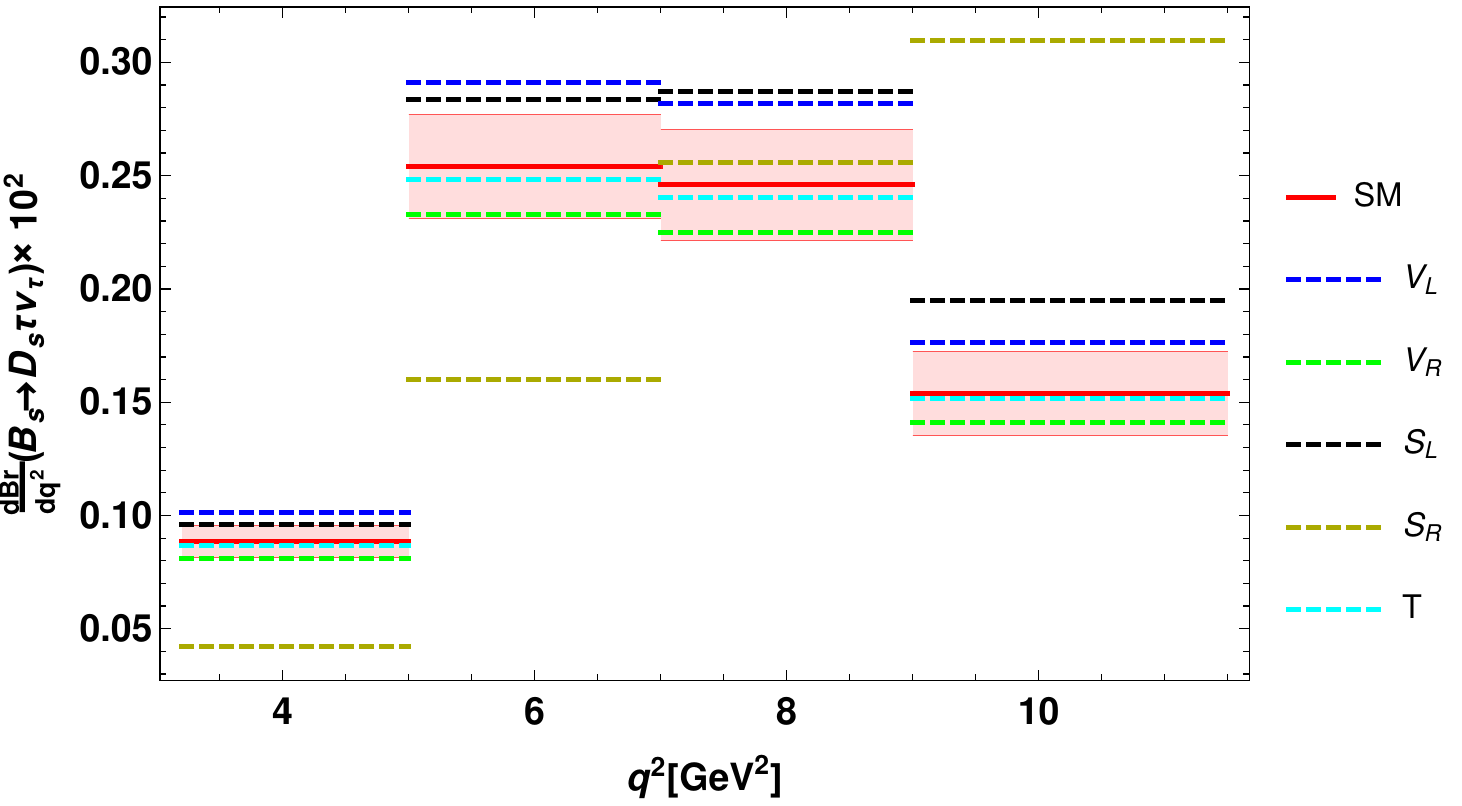}
\quad
\includegraphics[scale=0.5]{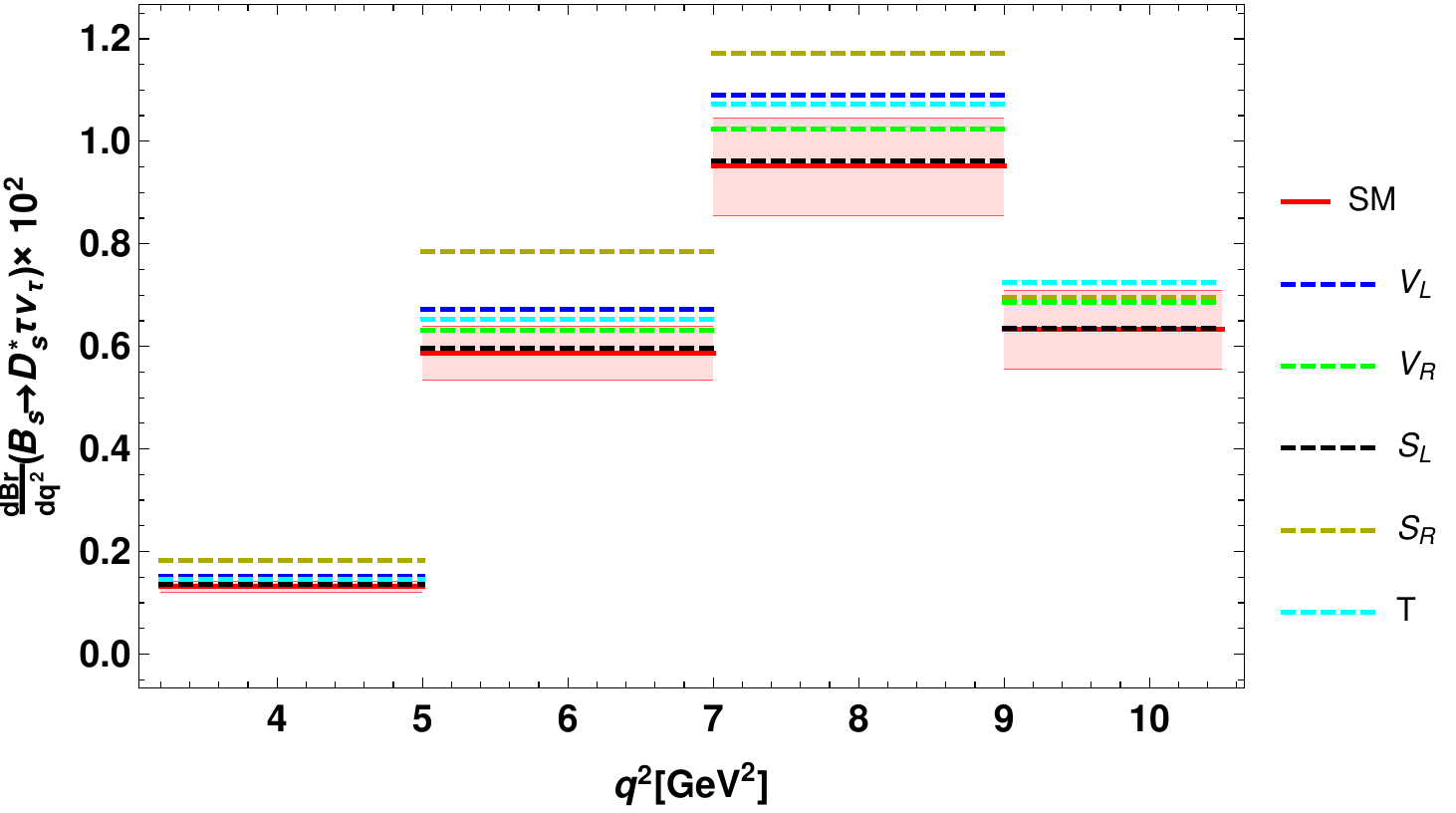}
\caption{ The bin-wise branching ratios of $\bar B \to D \tau \bar \nu_\tau$ (top-left panel), $\bar B \to D^* \tau \bar \nu_\tau$ (top-right panel), $B_c^+ \to \eta_c \tau^+   \nu_\tau$ (middle-left panel), $ B_c^+ \to J/\psi \tau^+  \nu_\tau$ (middle-right panel), $ B_s \to D_s \tau \bar \nu_\tau$ (bottom-left panel) and $ B_s \to D_s^* \tau \bar \nu_\tau$ (bottom-right panel) processes in four $q^2$ bins  for case A of our analysis. Here the solid red lines (light red bands) stand for the  central values (1-$\sigma$ uncertainties) in the SM. The blue, green, black, dark yellow and cyan dashed lines are drawn by using the best-fit values of $V_L$, $V_R$, $S_L$, $S_R$ and $T$  coefficients, respectively. } \label{Fig:CA-BR}
\end{figure}

Here we consider the presence of  a  single  real new Wilson coefficient  at a time, whose predicted best-fit values are presented in Table \ref{Tab:Best-fit}\,. Using these values, the branching ratios of $\bar B \to D \tau \bar \nu_\tau$ (top-left panel), $\bar B \to D^* \tau \bar \nu_\tau$ (top-right panel),  $B_c^+  \to \eta_c \tau^+  \nu_\tau$ (middle-left panel), $B_c^+ \to J/\psi \tau^+  \nu_\tau$ (middle-right panel), $B_s \to D_s \tau \bar \nu_\tau$ (bottom-left panel) and $B_s \to D_s^* \tau \bar \nu_\tau$ (bottom-right panel) processes in four $q^2$ bins are presented in Fig. \ref{Fig:CA-BR}\,. The solid red lines in these plots represent the SM predictions and the light red  bands represent the SM $1\sigma$ uncertainties, which  arise due to the uncertainties in the input parameters such as CKM matrix elements and hadronic form factors. The dashed blue, green, black, dark yellow and cyan lines are obtained by using the best-fit values of $V_L$, $V_R$, $S_L$, $S_R$ and $T$ coefficients, respectively. 
The following observations are inferred from these plots.
\begin{itemize}

\item  In $\bar B \to  D$ decays, the  $S_R$ contribution gives significant deviation from the SM in first, second and fourth $q^2$ bins. The contributions of  $V_L$ and $S_L$ couplings show reasonable deviations  from the SM central values  in the second, third and fourth $q^2$ bins. The predicted branching ratio with $V_R$ coupling lies within $1\sigma$ range of SM, while the tensor coefficient has negligible impact on the entire  $q^2$ region.


\item For $\bar B \to  D^* \tau \bar \nu_\tau$ process, the branching ratio show significant deviation in the first two bins  due to the contribution of additional $T$ coefficient, whereas the effect of other contributions are rather marginal. The predicted bin-wise branching ratios for $V_R$ coupling in all bins are found to be within the $1\sigma$ SM uncertainties.  The $S_L$ coefficient has negligible impact in all four $q^2$ bins.

\item 
 The deviation in the branching fraction of $B_c^+ \to \eta_c \tau^+  \nu_\tau$ and $B_s \to D_s \tau \bar \nu_\tau$  processes due to the contribution of $S_R$ coefficient is significantly large in the first, second and fourth bins whereas the impact of other coefficients are quite marginal in all the bins. 
 
 \item For  $B_c^+ \to J/\psi \tau^+  \nu_\tau$  ($B_s \to D_s^* \tau \bar \nu_\tau$) process, the effect of $S_R$ is significantly large in the second (second and third)  bin(s), whereas the impact of other couplings are rather nominal. 

\end{itemize}
 In order to quantify  the above discussed results, we define the pull metric at the  observable level as
\bea
{\rm Pull}_i=\frac{{\cal O}_i^{\rm NP} - {\cal O}_i^{\rm SM}}{\sqrt{{\Delta 
{\cal O}_i^{\rm NP}}^2 +\Delta {{\cal O}_i^{\rm SM}}^2}}\;,
\eea
where  $i$ represents all observables, ${\cal O}_i^{\rm SM}$ and  ${\cal O}_i^{\rm NP}$ stand for  the values of the observables  in the SM and NP scenarios and   
${\Delta {\cal O}_i^{\rm SM}}$,  $\Delta {{\cal O}_i^{\rm NP}}$ are the corresponding $1\sigma$ uncertainties.  The pull values of the branching fractions of all the decay modes in the presence of individual  new coefficients in all the four $q^2$ bins are presented in Table \ref{Tab:pull-CaseA}\,.  Though the presence of only real $S_R$ coupling provide comparatively large  deviation from the SM results, the corresponding $\chi^2_{\rm min}/{\rm d.o.f}$  rather large, which implies the fit is not good enough to accommodate all the observed $b \to c \tau \bar \nu_\tau$ anomalies. 
\begin{table}[htb]
\caption{Pull values of the branching ratios of all the decay modes in the presence of individual  new coefficients in all the four $q^2$ bins for case A. Here the first row contains  the pull values for only $V_L,~V_R,~T$ and second row represents the $S_L,~S_R$ pull values.}\label{Tab:pull-CaseA}
\begin{center}
\begin{tabular}{|c|c|c|c|}
\hline
~modes ~&~Only $V_L$~&~Only $V_R$~&~Only $T$~\\
~~&~Only $S_L$~&~Only $S_R$~&~~\\
\hline \hline
$\bar B \to D $~&~$(1.192,1.06,0.953,0.794)$~&~$(0.783,0.696,0.627,0.522)$~&~$(0.24,0.23,0.197,0.112)$~\\

~&~$(0.72,0.856,1.074,1.364)$~&~$(5.9,3.45,0.285,3.78)$~&~~\\

\hline

$\bar B \to D^* $~&~$(1.192,1.06,0.953,0.794)$~&~$(0.648,0.574,0.535,0.475)$~&~$(1.1,1.03,0.954,0.82)$~\\

~&~$(0.165,0.123,0.072,0.024)$~&~$(2.65,2.15,1.423,0.544)$~&~~\\

\hline

$B_c \to \eta_c$~&~$(1.192,1.06,0.953,0.794)$~&~$(0.783,0.696,0.627,0.522)$~&~$(0.206,0.194,0.16,0.082)$~\\

~&~$(0.71,0.87,1.125,1.404)$~&~$(5.925,3.518,0.281,3.482)$~&~~\\

\hline

$B_c \to J/\psi $~&~$(1.192,1.06,0.953,0.794)$~&~$(0.714,0.635,0.58,0.495)$~&~$(1.143,1.13,1.053,0.88)$~\\

~&~$(0.188,0.124,0.057,0.014)$~&~$(2.95,2.17,1.16,0.318)$~&~~\\

\hline

$\bar B \to D_s$~&~$(1.192,1.06,0.953,0.794)$~&~$(0.783,0.696,0.625,0.522)$~&~$(0.197,0.191,0.166,0.096)$~\\

~&~$(0.72,0.861,1.09,1.372)$~&~$(5.92,3.485,0.273,3.75)$~&~\\

\hline

$\bar B \to D_s^* $~&~$(1.192,1.06,0.953,0.794)$~&~$(0.63,0.555,0.52,0.469)$~&~$(0.813,0.829,0.843,0.785)$~\\

~&~$(0.175,0.128,0.074,0.025)$~&~$(2.78,2.234,1.46,0.55)$~&~\\
\hline
\end{tabular}
\end{center}
\end{table}
\begin{figure}[htb]
\includegraphics[scale=0.5]{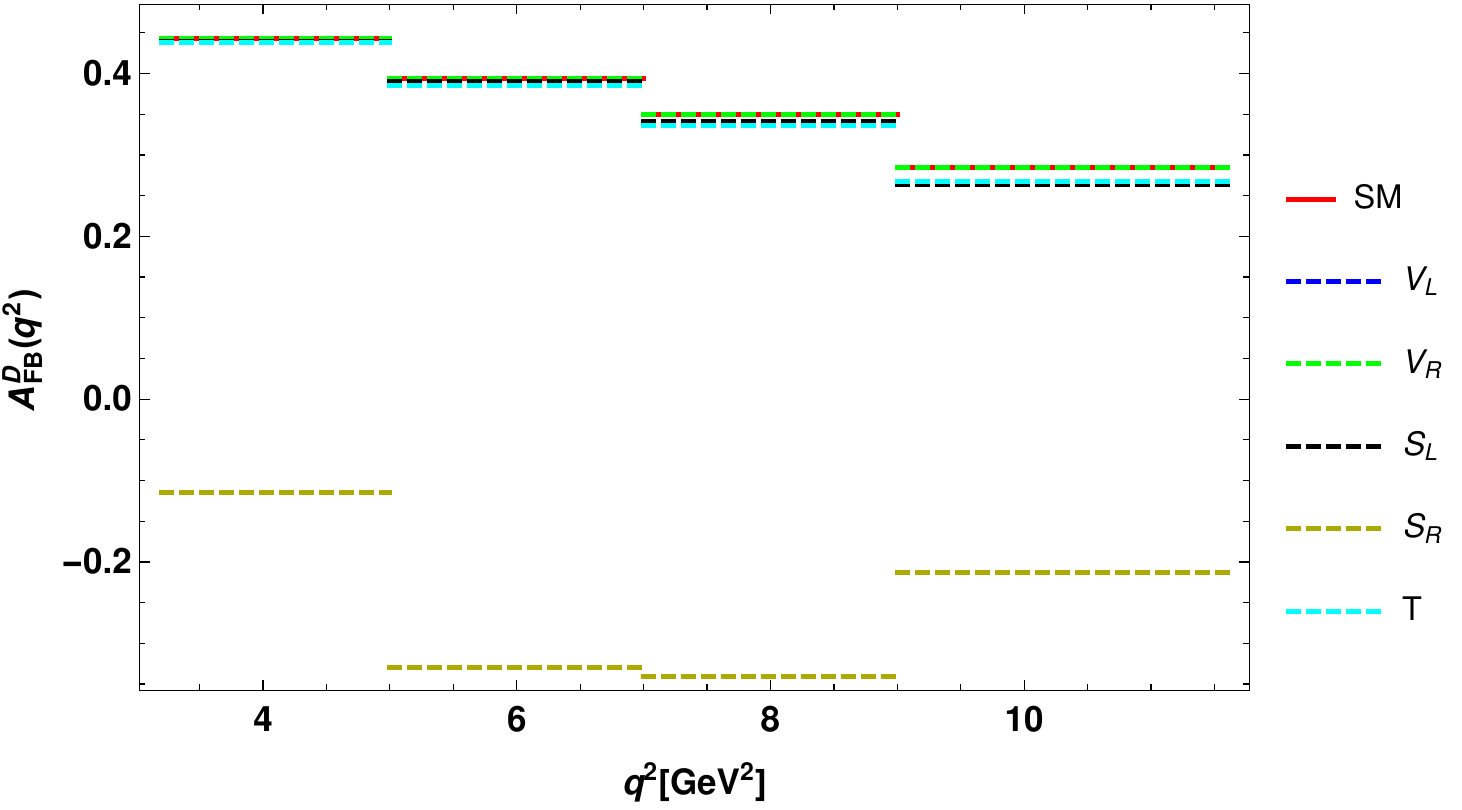}
\quad
\includegraphics[scale=0.5]{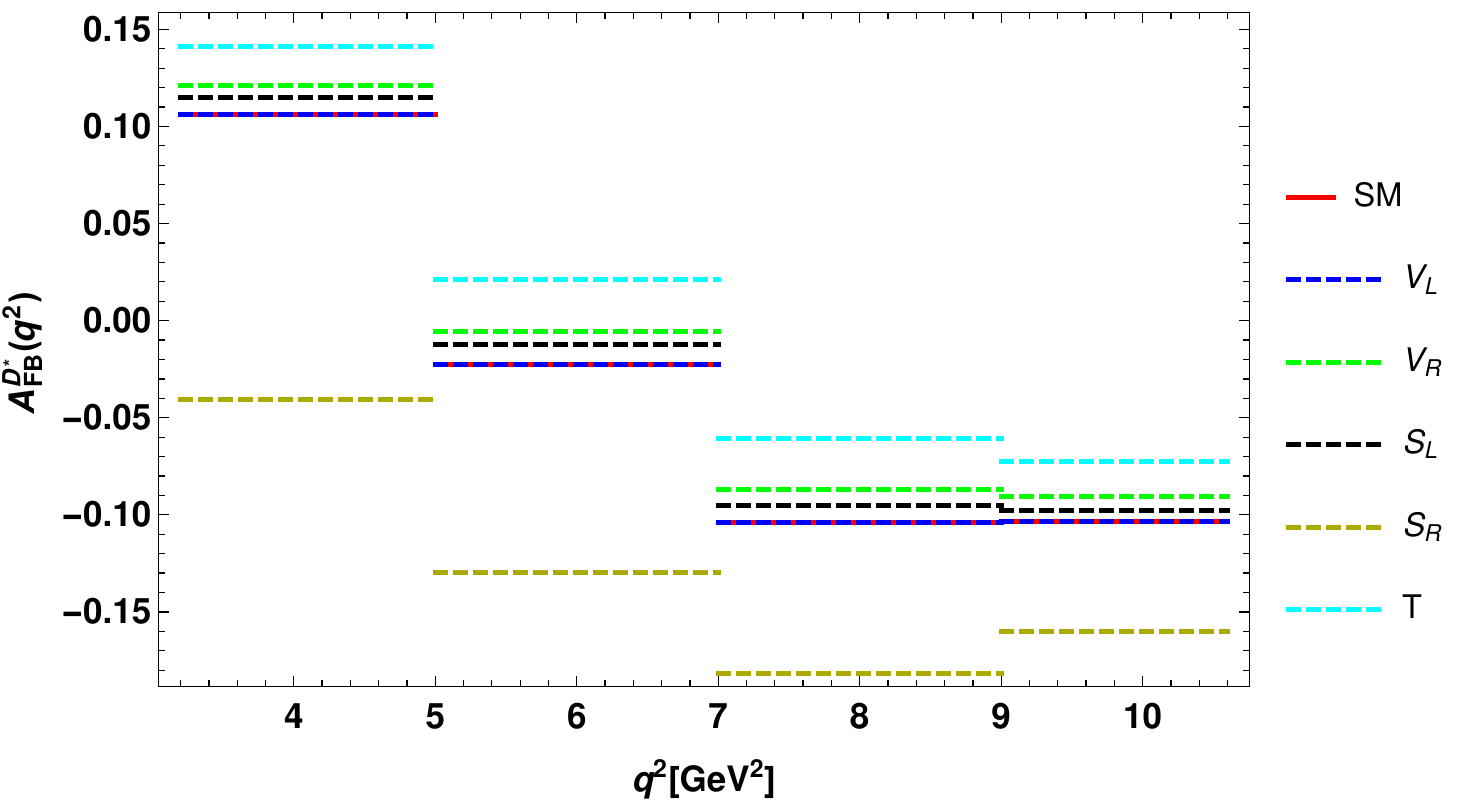}
\quad
\includegraphics[scale=0.5]{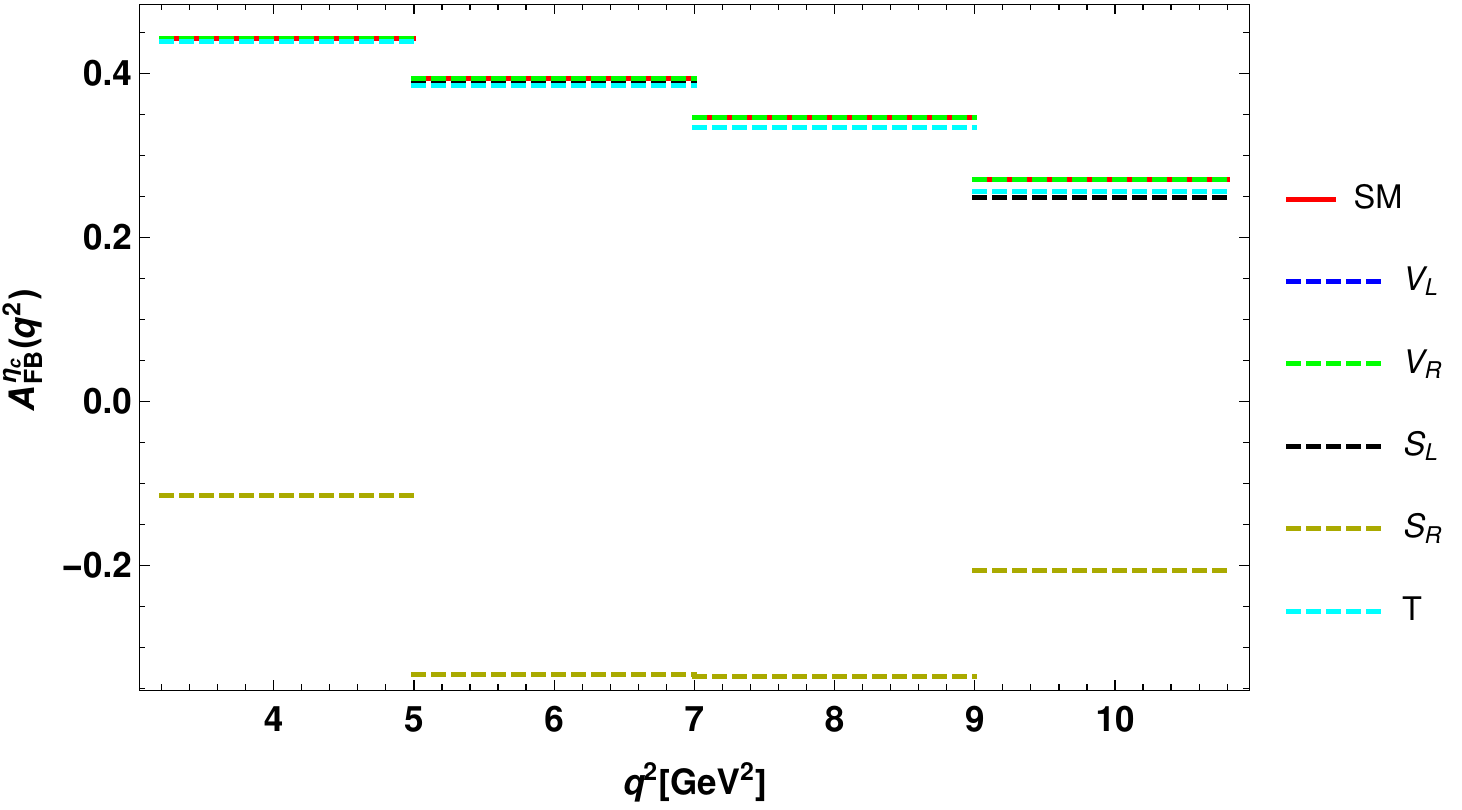}
\quad
\includegraphics[scale=0.5]{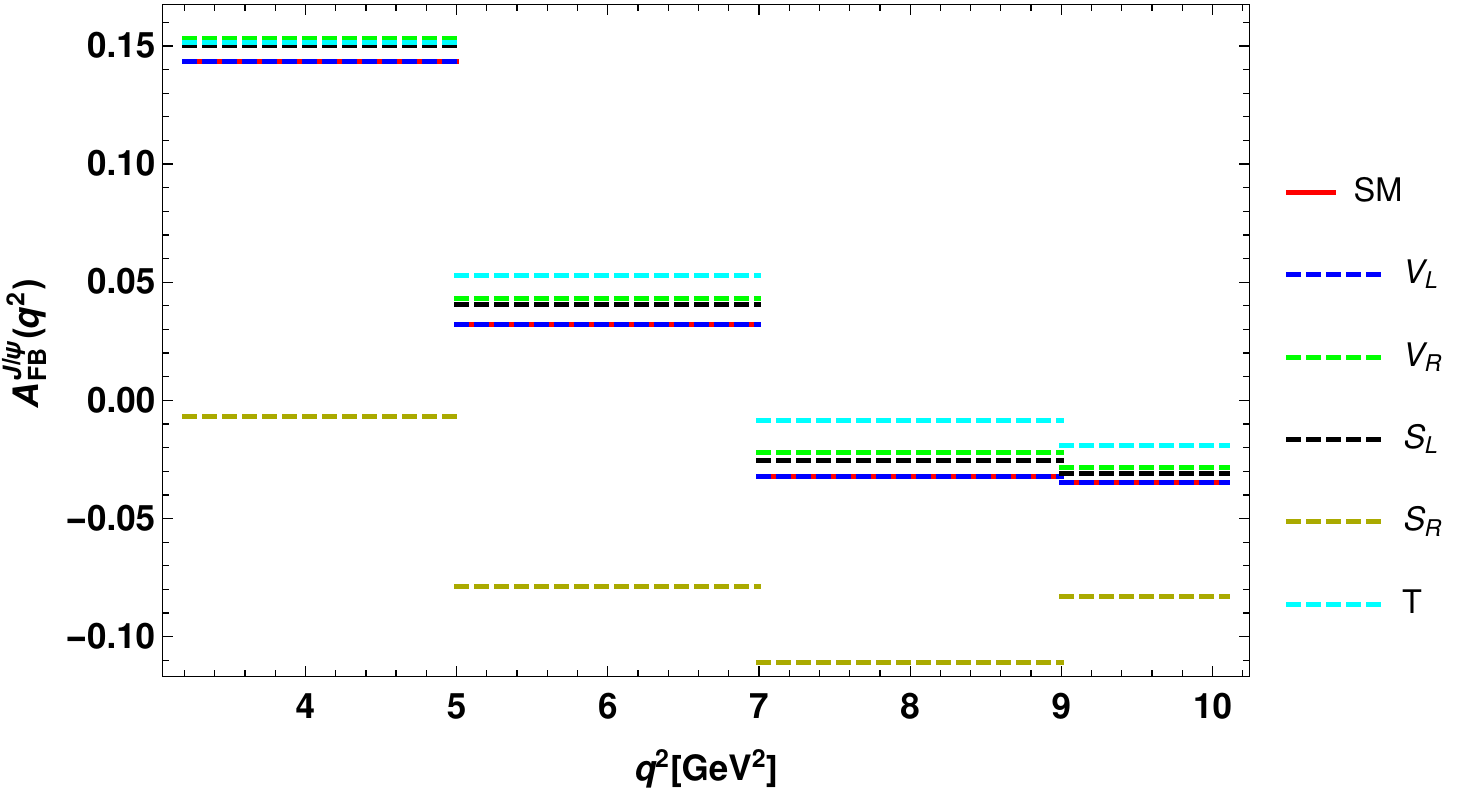}
\quad
\includegraphics[scale=0.5]{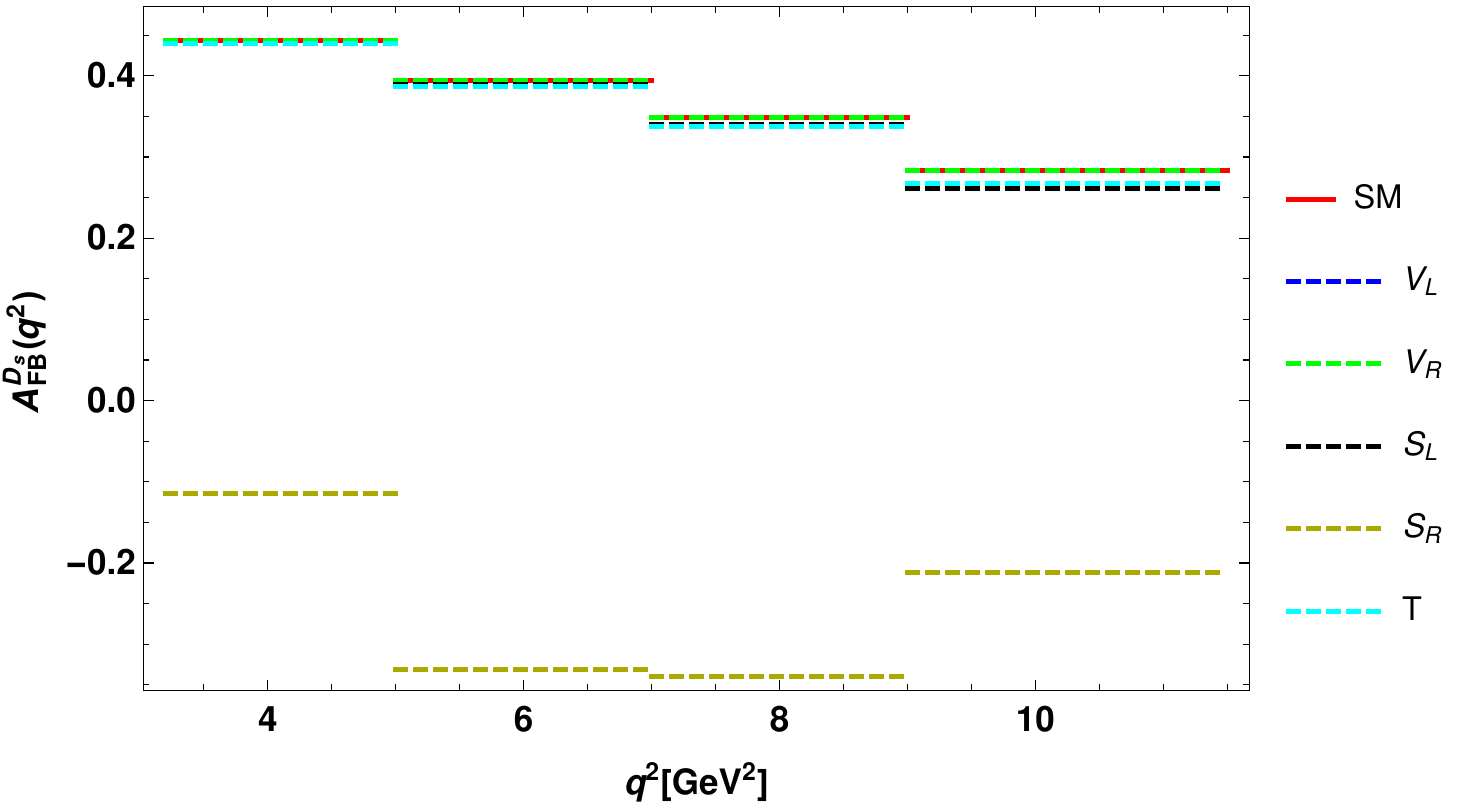}
\quad
\includegraphics[scale=0.5]{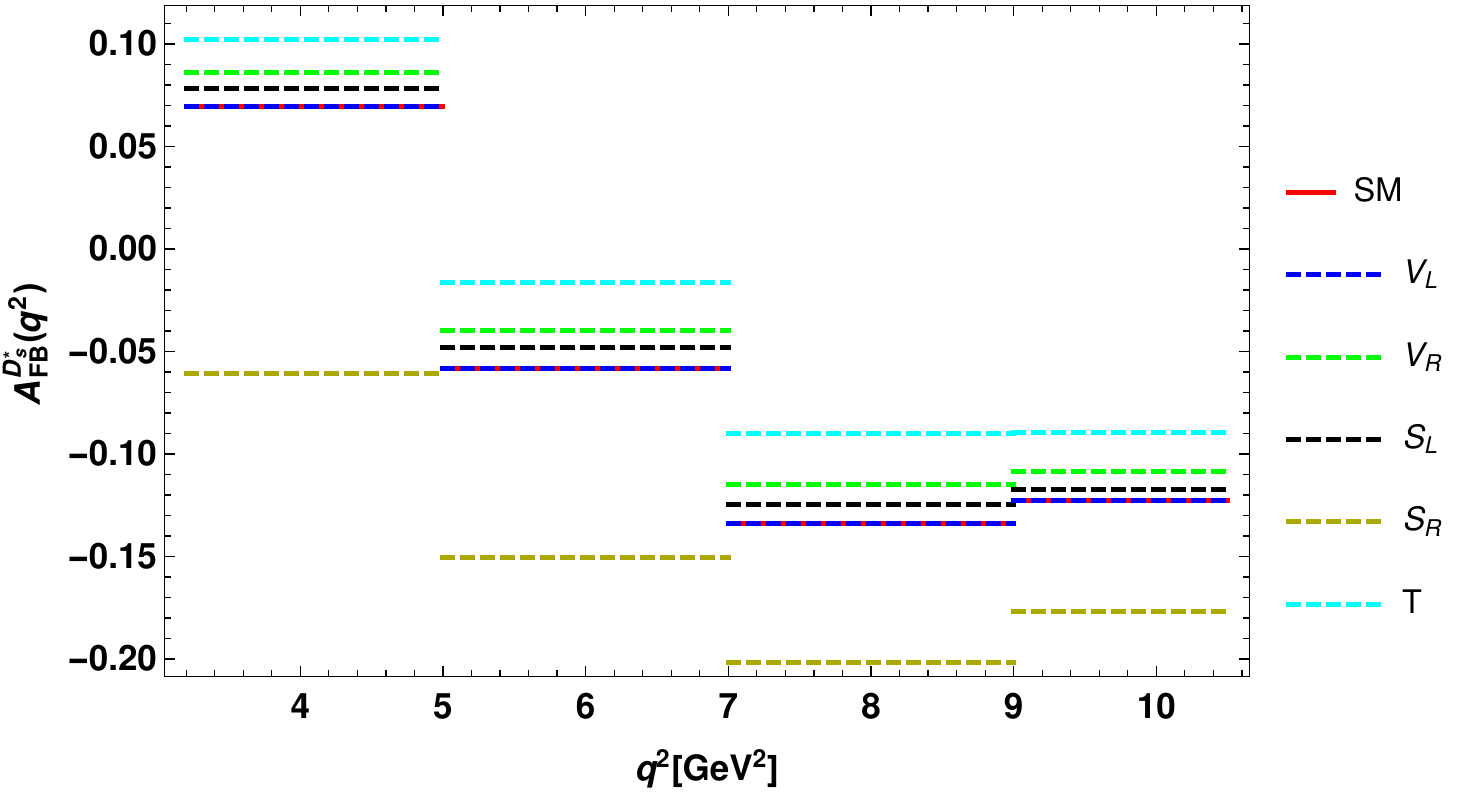}
\caption{ The bin-wise forward-backward asymmetry of $\bar B \to D\tau \bar \nu_\tau$ (top-left panel), $\bar B\to D^* \tau \bar \nu_\tau$ (top-right panel), $B_c^+ \to \eta_c \tau^+  \nu_\tau$ (middle-left panel), $ B_c^+ \to J/\psi \tau^+  \nu_\tau$ (middle-right panel), $ B_s \to D_s \tau \bar \nu_\tau$ (bottom-left panel) and $ B_s \to D_s^* \tau \bar \nu_\tau$ (bottom-right panel) processes in four $q^2$ bins for case A. }\label{Fig:CA-AFB}
\end{figure}
\begin{figure}[htb]
\includegraphics[scale=0.5]{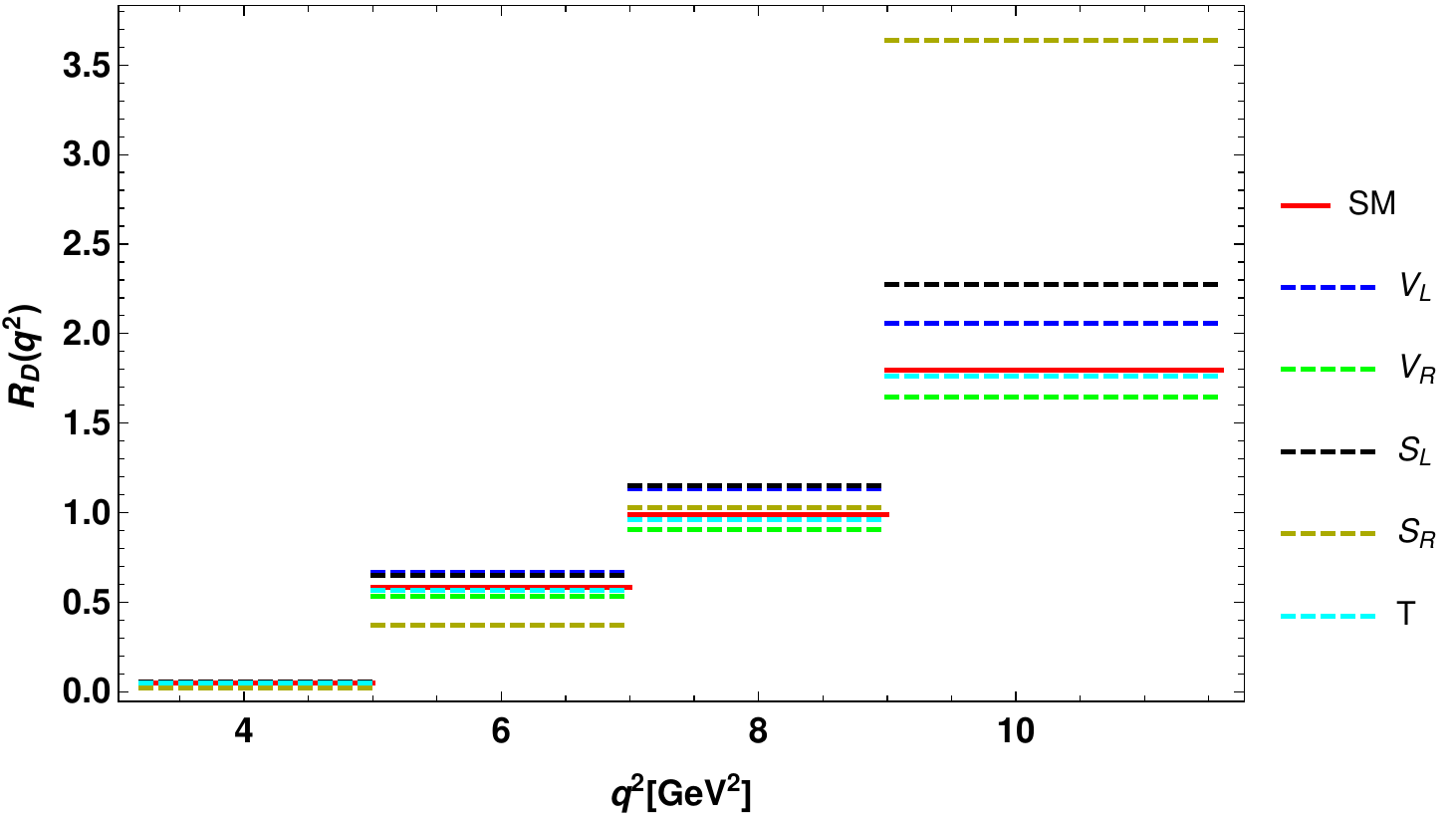}
\quad
\includegraphics[scale=0.5]{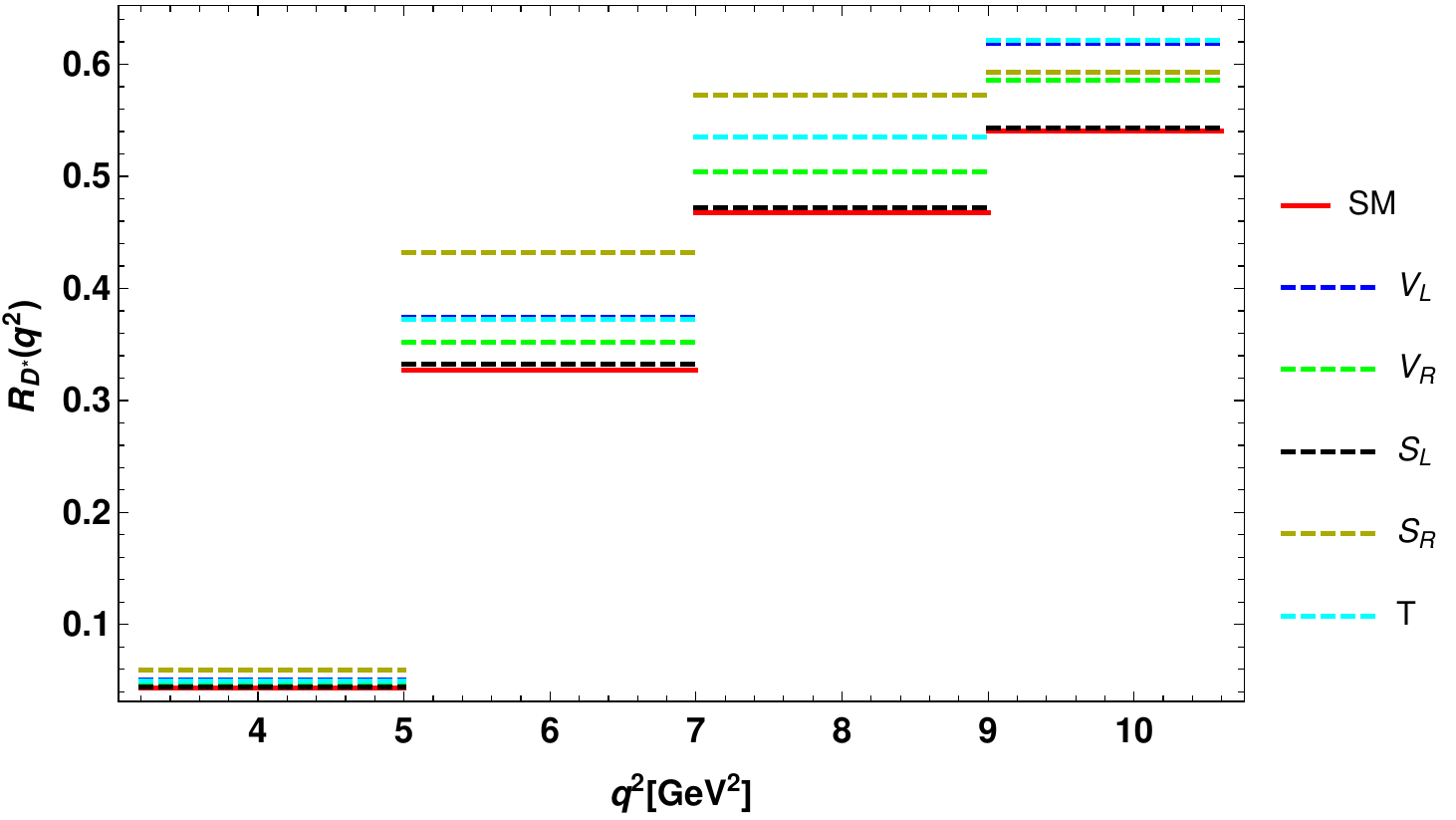}
\quad
\includegraphics[scale=0.5]{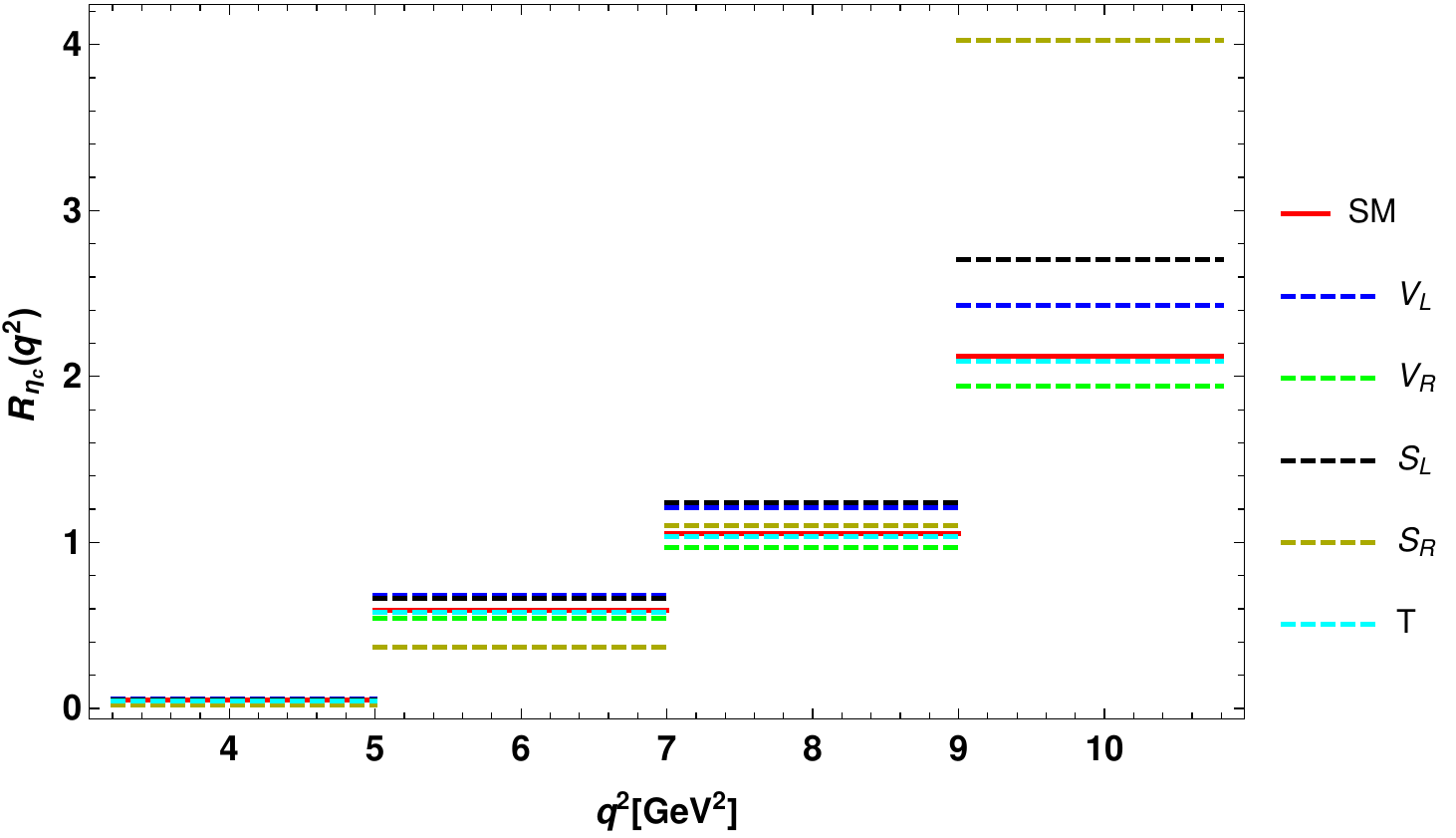}
\quad
\includegraphics[scale=0.5]{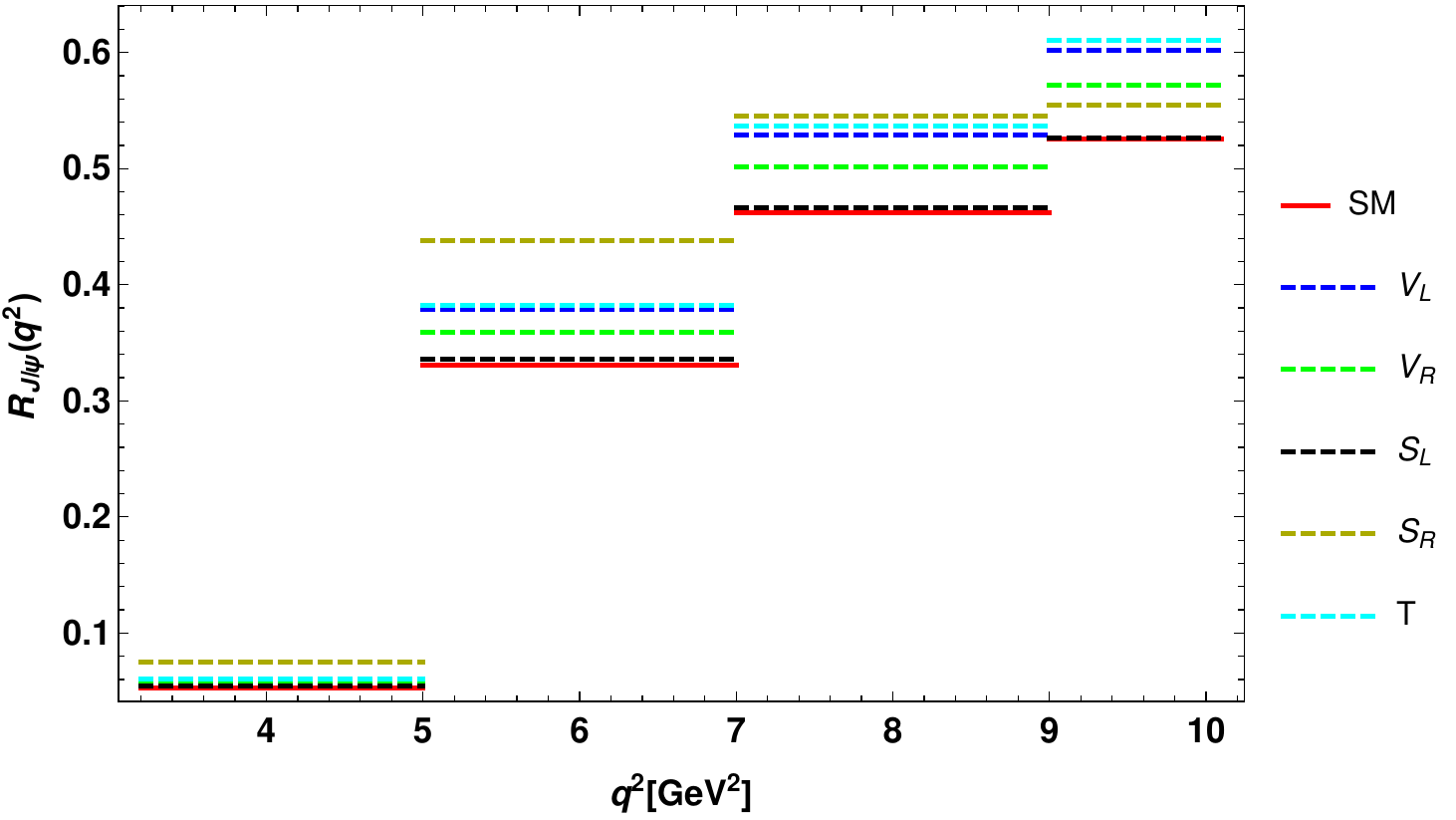}
\quad
\includegraphics[scale=0.5]{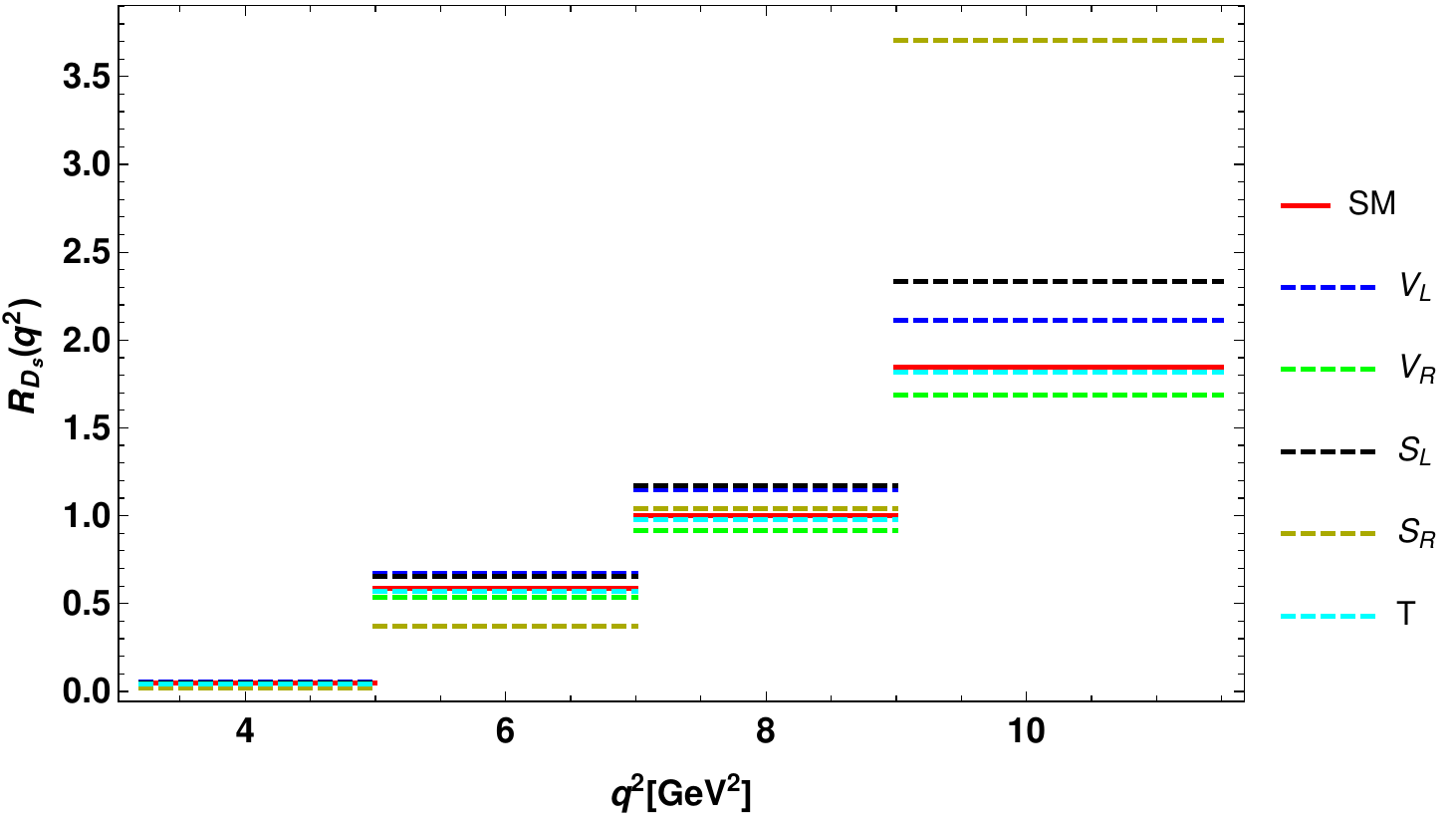}
\quad
\includegraphics[scale=0.5]{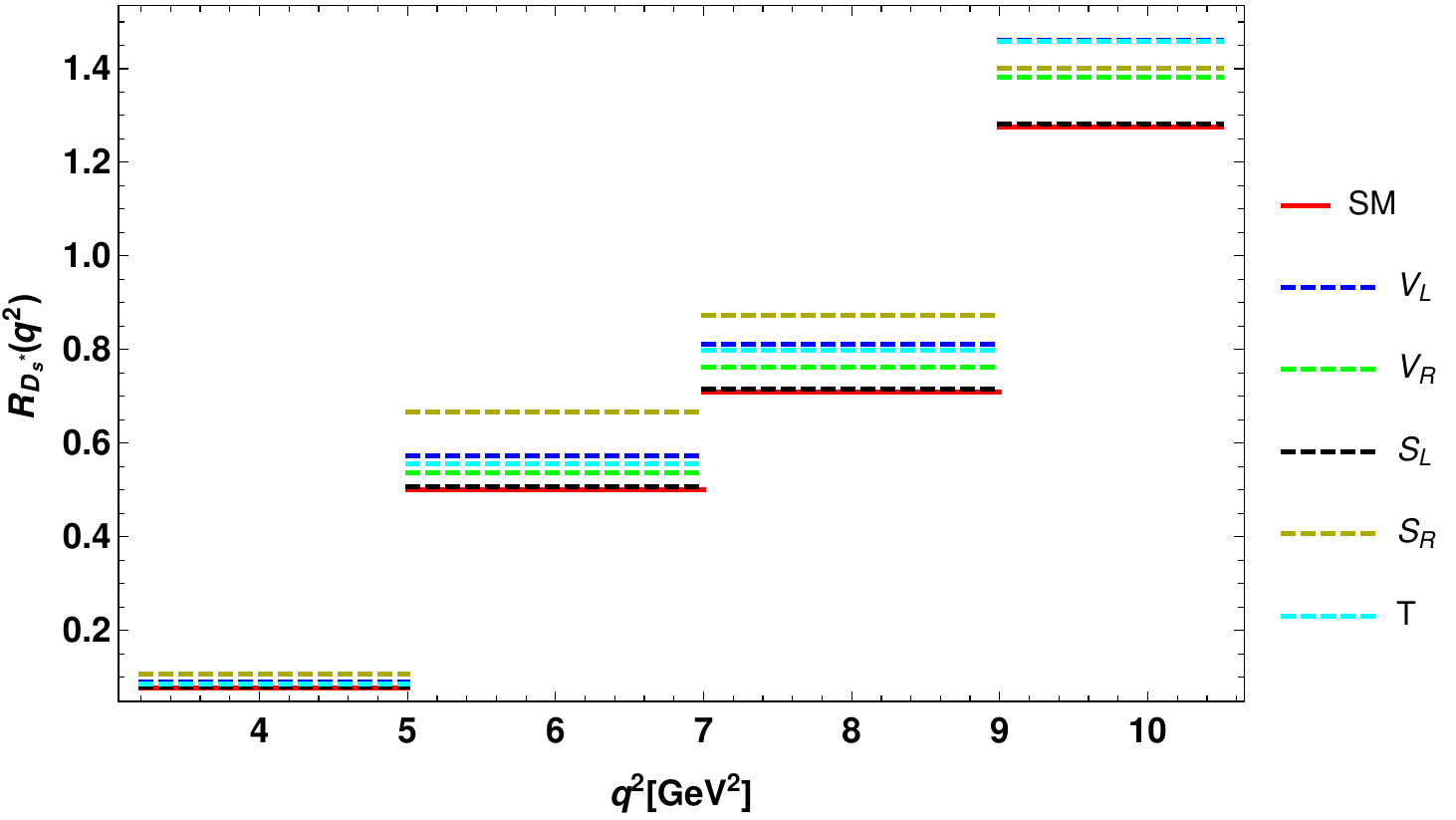}
\caption{ The bin-wise  $R_D$ (top-left panel), $R_{D^*}$ (top-right panel), $R_{\eta_c}$ (middle-left panel), $R_{J/\psi}$ (middle-right panel), $R_{D_s}$ (bottom-left panel) and $R_{D_s^*}$ (bottom-right panel)  in four $q^2$ bins for case A. }\label{Fig:CA-LNU}
\end{figure}
\begin{figure}[htb]
\includegraphics[scale=0.5]{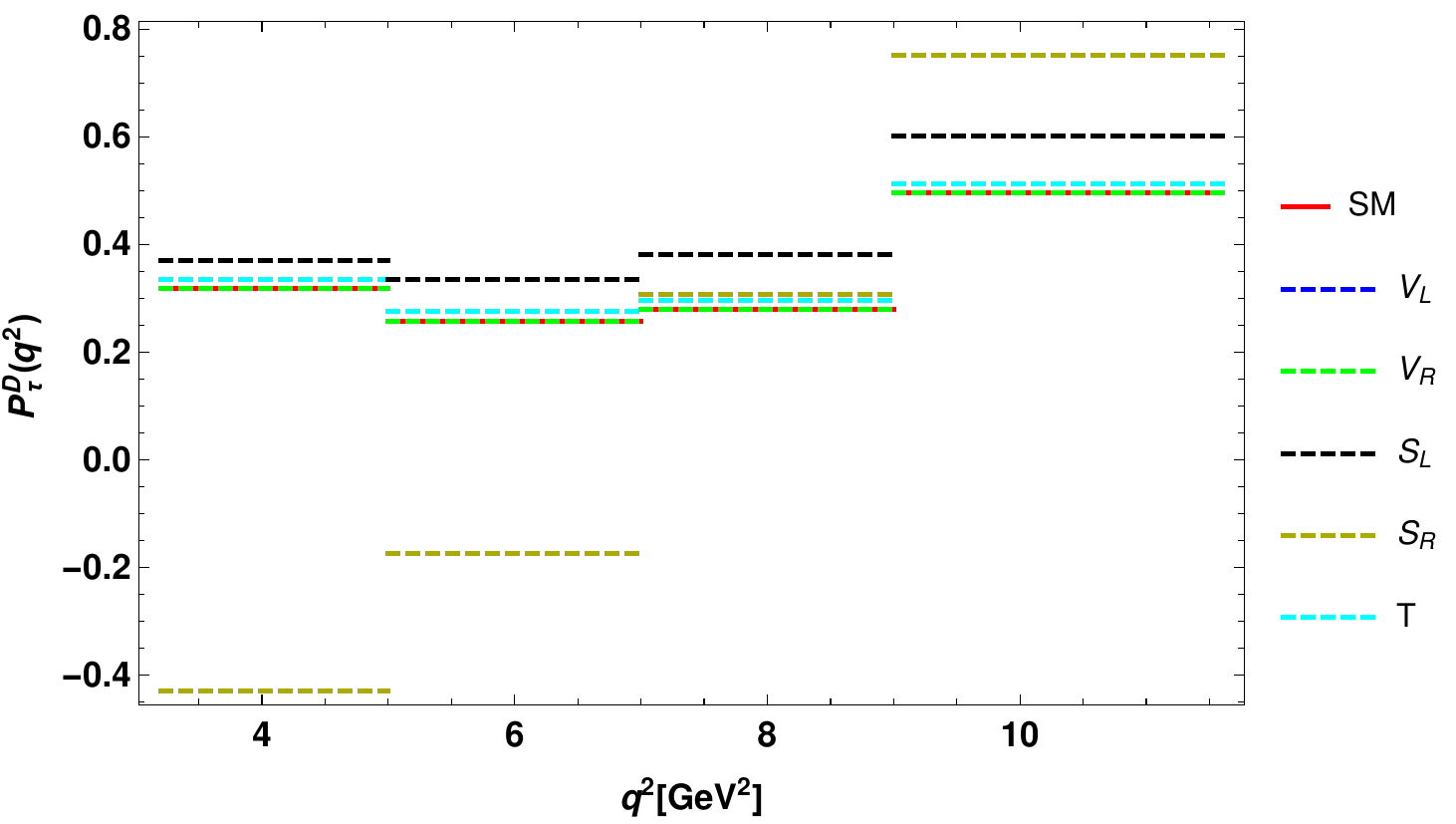}
\quad
\includegraphics[scale=0.5]{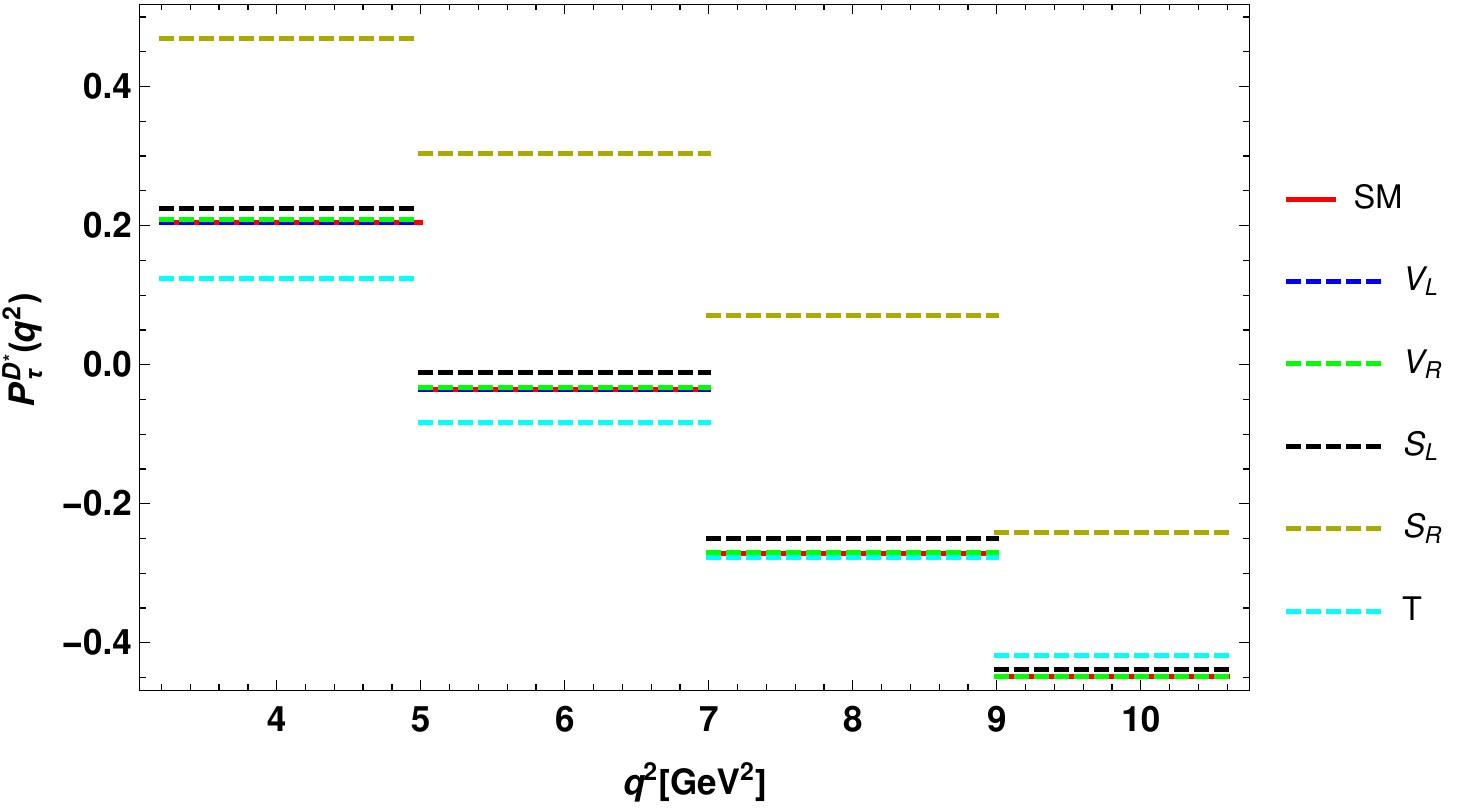}
\quad
\includegraphics[scale=0.5]{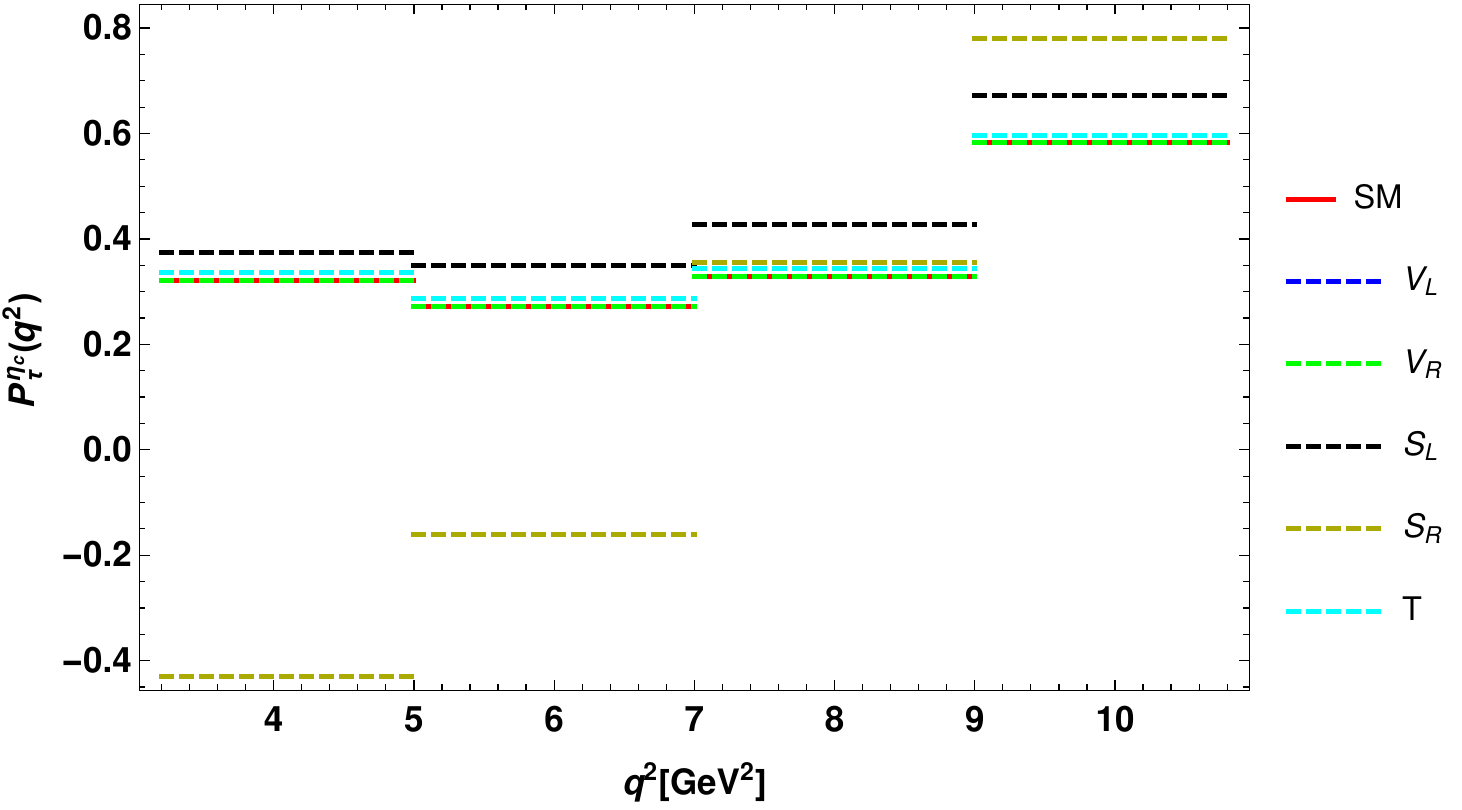}
\quad
\includegraphics[scale=0.5]{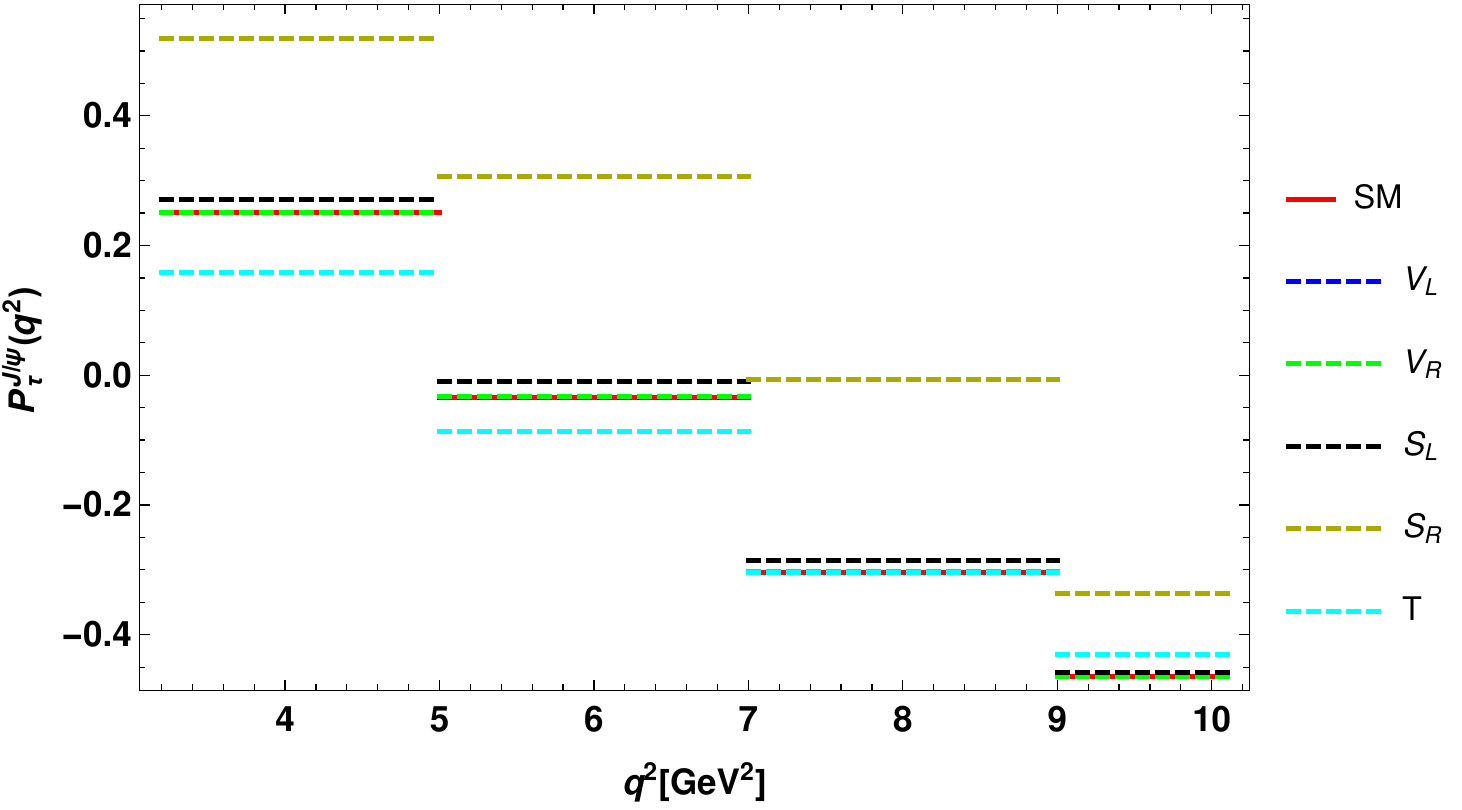}
\quad
\includegraphics[scale=0.5]{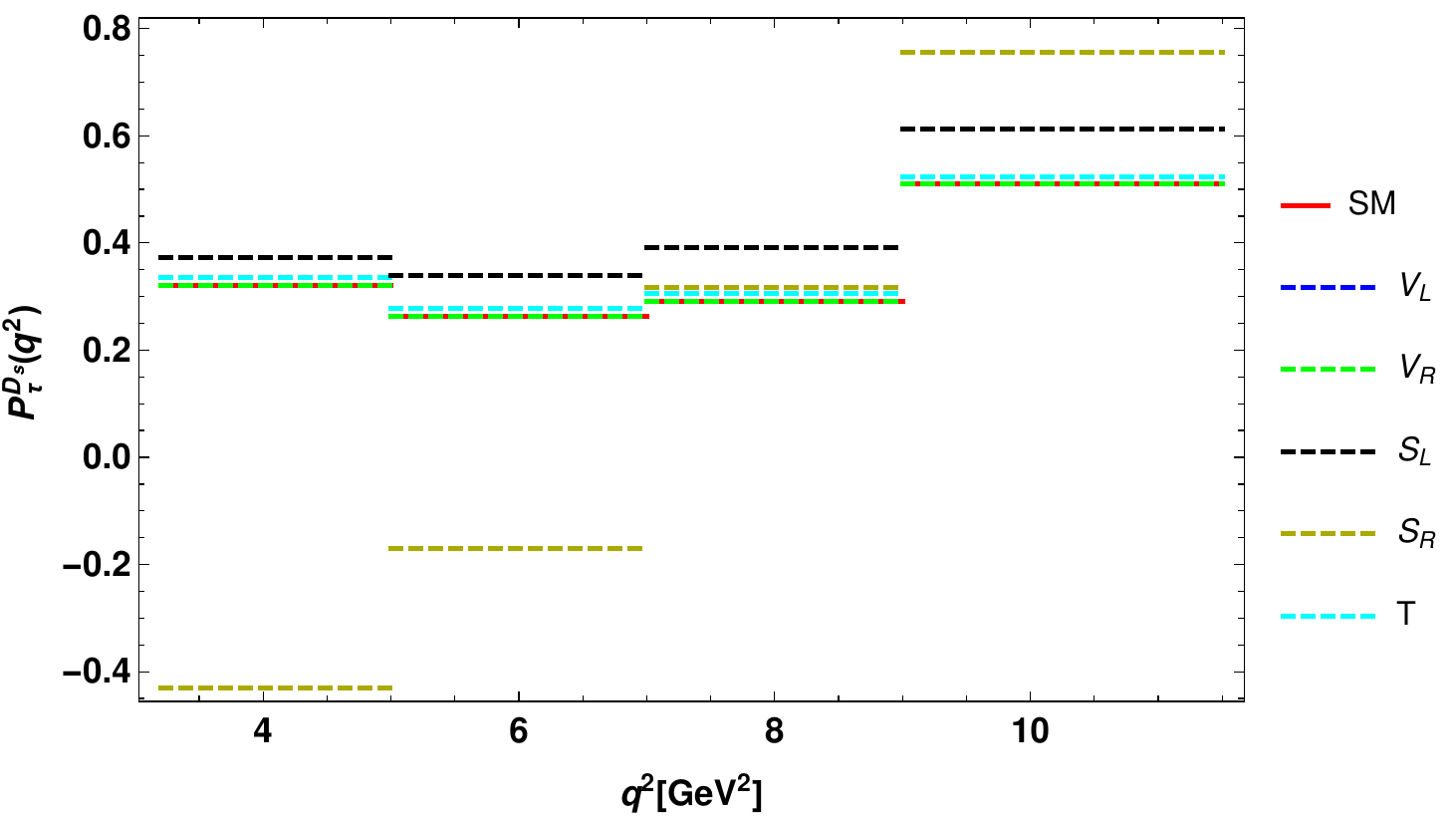}
\quad
\includegraphics[scale=0.5]{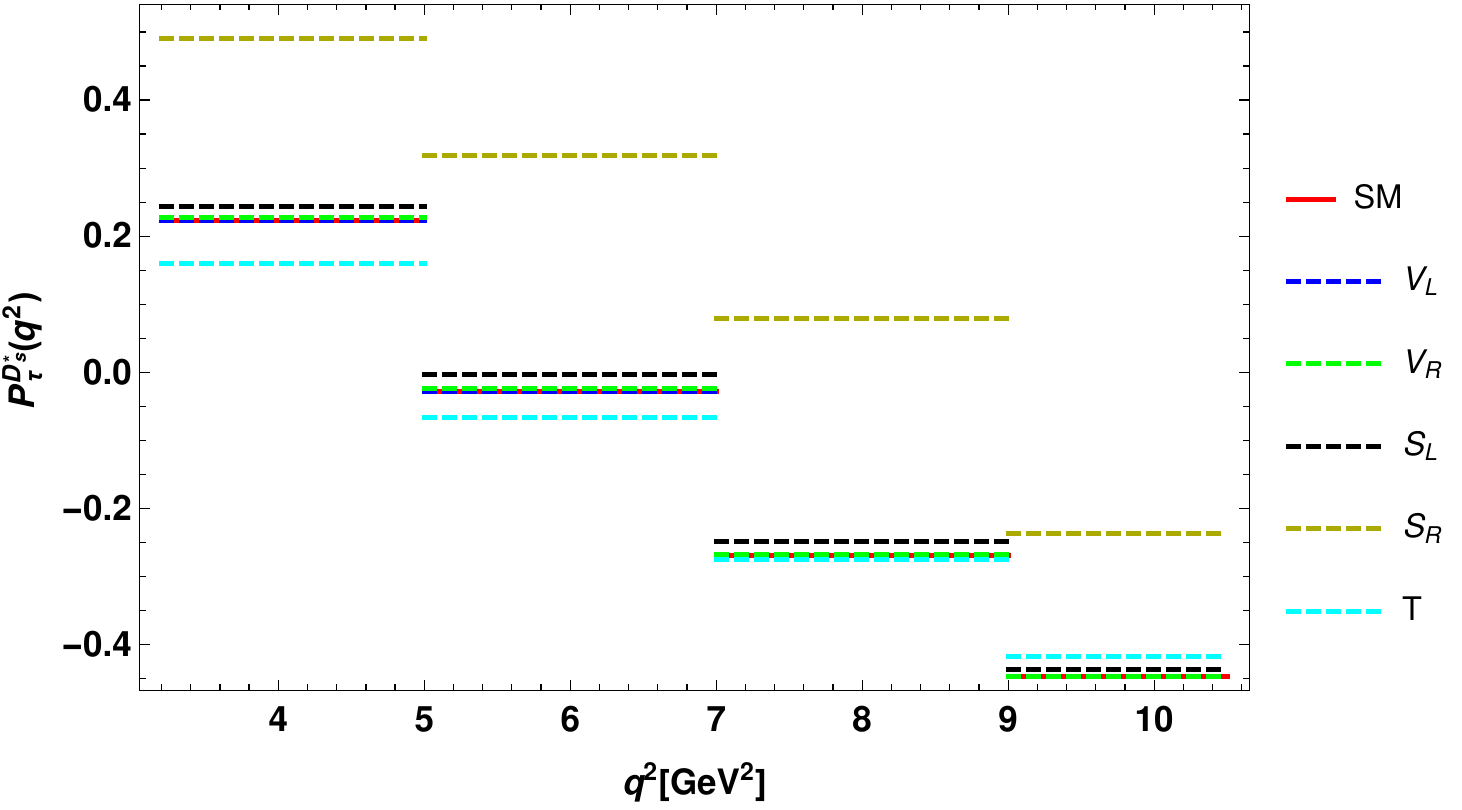}
\caption{ The bin-wise $\tau$-polarization asymmetry of $\bar B \to D \tau \bar \nu_\tau$ (top-left panel), $\bar B \to D^* \tau \bar \nu_\tau$ (top-right panel), $B_c^+ \to \eta_c \tau^+  \nu_\tau$ (middle-left panel), $ B_c^+ \to J/\psi \tau^+  \nu_\tau$ (middle-right panel), $\bar B_s \to\bar D_s \tau \bar \nu_\tau$ (bottom-left panel) and $\bar B_s \to\bar D_s^* \tau \bar \nu_\tau$ (bottom-right panel) processes in four $q^2$ bins for case A. } \label{Fig:CA-ptau}
\end{figure}
\begin{figure}[htb]
\includegraphics[scale=0.5]{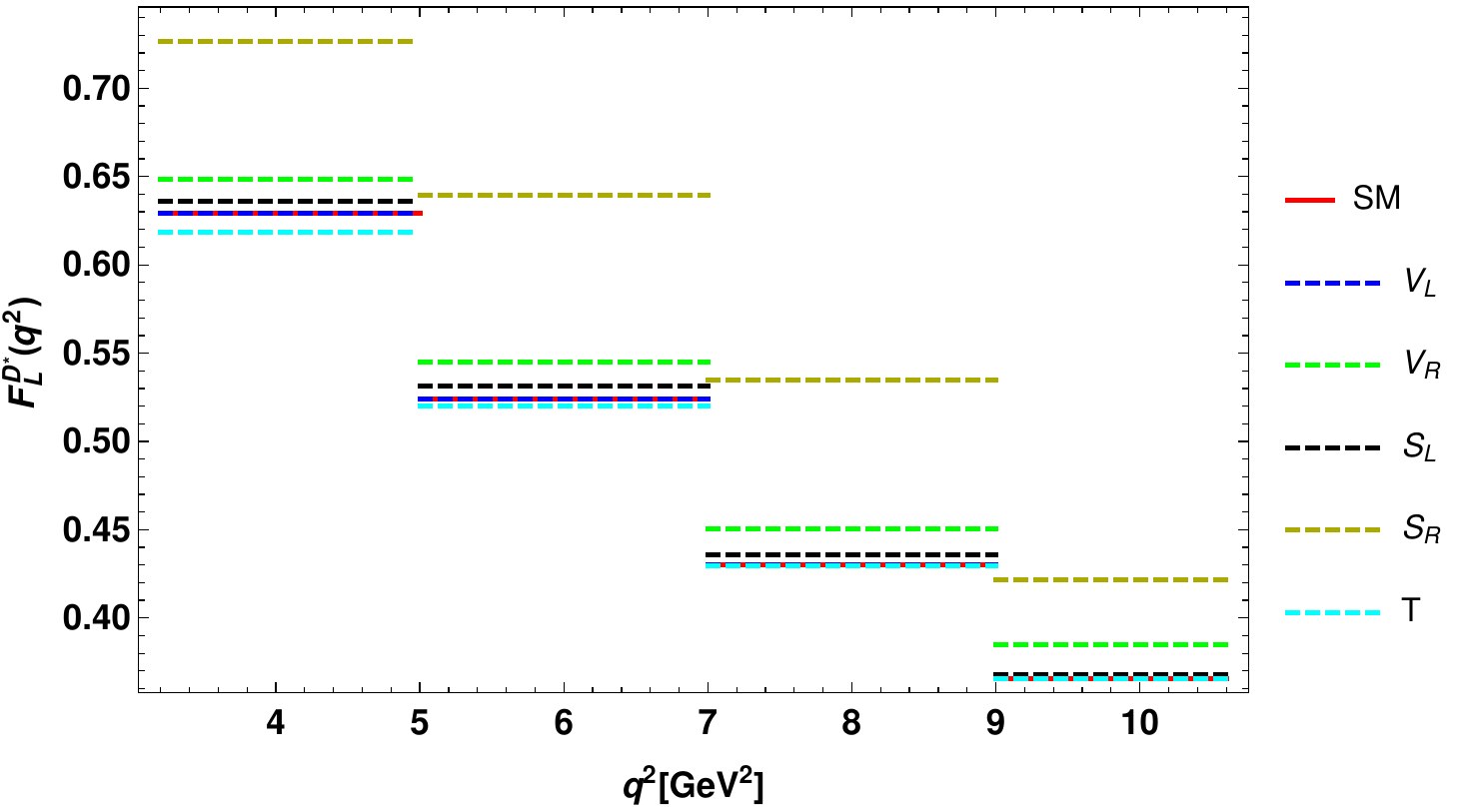}
\quad
\includegraphics[scale=0.5]{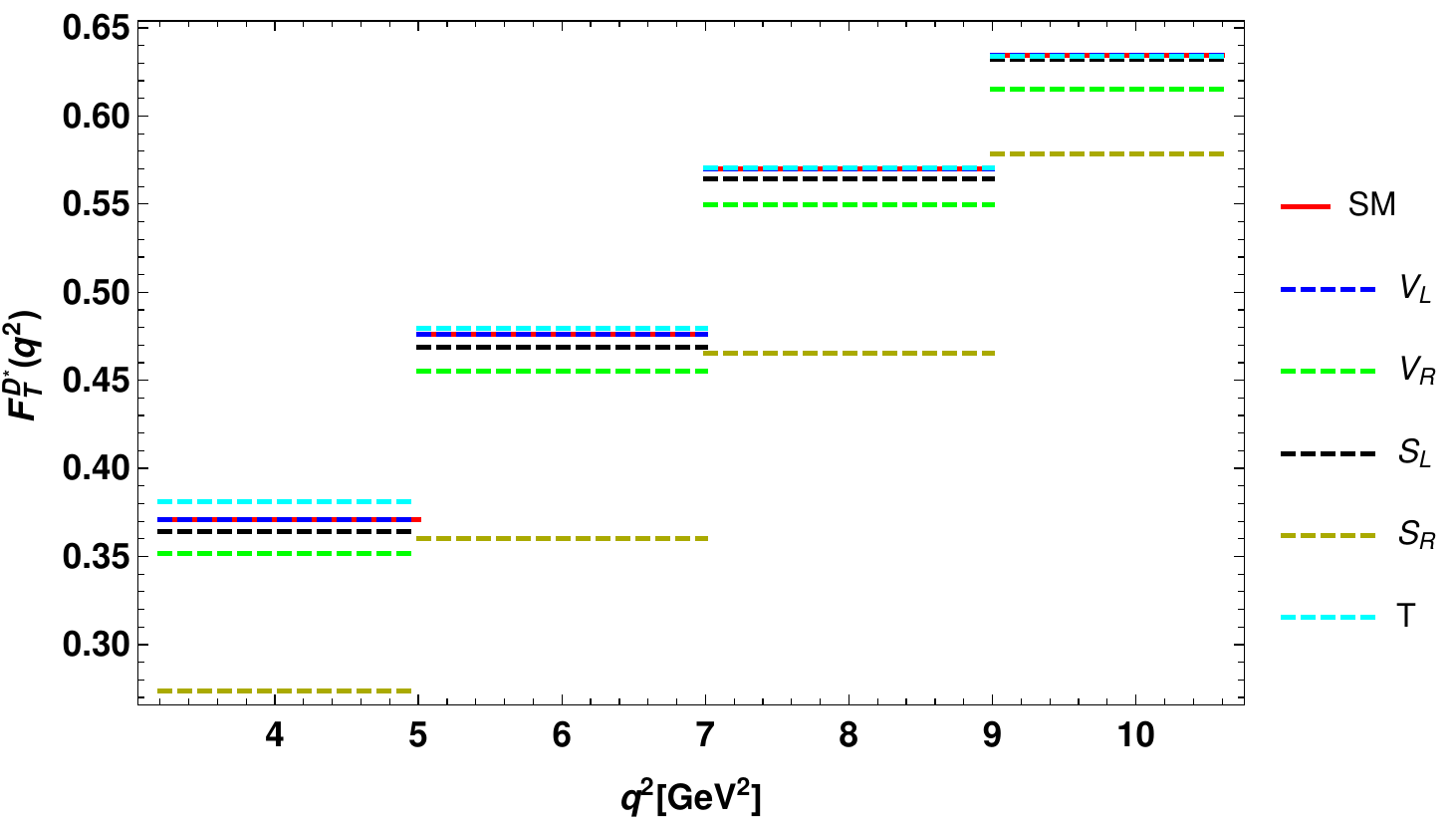}
\quad
\includegraphics[scale=0.5]{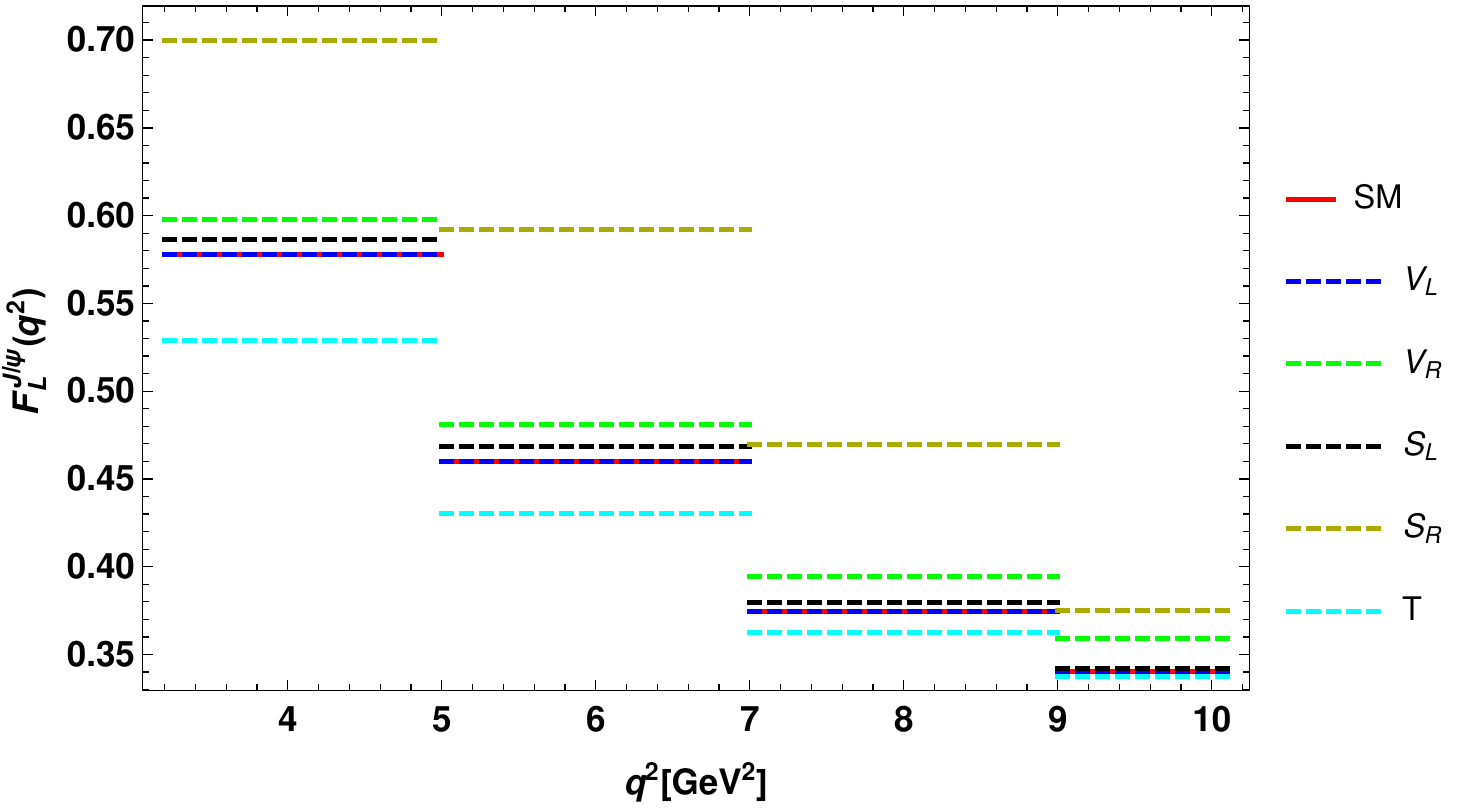}
\quad
\includegraphics[scale=0.5]{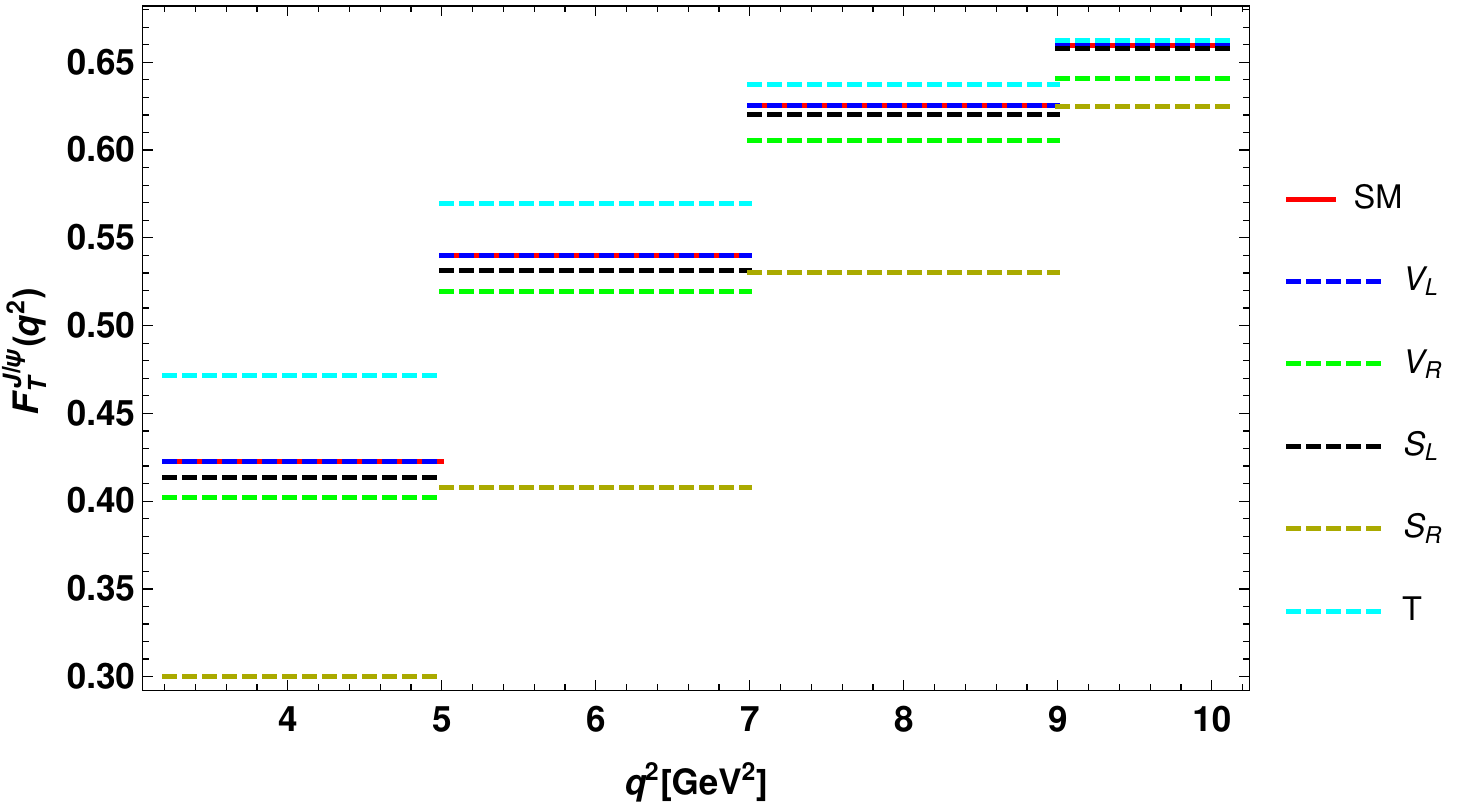}
\quad
\includegraphics[scale=0.5]{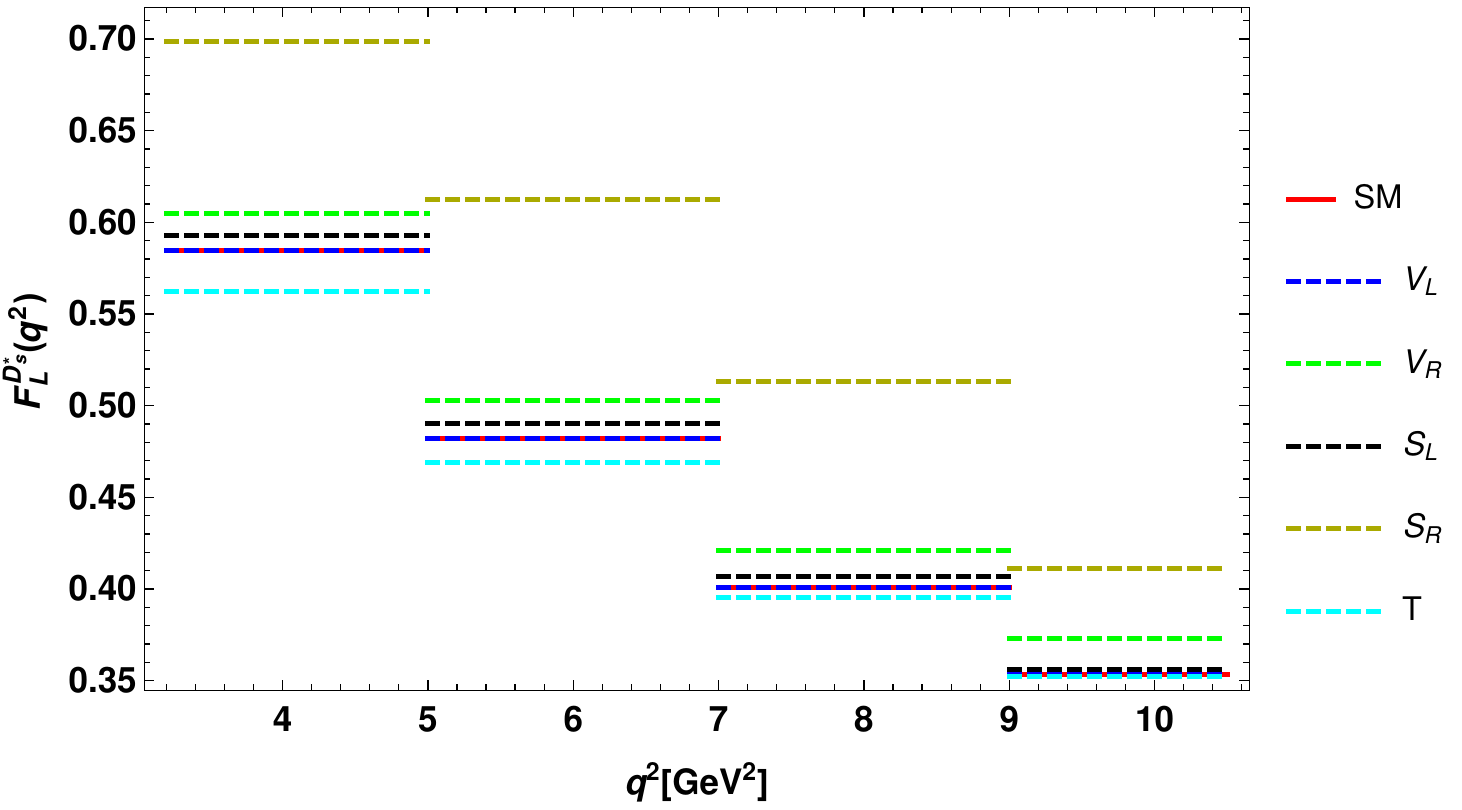}
\quad
\includegraphics[scale=0.5]{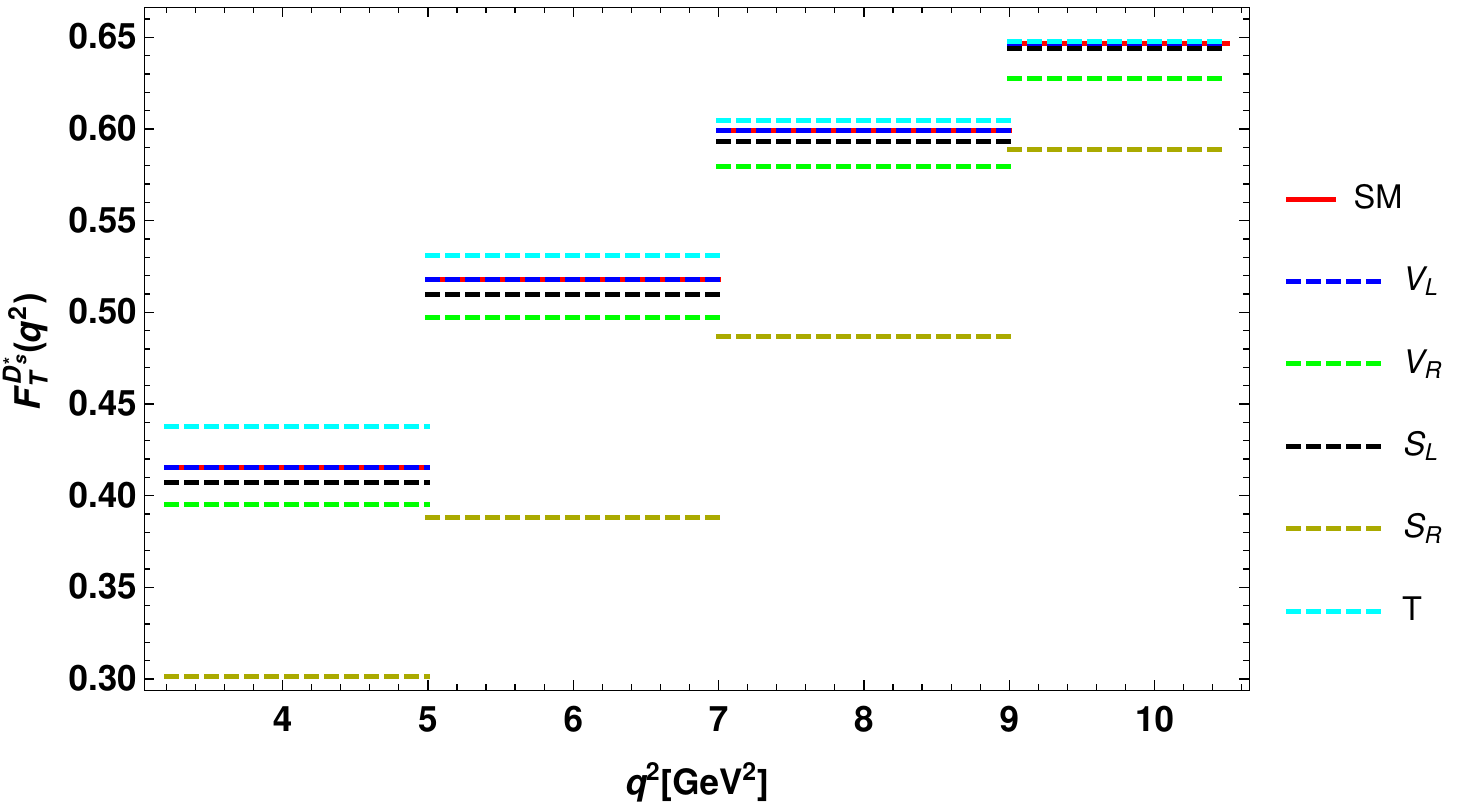}
\caption{ The bin-wise longitudinal (left panel) and transverse (right panel) polarization asymmetry of daughter vector meson of  $\bar B \to D^* \tau \bar \nu_\tau$ (top),  $ B_c^+ \to J/\psi \tau^+  \nu_\tau$ (middle) and $ B_s \to D_s^* \tau \bar \nu_\tau$ (bottom) processes in four $q^2$ bins for case A. } \label{Fig:CA-polarization}
\end{figure}

The bin-wise values of forward-backward asymmetry for $\bar B \to  D \tau \bar \nu_\tau$ (top-left panel), $\bar B \to  D^* \tau  \bar \nu_\tau$ (top-right panel), $B_c^+ \to \eta_c \tau^+  \nu_\tau$ (middle-left panel), $B_c^+ \to J/\psi \tau^+  \nu_\tau$ (middle-right  panel), $B_s \to D_s \tau \bar \nu_\tau$ (bottom-left panel) and $B_s \to D_s^* \tau \bar \nu_\tau$ (bottom-right panel) decay modes are  represented in Fig. \ref{Fig:CA-AFB}\,.
Since the forward-backward asymmetry is independent of $V_L$ coupling,
the presence of  only $V_L$   coefficient 
 does not have any impact on this observable. The forward-backward asymmetry obtained by using the best-fit value of only $S_R$ coupling has  significant deviation from SM results for $\bar B\to  D$, $B_c^+ \to \eta_c$ and $B_s \to D_s$ processes. The remaining coefficients have  negligible effect on $A_{FB}$ of these decay modes. Either $S_R$ or $T$ coefficient also provide maximum deviation in all the $q^2$ bins of $ B_{(s)}\to D_{(s)}^*$ and $B_c^+ \to J/\psi$ processes, except negligible impact of tensor coefficient in the first bin of $B_c^+ \to J/\psi$. The $V_R$ coefficient also affects this observable, but rather weakly in the last bin. 

The graphical representation of  lepton non-universality ratios, $R_D$ (top-left panel), $R_{D^*}$ (top-right panel), $R_{\eta_c}$ (middle-left panel), $R_{J/\psi}$ (middle-right panel), $R_{D_s}$ (bottom-left panel) and $R_{D_s^*}$ (bottom-right panel) in four  $q^2$ bins are depicted in Fig. \ref{Fig:CA-LNU}\,. The presence of either $V_L$,  $S_R$ or $S_L$ coefficients show significant deviation (the $V_R$ coupling has comparatively less impact) from the SM  values of $R_{D_{(s)}}$ and  $R_{\eta_c}$ observables in the fourth $q^2$ bin, whereas tensor coupling has vanishingly small effect.   All the coefficients except $S_L$ affect  the $R_{D_{(s)}^*}$ and $R_{J/\psi}$  observables
 remarkably in the last three $q^2$ bins. 
 
 The bin-wise plots for the $\tau$-polarization asymmetry of $\bar B \to  D \tau \bar \nu_\tau$ (top-left panel), $\bar B \to  D^*  \tau \bar \nu_\tau$ (top-right panel), $B_c^+ \to \eta_c \tau^+  \nu_\tau$ (middle-left panel), $B_c^+ \to J/\psi \tau^+  \nu_\tau$ (middle-right  panel), $B_s \to D_s \tau \bar \nu_\tau$ (bottom-left panel) and $B_s \to D_s^* \tau \bar \nu_\tau$ (bottom-right panel) decay modes are shown in Fig. \ref{Fig:CA-ptau}\,. Since the dependence of vector couplings drops out in $P_\tau$, so only scalar type couplings provide deviation from SM results for $B_{(s)}\to D_{(s)}$ and $B_c^+ \to \eta_c$ transitions.   We observe that, the contributions from  additional $S_L~(S_R)$ coefficient affect the $P_\tau$ observable significantly except in the  first (third) bin, whereas tensor coupling  has negligible effect in all $q^2$ bins.  In the $P_\tau^V$ observables ($V=D_{(s)}^*, J/\psi$), the $S_L$ and $T$   contributions are dominant in the first two bins and $S_R$ has significant impact in full $q^2$ regions. In Fig. \ref{Fig:CA-polarization}\,, the bin-wise plots for  longitudinal (left) and transverse (right) polarization asymmetry of  $\bar B \to  D^* \tau \bar \nu_\tau$,  $B_c^+ \to J/\psi \tau^+  \nu_\tau$ and $B_s \to D_s^* \tau \bar \nu_\tau$ processes  are shown in the top, middle and bottom panels, respectively. Both the longitudinal and transverse polarization asymmetry of all the decay modes shift from their respective SM predictions due to the presence of either $V_R$ or $S_R$ couplings. The tensor coefficient affects the first three (two) bins of $B_c^+ \to  J/\psi~ (B_s \to D_s^*)$ decay mode. 
\subsection{Case B}
\begin{figure}[htb]
\includegraphics[scale=0.5]{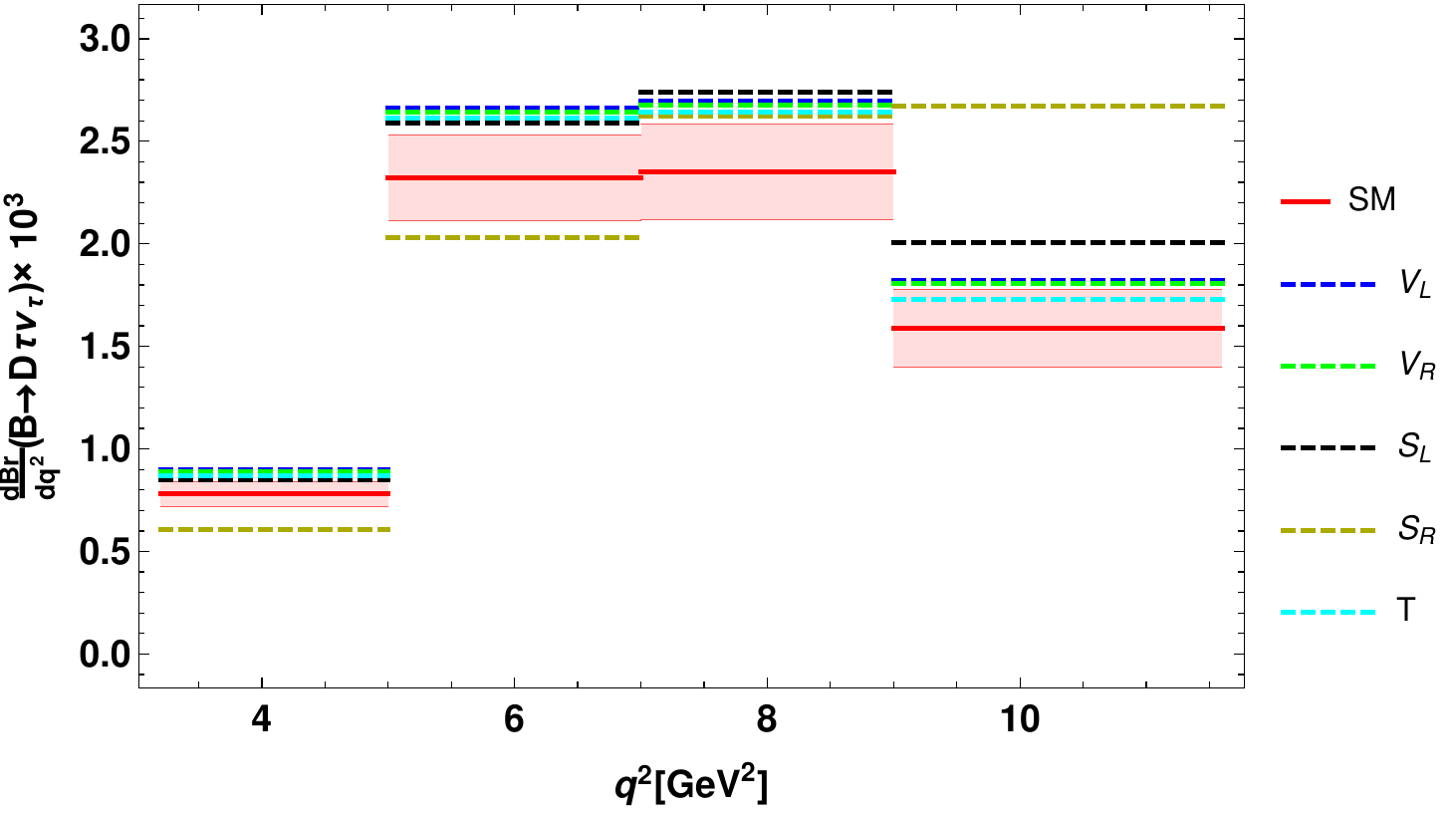}
\quad
\includegraphics[scale=0.5]{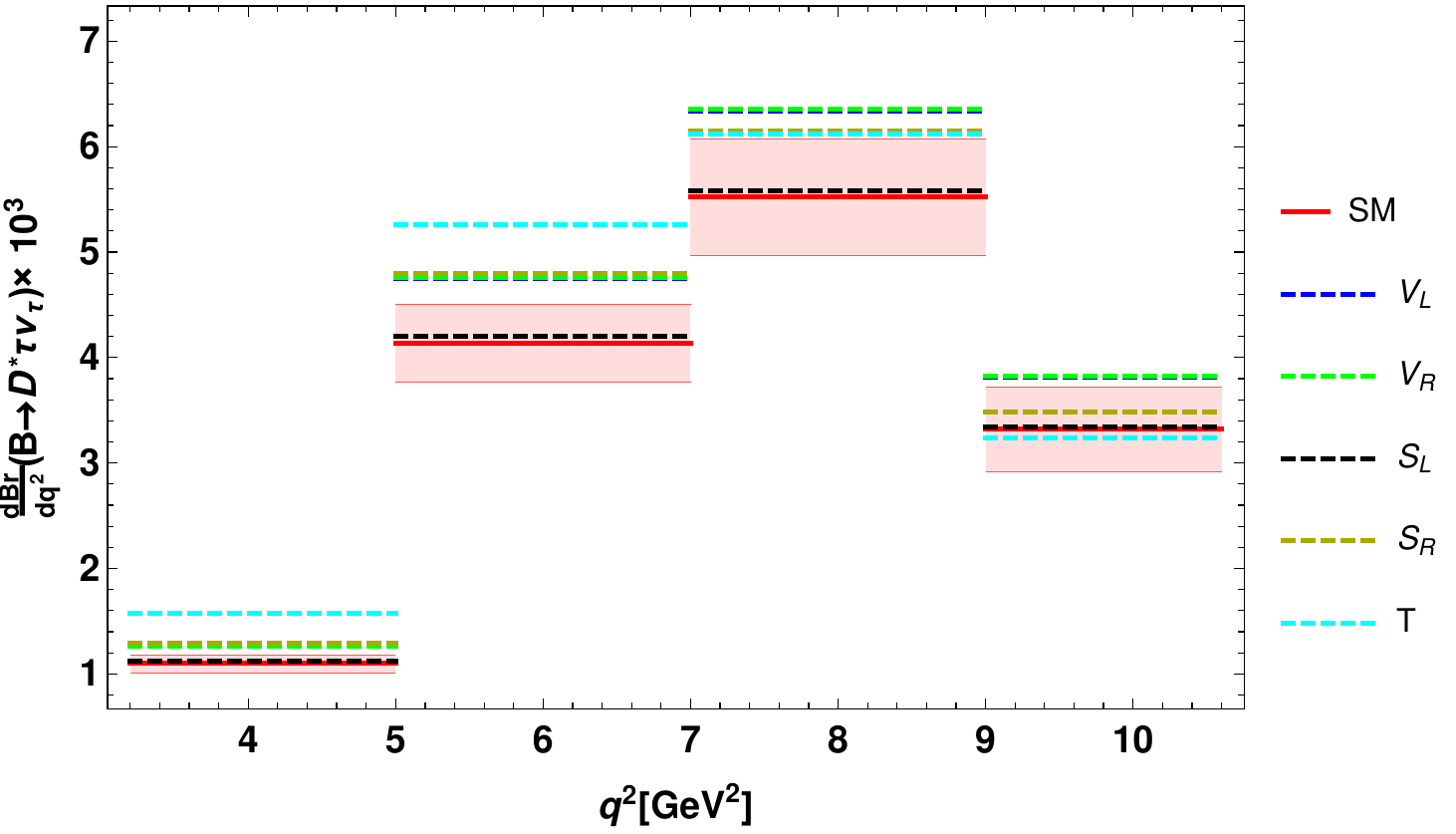}
\quad
\includegraphics[scale=0.5]{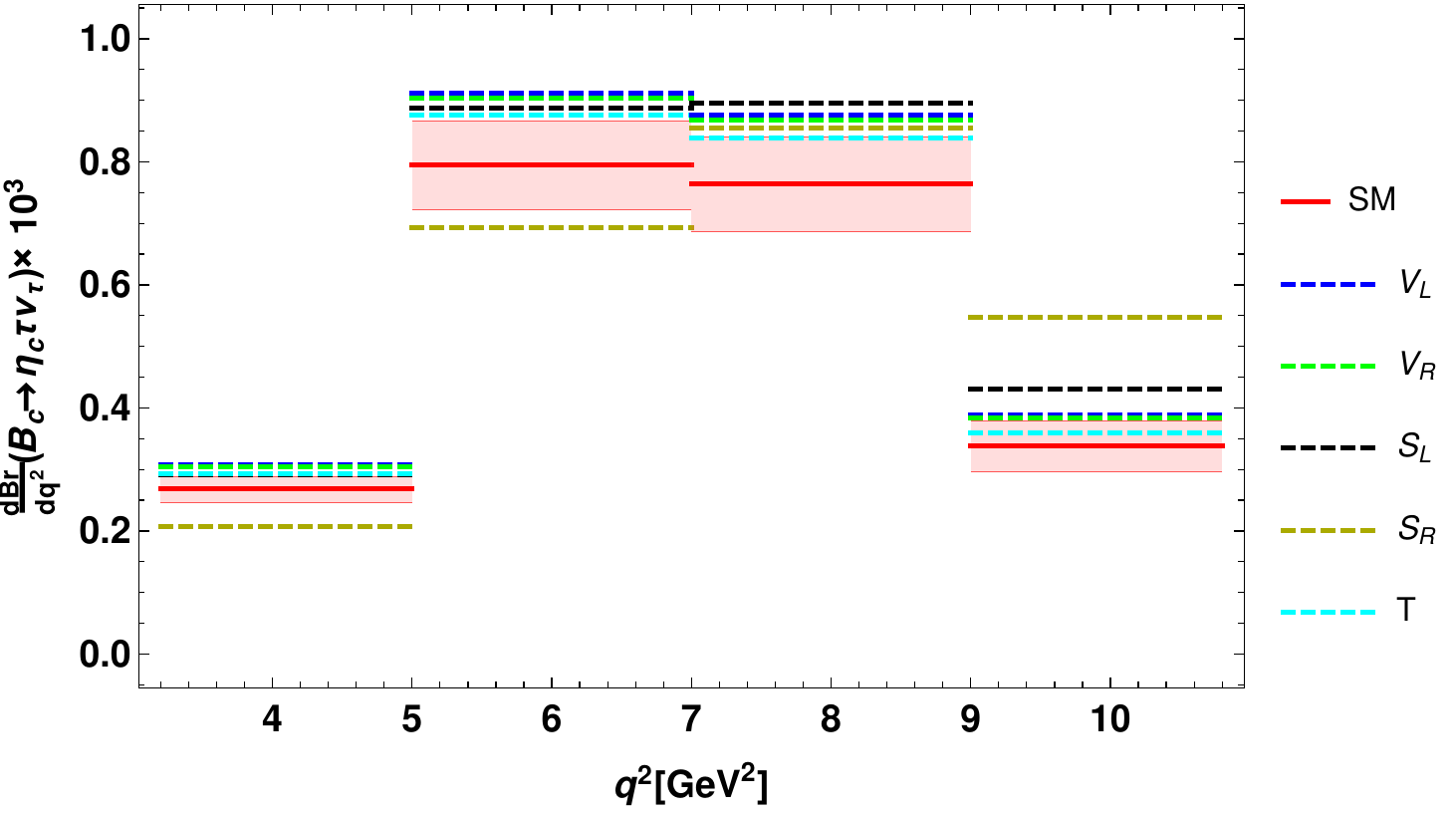}
\quad
\includegraphics[scale=0.5]{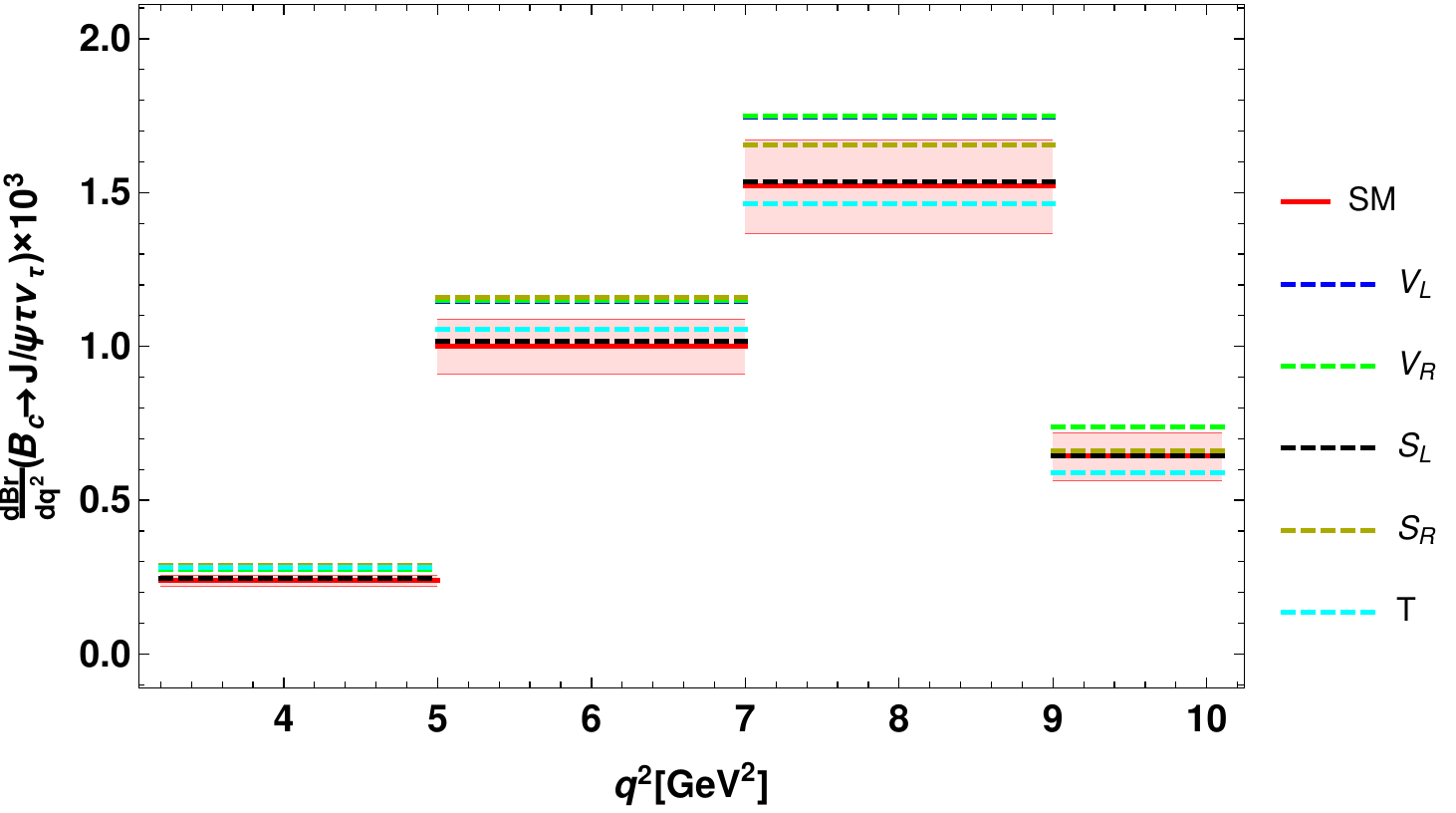}
\quad
\includegraphics[scale=0.5]{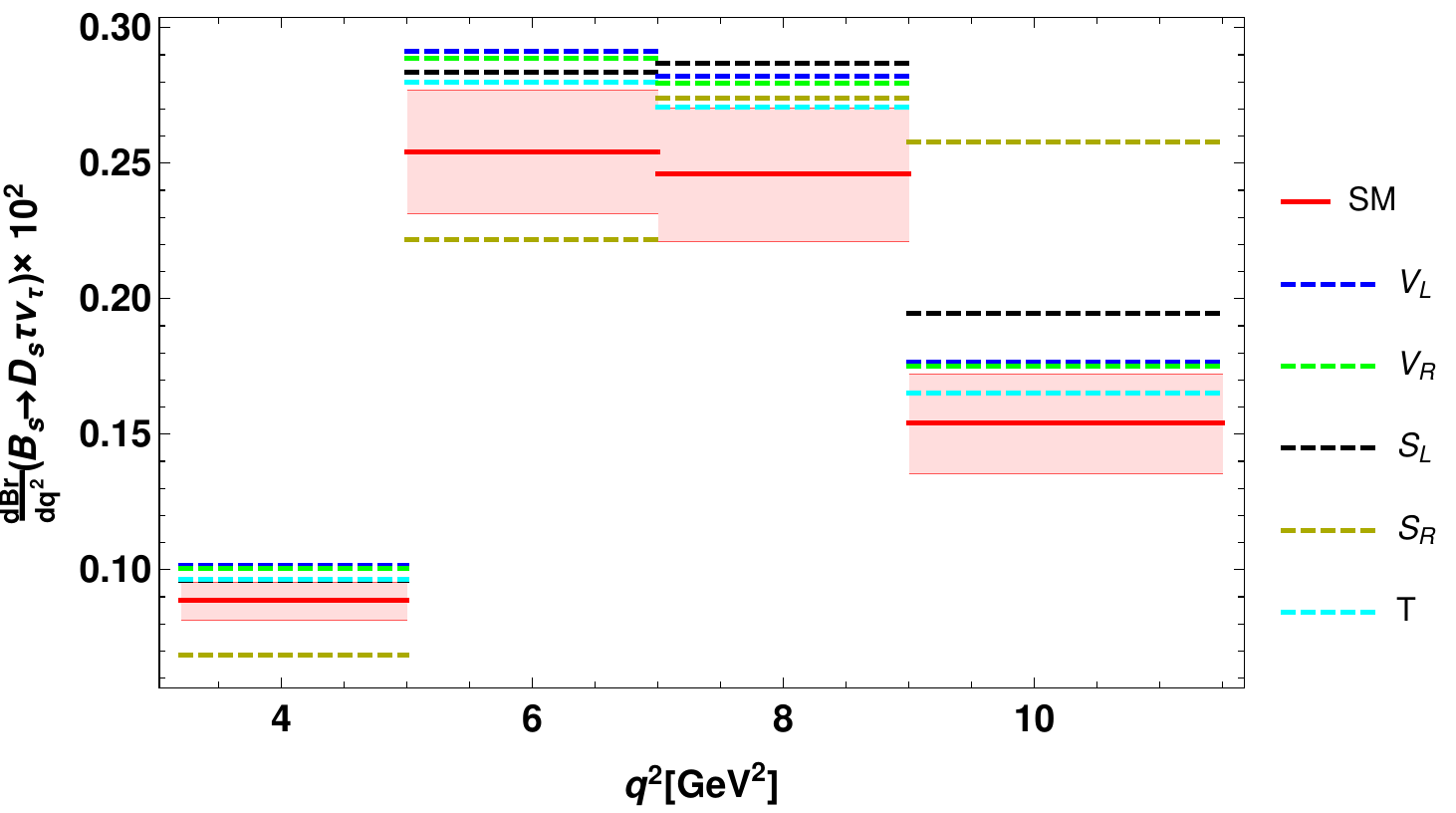}
\quad
\includegraphics[scale=0.5]{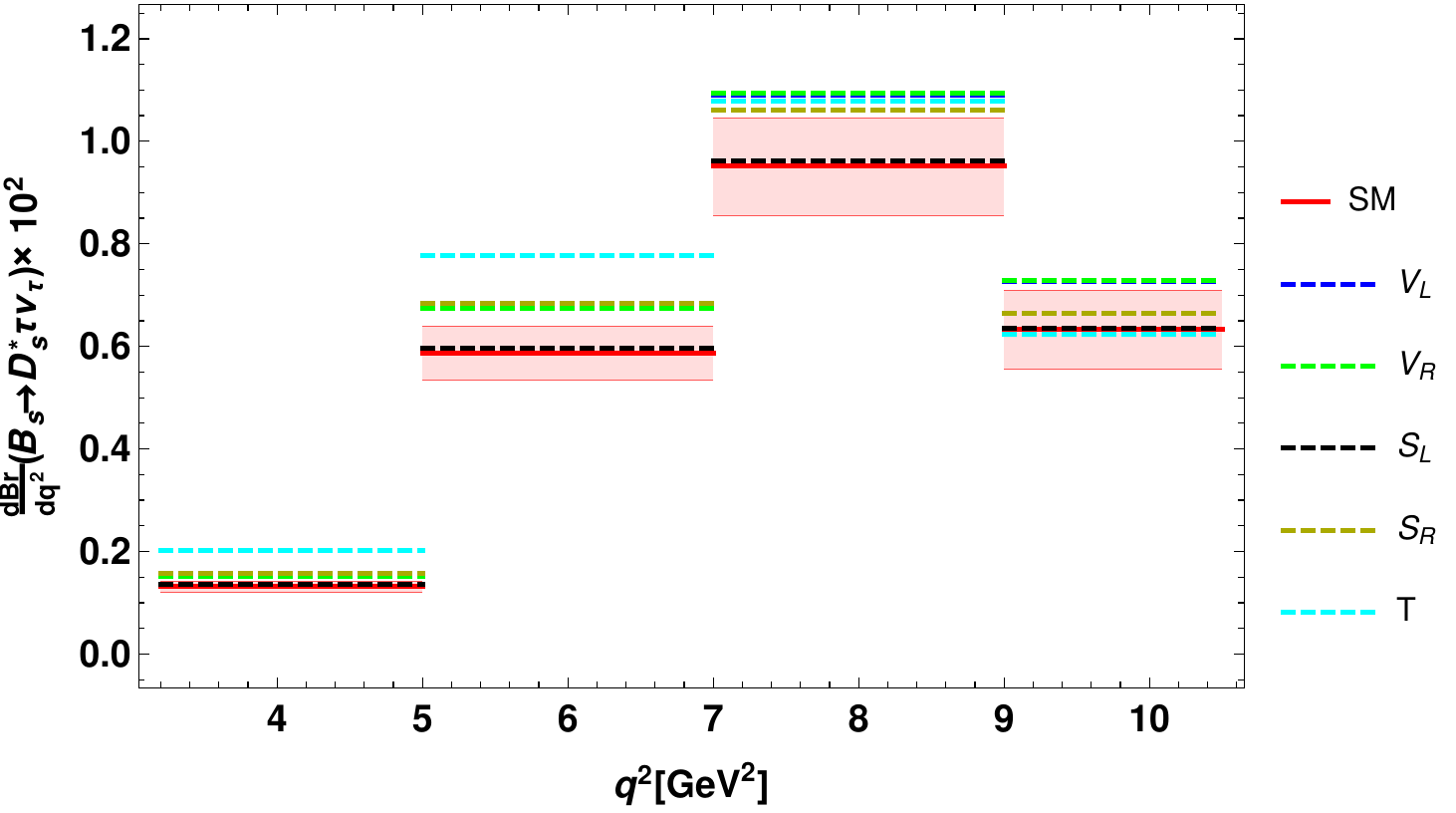}
\caption{ The bin-wise branching ratios of $\bar B \to D \tau \bar \nu_\tau$ (top-left panel), $\bar B \to D^* \tau \bar \nu_\tau$ (top-right panel), $B_c^+ \to \eta_c \tau^+  \nu_\tau$ (middle-left panel), $ B_c^+ \to J/\psi \tau^+   \nu_\tau$ (middle-right panel), $\bar B_s \to\bar D_s \tau \bar \nu_\tau$ (bottom-left panel) and $\bar B_s \to\bar D_s^* \tau \bar \nu_\tau$ (bottom-right panel) processes in four $q^2$ bins for case B. }\label{Fig:CB-BR}
\end{figure}

In this case, the new coefficients are assumed to be complex and considering the presence of one coefficient at a time, the best-fit values of the real and imaginary parts of these couplings are presented in Table \ref{Tab:Best-fit}\,. Using the best-fit values, the bin-wise plots for the branching ratio of $\bar B \to  D \tau \bar \nu_\tau$ (top-left panel), $\bar B \to  D^* \tau \bar \nu_\tau$ (top-right panel), $B_c^+ \to \eta_c \tau^+  \nu_\tau$ (middle-left panel), $B_c^+ \to J/\psi \tau^+  \nu_\tau$ (middle-right  panel), $B_s \to D_s \tau \bar \nu_\tau$ (bottom-left panel) and $B_s \to D_s^* \tau \bar \nu_\tau$ (bottom-right panel) decay processes are shown in Fig. \ref{Fig:CB-BR}\,. 
In the presence of either $V_L$ or $V_R$  coefficient, the branching ratios of all the decay modes  deviate from their  SM values in all the bins except the first bin (where the deviation is minimal). Due to the additional $S_L$ coupling, the branching fractions of $B_{(s)} \to  D_{(s)}$ and $B_c^+ \to \eta_c$ show significant deviations in all bins except in the first one, whereas $B_{(s)} \to D_{(s)}^*$ and $B_c^+ \to J/\psi$ have negligibly effect from $S_L$.  The branching ratios of all processes deviate maximally due to the new contribution from $S_R$ coefficient. The tensor coefficient affects $2^{\rm nd}$ and $3^{\rm rd}$ bins of $ B_{(s)} \to  D_{(s)} $ and $B_c^+ \to \eta_c$, first three bins of $ B_{(s)} \to  D_{(s)}^*$ and no effect on $B_c^+ \to J/\psi$. For this case, the pull values of the branching ratios of all these channels in the presence of individual  complex  coefficient, in all the four $q^2$ bins are given  in Table \ref{Tab:pull-CaseB}\,. 
\begin{table}[htb]
\caption{Pull values of the branching ratios of all the decay modes in the presence of individual  new coefficients in all the four $q^2$ bins for case B. Here the first row contains  the pull values for only $V_L,~V_R,~T$ and second row represents the $S_L,~S_R$ pull values.}\label{Tab:pull-CaseB}
\begin{center}
\begin{tabular}{|c|c|c|c|c|c|}
\hline
~modes ~&~Only $V_L$~&~Only $V_R$~&~Only $T$~\\
~ ~&~Only $S_L$~&~Only $S_R$~&~\\
\hline \hline
$\bar B \to D $~&~$(1.2,1.07,0.962,0.802)$~&~$(1.13,1.0,0.9,0.751)$~&~$(0.942,0.921,0.818,0.49)$~\\

~&~$(0.712,0.853,1.07,1.36)$~&~$(2.215,1.053,0.762,2.9)$~&~~\\

\hline
$\bar B \to D^* $~&~$(1.2,1.07,0.962,0.802)$~&~$(1.23,1.086,0.979,0.818)$~&~$(3.102,1.86,0.722,0.147)$~\\

&~$(0.164,0.122,0.072,0.242)$~&~$(1.42,1.149,0.746,0.277)$~&~~\\

\hline
$B_c \to \eta_c$~&~$(1.2,1.07,0.962,0.802)$~&~$(1.13,1.0,0.9,0.751)$~&~$(0.797,0.797,0.655,0.35)$~\\

&~$(0.715,0.867,1.121,1.4)$~&~$(2.23,1.07,0.79,2.71)$~&~~\\
\hline
$B_c \to J/\psi $~&~$(1.2,1.07,0.962,0.802)$~&~$(1.23,1.1,0.983,0.82)$~&~$(1.48,0.41,0.274,0.512)$~\\

~&~$(0.187,0.123,0.057,0.014)$~&~$(1.6,1.16,0.6,0.16)$~&~~\\
\hline
$\bar B \to D_s$~&~$(1.2,1.07,0.962,0.802)$~&~$(1.13,1.0,0.9,0.751)$~&~$(0.762,0.753,0.677,0.41)$~\\

~&~$(0.713,0.857,1.08,1.37)$~&~$(2.23,1.07,0.763,2.879)$~&~~\\
\hline
$\bar B \to D_s^* $~&~$(1.2,1.07,0.962,0.802)$~&~$(1.22,1.085,0.978,0.818)$~&~$(3.587,2.16,0.89,0.082)$~\\

~&~$(0.174,0.127,0.074,0.025)$~&~$(1.5,1.197,0.766,0.281)$~&~~\\
\hline
\end{tabular}
\end{center}
\end{table}
\begin{figure}[htb]
\includegraphics[scale=0.5]{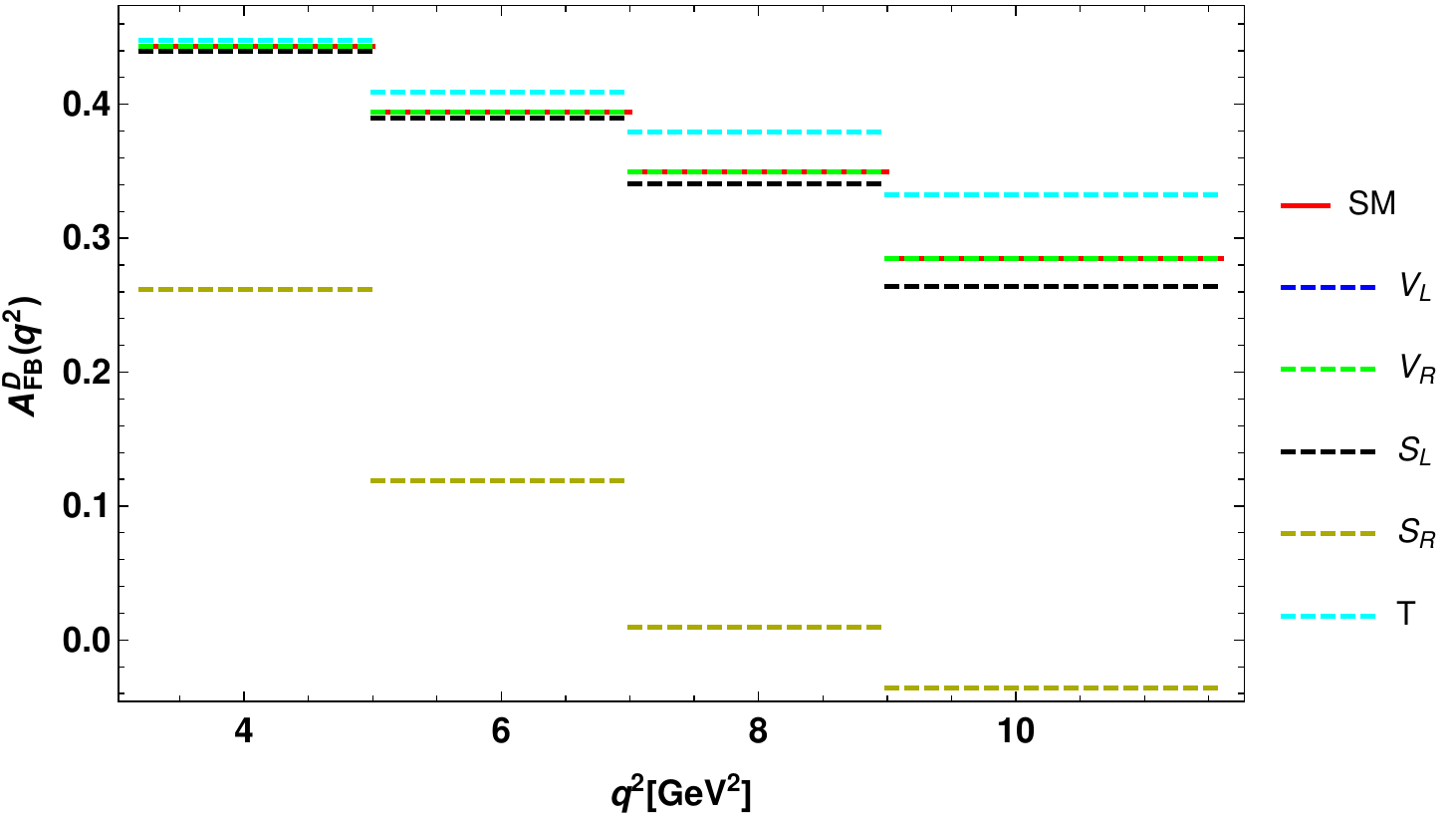}
\quad
\includegraphics[scale=0.5]{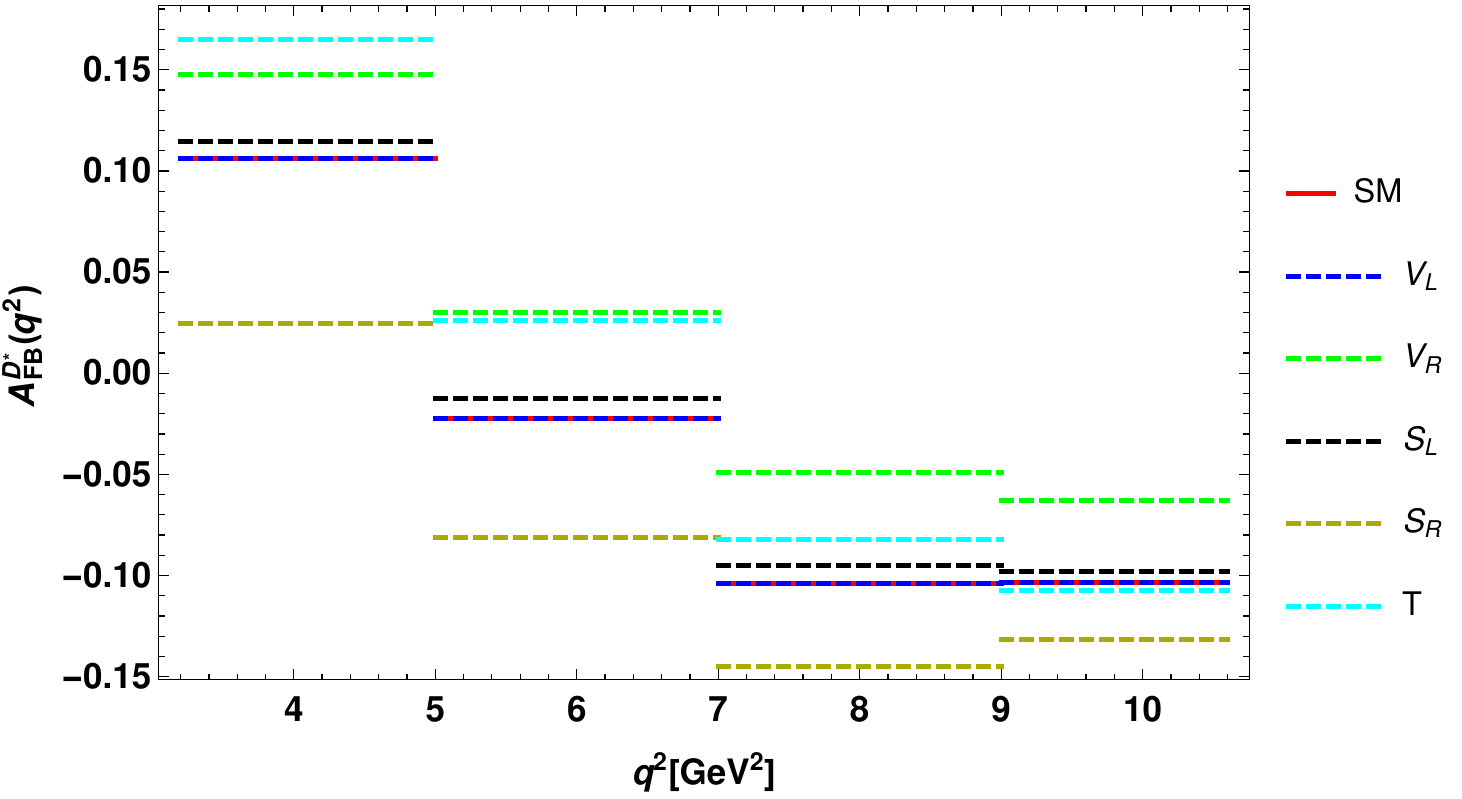}
\quad
\includegraphics[scale=0.5]{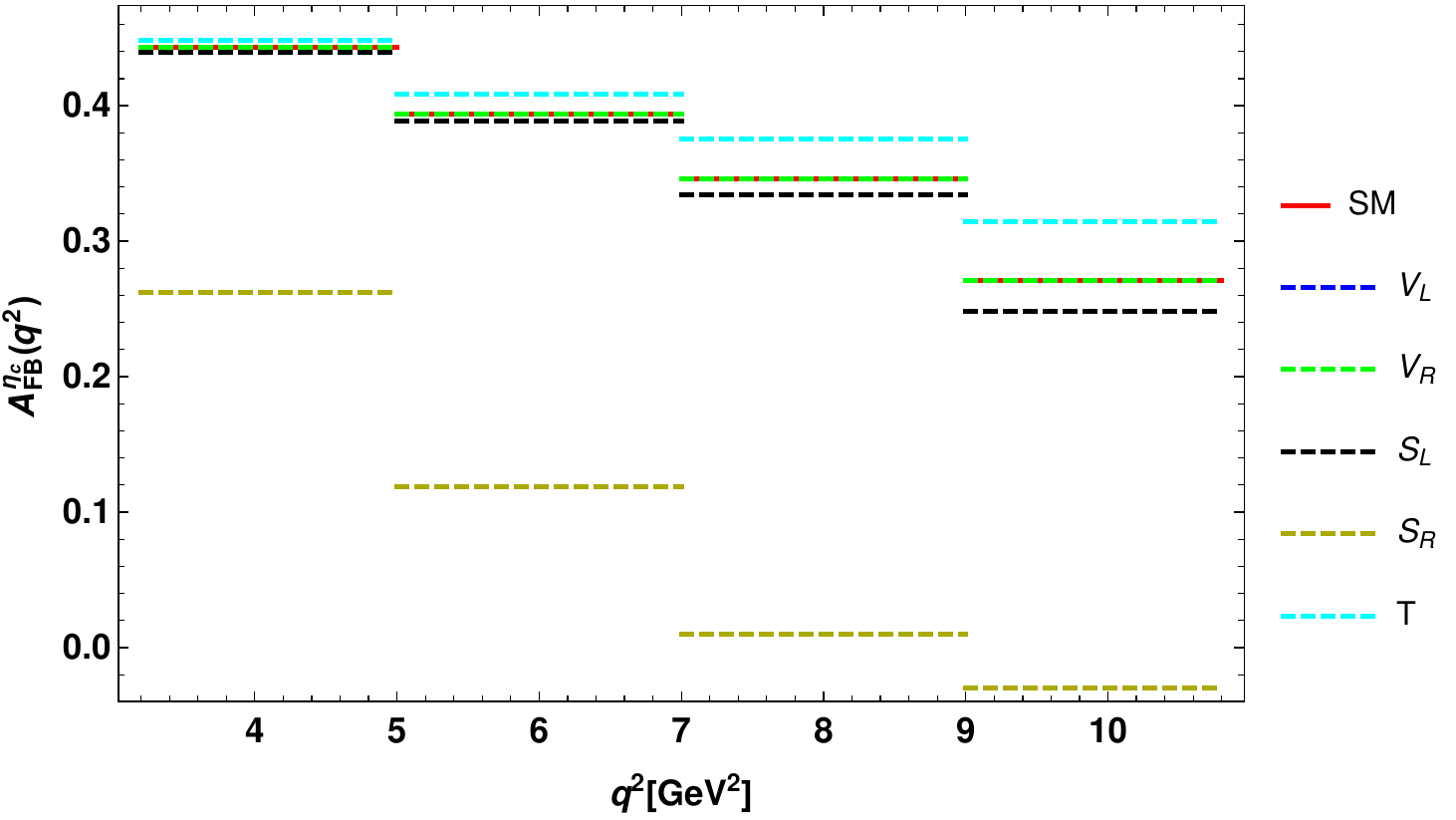}
\quad
\includegraphics[scale=0.5]{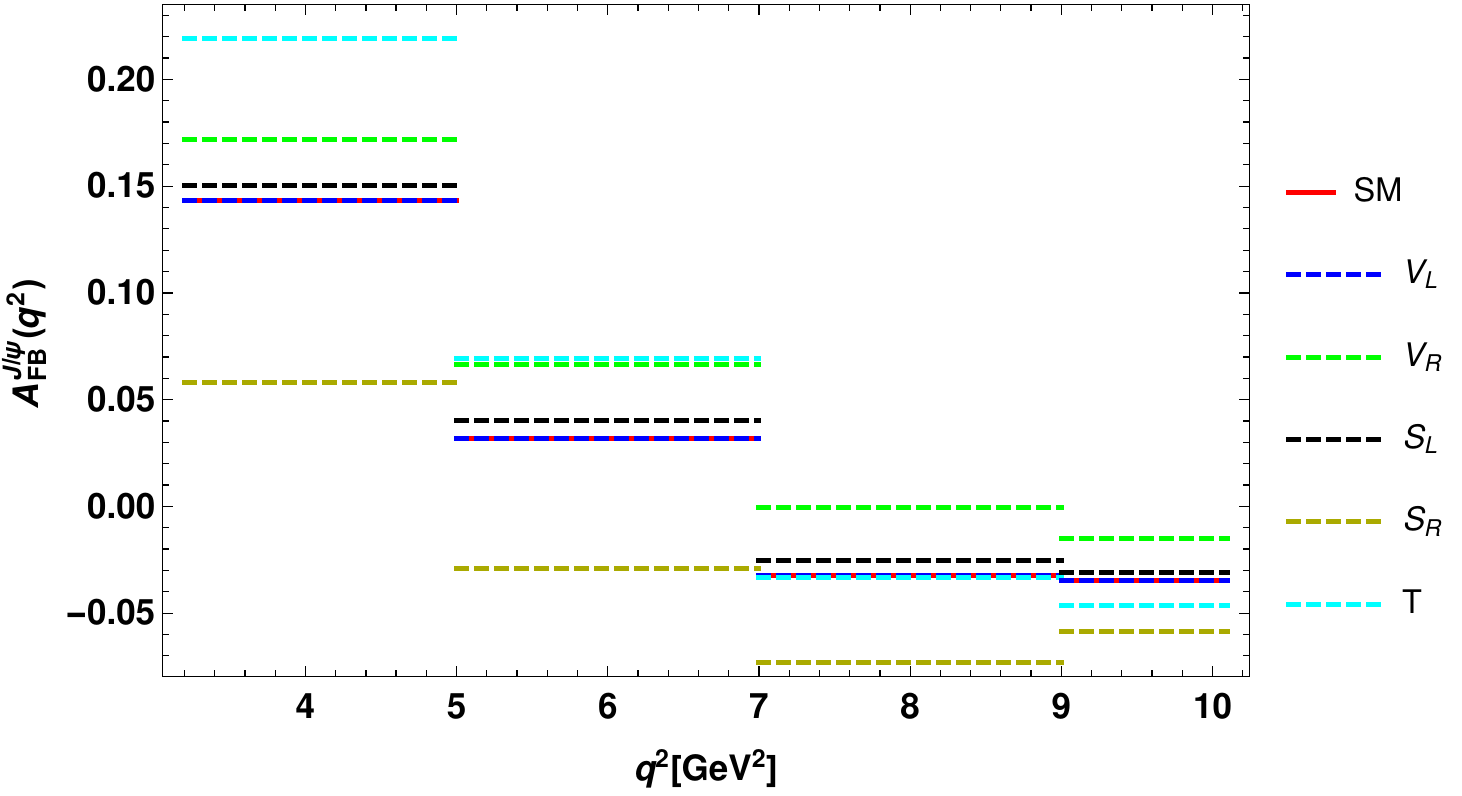}
\quad
\includegraphics[scale=0.5]{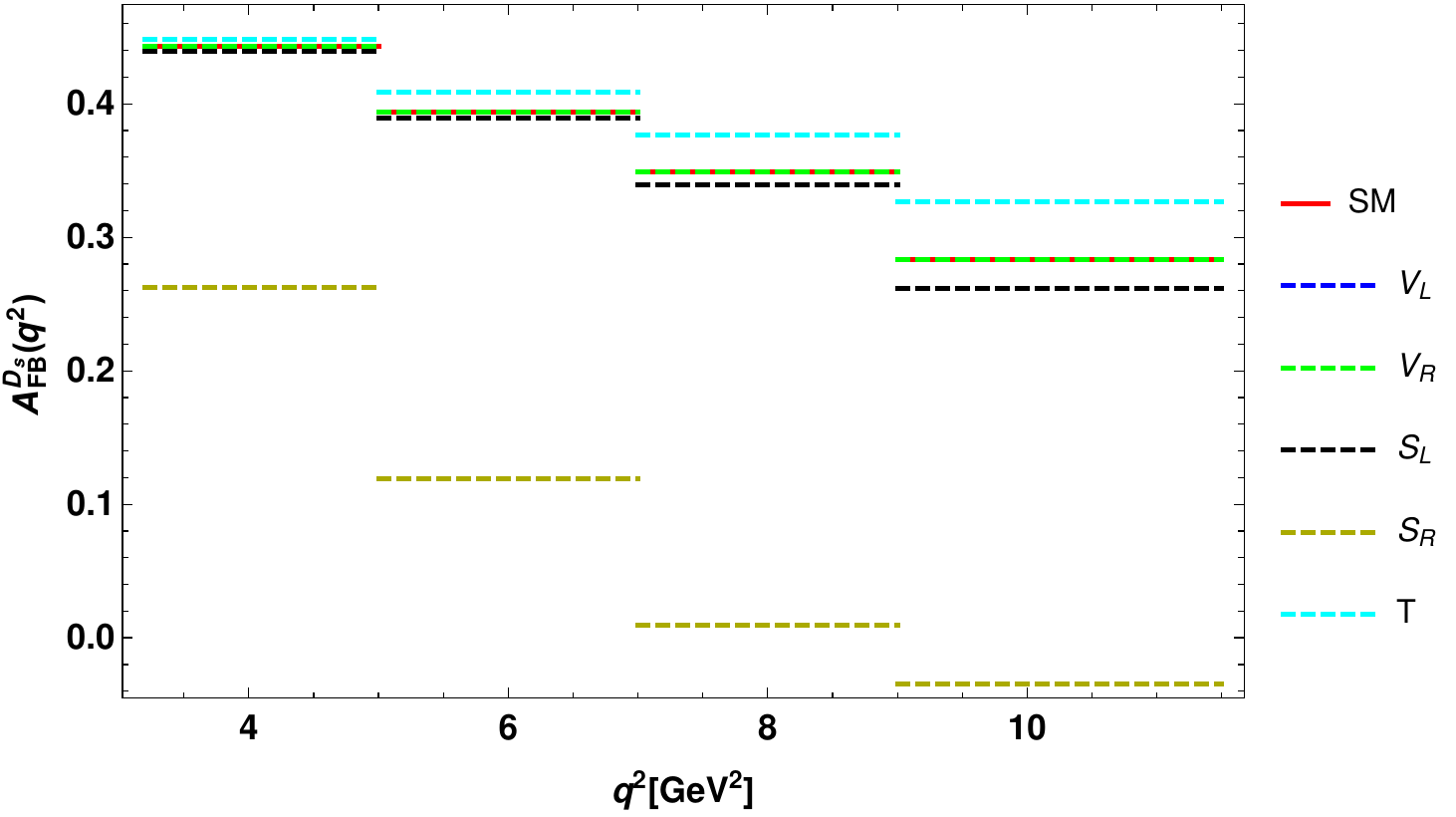}
\quad
\includegraphics[scale=0.5]{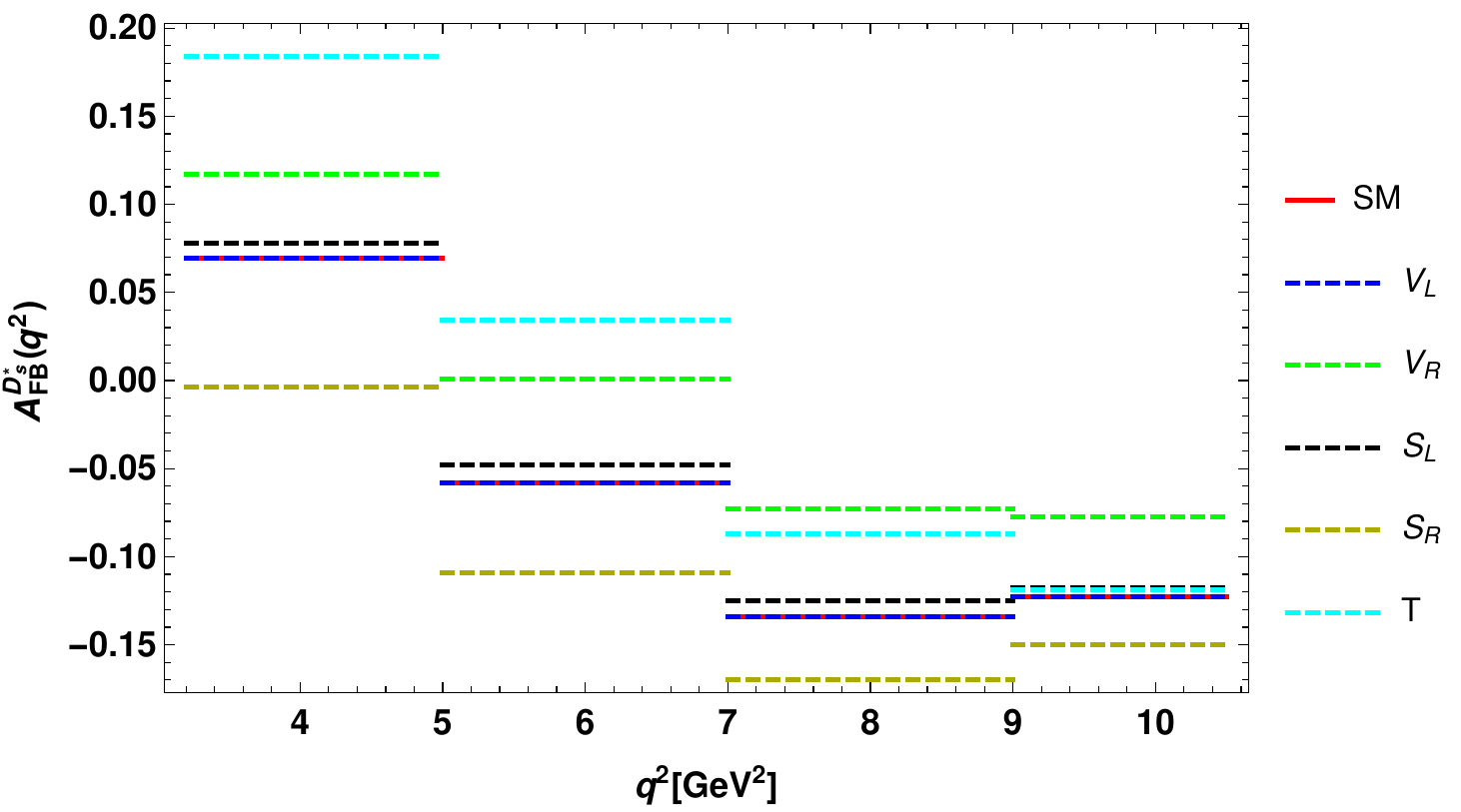}
\caption{ The bin-wise forward-backward asymmetry of $\bar B \to D \tau \bar \nu_\tau$ (top-left panel), $\bar B \to D^* \tau \bar \nu_\tau$ (top-right panel), $B_c^+ \to \eta_c \tau^+  \nu_\tau$ (middle-left panel), $ B_c^+ \to J/\psi \tau^+  \nu_\tau$ (middle-right panel), $\bar B_s \to\bar D_s \tau \bar \nu_\tau$ (bottom-left panel) and $\bar B_s \to\bar D_s^* \tau \bar \nu_\tau$ (bottom-right panel) processes in four $q^2$ bins for case B. }\label{Fig:CB-AFB}
\end{figure}

The graphical representation of forward-backward asymmetry of $\bar B \to  D \tau \bar \nu_\tau$ (top-left panel), $\bar B \to  D^* \tau \bar \nu_\tau$ (top-right panel), $B_c^+ \to \eta_c \tau^+  \nu_\tau$ (middle-left panel), $B_c^+ \to J/\psi \tau^+ \nu_\tau$ (middle-right  panel), $B_s \to D_s \tau \bar \nu_\tau$ (bottom-left panel) and $B_s \to D_s^* \tau \bar \nu_\tau$ (bottom-right panel) decay processes are shown in Fig. \ref{Fig:CB-AFB}\,. The  presence of $V_R$ coefficient affects the forward-backward asymmetry of only $B \to V$ processes. The $S_L$ coupling shows vanishing results.  The $S_R$ coefficients contributes maximally to the forward-backward asymmetry of all these decay modes in all the four $q^2$ bin. The tensor coefficients show deviation in the last three bins of $B_{(s)} \to D_{(s)}~(B_c^+ \to \eta_c)$, first three bins of $B_{(s)} \to D_{(s)}^*$ and first two bins of $B_c^+ \to J/\psi$ process. The bin-wise numerical values of branching ratios and forward-backward asymmetries are given in Table \ref{Tab:BR-AFB-CaseB}\,.

\begin{figure}[htb]
\includegraphics[scale=0.5]{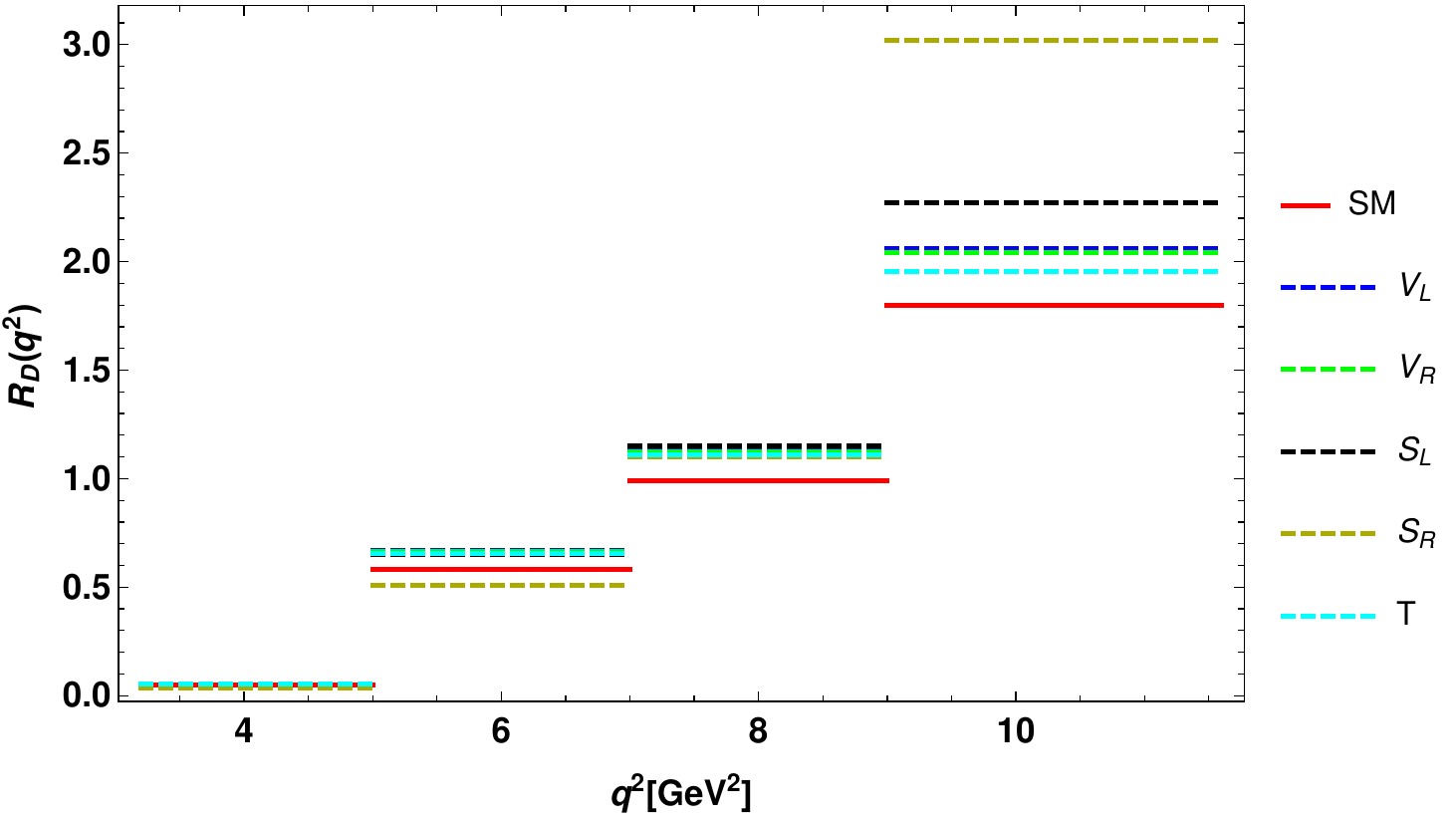}
\quad
\includegraphics[scale=0.5]{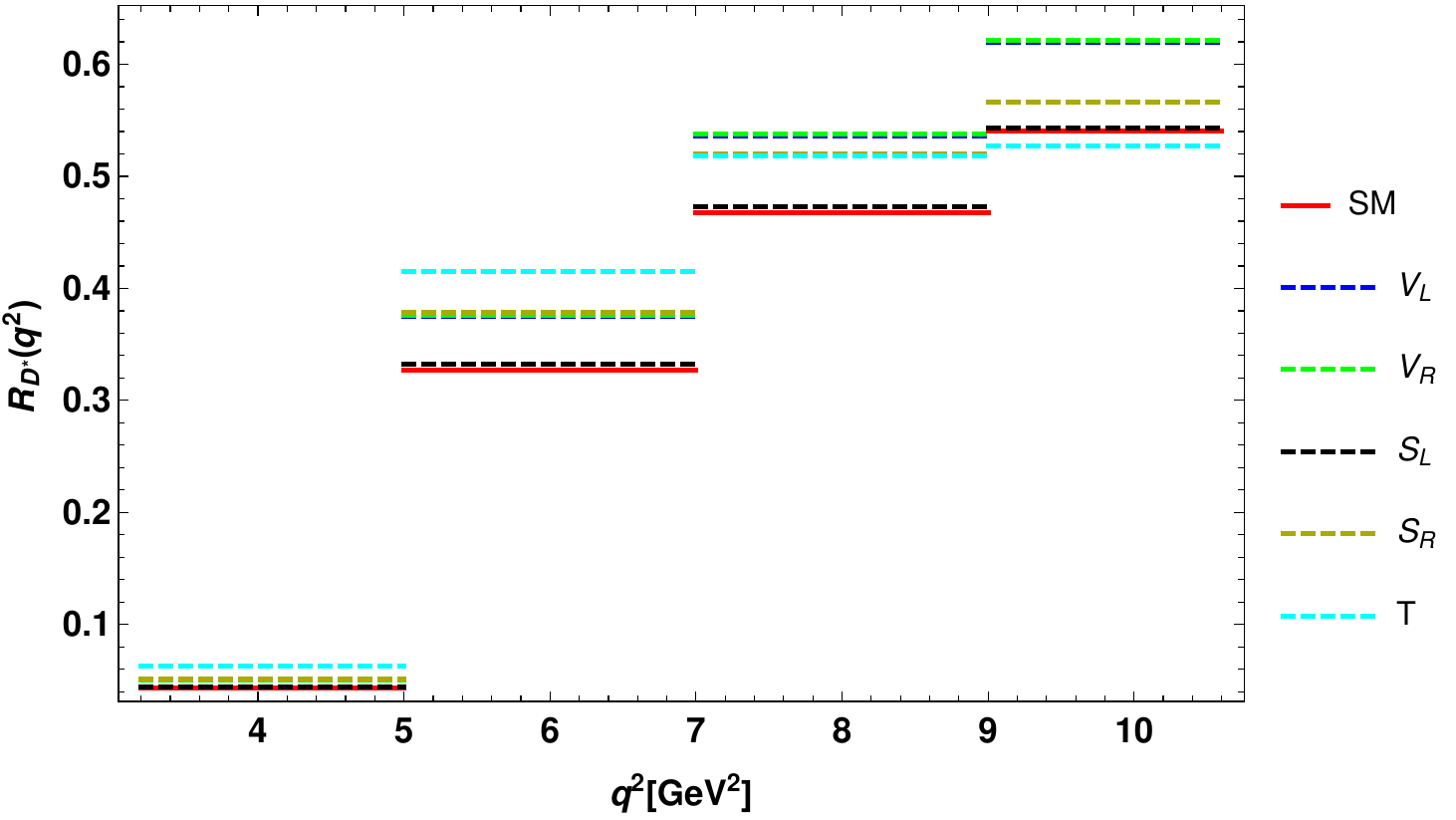}
\quad
\includegraphics[scale=0.5]{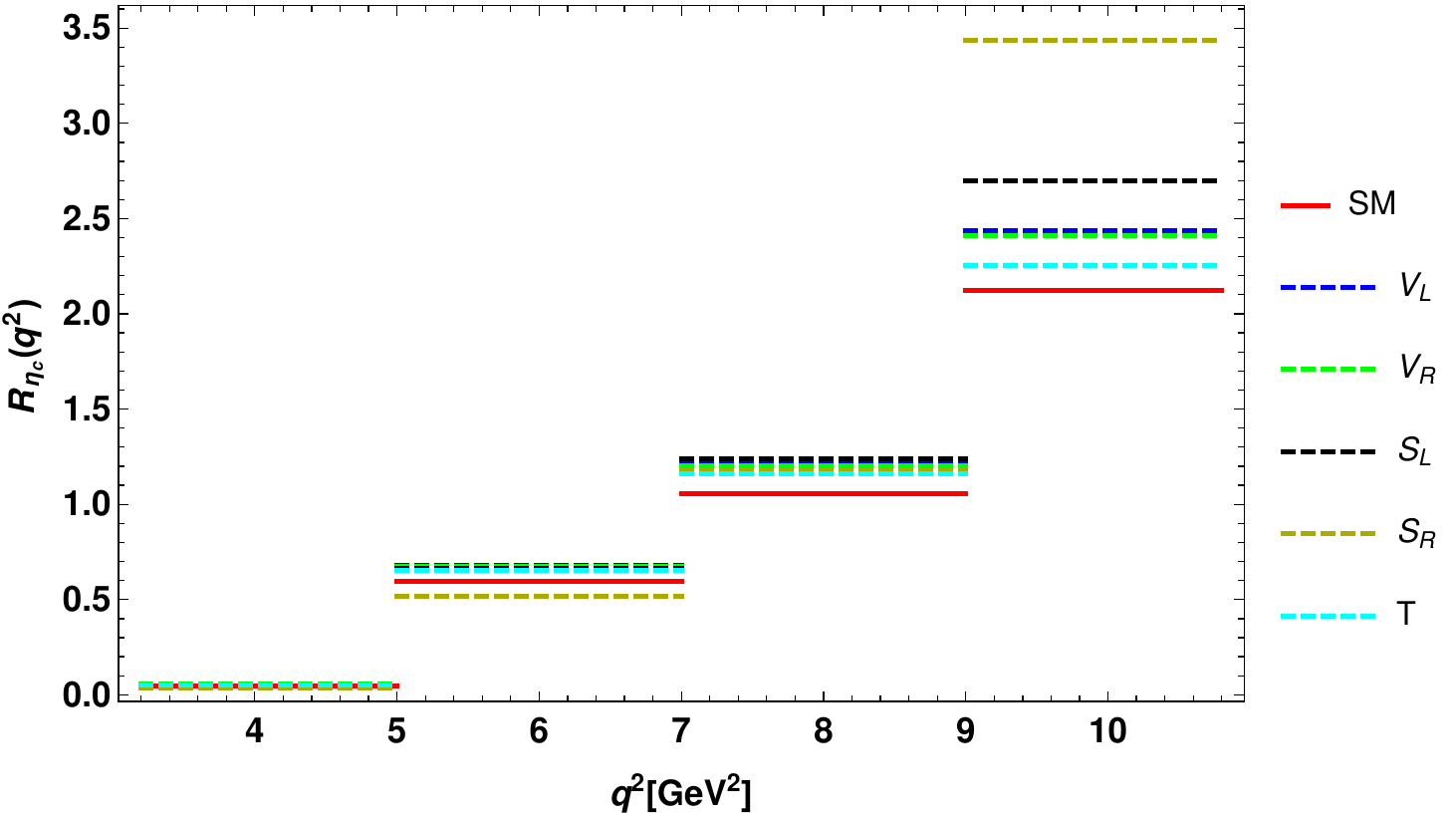}
\quad
\includegraphics[scale=0.5]{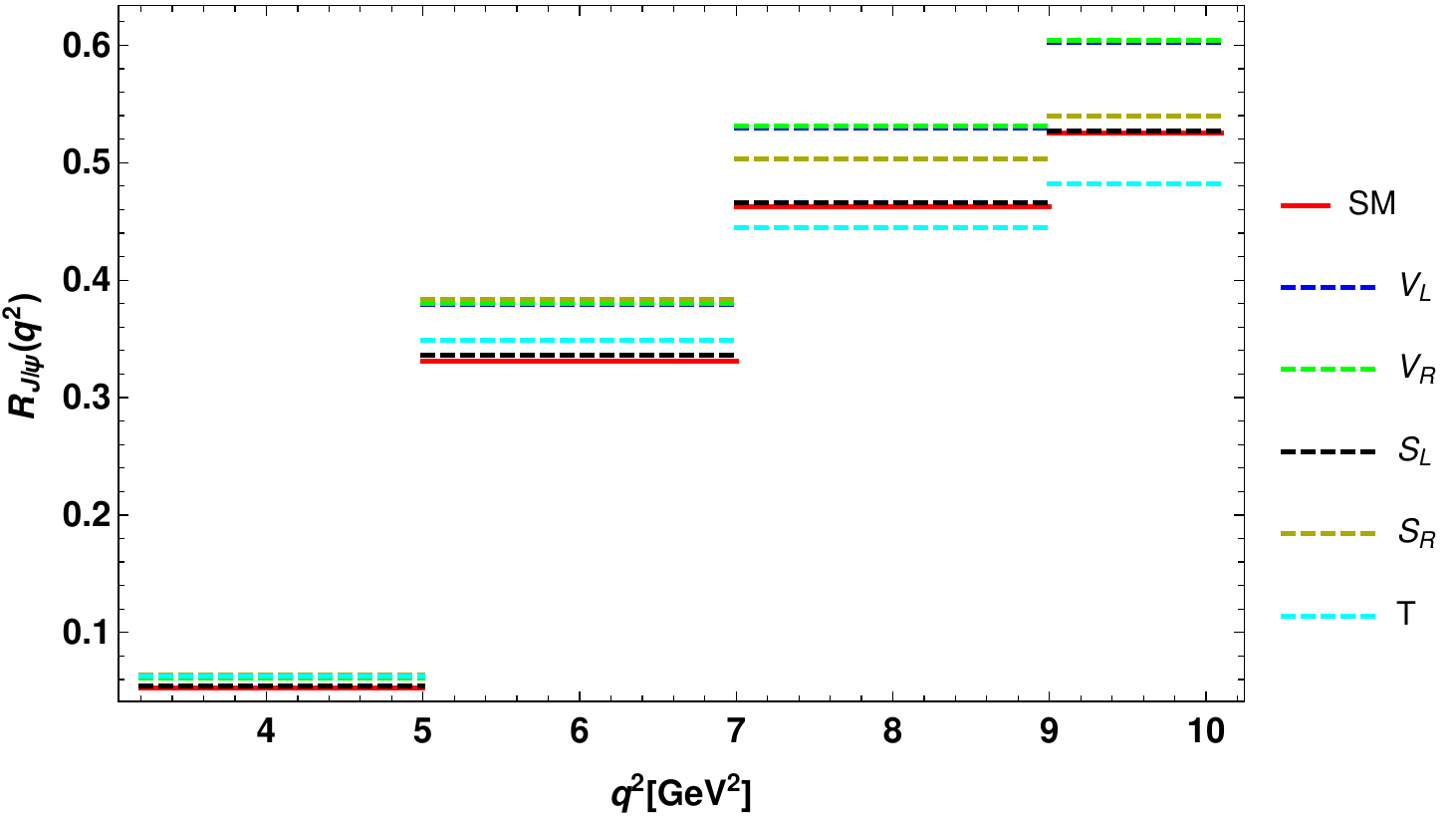}
\quad
\includegraphics[scale=0.5]{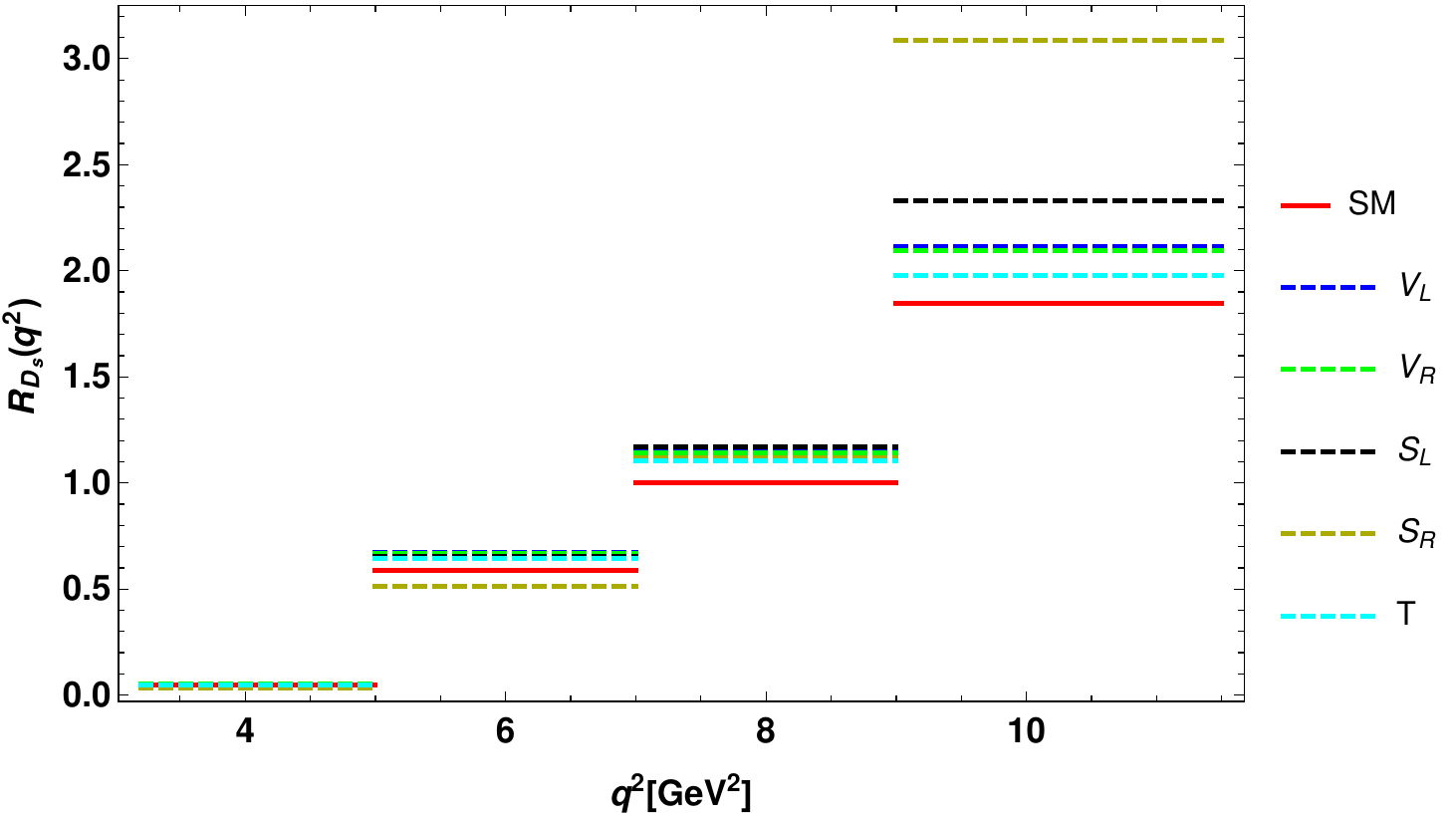}
\quad
\includegraphics[scale=0.5]{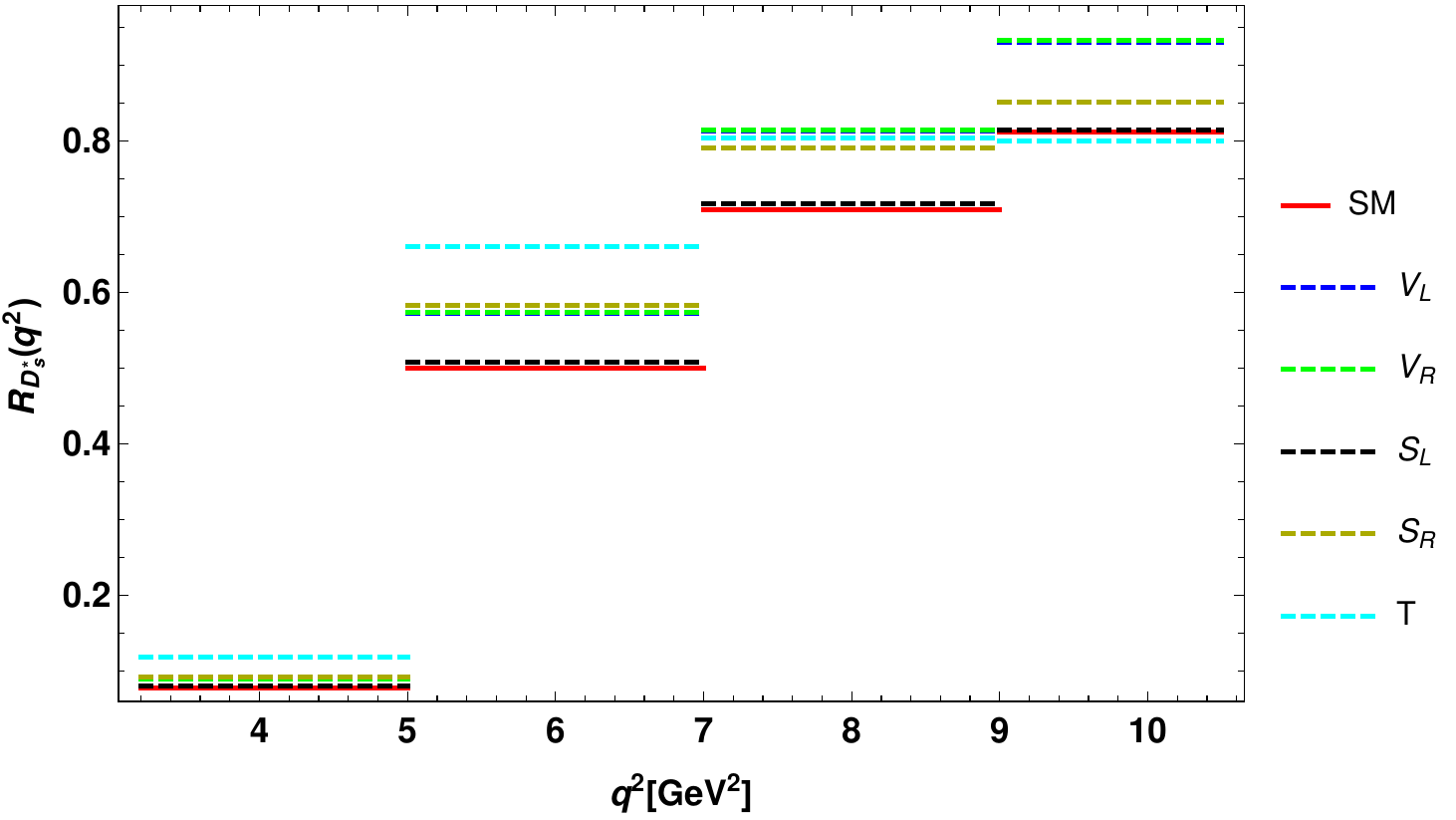}
\caption{ The bin-wise  $R_D$ (top-left panel), $R_{D^*}$ (top-right panel), $R_{\eta_c}$ (middle-left panel), $R_{J/\psi}$ (middle-right panel), $R_{D_s}$ (bottom-left panel) and $R_{D_s^*}$ (bottom-right panel)  in four $q^2$ bins for case B. }\label{Fig:CB-LNU}
\end{figure}

The bin-wise values of $R_D$ (top-left panel), $R_{D^*}$ (top-right panel), $R_{\eta_c}$ (middle-left panel), $R_{J/\psi}$ (middle-right panel), $R_{D_s}$ (bottom-left panel) and $R_{D_s^*}$ (bottom-right panel) are presented graphically in Fig. \ref{Fig:CB-LNU}\,. All the coefficients  provide profound deviations in the last  $q^2$ bin of $R_{D_{(s)}}$ and $R_{\eta_c}$ observables. Including either $V_L$ or $V_R$ coefficient, we find significant deviation in the last three bins of  the $R_V$ observable and the additional $S_R$ coupling maximally affect  the two middle bins. The $S_L$ coefficient has negligible impact on these LNU parameters, whereas tensor coupling affects the second and third bins of $R_{D_{(s)}^*}$. Though $T$ coupling has impact on last  bin of $B_c^+ \to J/\psi$, the deviation is minimal. 

\begin{figure}[htb]
\includegraphics[scale=0.5]{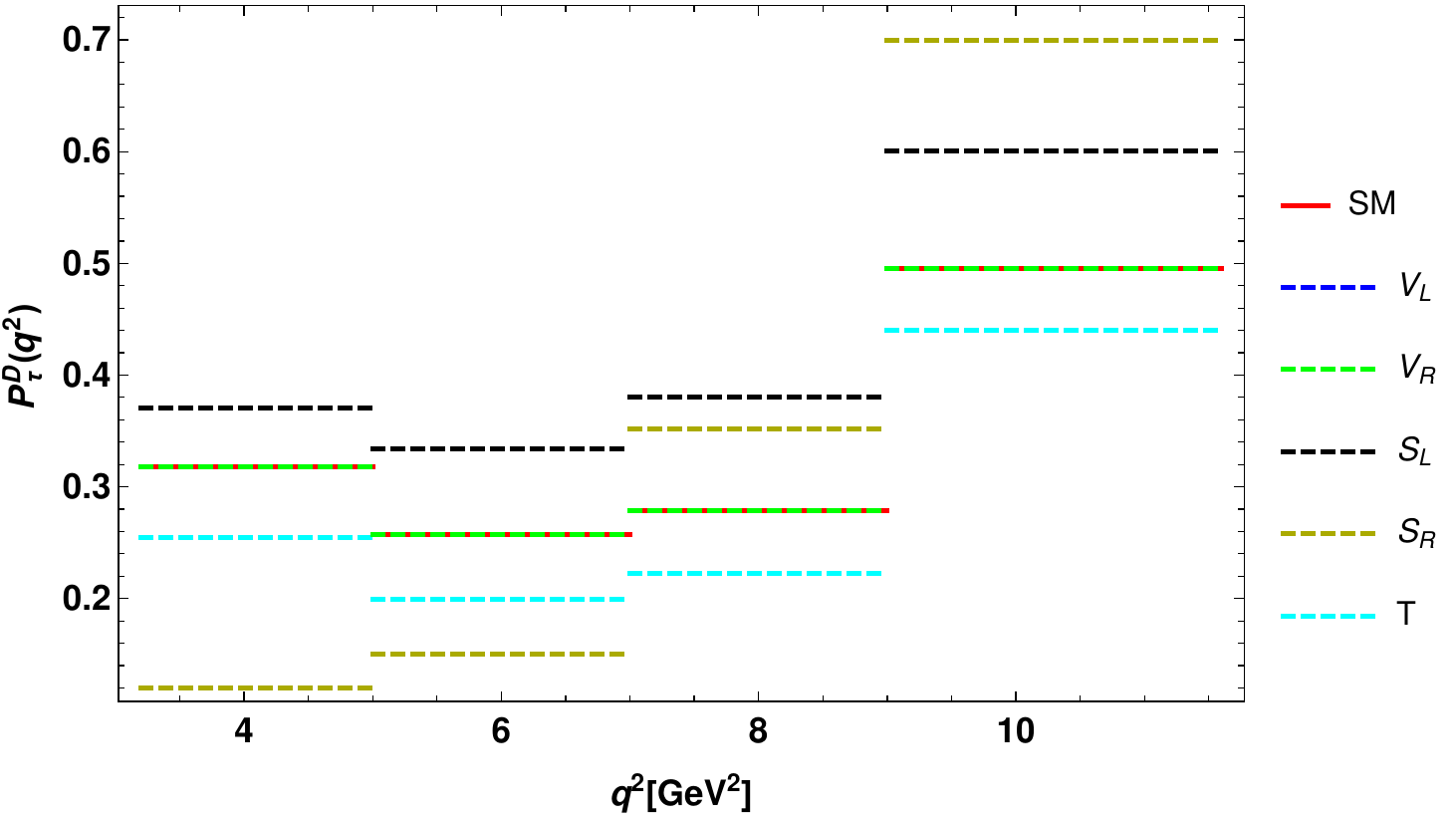}
\quad
\includegraphics[scale=0.5]{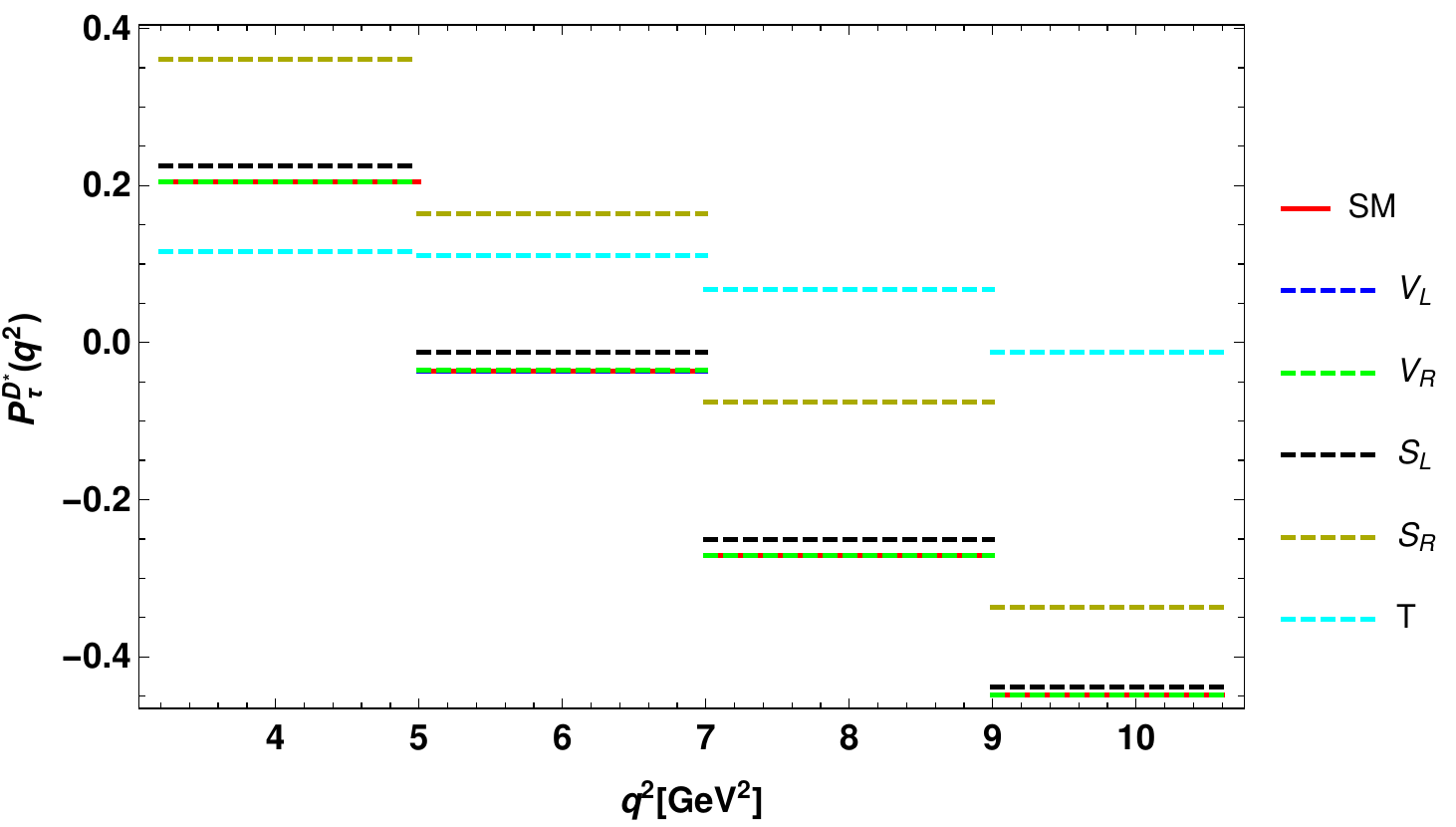}
\quad
\includegraphics[scale=0.5]{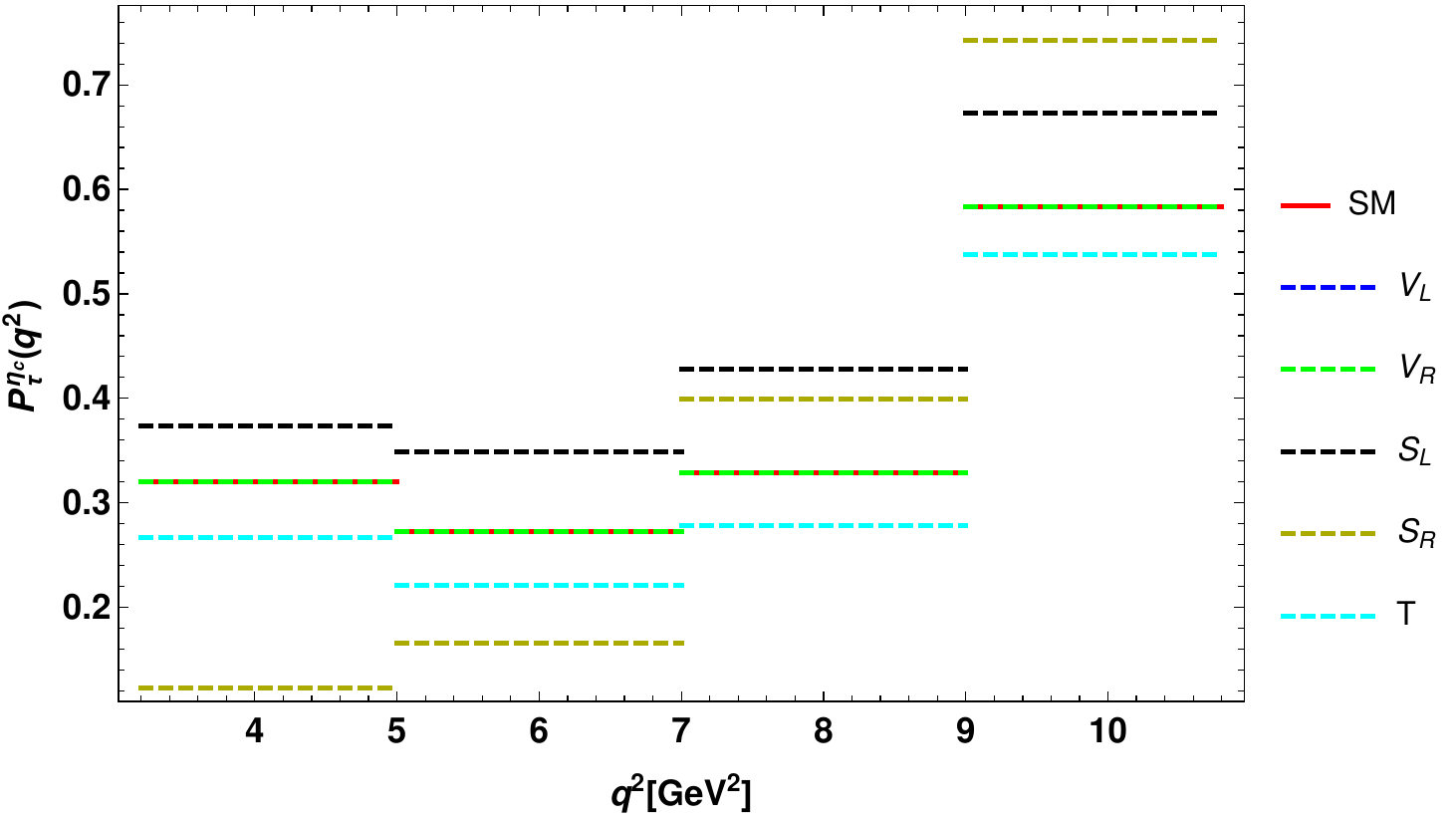}
\quad
\includegraphics[scale=0.5]{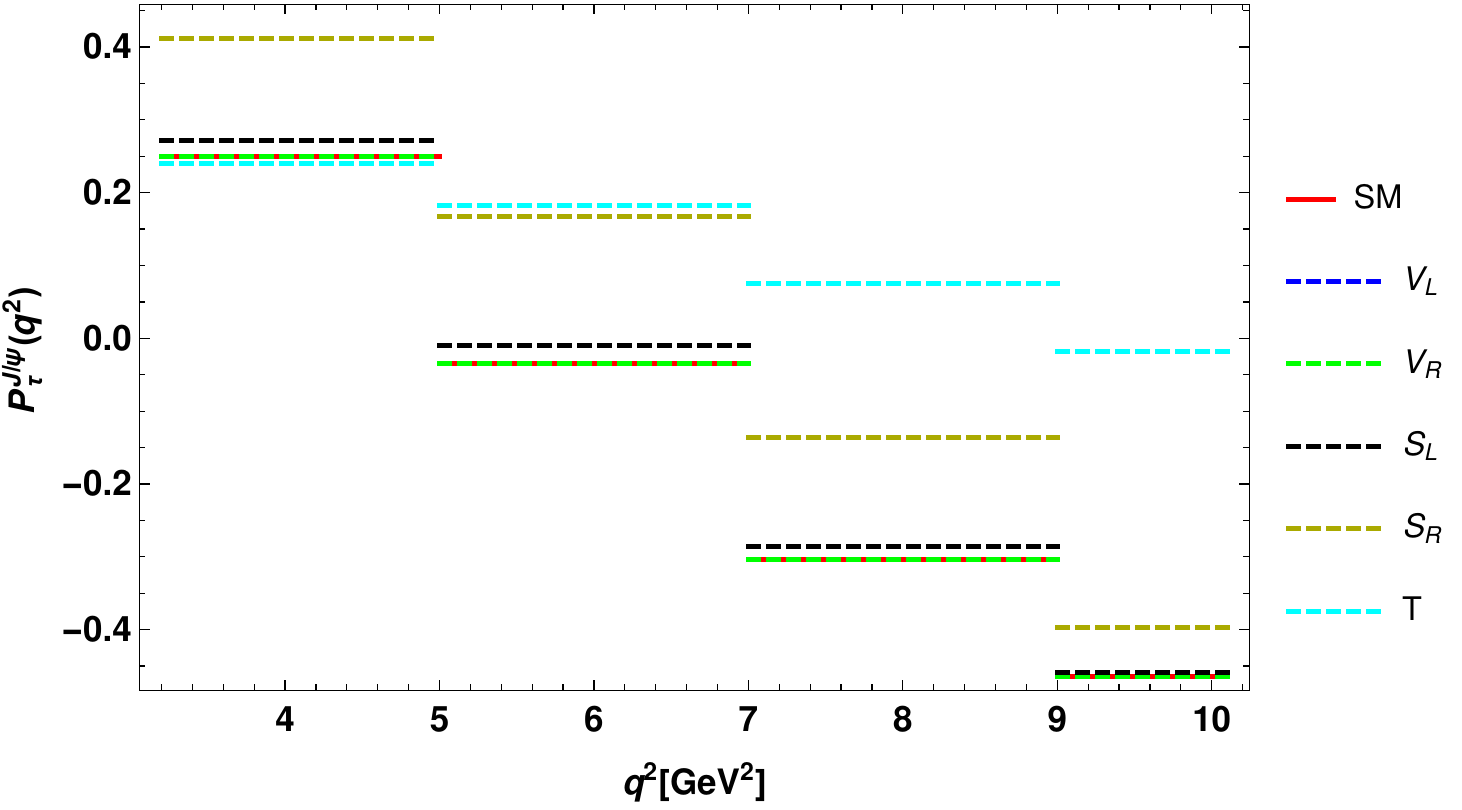}
\quad
\includegraphics[scale=0.5]{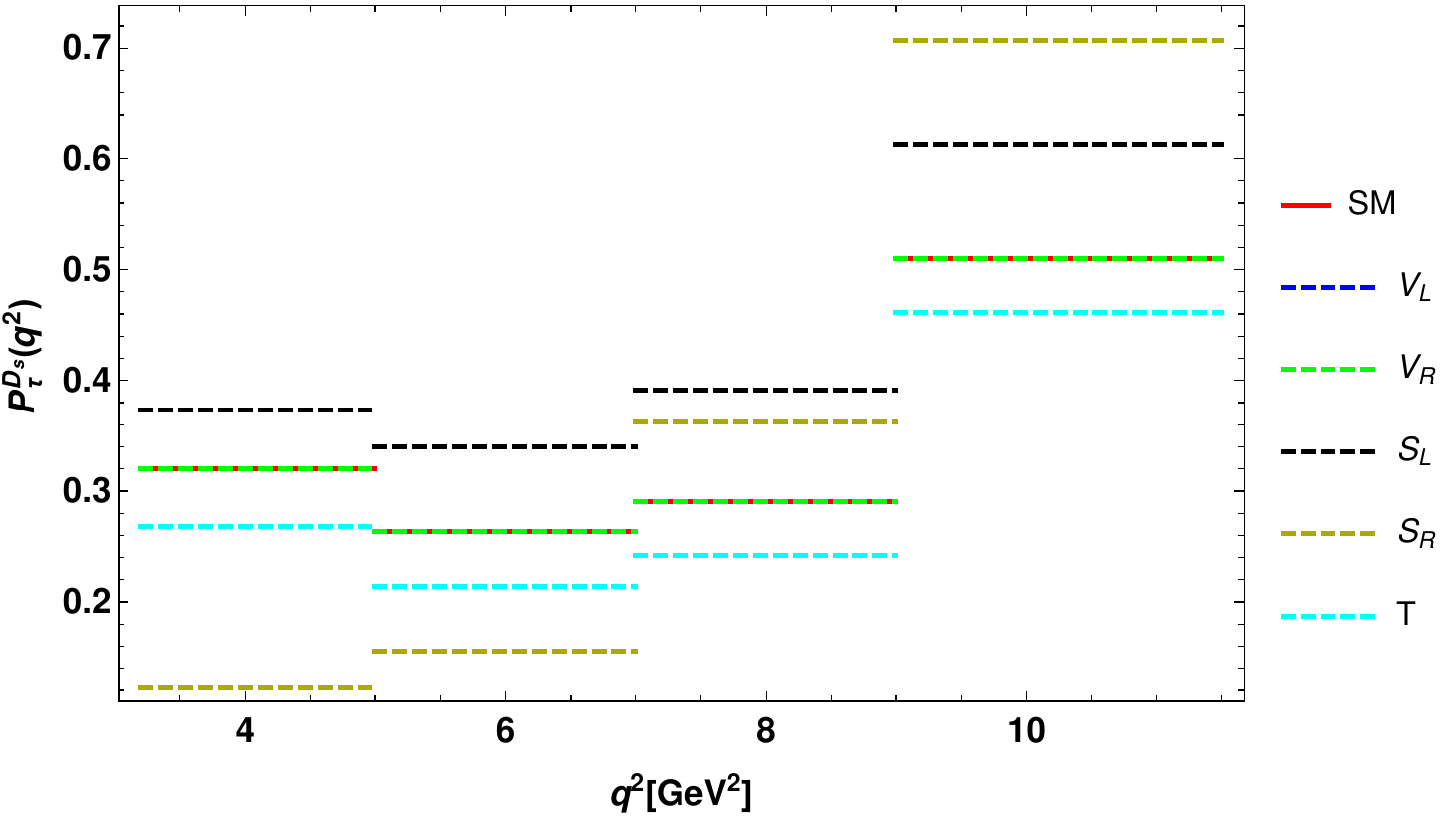}
\quad
\includegraphics[scale=0.5]{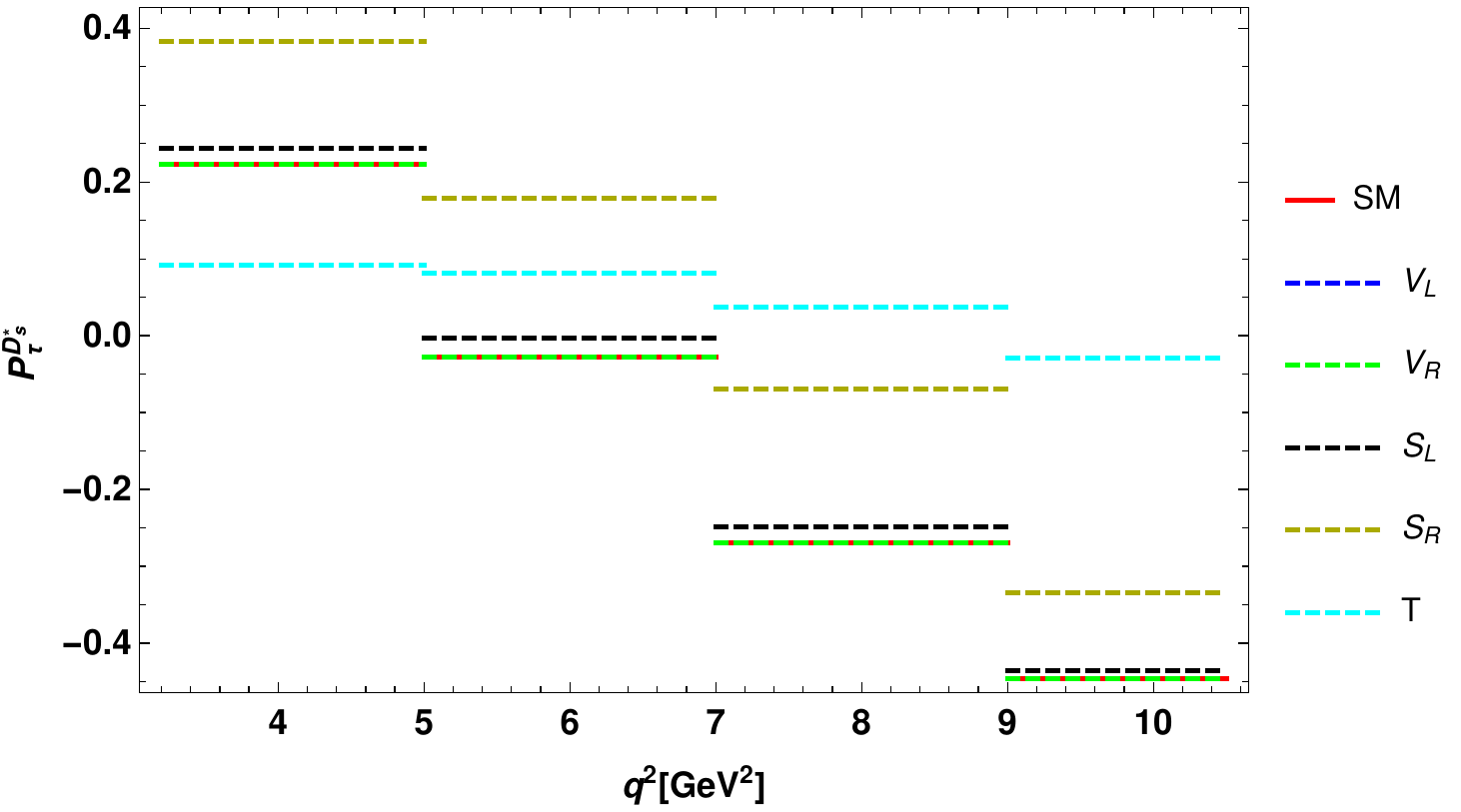}
\caption{ The bin-wise $\tau$-polarization asymmetry of $\bar B \to D \tau \bar \nu_\tau$ (top-left panel), $\bar B \to\bar D^* \tau  \nu_\tau$ (top-right panel), $B_c^+ \to \eta_c \tau^+  \nu_\tau$ (middle-left panel), $ B_c^+ \to J/\psi \tau^+  \nu_\tau$ (middle-right panel), $\bar B_s \to\bar D_s \tau \bar \nu_\tau$ (bottom-left panel) and $\bar B_s \to\bar D_s^* \tau \bar \nu_\tau$ (bottom-right panel) processes in four $q^2$ bins for case B. }\label{Fig:CB-ptau}
\end{figure}

The graphical representation of tau polarization observable of $\bar B \to  D \tau \bar \nu_\tau$ (top-left panel), $\bar B \to  D^* \tau \bar \nu_\tau$ (top-right panel), $B_c^+ \to \eta_c \tau^+  \nu_\tau$ (middle-left panel), $B_c^+ \to J/\psi \tau^+  \nu_\tau$ (middle-right  panel), $B_s \to D_s \tau \bar \nu_\tau$ (bottom-left panel) and $B_s \to D_s^* \tau \bar \nu_\tau$ (bottom-right panel) decay processes are shown in Fig. \ref{Fig:CB-ptau}\,. The $S_L$ coefficient has significant impact on $P_\tau$ observable of $B \to P$  and almost null  effect on $B \to V$ decay modes. The inclusion of either $S_R$ or $T$ coefficient has shifted $P_\tau$ significantly from the SM values of all decay modes. The bin-wise numerical values of the LNU  ratios and $\tau$ polarization asymmetries are presented in Table \ref{Tab:LNU-Ptau-CaseB}\,. 

\begin{figure}[htb]
\includegraphics[scale=0.5]{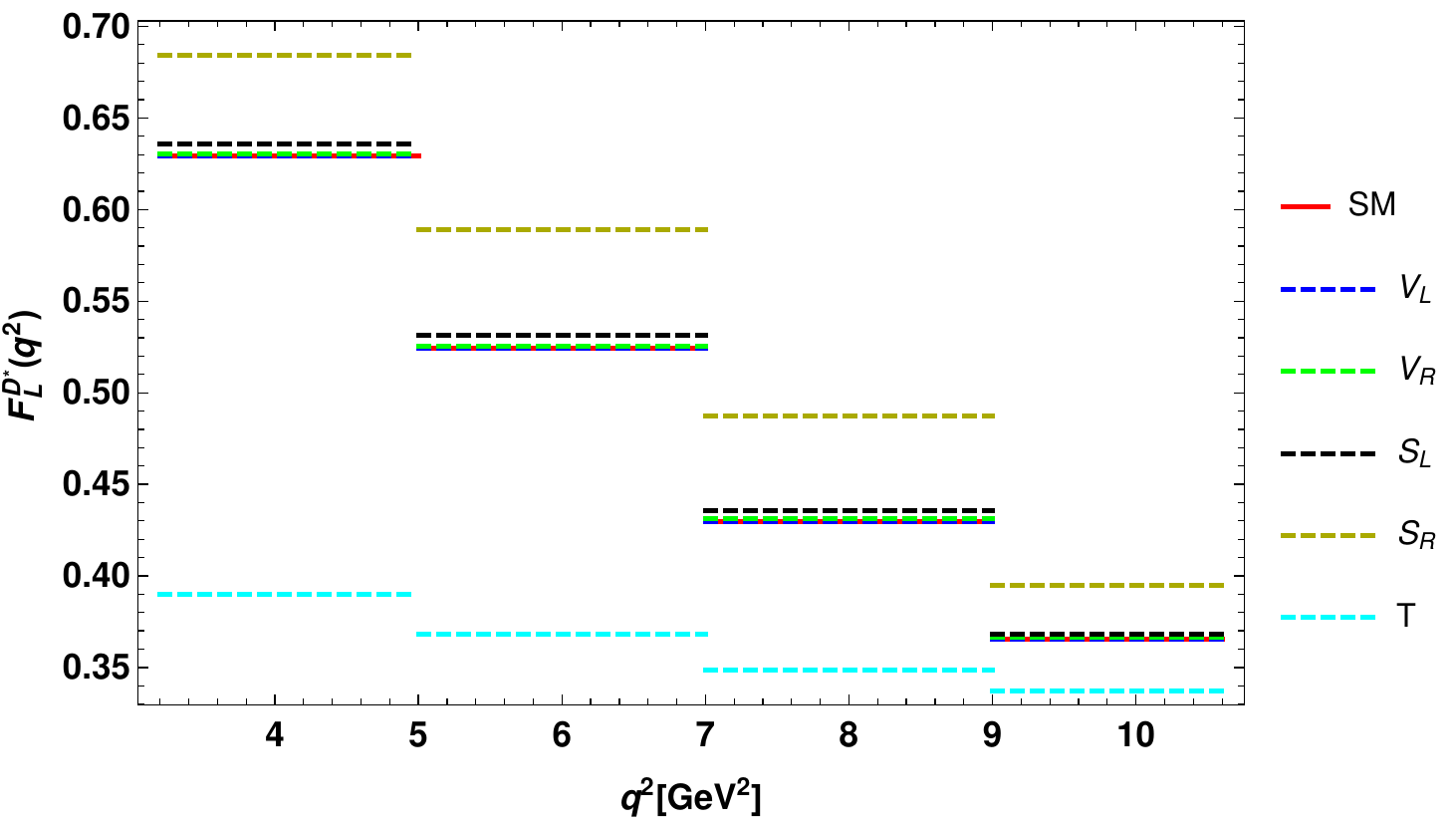}
\quad
\includegraphics[scale=0.5]{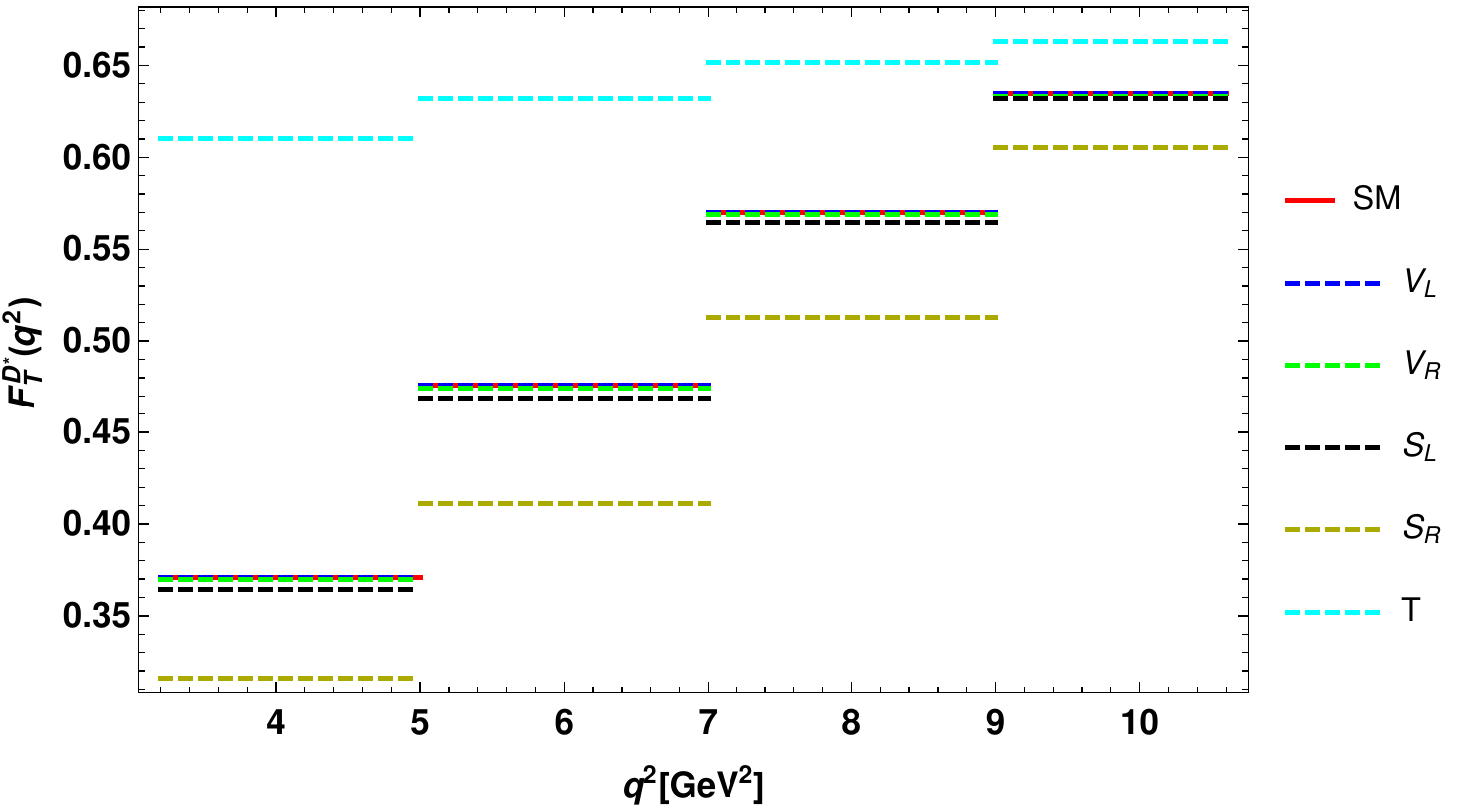}
\quad
\includegraphics[scale=0.5]{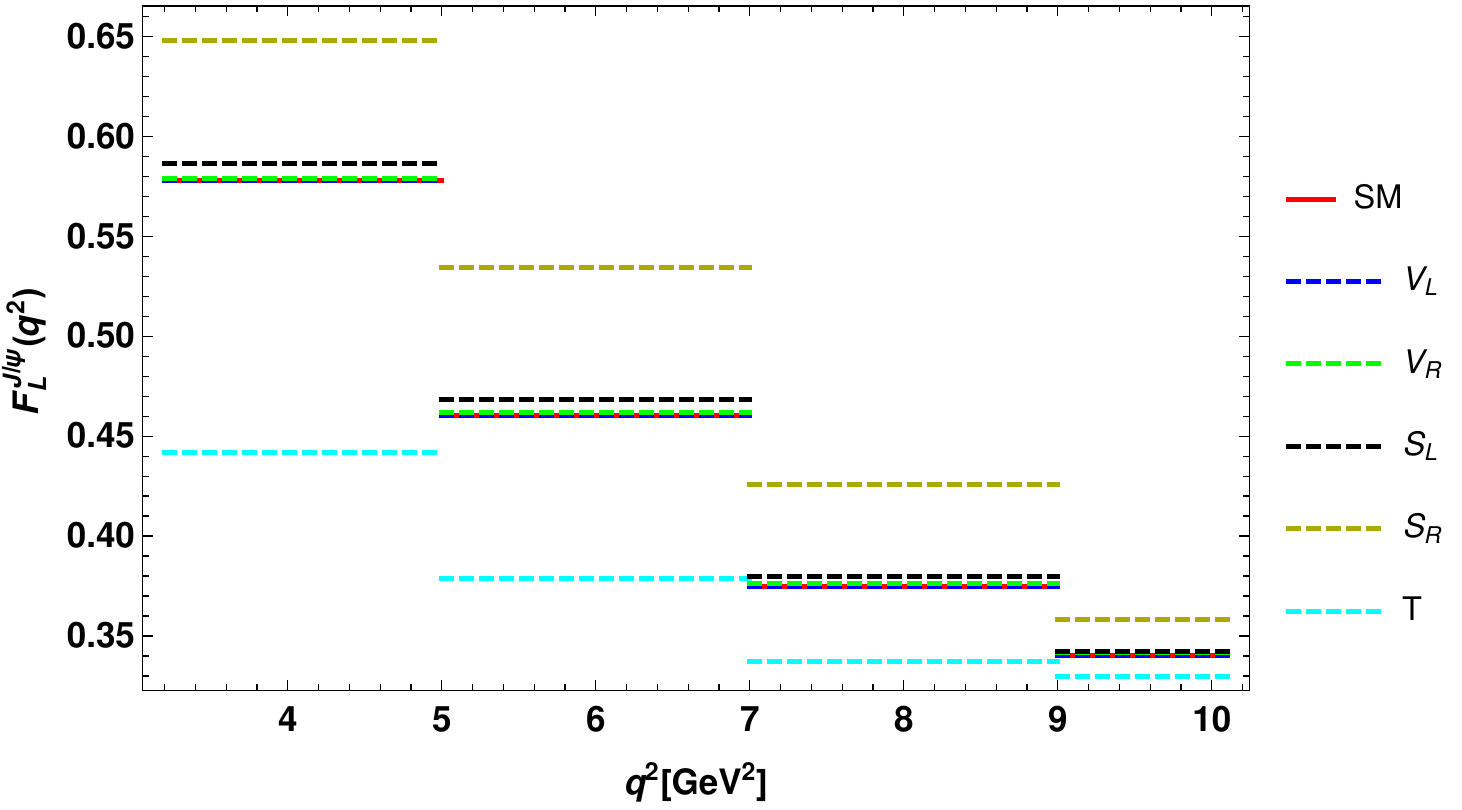}
\quad
\includegraphics[scale=0.5]{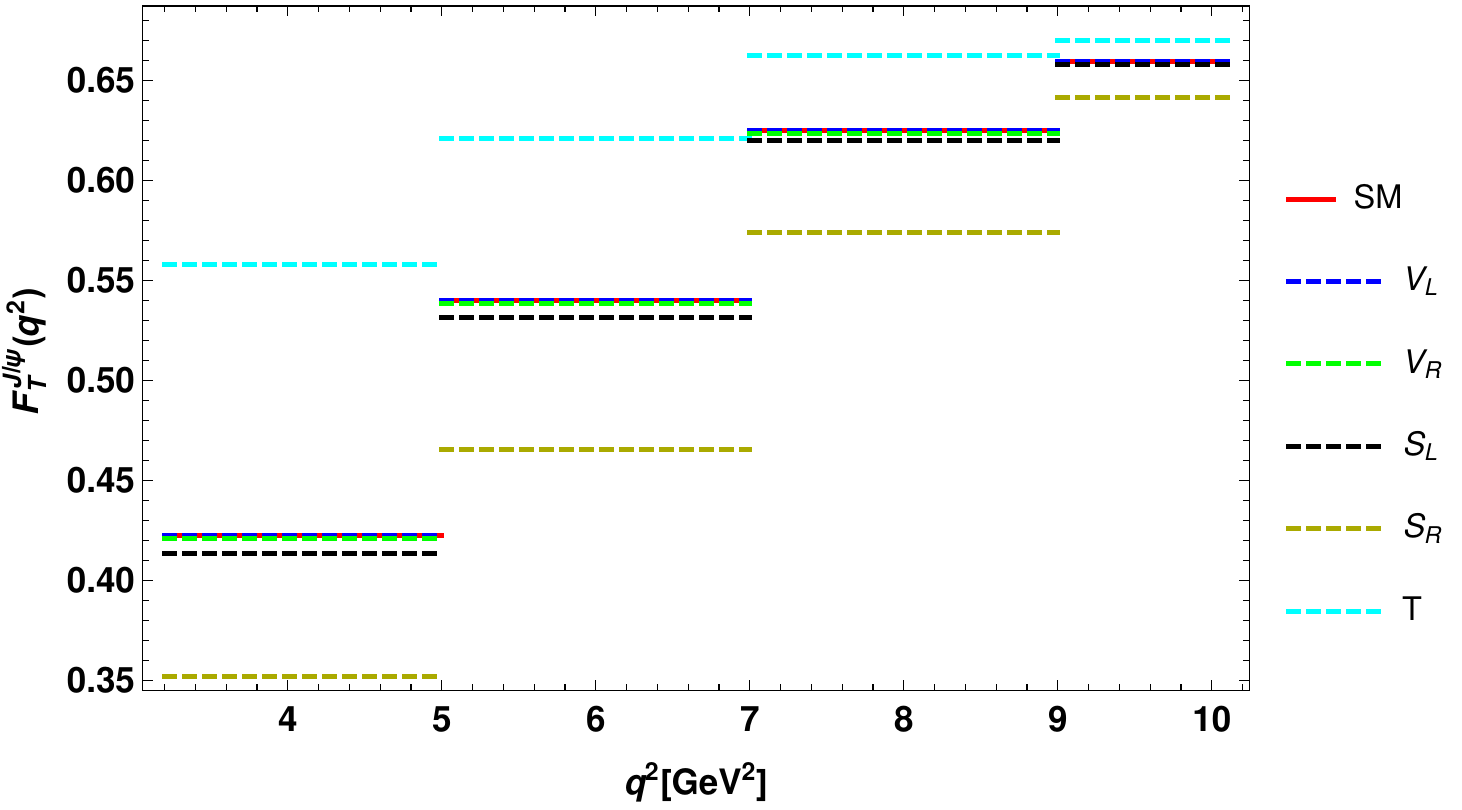}
\quad
\includegraphics[scale=0.5]{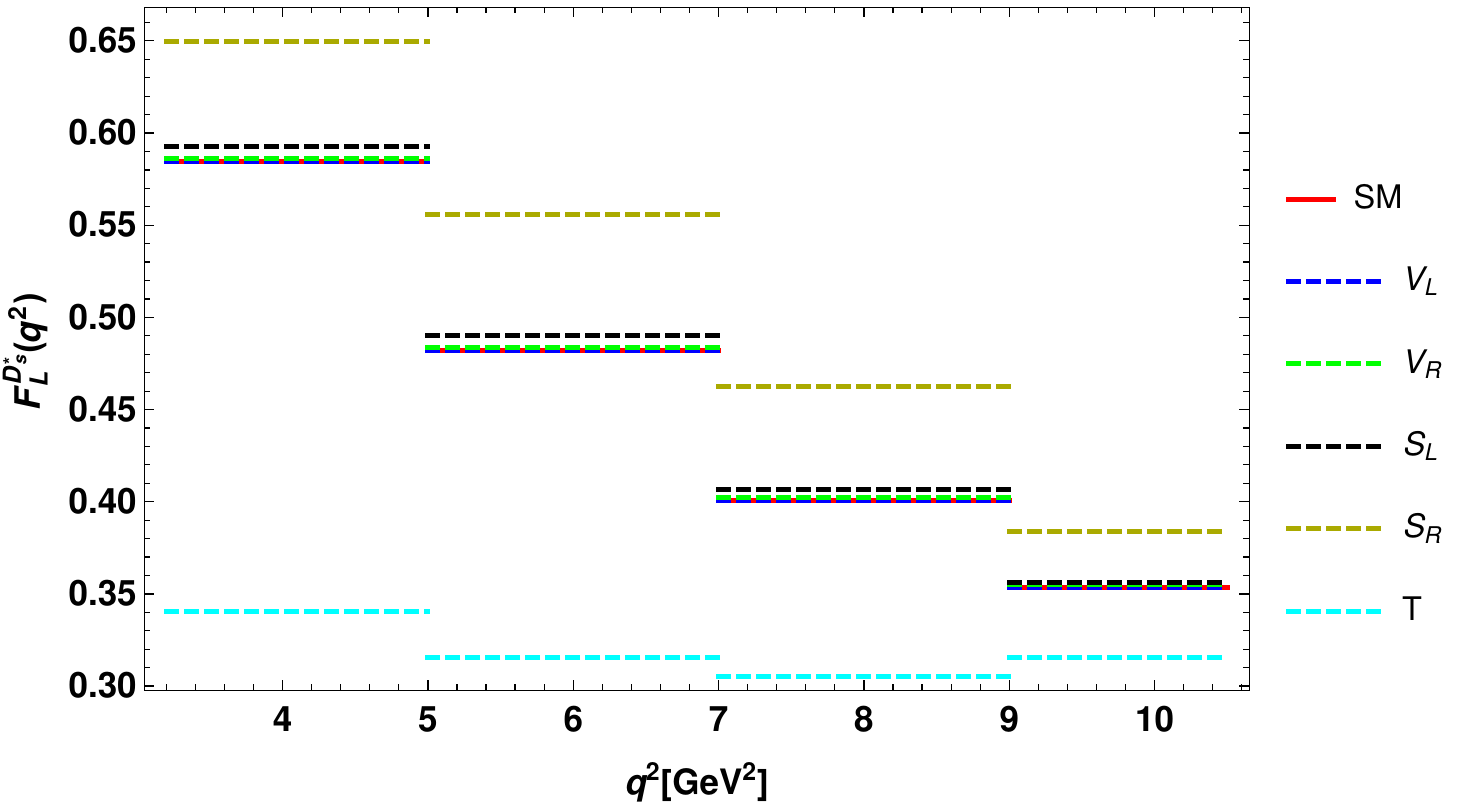}
\quad
\includegraphics[scale=0.5]{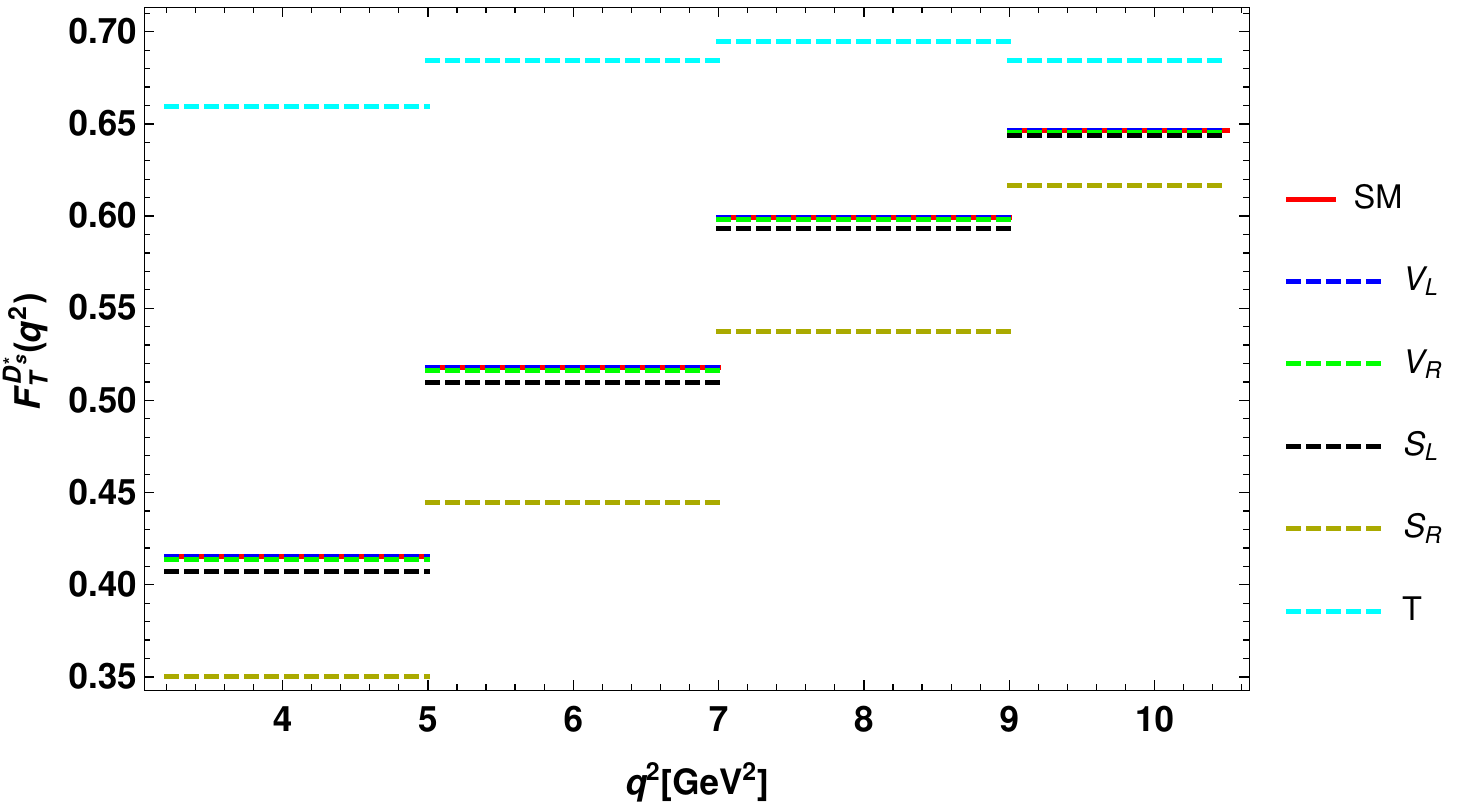}
\caption{ The bin-wise longitudinal (left panel) and transverse (right panel) polarization asymmetry of daughter vector meson of  $\bar B \to D^* \tau \bar \nu_\tau$ (top),  $ B_c^+ \to J/\psi \tau^+ \nu_\tau$ (middle) and $\bar B_s \to\bar D_s^* \tau \bar \nu_\tau$ (bottom) processes in four $q^2$ bin for case B. }\label{Fig:CB-polarization} 
\end{figure}

Fig. \ref{Fig:CB-polarization} depicts the bin-wise plots for  longitudinal (left) and transverse (right) polarization asymmetry parameters of  $\bar B \to  D^* \tau \bar \nu_\tau$ (top),  $B_c^+ \to J/\psi \tau^+  \nu_\tau$ (middle) and $B_s \to D_s^* \tau \bar \nu_\tau$ (bottom) processes for case B. The $F_L$ and $F_T$ parameters of none of the decay modes are affected by the $V_L/V_R/S_L$ coefficients, whereas $S_R/T$ have significant impact on the decay modes in all the $q^2$ bins.  The bin-wise numerical values of longitudinal  polarization asymmetries of $D_{(s)}^*,~J/\psi$ vector mesons are presented in Table \ref{Tab:FL-FT-CaseB}\,. The transverse polarization asymmetries can be obtained from Table \ref{Tab:FL-FT-CaseB} by using the relation $F_T=1-F_L$. Fig. \ref{Fig:correlation} depicts the correlations between lepton non-universality, lepton and hadron longitudinal polarization asymmetries of all the above discussed decay modes in their corresponding full $q^2$ range that obtained by using the $1\sigma$ range of new coefficients for case B. The $1\sigma$ range of new vector and scalar complex coefficients obtained from the joint confidence regions of the real and imaginary planes of these new couplings are given in Table III of our previous work \cite{Ray:2019gkv}\,. The $1\sigma$ ranges of real and imaginary parts of the tensor coefficient are $({\rm Re}[T], {\rm Im}[T])=([0.066,0.12],[-0.18,-0.16])$.


\begin{table}[htb]
\scriptsize
\caption{\scriptsize Predicted bin-wise values of branching ratios and forward-backward asymmetries of  $ B_{(s)} \to D^{(*)}_{(s)} \tau \bar \nu_\tau$ and $B_c^+ \to (\eta_c, J/\psi) \tau^+  \nu_\tau$ processes in the SM and in the presence of new complex Wilson coefficients  (case B).}\label{Tab:BR-AFB-CaseB}
\begin{center}
\begin{tabular}{|c|c|c|c|c|c|c|}
\hline
 ~Observables~&~Values for SM~&~values for $V_L$~&~Values for $V_R$~&~values for $S_L$~&~Values for $S_R$~&~Values for $T$~\\
\hline\hline
~$10^3$ Br$(\bar B \to\bar D)|_{q^2\in [m_\tau^2,5]}$~&~$(0.782\pm 0.03)$~&~$0.896$~&~$0.888$~~&~$0.848$~&~$0.61$~&~$0.87$\\

~$10^3$ Br$(\bar B \to\bar D )|_{q^2\in [5,7]}$~&~$(2.324\pm 0.21)$~&~$2.66$~&~$2.64$~&~$2.591$~&~$2.03$~&~$2.62$\\

~$10^3$ Br$(\bar B \to\bar D )|_{q^2\in [7,9]}$~&~$(2.36\pm  0.24)$~&~$2.7$~&~$2.674$~&~$2.74$~&~$2.62$~&~$2.643$\\

~$10^3$ Br$(\bar B \to\bar D )|_{q^2\in [9,11.6]}$~&~$(1.59\pm  0.191)$~&~$1.82$~&~$1.81$~&~$2.0$~&~$2.671$~&~$1.73$\\

~$\langle A_{FB}^D \rangle|_{q^2\in [m_\tau^2,5]}$~&~$(0.443\pm  0.0354)$~&~$0.443$~&~$0.443$~&~$0.4395$~&~$0.262$~&~$0.448$~\\

~$\langle A_{FB}^D \rangle|_{q^2\in [5,7]}$~&~$(0.394\pm 0.035)$~&~$0.394$~&~$0.394$~&~$0.3896$~&~$0.12$~&~$0.41$~\\

~$\langle A_{FB}^D \rangle|_{q^2\in [7,9]}$~&~$(0.3494\pm  0.33)$~&~$0.3494$~&~$0.3494$~&~$0.3406$~&~$0.0093$~&~$0.379$~\\

~$\langle A_{FB}^D \rangle|_{q^2\in [9,11.6]}$~&~$(0.2851 \pm 0.034)$~&~$0.2851$~&~$0.2851$~&~$0.2638$~&~$-0.0356$~&~$0.3326$~\\

\hline
~$10^3$ Br$(\bar B \to\bar D^*)|_{q^2\in [m_\tau^2,5]}$~&~$(1.1 \pm 0.088)$~&~$1.258$~&~$1.261$~&~$1.12$~&~$1.29$~&~$1.574$~\\

~$10^3$ Br$(\bar B \to\bar D^* )|_{q^2\in [5,7]}$~&~$(4.14\pm 0.373)$~&~$4.75$~&~$4.76$~&~$4.2$~&~$4.8$~&~$5.26$~\\

~$10^3$ Br$(\bar B \to\bar D^*)|_{q^2\in [7,9]}$~&~$(5.53\pm 0.55)$~&~$6.34$~&~$6.352$~&~$5.584$~&~$6.144$~&~$6.123$~\\

~$10^3$ Br$(\bar B \to\bar D^* )|_{q^2\in [9,11.6]}$~&~$(3.32\pm 0.4)$~&~$3.81$~&~$3.82$~&~$3.34$~&~$3.483$~&~$3.242$~\\

~$\langle A_{FB}^{D^*} \rangle|_{q^2\in [m_\tau^2,5]}$~&~$(0.106\pm 0.009)$~&~$0.106$~&~$0.148$~&~$0.115$~&~$0.025$~&~$0.165$~\\

~$\langle A_{FB}^{D^*} \rangle|_{q^2\in [5,7]}$~&~$(-0.0225\pm 0.002)$~&~$-0.0225$~&~$0.03$~&~$-0.0123$~&~$-0.081$~&~$0.0261$~\\

~$\langle A_{FB}^{D^*} \rangle|_{q^2\in [7,9]}$~&~$(-0.104\pm 0.008)$~&~$-0.104$~&~$-0.0493$~&~$-0.095$~&~$-0.145$~&~$-0.0824$~\\

~$\langle A_{FB}^{D^*} \rangle|_{q^2\in [9,11.6]}$~&~$(-0.103\pm 0.013)$~&~$-0.103$~&~$-0.063$~&~$-0.098$~&~$-0.132$~&~$-0.11$~\\

\hline
~$10^4$ Br$( B_c \to \eta_c)|_{q^2\in [m_\tau^2,5]}$~&~$(2.7\pm 0.22)$~&~$3.1$~&~$3.05$~&~$2.91$~&~$2.08$~&~$2.94$~\\

~$10^4$ Br$(B_c \to \eta_c)|_{q^2\in [5,7]}$~&~$(7.95\pm 0.72)$~&~$9.11$~&~$9.03$~&~$8.877$~&~$6.93$~&~$8.77$~\\

~$10^4$ Br$(B_c \to \eta_c)|_{q^2\in [7,9]}$~&~$(7.64\pm 0.611)$~&~$8.76$~&~$8.682$~&~$8.96$~&~$8.545$~&~$8.382$~\\

~$10^4$ Br$(B_c \to \eta_c)|_{q^2\in [9,11.6]}$~&~$(3.384\pm 0.41)$~&~$3.88$~&~$3.85$~&~$4.3$~&~$5.5$~&~$3.6$~\\

~$\langle A_{FB}^{\eta_c} \rangle|_{q^2\in [m_\tau^2,5]}$~&~$(0.4429\pm 0.035)$~&~$0.4429$~&~$0.4429$~&~$0.4392$~&~$0.262$~&~$0.4483$~\\

~$\langle A_{FB}^{\eta_c} \rangle|_{q^2\in [5,7]}$~&~$(0.394\pm 0.035)$~&~$0.394$~&~$0.394$~&~$0.3884$~&~$0.12$~&~$0.41$~\\

~$\langle A_{FB}^{\eta_c} \rangle|_{q^2\in [7,9]}$~&~$(0.3458\pm 0.028)$~&~$0.3458$~&~$0.3458$~&~$0.3346$~&~$-0.01$~&~$0.375$~\\

~$\langle A_{FB}^{\eta_c} \rangle|_{q^2\in [9,11.6]}$~&~$(0.27\pm 0.032)$~&~$0.27$~&~$0.27$~&~$0.2483$~&~$-0.03$~&~$0.315$~\\
\hline
~$10^3$ Br$( B_c \to J/\psi)|_{q^2\in [m_\tau^2,5]}$~&~$(0.24\pm 0.019)$~&~$0.274$~&~$0.275$~&~$0.245$~&~$0.287$~&~$0.283$~\\

~$10^3$ Br$(B_c \to J/\psi)|_{q^2\in [5,7]}$~&~$(1.0\pm 0.09)$~&~$1.15$~&~$1.15$~&~$1.02$~&~$1.16$~&~$1.1$~\\

~$10^3$  Br$(B_c \to J/\psi)|_{q^2\in [7,9]}$~&~$(1.52\pm 0.132)$~&~$1.74$~&~$1.75$~&~$1.534$~&~$1.66$~&~$1.463$~\\

~$10^3$  Br$(B_c \to J/\psi)|_{q^2\in [9,11.6]}$~&~$(0.643\pm 0.077)$~&~$0.737$~&~$0.74$~&~$0.644$~&~$0.661$~&~$0.589$~\\

~$\langle A_{FB}^{J/\psi} \rangle|_{q^2\in [m_\tau^2,5]}$~&~$(0.1434\pm 0.012)$~&~$0.1434$~&~$0.172$~&~$0.15$~&~$0.058$~&~$0.219$~\\

~$\langle A_{FB}^{J/\psi} \rangle|_{q^2\in [5,7]}$~&~$(0.032\pm 0.003)$~&~$0.032$~&~$0.066$~&~$0.04$~&~$-0.029$~&~$0.07$~\\

~$\langle A_{FB}^{J/\psi} \rangle|_{q^2\in [7,9]}$~&~$(-0.032\pm 0.0026)$~&~$-0.032$~&~$-4.9\times 10^{-4}$~&~$-0.0254$~&~$-0.073$~&~$-0.0334$~\\

~$\langle A_{FB}^{J/\psi} \rangle|_{q^2\in [9,11.6]}$~&~$(-0.0345\pm 0.0042)$~&~$-0.0345$~&~$-0.015$~&~$-0.031$~&~$-0.059$~&~$-0.046$~\\

\hline
~$10^3$ Br$(B_s \to D_s)|_{q^2\in [m_\tau^2,5]}$~&~$(0.86\pm 0.69)$~&~$1.02$~&~$1.0$~&~$0.96$~&~$0.687$~&~$0.966$~\\

~$10^3$ Br$(B_s \to D_s )|_{q^2\in [5,7]}$~&~$(2.542\pm 0.23)$~&~$2.92$~&~$2.89$~&~$2.836$~&~$2.22$~&~$2.8$~\\

~$10^3$ Br$(B_s \to D_s )|_{q^2\in [7,9]}$~&~$(2.46\pm 0.23)$~&~$2.82$~&~$2.8$~&~$2.87$~&~$2.74$~&~$2.71$~\\

~$10^3$ Br$( B_s \to D_s )|_{q^2\in [9,11.6]}$~&~$(1.54\pm 0.185)$~&~$1.77$~&~$1.75$~&~$1.95$~&~$2.58$~&~$1.65$~\\

~$\langle A_{FB}^{D_s} \rangle|_{q^2\in [m_\tau^2,5]}$~&~$(0.443\pm 0.0354)$~&~$0.443$~&~$0.443$~&~$0.44$~&~$0.263$~&~$0.449$~\\

~$\langle A_{FB}^{D_s} \rangle|_{q^2\in [5,7]}$~&~$(0.4\pm 0.036)$~&~$0.4$~&~$0.4$~&~$0.3893$~&~$0.12$~&~$0.41$~\\

~$\langle A_{FB}^{D_s} \rangle|_{q^2\in [7,9]}$~&~$(0.349\pm 0.032)$~&~$0.349$~&~$0.349$~&~$0.34$~&~$0.0096$~&~$0.377$~\\

~$\langle A_{FB}^{D_s} \rangle|_{q^2\in [9,11.6]}$~&~$(0.2833\pm 0.034)$~&~$0.2833$~&~$0.2833$~&~$0.262$~&~$-0.035$~&~$0.327$~\\

\hline
~$10^3$ Br$( B_s \to  D_s^*)|_{q^2\in [m_\tau^2,5]}$~&~$(1.325\pm 0.1)$~&~$1.52$~&~$1.53$~&~$1.35$~&~$1.572$~&~$2.02$~\\

~$10^3$ Br$( B_s \to D_s^* )|_{q^2\in [5,7]}$~&~$(5.88\pm 0.53)$~&~$6.74$~&~$6.75$~&~$5.97$~&~$6.85$~&~$7.77$~\\

~$10^3$ Br$( B_s \to D_s^*)|_{q^2\in [7,9]}$~&~$(9.515\pm 0.86)$~&~$10.1$~&~$10.1$~&~$9.614$~&~$10.6$~&~$10.8$~\\

~$10^3$ Br$( B_s \to  D_s^* )|_{q^2\in [9,11.6]}$~&~$(6.33\pm 0.76)$~&~$7.2$~&~$7.28$~&~$6.36$~&~$6.64$~&~$6.25$~\\

~$\langle A_{FB}^{D_s^*} \rangle|_{q^2\in [m_\tau^2,5]}$~&~$(0.07\pm 0.006)$~&~$0.07$~&~$0.12$~&~$0.078$~&~$-0.004$~&~$0.184$~\\

~$\langle A_{FB}^{D_s^*} \rangle|_{q^2\in [5,7]}$~&~$(-0.0584\pm 0.0053)$~&~$-0.0584$~&~$7.88\times 10^{-4}$~&~$-0.048$~&~$-0.11$~&~$0.0342$~\\

~$\langle A_{FB}^{D_s^*} \rangle|_{q^2\in [7,9]}$~&~$(-0.134\pm 0.12)$~&~$-0.134$~&~$-0.073$~&~$-0.125$~&~$-0.17$~&~$-0.0872$~\\

~$\langle A_{FB}^{D_s^*} \rangle|_{q^2\in [9,11.6]}$~&~$(-0.123\pm 0.015)$~&~$-0.123$~&~$-0.077$~&~$-0.1174$~&~$-0.15$~&~$-0.119$~\\

\hline
\end{tabular}
\end{center}
\end{table} 

\begin{table}[htb]
\scriptsize
\caption{\scriptsize Predicted bin-wise values of LNU ratios and $\tau$-polarization asymmetries of $ B_{(s)} \to  D_{(s)}^{(*)} \tau \bar \nu_\tau$ and $B_c^+ \to (\eta_c, J/\psi) \tau^+  \nu_\tau$ processes in the SM and in the presence of new complex Wilson coefficients  (case B).}\label{Tab:LNU-Ptau-CaseB}
\begin{center}
\begin{tabular}{|c|c|c|c|c|c|c|}
\hline
 ~Observables~&~Values for SM~&~values for $V_L$~&~Values for $V_R$~&~values for $V_L$~&~Values for $V_R$~&~Values for $T$~\\
\hline\hline

~$\langle R_D \rangle|_{q^2\in [m_\tau^2,5]}$~&~$0.048$~&~$0.055$~&~$0.054$~&~$0.052$~&~$0.037$~&~$0.053$~\\

~$\langle R_D \rangle|_{q^2\in [5,7]}$~&~$0.583$~&~$0.668$~&~$0.662$~&~$0.65$~&~$0.51$~&~$0.655$~\\

~$\langle R_D \rangle|_{q^2\in [7,9]}$~&~$0.989$~&~$1.133$~&~$1.123$~&~$1.151$~&~$1.101$~&~$1.11$~\\

~$\langle R_D \rangle|_{q^2\in [9,11.6]}$~&~$1.8$~&~$2.061$~&~$2.043$~&~$2.27$~&~$3.02$~&~$1.954$~\\

~$\langle P_\tau^D \rangle|_{q^2\in [m_\tau^2,5]}$~&~$0.318$~&~$0.318$~&~$0.318$~&~$0.37$~&~$0.12$~&~$0.26$~\\

~$\langle P_\tau^D \rangle|_{q^2\in [5,7]}$~&~$0.2572$~&~$0.2572$~&~$0.2572$~&~$0.334$~&~$0.15$~&~$0.199$~\\

~$\langle P_\tau^D \rangle|_{q^2\in [7,9]}$~&~$0.2783$~&~$0.2783$~&~$0.2783$~&~$0.38$~&~$0.352$~&~$0.223$~\\

~$\langle P_\tau^D \rangle|_{q^2\in [9,11.6]}$~&~$0.4955$~&~$0.4955$~&~$0.4955$~&~$0.6$~&~$0.7$~&~$0.44$~\\

\hline

~$\langle R_{D^*} \rangle|_{q^2\in [m_\tau^2,5]}$~&~$0.044$~&~$0.05$~&~$0.0502$~&~$0.045$~&~$0.052$~&~$0.063$~\\

~$\langle R_{D^*}\rangle|_{q^2\in [5,7]}$~&~$0.33$~&~$0.375$~&~$0.376$~&~$0.332$~&~$0.379$~&~$0.416$~\\

~$\langle R_{D^*} \rangle|_{q^2\in [7,9]}$~&~$0.468$~&~$0.536$~&~$0.538$~&~$0.472$~&~$0.52$~&~$0.518$~\\

~$\langle R_{D^*} \rangle|_{q^2\in [9,11.6]}$~&~$0.54$~&~$0.62$~&~$0.622$~&~$0.543$~&~$0.567$~&~$0.527$~\\

~$\langle P_\tau^{D^*} \rangle|_{q^2\in [m_\tau^2,5]}$~&~$0.205$~&~$0.205$~&~$0.206$~&~$0.225$~&~$0.361$~&~$0.116$~\\

~$\langle P_\tau^{D^*} \rangle|_{q^2\in [5,7]}$~&~$-0.036$~&~$-0.036$~&~$-0.036$~&~$-0.0119$~&~$0.165$~&~$0.11$~\\

~$\langle P_\tau^{D^*} \rangle|_{q^2\in [7,9]}$~&~$-0.2716$~&~$-0.2716$~&~$-0.271$~&~$-0.2508$~&~$-0.0754$~&~$0.682$~\\

~$\langle P_\tau^{D^*} \rangle|_{q^2\in [9,11.6]}$~&~$-0.4484$~&~$-0.4484$~&~$-0.4483$~&~$-0.4382$~&~$-0.338$~&~$-0.013$~\\
\hline

~$\langle R_{\eta_c} \rangle|_{q^2\in [m_\tau^2,5]}$~&~$0.049$~&~$0.056$~&~$0.0556$~&~$0.053$~&~$0.038$~&~$0.0381$~\\

~$\langle R_{\eta_c} \rangle|_{q^2\in [5,7]}$~&~$0.593$~&~$0.68$~&~$0.674$~&~$0.663$~&~$0.517$~&~$0.654$~\\

~$\langle R_{\eta_c} \rangle|_{q^2\in [7,9]}$~&~$1.06$~&~$1.211$~&~$1.2$~&~$1.24$~&~$1.181$~&~$1.16$~\\

~$\langle R_{\eta_c} \rangle|_{q^2\in [9,11.6]}$~&~$2.13$~&~$2.44$~&~$2.41$~&~$2.7$~&~$3.436$~&~$2.254$~\\

~$\langle P_\tau^{\eta_c} \rangle|_{q^2\in [m_\tau^2,5]}$~&~$0.3205$~&~$0.3205$~&~$0.3205$~&~$0.3733$~&~$0.123$~&~$0.266$~\\

~$\langle P_\tau^{\eta_c} \rangle|_{q^2\in [5,7]}$~&~$0.2725$~&~$0.2725$~&~$0.2725$~&~$0.349$~&~$0.166$~&~$0.2214$~\\

~$\langle P_\tau^{\eta_c} \rangle|_{q^2\in [7,9]}$~&~$0.3287$~&~$0.3287$~&~$0.3287$~&~$0.428$~&~$0.4$~&~$0.278$~\\

~$\langle P_\tau^{\eta_c} \rangle|_{q^2\in [9,11.6]}$~&~$0.584$~&~$0.584$~&~$0.584$~&~$0.673$~&~$0.743$~&~$0.538$~\\
\hline
~$\langle R_{J/\psi} \rangle|_{q^2\in [m_\tau^2,5]}$~&~$0.053$~&~$0.061$~&~$0.061$~&~$0.054$~&~$0.064$~&~$0.063$~\\

~$\langle R_{J/\psi} \rangle|_{q^2\in [5,7]}$~&~$0.331$~&~$0.38$~&~$0.38$~&~$0.336$~&~$0.384$~&~$0.349$~\\

~$\langle R_{J/\psi} \rangle|_{q^2\in [7,9]}$~&~$0.4622$~&~$0.53$~&~$0.531$~&~$0.466$~&~$0.503$~&~$0.445$~\\

~$\langle R_{J/\psi} \rangle|_{q^2\in [9,11.6]}$~&~$0.526$~&~$0.602$~&~$0.604$~&~$0.527$~&~$0.54$~&~$0.482$~\\

~$\langle P_\tau^{J/\psi} \rangle|_{q^2\in [m_\tau^2,5]}$~&~$0.25$~&~$0.25$~&~$0.25$~&~$0.272$~&~$0.412$~&~$0.24$~\\

~$\langle P_\tau^{J/\psi} \rangle|_{q^2\in [5,7]}$~&~$-0.0343$~&~$-0.0343$~&~$-0.0339$~&~$-0.01$~&~$0.168$~&~$0.182$~\\

~$\langle P_\tau^{J/\psi} \rangle|_{q^2\in [7,9]}$~&~$-0.3034$~&~$-0.3034$~&~$-0.3032$~&~$-0.286$~&~$-0.137$~&~$0.076$~\\

~$\langle P_\tau^{J/\psi} \rangle|_{q^2\in [9,11.6]}$~&~$-0.4644$~&~$-0.4644$~&~$-0.4644$~&~$-0.4584$~&~$-0.4$~&~$-0.02$~\\

\hline

~$\langle R_{D_s} \rangle|_{q^2\in [m_\tau^2,5]}$~&~$0.046$~&~$0.053$~&~$0.052$~&~$0.05$~&~$0.0354$~&~$0.05$~\\

~$\langle R_{D_s} \rangle|_{q^2\in [5,7]}$~&~$0.586$~&~$0.672$~&~$0.666$~&~$0.653$~&~$0.511$~&~$0.645$~\\

~$\langle R_{D_s} \rangle|_{q^2\in [7,9]}$~&~$1.0$~&~$1.15$~&~$1.14$~&~$1.17$~&~$1.12$~&~$1.1$~\\

~$\langle R_{D_s} \rangle|_{q^2\in [9,11.6]}$~&~$1.844$~&~$2.114$~&~$2.1$~&~$2.332$~&~$3.1$~&~$1.977$~\\

~$\langle P_\tau^{D_s} \rangle|_{q^2\in [m_\tau^2,5]}$~&~$0.32$~&~$0.32$~&~$0.32$~&~$0.373$~&~$0.123$~&~$0.268$~\\

~$\langle P_\tau^{D_s} \rangle|_{q^2\in [5,7]}$~&~$0.2634$~&~$0.2634$~&~$0.2634$~&~$0.3397$~&~$0.156$~&~$0.214$~\\

~$\langle P_\tau^{D_s} \rangle|_{q^2\in [7,9]}$~&~$0.29$~&~$0.29$~&~$0.29$~&~$0.39$~&~$0.363$~&~$0.242$~\\

~$\langle P_\tau^{D_s} \rangle|_{q^2\in [9,11.6]}$~&~$0.51$~&~$0.51$~&~$0.51$~&~$0.612$~&~$0.71$~&~$0.462$~\\
\hline

~$\langle R_{D_s^*} \rangle|_{q^2\in [m_\tau^2,5]}$~&~$0.078$~&~$0.0894$~&~$0.0896$~&~$0.08$~&~$0.093$~&~$0.119$~\\

~$\langle R_{D_s^*} \rangle|_{q^2\in [5,7]}$~&~$0.5$~&~$0.573$~&~$0.574$~&~$0.51$~&~$0.583$~&~$0.661$~\\

~$\langle R_{D_s^*} \rangle|_{q^2\in [7,9]}$~&~$0.71$~&~$0.813$~&~$0.815$~&~$0.717$~&~$0.79$~&~$0.81$~\\

~$\langle R_{D_s^*} \rangle|_{q^2\in [9,11.6]}$~&~$0.812$~&~$0.931$~&~$0.933$~&~$0.82$~&~$0.852$~&~$0.8$~\\

~$\langle P_\tau^{D_s^*} \rangle|_{q^2\in [m_\tau^2,5]}$~&~$0.223$~&~$0.223$~&~$0.225$~&~$0.2435$~&~$0.382$~&~$0.092$~\\

~$\langle P_\tau^{D_s^*} \rangle|_{q^2\in [5,7]}$~&~$-0.028$~&~$-0.028$~&~$-0.0273$~&~$-0.0028$~&~$0.178$~&~$0.082$~\\

~$\langle P_\tau^{D_s^*} \rangle|_{q^2\in [7,9]}$~&~$-0.27$~&~$-0.27$~&~$-0.27$~&~$-0.248$~&~$-0.07$~&~$0.038$~\\

~$\langle P_\tau^{D_s^*} \rangle|_{q^2\in [9,11.6]}$~&~$-0.447$~&~$-0.47$~&~$-0.446$~&~$-0.436$~&~$-0.334$~&~$-0.03$~\\
\hline
\end{tabular}
\end{center}
\end{table}
\thispagestyle{empty}
\begin{table}[!htbp]
\scriptsize
\caption{ Predicted bin-wise values of  longitudinal  polarization asymmetries of daughter meson of $B_{(s)} \to D_{(s)}^{*} \tau \bar \nu_\tau$ and  $B_c^+ \to J/\psi \tau^+  \nu_\tau$ processes in the SM and in the presence of new complex Wilson coefficients (case B).}\label{Tab:FL-FT-CaseB}
\begin{center}
\begin{tabular}{|c|c|c|c|c|c|c|}
\hline
 ~Observables~&~Values for SM~&~values for $V_L$~&~Values for $V_R$~&~values for $V_L$~&~Values for $V_R$~&~Values for $T$~\\
\hline\hline

~$\langle F_L^{D^*} \rangle|_{q^2\in [m_\tau^2,5]}$~&~$(0.629\pm 0.023)$~&~$0.629$~&~$0.63$~&~$0.6359$~&~$0.6843$~&~$0.39$~\\

~$\langle F_L^{D^*} \rangle|_{q^2\in [5,7]}$~&~$(0.5242\pm 0.047)$~&~$0.5242$~&~$0.526$~&~$0.5315$~&~$0.589$~&~$0.368$~\\

~$\langle F_L^{D^*} \rangle|_{q^2\in [7,9]}$~&~$(0.43\pm 0.04)$~&~$0.43$~&~$0.432$~&~$0.436$~&~$0.4871$~&~$0.35$~\\

~$\langle F_L^{D^*} \rangle|_{q^2\in [9,11.6]}$~&~$(0.3654\pm 0.044)$~&~$0.3654$~&~$0.37$~&~$0.368$~&~$0.395$~&~$0.337$~\\
\hline
~$\langle F_L^{J/\psi} \rangle|_{q^2\in [m_\tau^2,5]}$~&~$(0.5777\pm 0.046)$~&~$0.5777$~&~$0.5792$~&~$0.5865$~&~$0.648$~&~$0.442$~\\

~$\langle  F_L^{J/\psi}\rangle|_{q^2\in [5,7]}$~&~$(0.46\pm 0.041)$~&~$0.46$~&~$0.462$~&~$0.469$~&~$0.54$~&~$0.379$~\\

~$\langle  F_L^{J/\psi} \rangle|_{q^2\in [7,9]}$~&~$(0.3748\pm 0.032)$~&~$0.3748$~&~$0.386$~&~$0.38$~&~$0.426$~&~$0.338$~\\

~$\langle  F_L^{J/\psi}\rangle|_{q^2\in [9,11.6]}$~&~$(0.3405\pm 0.041)$~&~$0.3405$~&~$0.342$~&~$0.342$~&~$0.358$~&~$0.33$~\\

\hline

~$\langle F_L^{D_s^*} \rangle|_{q^2\in [m_\tau^2,5]}$~&~$(0.585\pm 0.047)$~&~$0.585$~&~$0.586$~&~$0.593$~&~$0.65$~&~$0.341$~\\

~$\langle F_L^{D_s^*} \rangle|_{q^2\in [5,7]}$~&~$(0.482\pm 0.044)$~&~$0.482$~&~$0.484$~&~$0.49$~&~$0.556$~&~$0.32$~\\

~$\langle F_L^{D_s^*} \rangle|_{q^2\in [7,9]}$~&~$(0.4\pm \pm 0.035)$~&~$0.4$~&~$0.402$~&~$0.41$~&~$0.463$~&~$0.31$~~\\

~$\langle F_L^{D_s^*} \rangle|_{q^2\in [9,11.6]}$~&~$(0.354\pm 0.043)$~&~$0.354$~&~$0.355$~&~$0.3562$~&~$0.384$~&~$0.32$~\\

\hline
\end{tabular}
\end{center}
\end{table}

\subsection{Case C}
\begin{figure}[htb]
\includegraphics[scale=0.38]{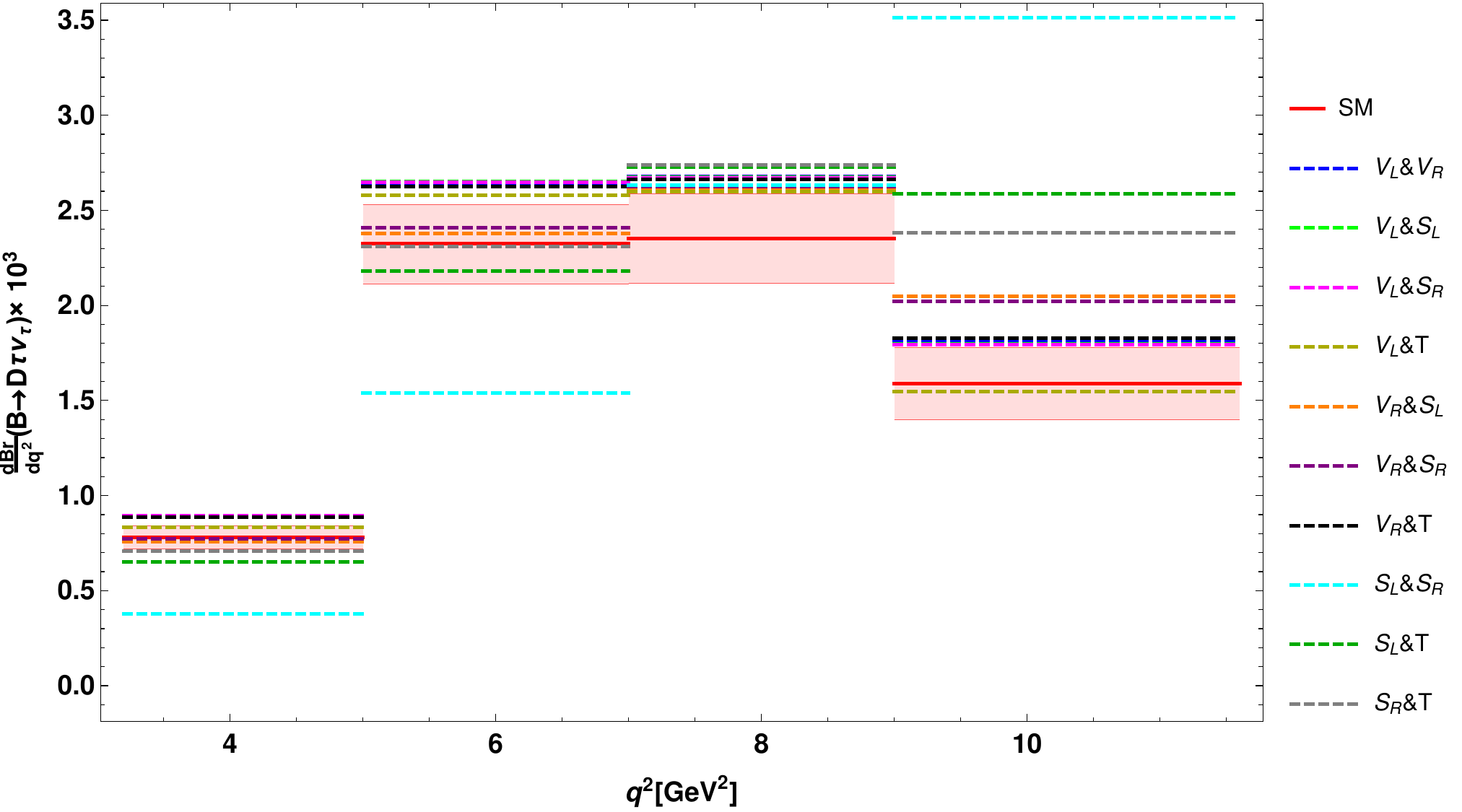}
\quad
\includegraphics[scale=0.38]{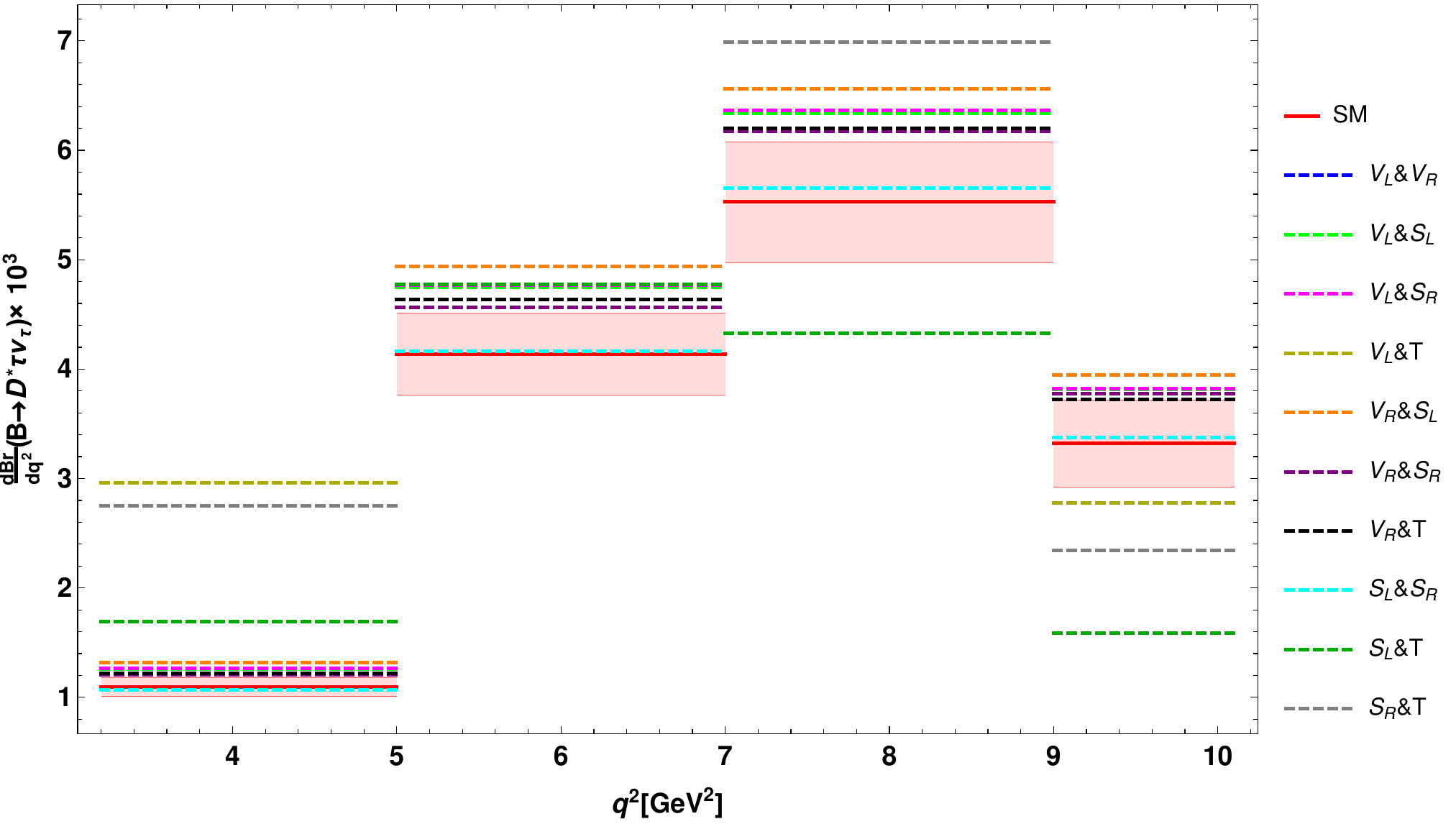}
\quad
\includegraphics[scale=0.38]{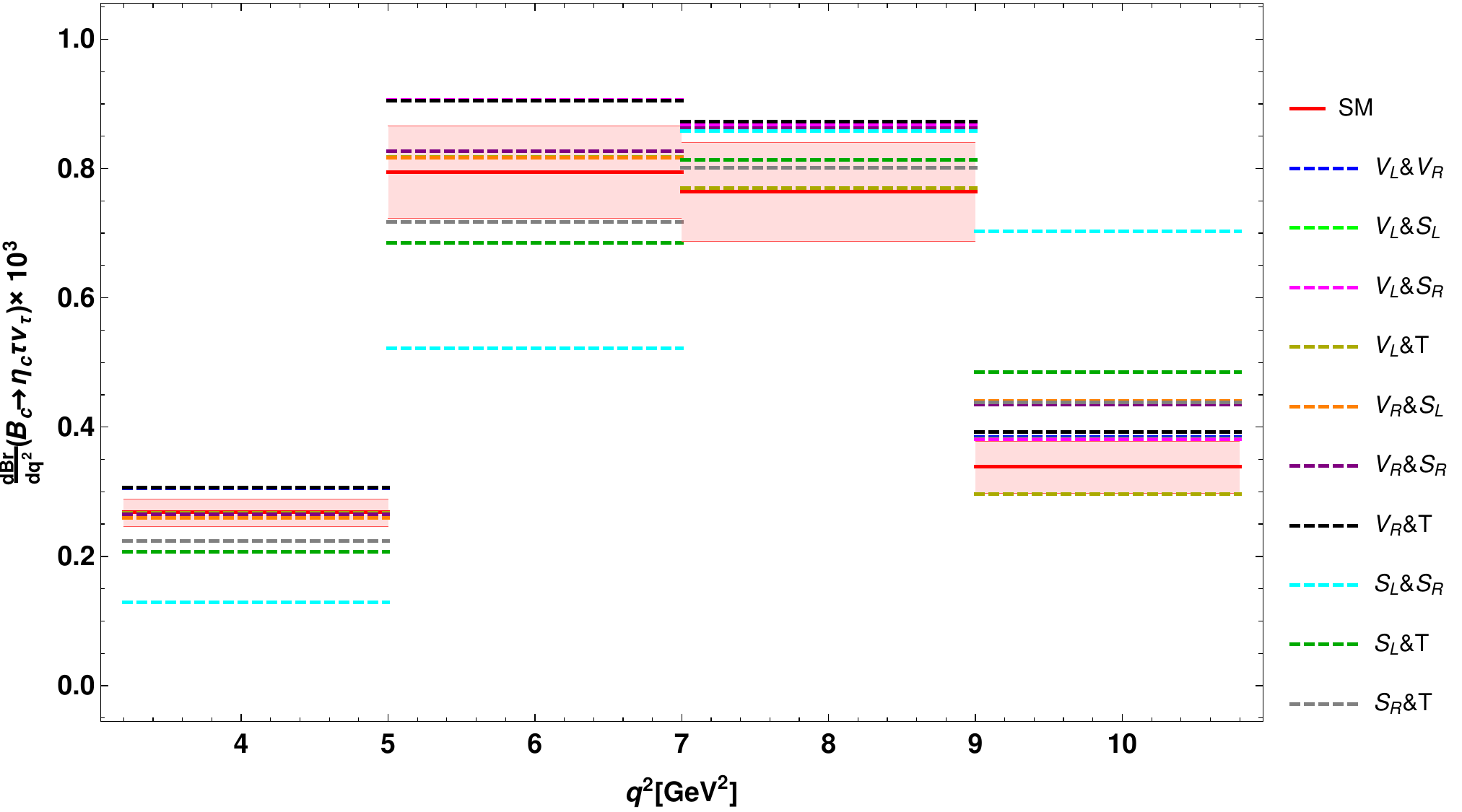}
\quad
\includegraphics[scale=0.38]{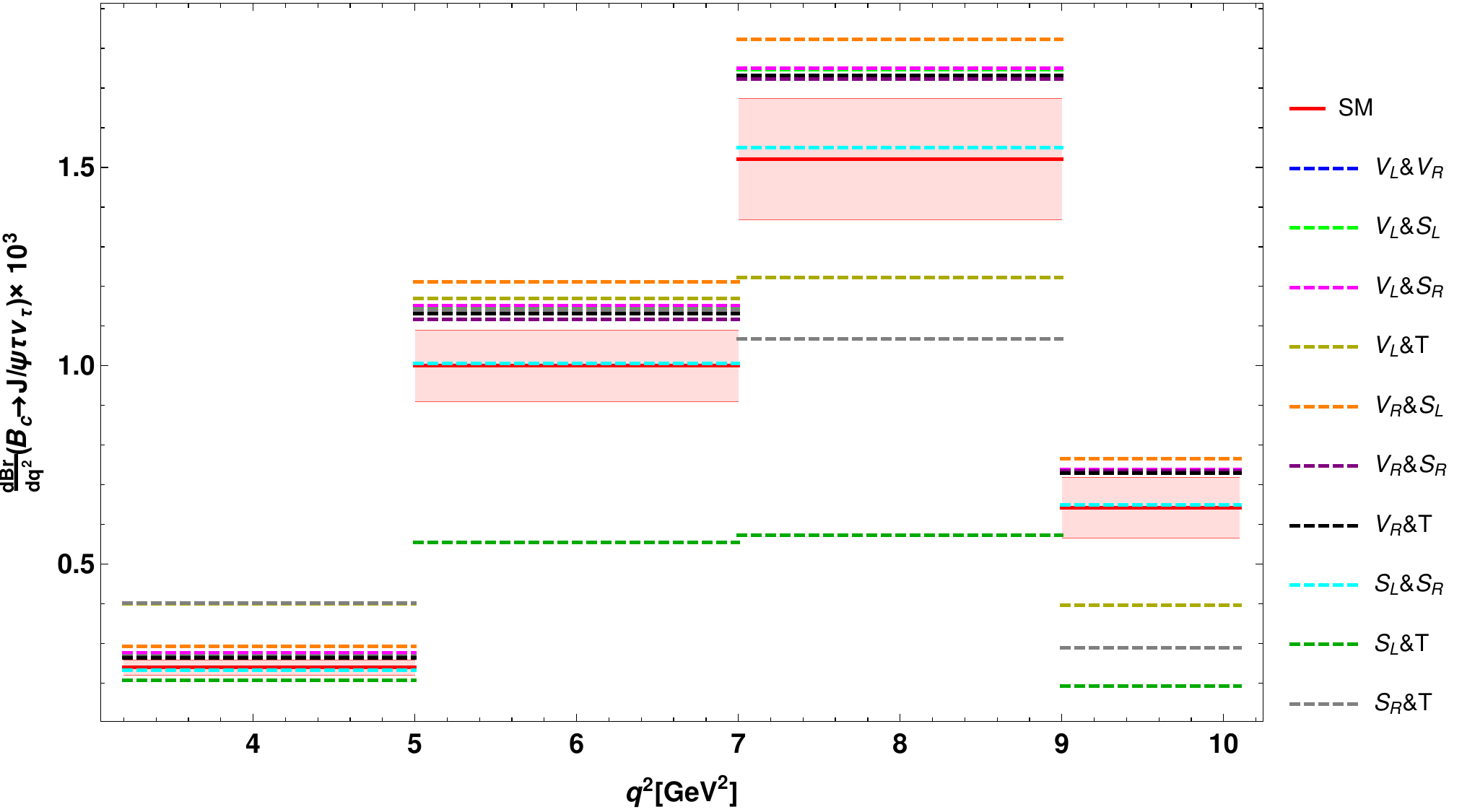}
\quad
\includegraphics[scale=0.38]{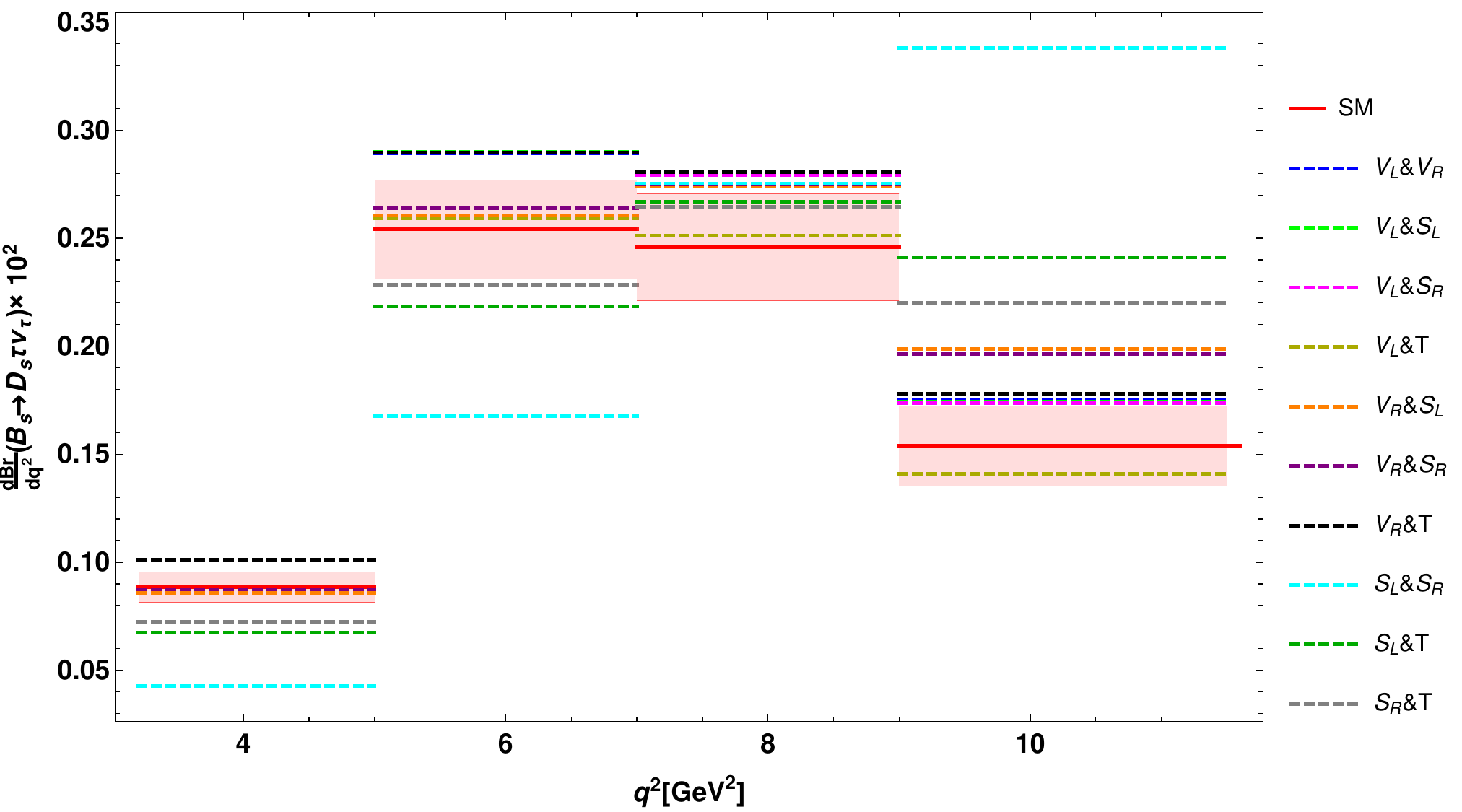}
\quad
\includegraphics[scale=0.38]{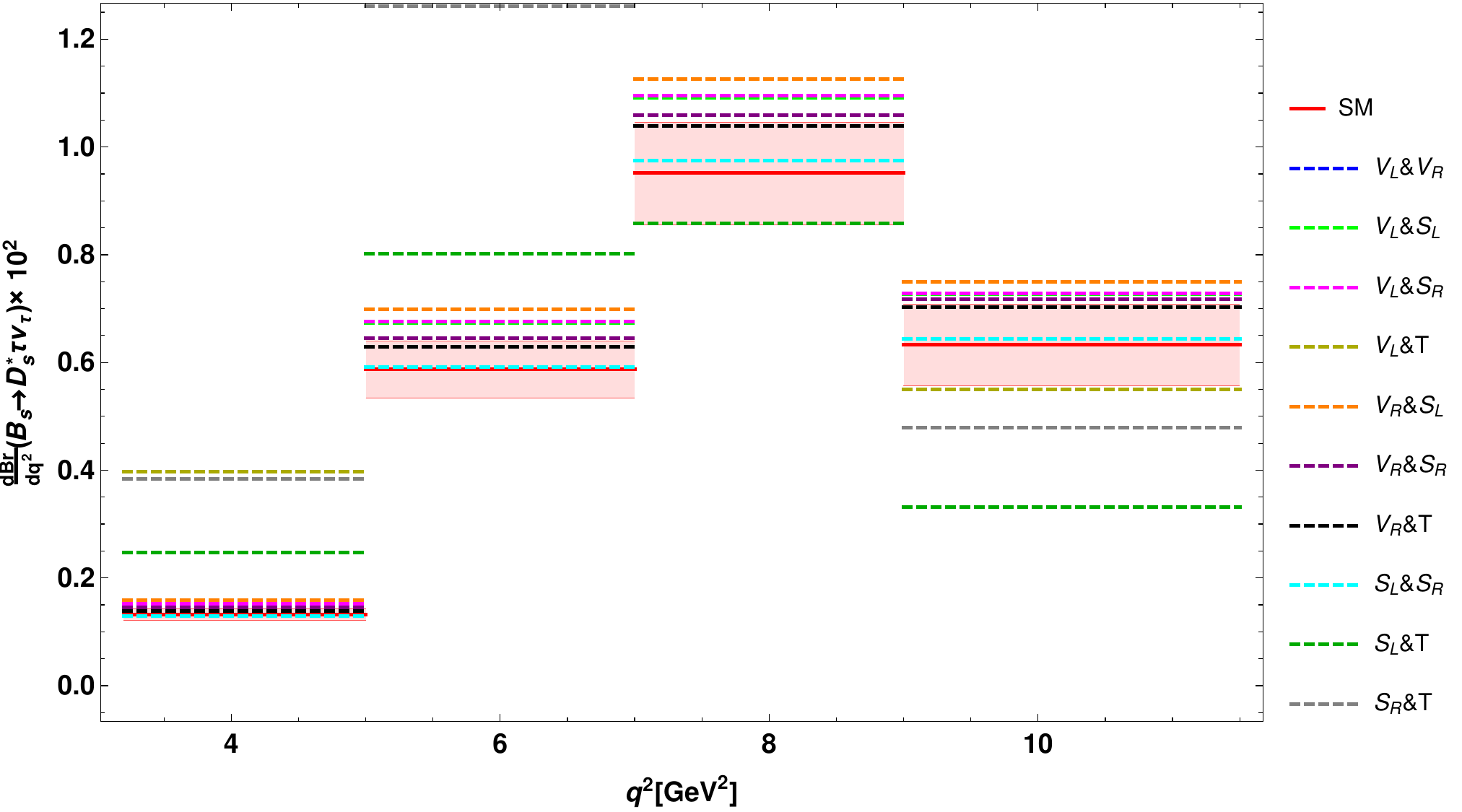}
\caption{ The bin-wise branching ratios of $\bar B \to  D \tau \bar \nu_\tau$ (top-left panel), $\bar B \to  D^* \tau \bar \nu_\tau$ (top-right panel), $B_c^+ \to \eta_c \tau^+  \nu_\tau$ (middle-left panel), $ B_c^+ \to J/\psi \tau^+  \nu_\tau$ (middle-right panel), $ B_s \to D_s \tau \bar \nu_\tau$ (bottom-left panel) and $\bar B_s \to\bar D_s^* \tau \bar \nu_\tau$ (bottom-right panel) processes in four $q^2$ bins for case C.}\label{Fig:CC-BR}
\end{figure}

In this subsection, we discuss the implications of two different types of real Wilson coefficients on the branching fractions, LNU ratios and angular observables of $ B_{(s)} \to D_{(s)}^{(*)} \tau \bar \nu_\tau$ and  $B_c^+ \to (\eta_c, J/\psi) \tau^+  \nu_\tau$  decay processes. We consider  $10$ possible combinations of new coefficients such as $(V_L, V_R)$, $(V_L, S_L)$, $(V_L, S_R)$, $(V_L, T)$, $(V_R, S_L)$, $(V_R, S_R)$, $(V_R, T)$, $(S_L, S_R)$, $(S_L, T)$ and $(S_R, T)$, whose best-fit values are presented in Table \ref{Tab:Best-fit}\,. Using the best-fit values, the branching ratios of  $\bar B \to  D \tau \bar \nu_\tau$ (top-left panel), $\bar B \to  D^* \tau \bar \nu_\tau$ (top-right panel), $B_c^+ \to \eta_c \tau^+  \nu_\tau$ (middle-left panel), $B_c^+ \to J/\psi \tau^+  \nu_\tau$ (middle-right  panel), $B_s \to D_s \tau \bar \nu_\tau$ (bottom-left panel) and $B_s \to D_s^* \tau \bar \nu_\tau$ (bottom-right panel)  processes in four $q^2$ bin for all possible set of coefficients are presented in Fig. \ref{Fig:CC-BR}\,. Here blue, green, magenta, dark yellow represents the $V_L\& V_R$, $V_L\&S_L$, $V_L\& S_R$, $V_L\& T$ sets respectively. The orange, purple, black colors are respectively stand for $V_R\& S_L$, $V_R\& S_R$, $V_R\& T$  and cyan, dark green, gray are for $S_L\& S_R$, $S_L\& T$, $S_R\& T$ combinations, respectively. In the first bin of Br($\bar B \to  D \tau \bar \nu_\tau$), the  $V_R\& S_L$, $V_R\& S_R$ and $S_R\& T$ have negligible deviation, slight deviation has been found for the sets $V_L\& V_R$, $V_L\& S_L$, $V_L\& S_R$, $V_L\& T$,  $V_R\& T$ and $S_L \& T$ and significant deviation from the SM predictions are noticed for the  $S_L\& S_R$ combination of Wilson coefficients. The $V_R\& S_L$ and $V_R\& S_R$ provide no deviation, whereas $V_L\& V_R$, $V_L\& S_L$,  $V_L\& S_R$,  $V_L\& T$ and   $S_L\& S_R$ give significant deviations from the SM results in the second bin and the $S_L \& S_R$ show maximum deviation. In the $3^{\rm rd}$ bin, all combinations of real coefficients provide significant contributions. Except $V_L\& T$, all the sets of Wilson coefficients provide large deviation in the last $q^2$ bin. The branching ratio of $\bar B \to D^* \tau \bar \nu_\tau$  is found to receive significant new physics contribution from all possible combinations of coefficients except $S_L\& S_R$ in all the $q^2$ bins. The  $V_R\& T$, $S_L\& S_R$, $S_L\& T$ and $S_R\& T$ sets provide  deviation (deviation is maximum for $S_L\& S_R$) in the first two bins of the branching ratios of $B_c \to \eta_c \tau \bar \nu_l$. Except  $V_L\& T$, $S_L\& T$ and $S_R\& T$, the Br($B_c \to \eta_c$) in $3^{\rm rd}$ bin deviate significantly due to the presence of rest combinations of Wilson coefficients.  All the sets show significant deviation in the last bin and the maximum deviation is observed for the  $S_L\& S_R$ set.  The bin-wise predictions on the branching ratio of $B_s \to D_s \tau \bar \nu_\tau$ behave almost similarly as $\bar B \to \bar D$. 

\begin{figure}[htb]
\includegraphics[scale=0.38]{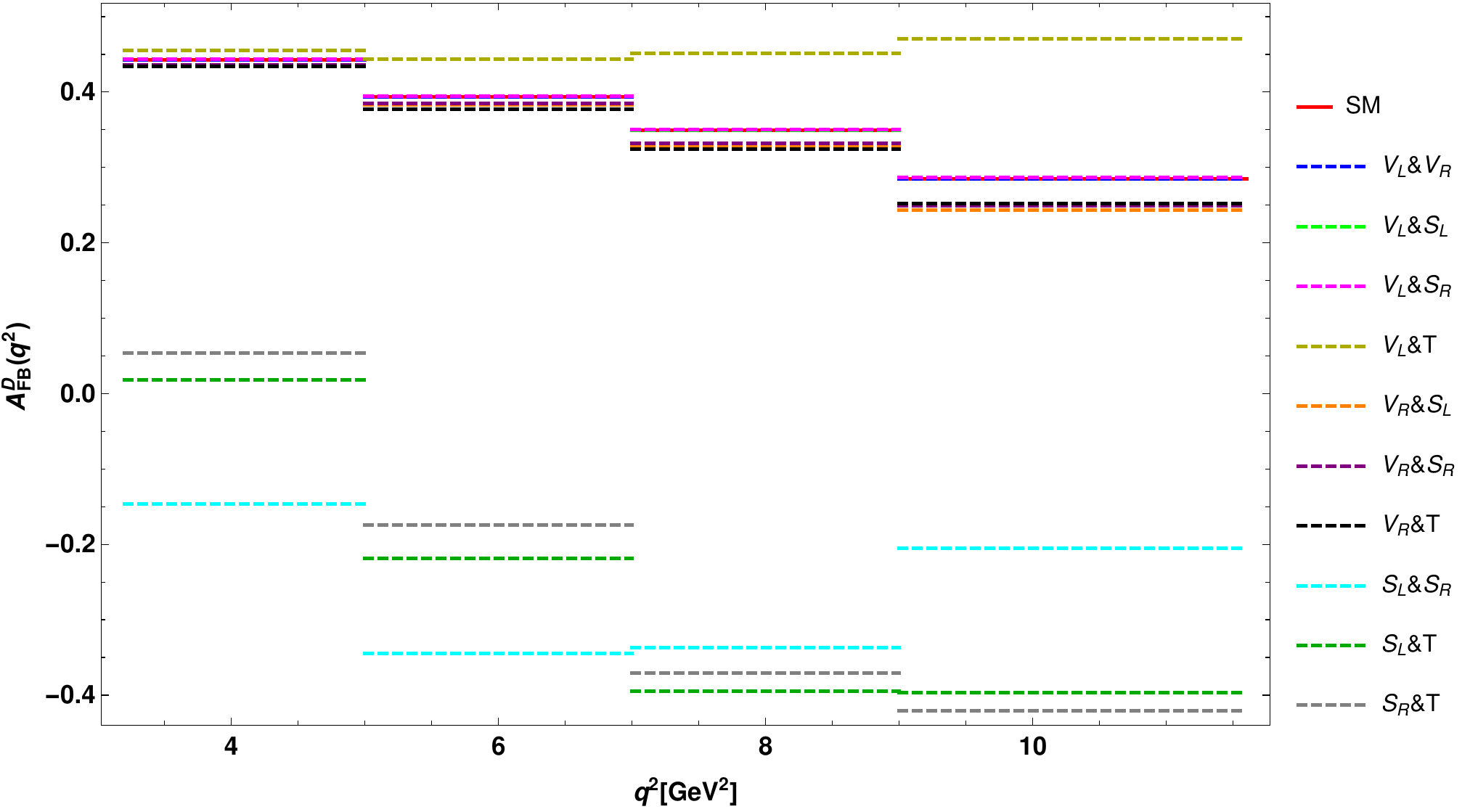}
\quad
\includegraphics[scale=0.38]{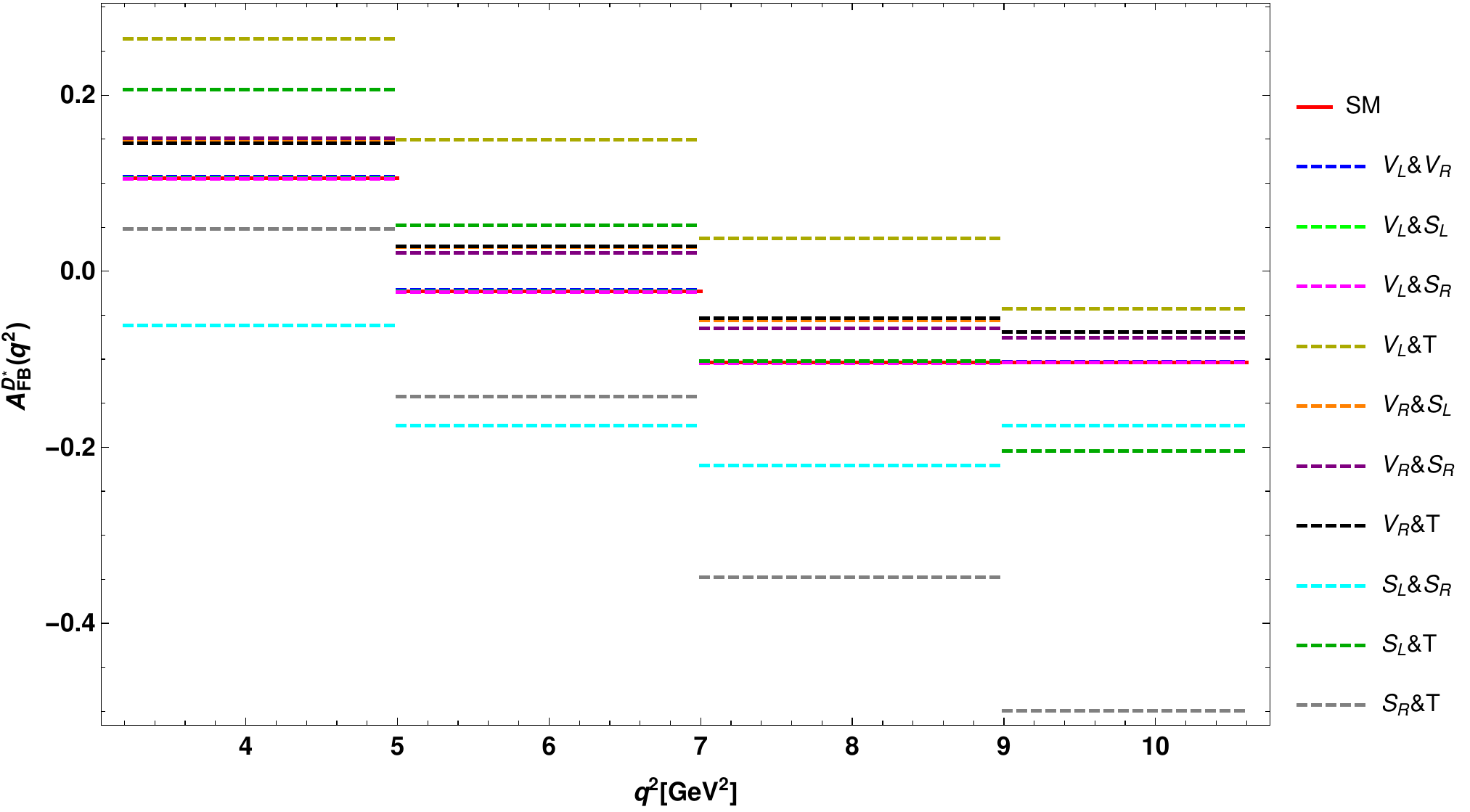}
\quad
\includegraphics[scale=0.38]{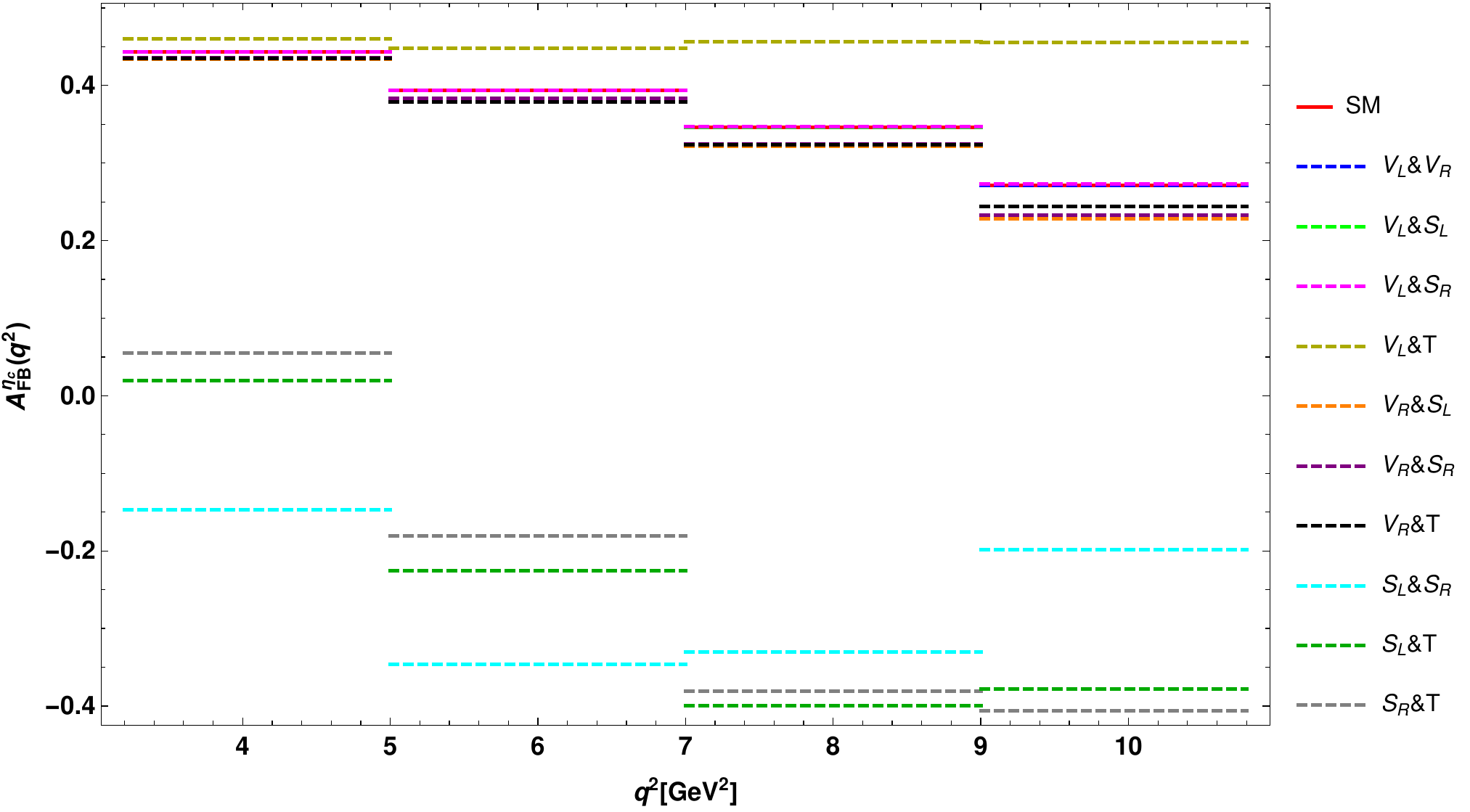}
\quad
\includegraphics[scale=0.38]{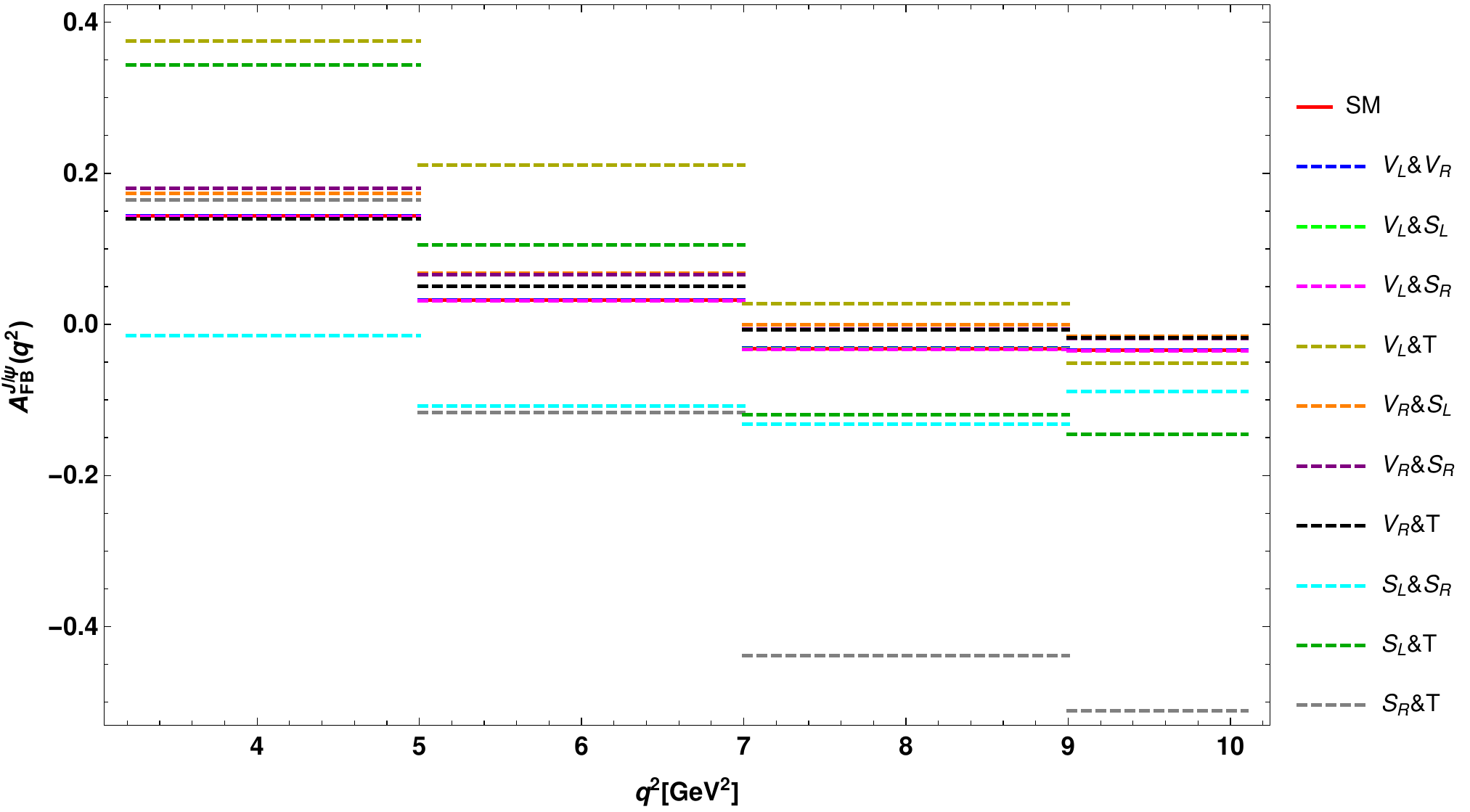}
\quad
\includegraphics[scale=0.38]{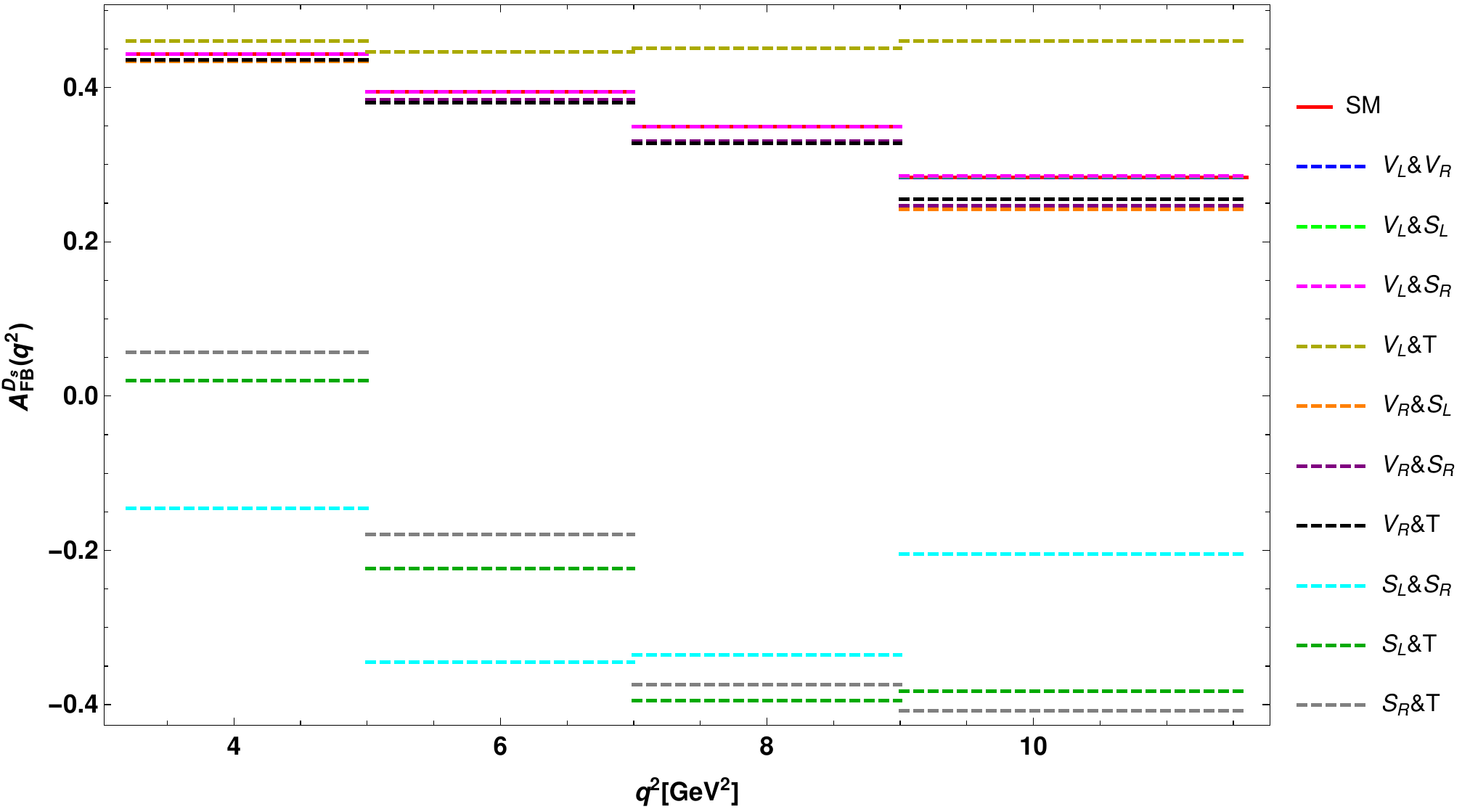}
\quad
\includegraphics[scale=0.38]{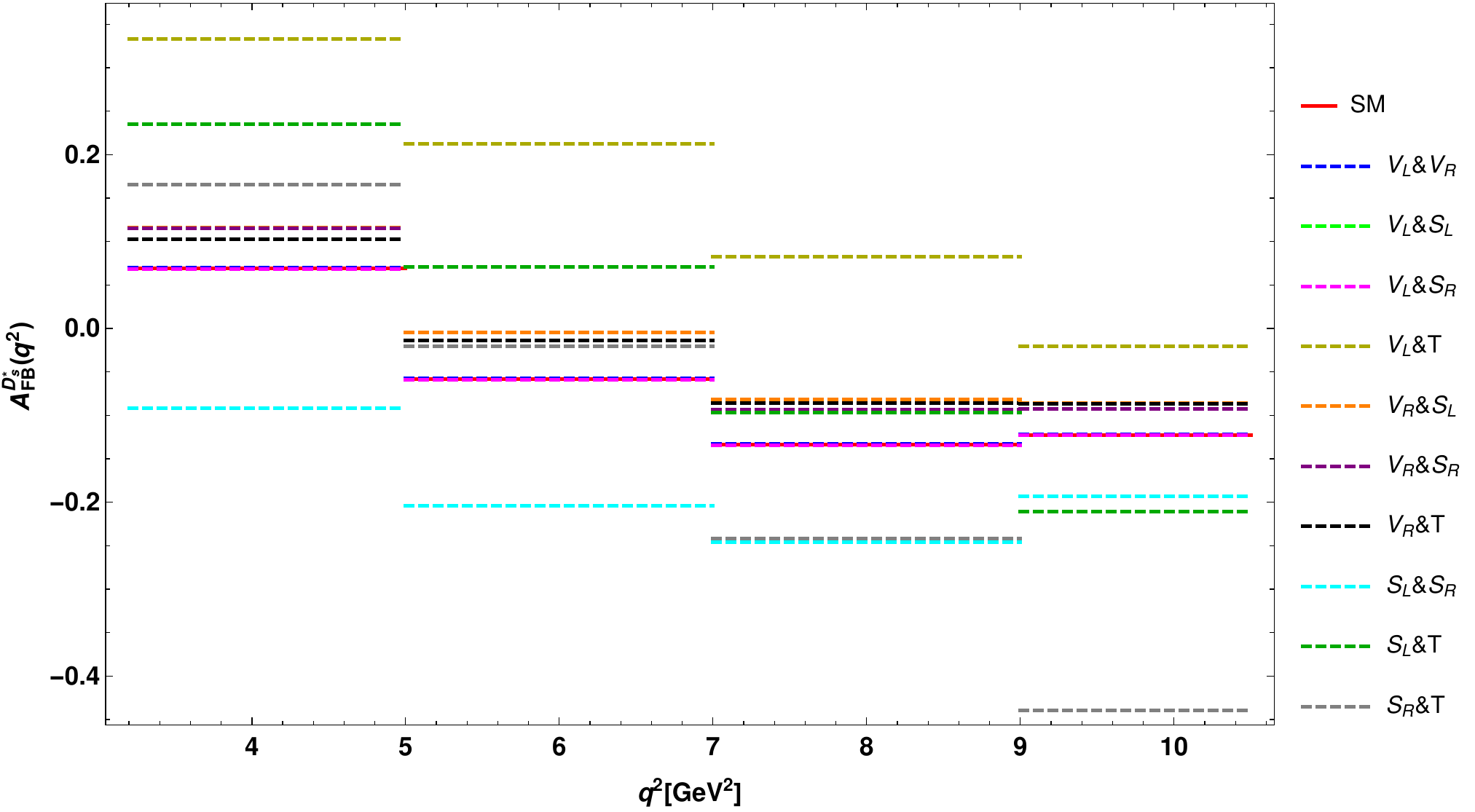}
\caption{ The bin-wise forward-backward asymmetry of  $\bar B \to  D \tau \bar \nu_\tau$ (top-left panel), $\bar B \to  D^* \tau \bar \nu_\tau$ (top-right panel), $B_c^+ \to \eta_c \tau^+  \nu_\tau$ (middle-left panel), $ B_c^+ \to J/\psi \tau^+  \nu_\tau$ (middle-right panel), $ B_s \to D_s \tau \bar \nu_\tau$ (bottom-left panel) and $ B_s \to D_s^* \tau \bar \nu_\tau$ (bottom-right panel) processes in four $q^2$ bins for case C. }\label{Fig:CC-AFB}
\end{figure}

In Fig. \ref{Fig:CC-AFB}\,, we show the bin-wise forward-backward asymmetry values of  $\bar B \to  D \tau \bar \nu_\tau$ (top-left panel), $\bar B \to  D^* \tau \bar \nu_\tau$ (top-right panel), $B_c^+ \to \eta_c \tau^+ \nu_\tau$ (middle-left panel), $B_c^+ \to J/\psi \tau^+  \nu_\tau$ (middle-right  panel), $B_s \to D_s \tau \bar \nu_\tau$ (bottom-left panel) and $B_s \to D_s^* \tau \bar \nu_\tau$ (bottom-right panel)  processes. We observe that the  $S_L\& S_R$, $S_L \& T$ and $S_R \& T$ coefficients provide significant deviation from the SM predictions of forward-backward asymmetry of $B_{(s)} \to D_{(s)}$ and $B_c^+ \to \eta_c$ in all $q^2$ bins, however  $V_L \& T$ affect the $A_{FB}$ parameter in the last bins. Except the $V_L\& V_R$, $V_L \& S_L$ and $V_L \& S_R$, all possible coefficients combinations shift the  $A_{FB}$ of $B_{(s)} \to  D_{(s)}^*$ and $B_c^+ \to J/\psi$ decay modes. The $S_L \& T$ set has no effect on the forward-backward asymmetry of  $\bar B \to D^*$ in the third bin. 

\begin{figure}[htb]
\includegraphics[scale=0.38]{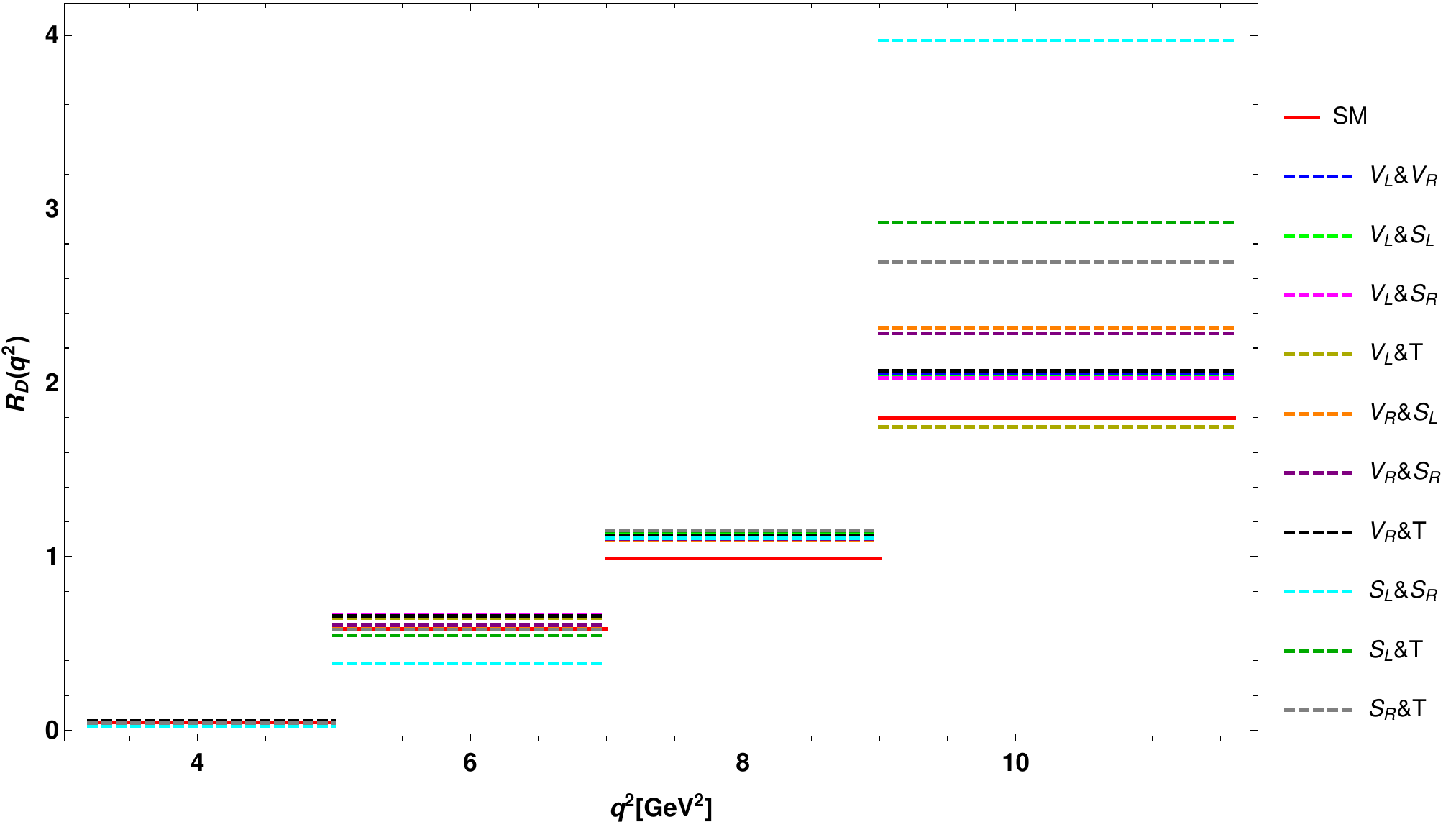}
\quad
\includegraphics[scale=0.38]{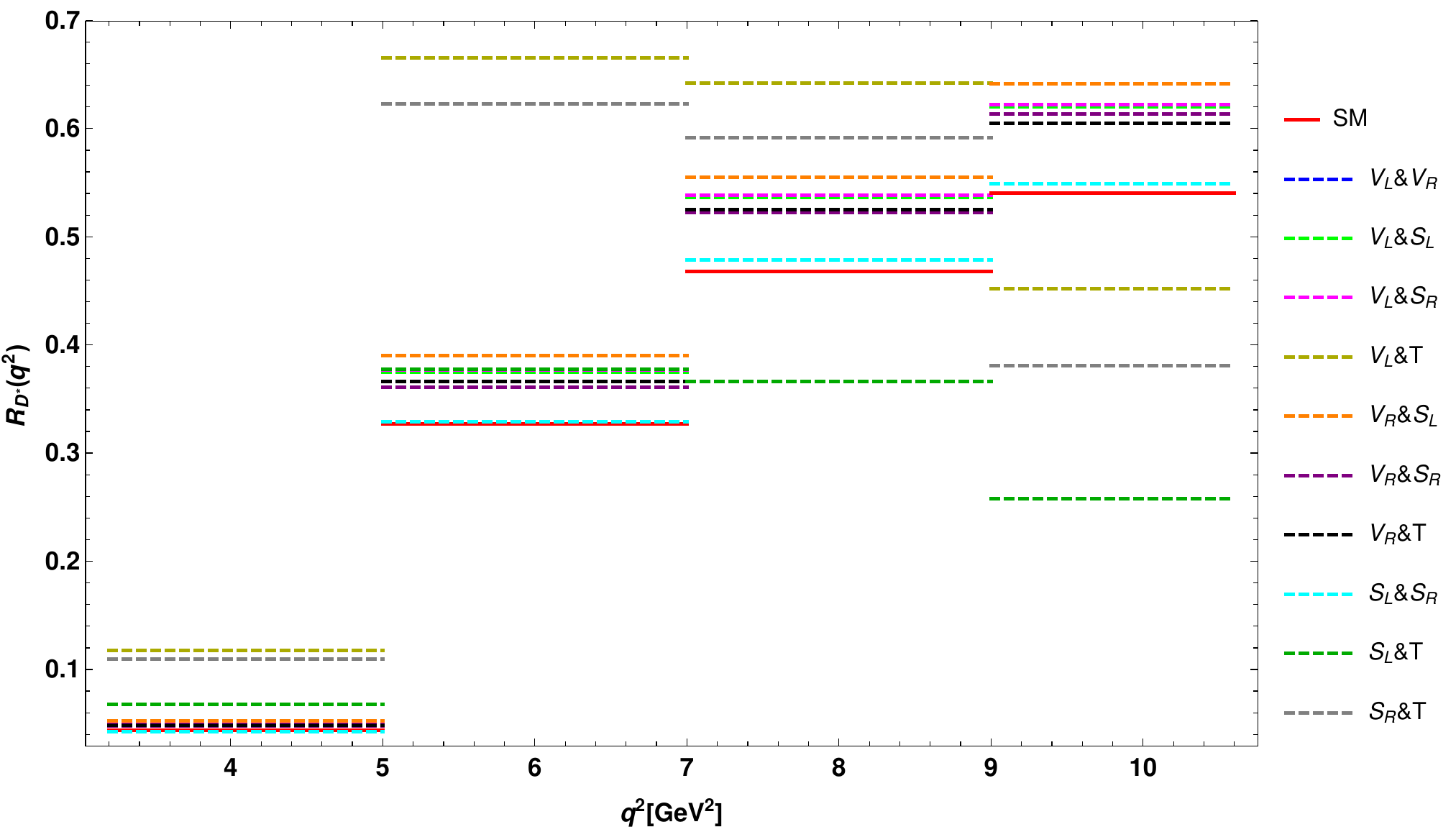}
\quad
\includegraphics[scale=0.38]{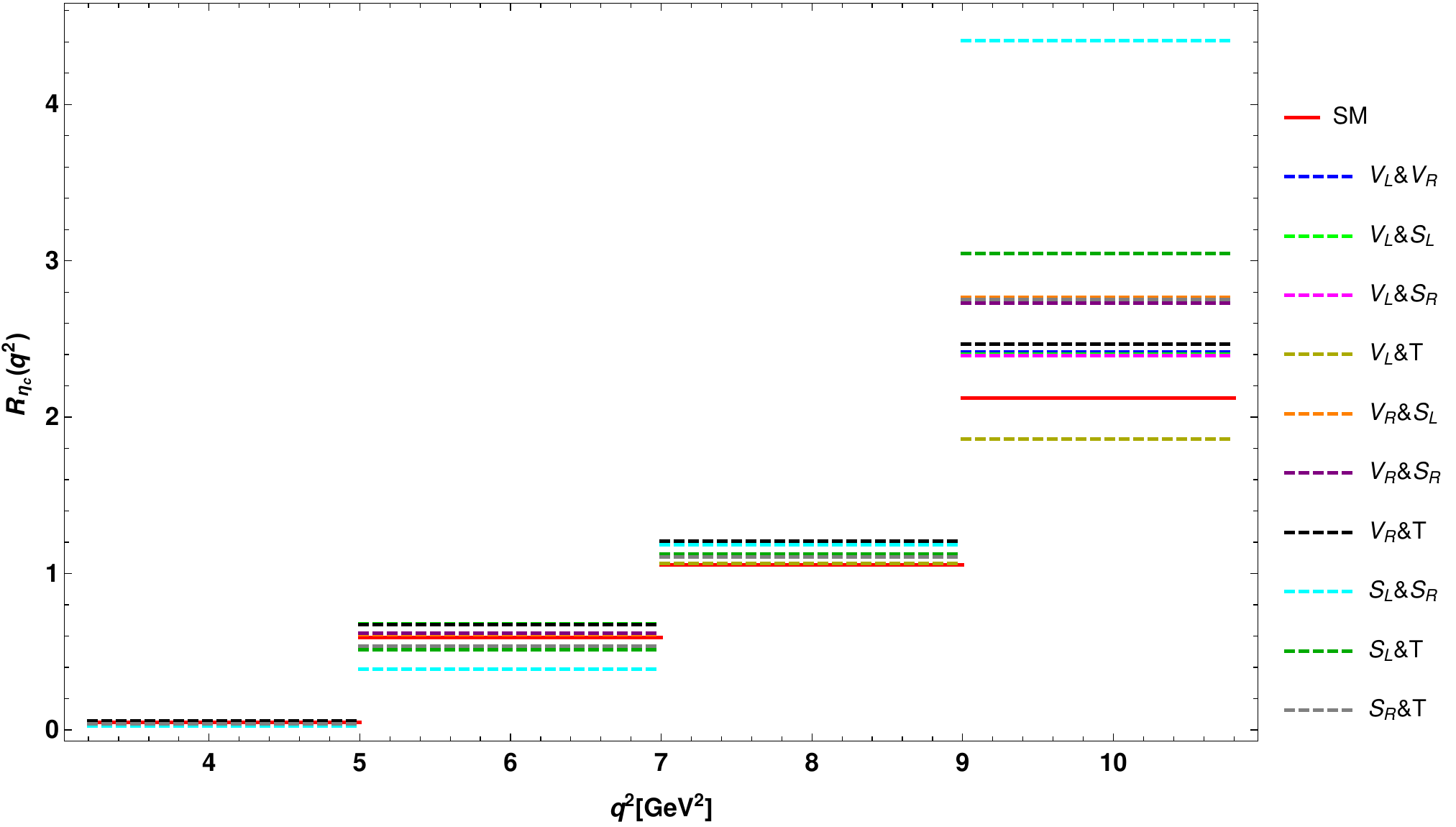}
\quad
\includegraphics[scale=0.38]{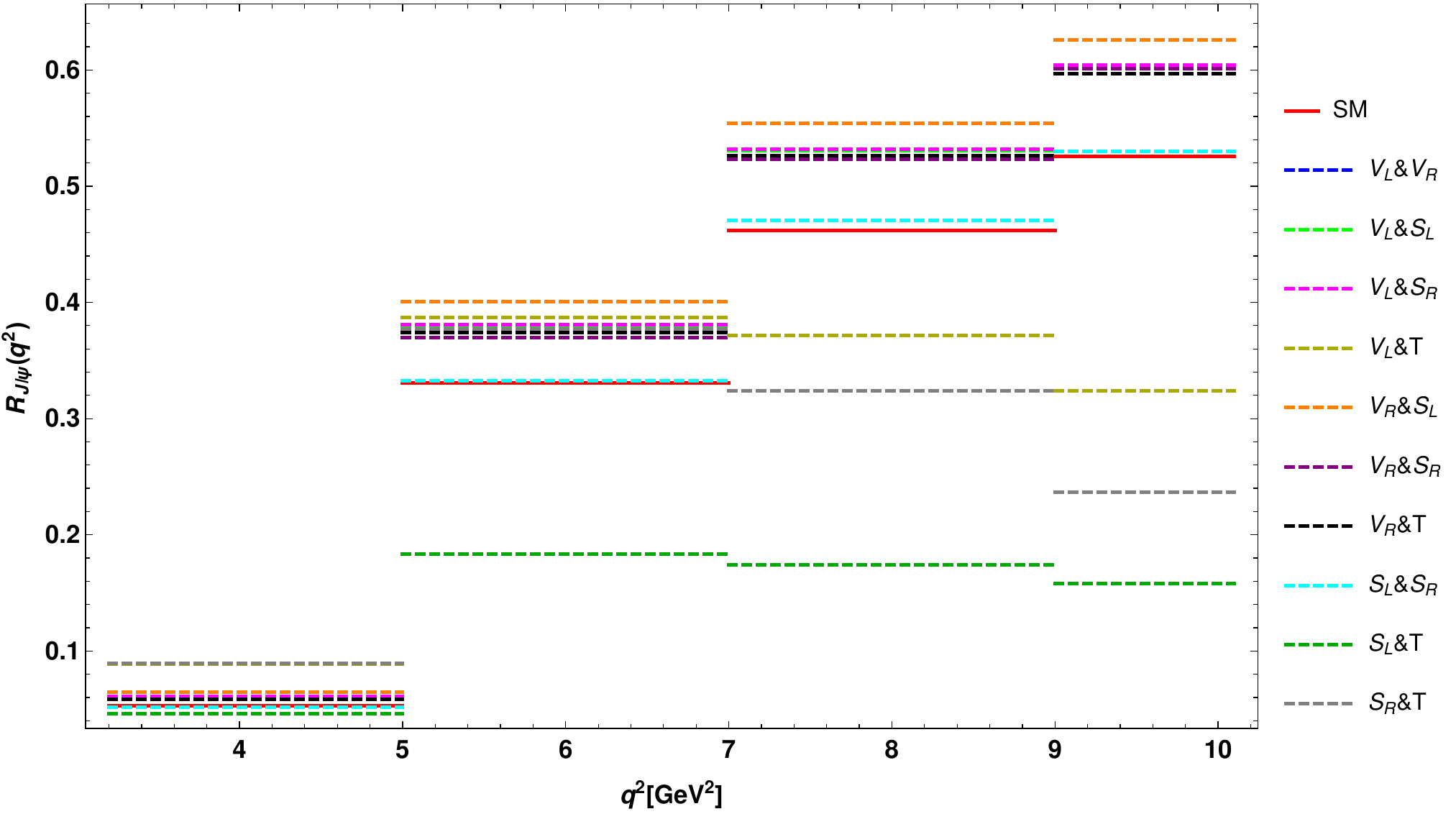}
\quad
\includegraphics[scale=0.38]{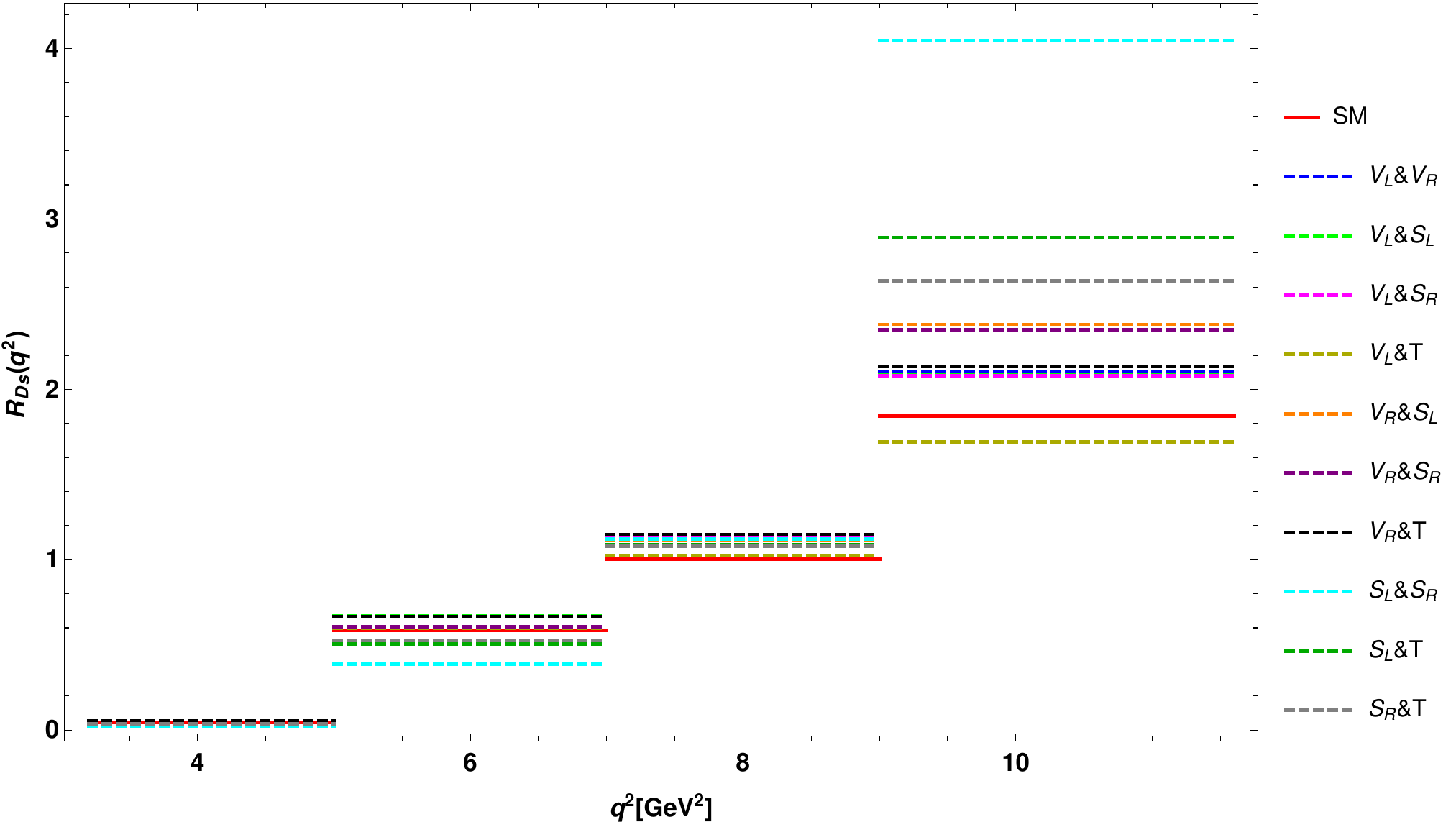}
\quad
\includegraphics[scale=0.38]{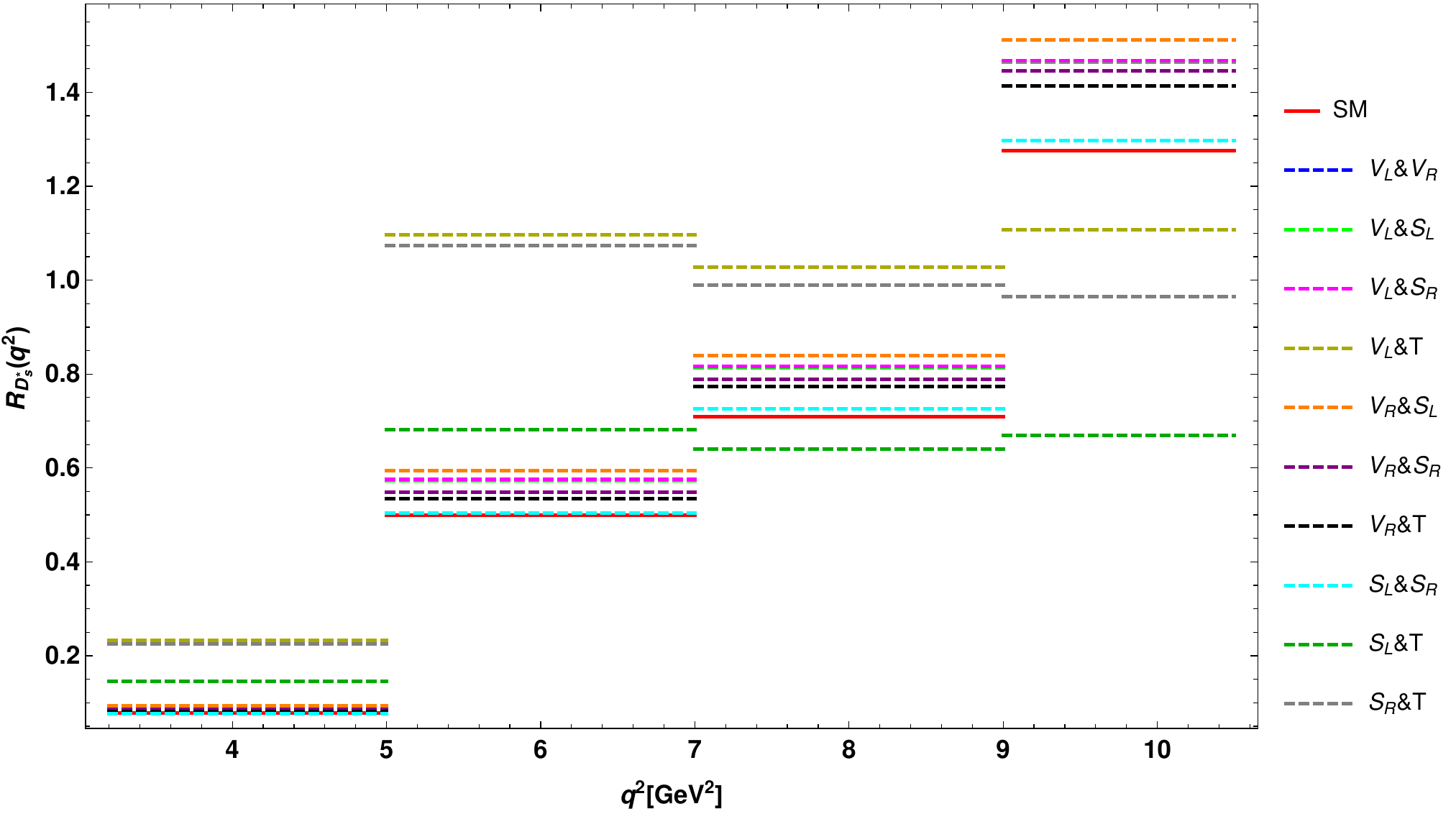}
\caption{ The bin-wise  $R_D$ (top-left panel), $R_{D^*}$ (top-right), $R_{\eta_c}$ (middle-left panel), $R_{J/\psi}$ (middle-right), $R_{D_s}$ (bottom-left panel) and $R_{D_s^*}$ (bottom-right panel)  in four $q^2$ bins for case C. }\label{Fig:CC-LNU}
\end{figure}

Fig. \ref{Fig:CC-LNU} depicts the bin-wise values of lepton non-universality observables of   $\bar B \to  D \tau \bar \nu_\tau$ (top-left panel), $\bar B \to  D^* \tau \bar \nu_\tau$ (top-right panel), $B_c^+ \to \eta_c \tau^+  \nu_\tau$ (middle-left panel), $B_c^+ \to J/\psi \tau^+  \nu_\tau$ (middle-right  panel), $B_s \to D_s \tau \bar \nu_\tau$ (bottom-left panel) and $B_s \to D_s^* \tau \bar \nu_\tau$ (bottom-right panel)  decay modes. All the sets of combinations have significant effects on the $R_{D_{(s)}}$, $R_{\eta_c}$ parameters in the last $q^2$ bin except  $V_L\& T$, which provides negligible deviation.  The $V_L\& T$ and $S_R\& T$ coefficients deviates significantly from SM results of $R_{D_{(s)}}$ ($R_{J/\psi}$) in the last three (two) bins.  The $S_L\& T$ coefficient show maximum deviation in $R_V$ ratios. 
\begin{figure}[htb]
\includegraphics[scale=0.38]{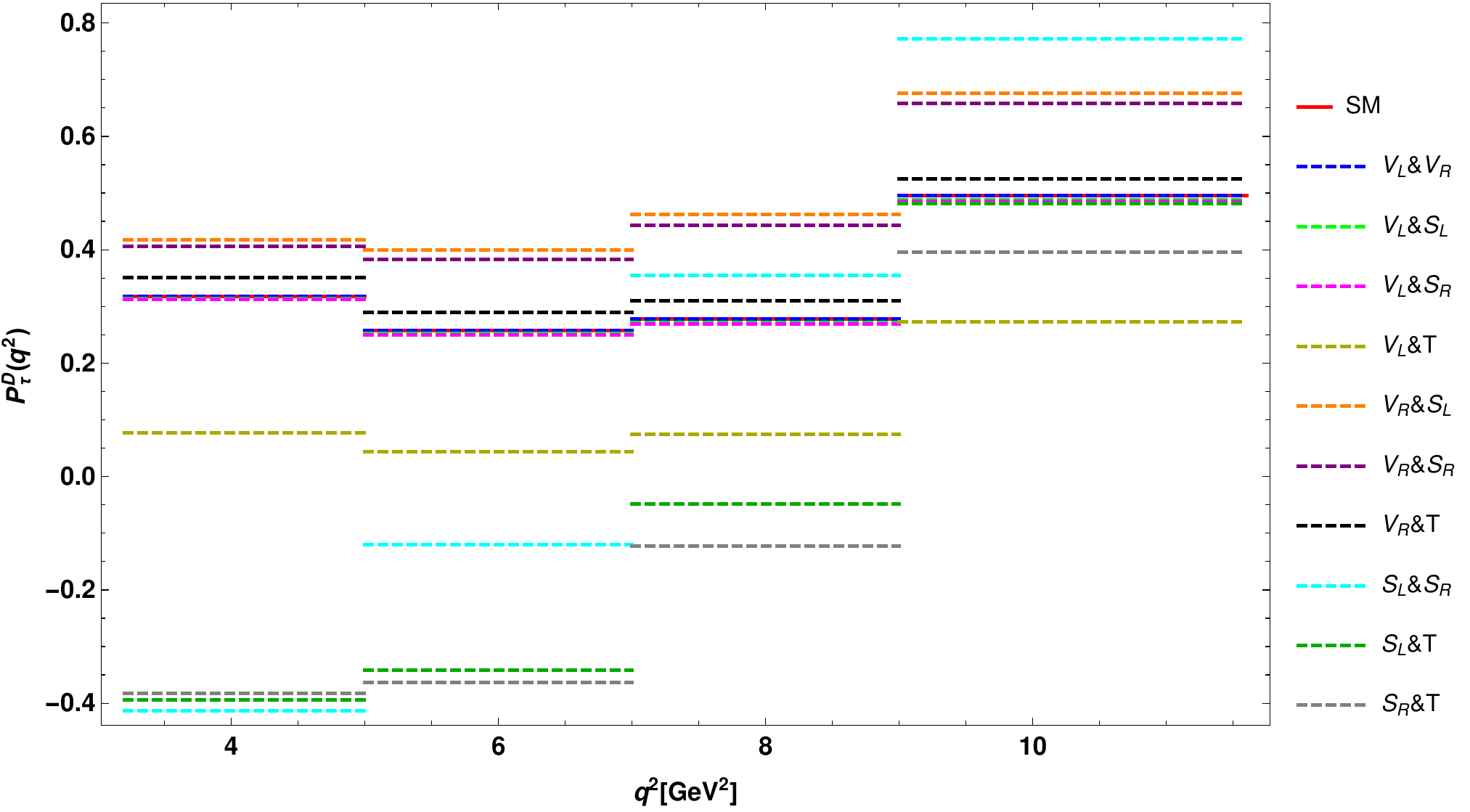}
\quad
\includegraphics[scale=0.38]{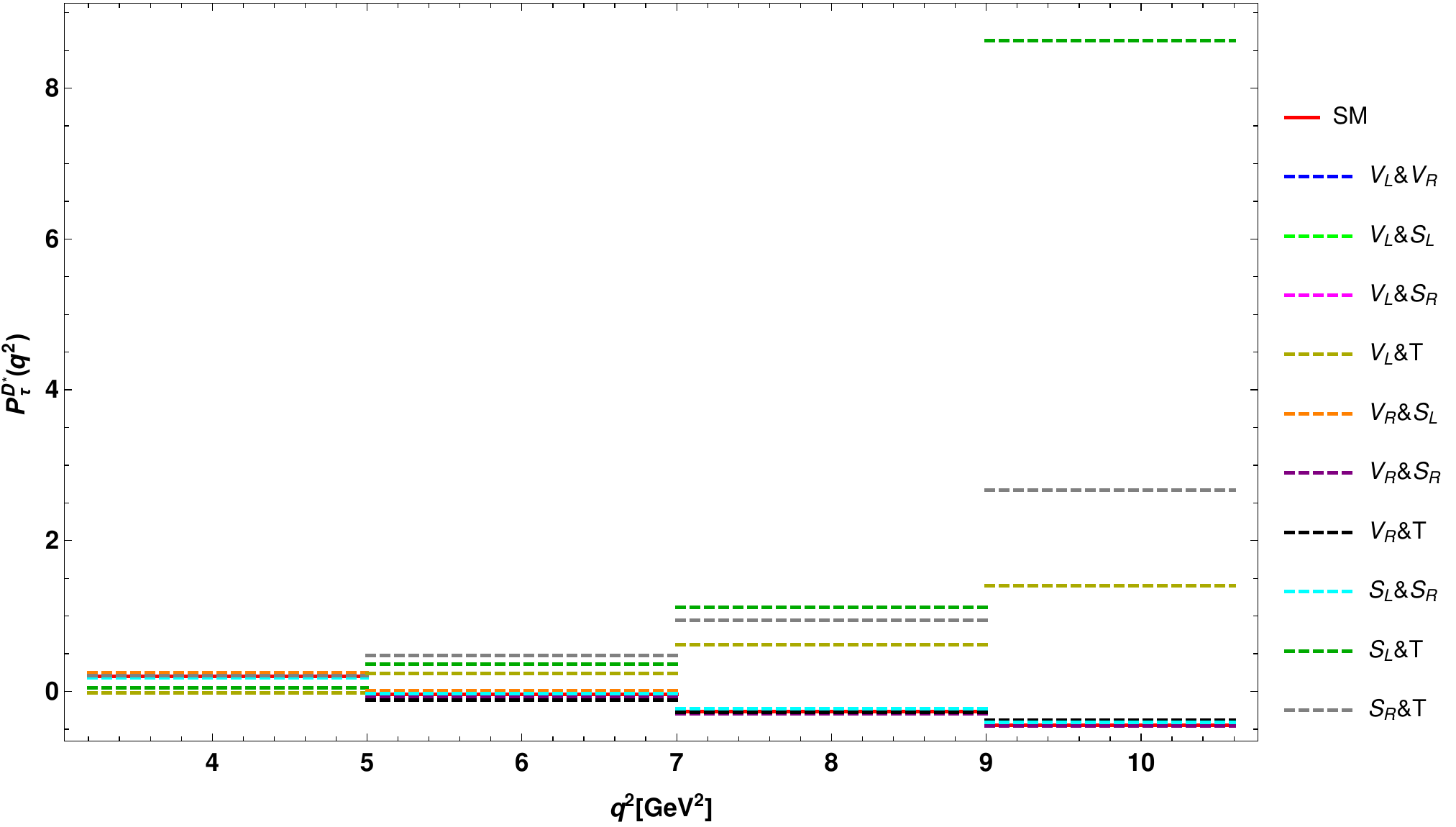}
\quad
\includegraphics[scale=0.38]{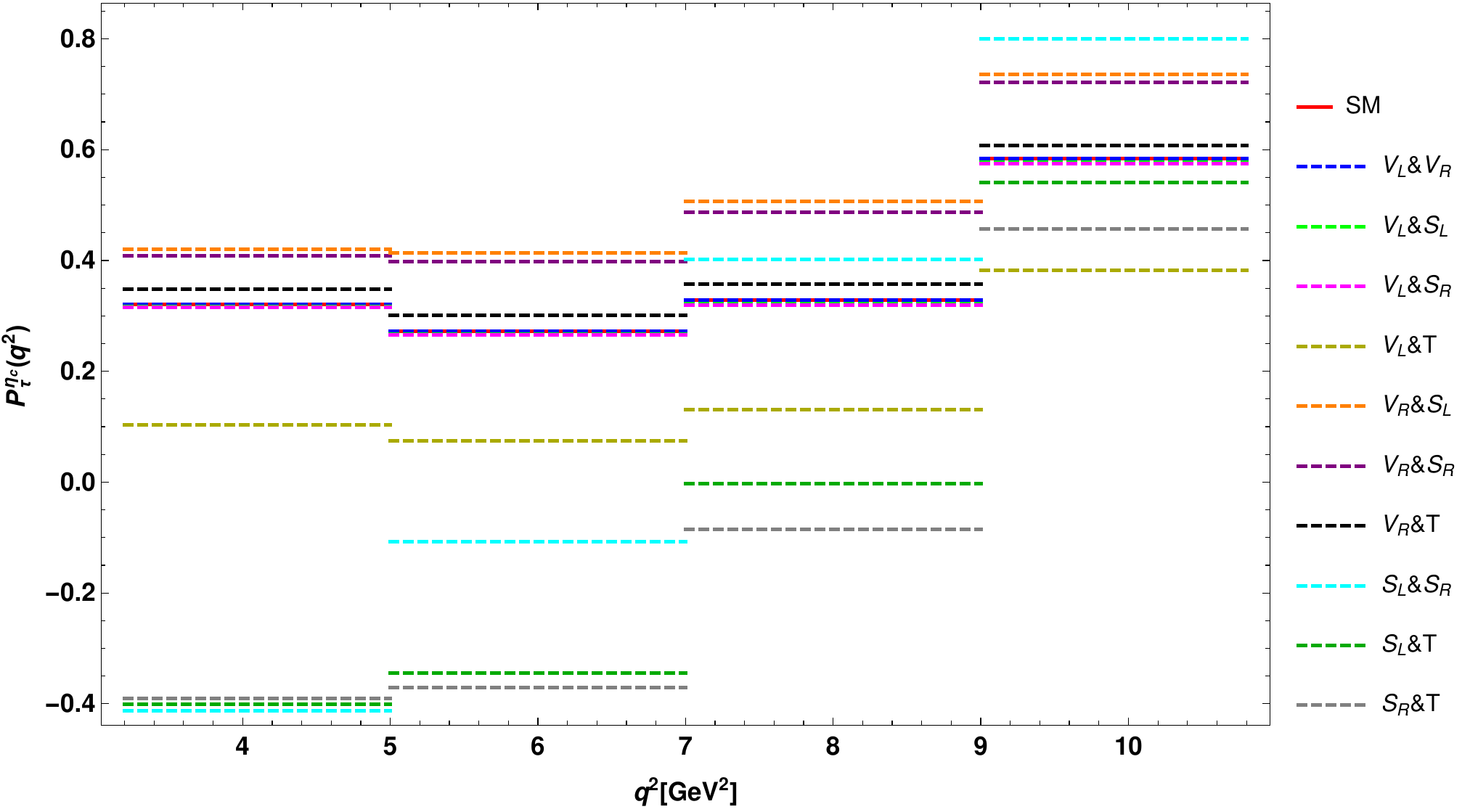}
\quad
\includegraphics[scale=0.38]{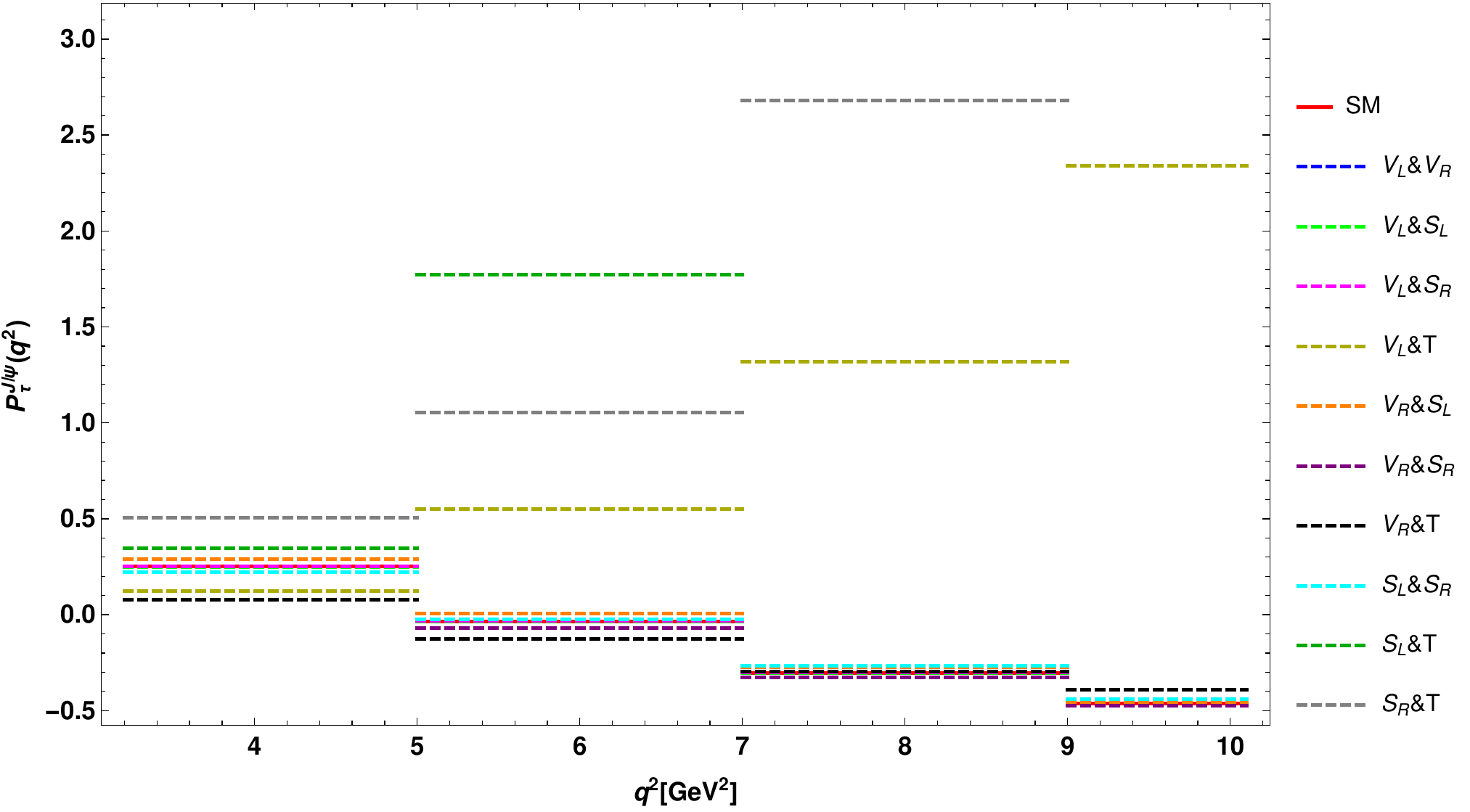}
\quad
\includegraphics[scale=0.38]{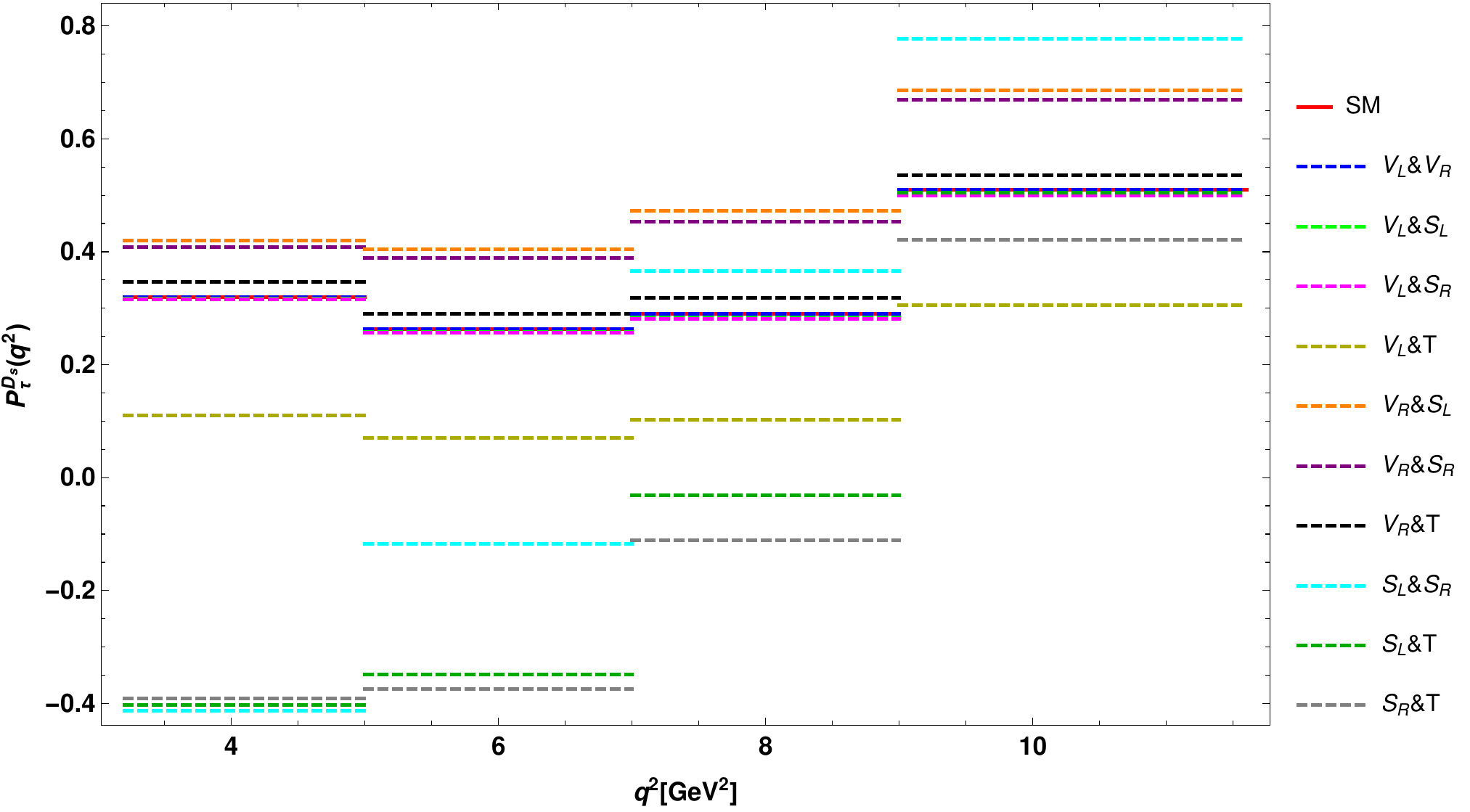}
\quad
\includegraphics[scale=0.38]{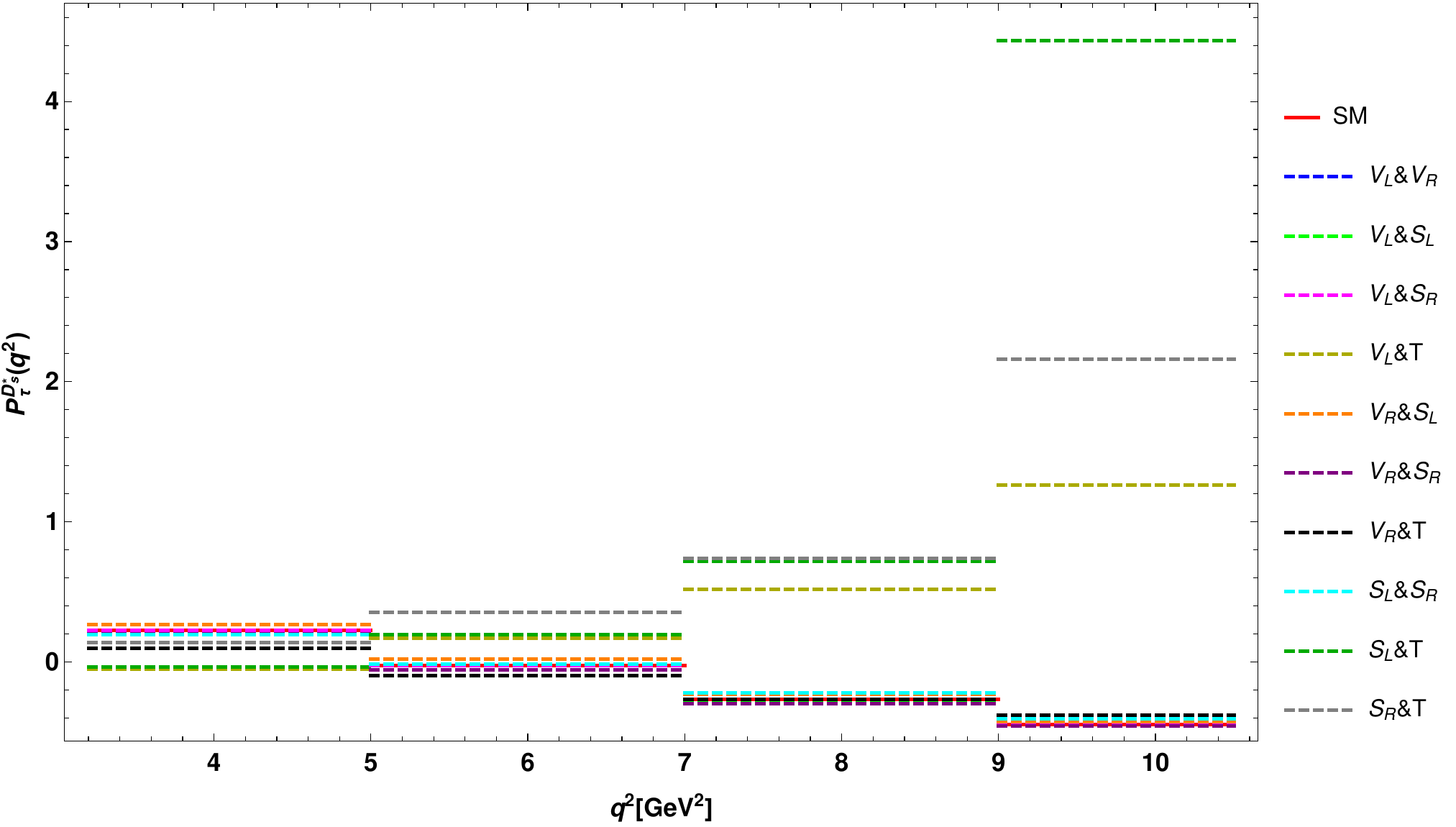}
\caption{ The bin-wise $\tau$-polarization asymmetry of $\bar B \to D \tau \bar \nu_\tau$ (top-left panel), $\bar B \to D^* \tau \bar \nu_\tau$ (top-right panel), $B_c^+ \to \eta_c \tau^+  \nu_\tau$ (middle-left panel), $ B_c^+ \to J/\psi \tau^+  \nu_\tau$ (middle-right panel), $B_s \to  D_s \tau \bar \nu_\tau$ (bottom-left panel) and $B_s \to D_s^* \tau \bar \nu_\tau$ (bottom-right panel) processes in four $q^2$ bins for case C. }\label{Fig:CC-ptau}
\end{figure}

The graphical representation of tau polarization asymmetry of all these decay modes are presented in Fig. \ref{Fig:CC-ptau}\,. For all $B \to P$ decay modes, the $V_L \& V_R$, $V_L \& S_L$ and $V_L\& S_R$ scenario provide no deviation from the SM predictions in all the $q^2$ bins and the $S_L\& T$ set has negligible effect on the $1^{\rm st}$ bin of $P_\tau^{D_{(s)}}$.   The impact of  $S_{L(R)}\& T$ and $V_L\& T$ scenarios on $\tau$-polarization asymmetries of $B \to V$ processes are far-reaching. Similarly, all possible combinations of Wilson coefficients have larger impact on $V$ polarization asymmetries of $B \to V$, except the $V_L\& V_R$, $V_L \& S_L$ and $V_L \& S_R$ scenarios. 
\begin{figure}[htb]
\includegraphics[scale=0.38]{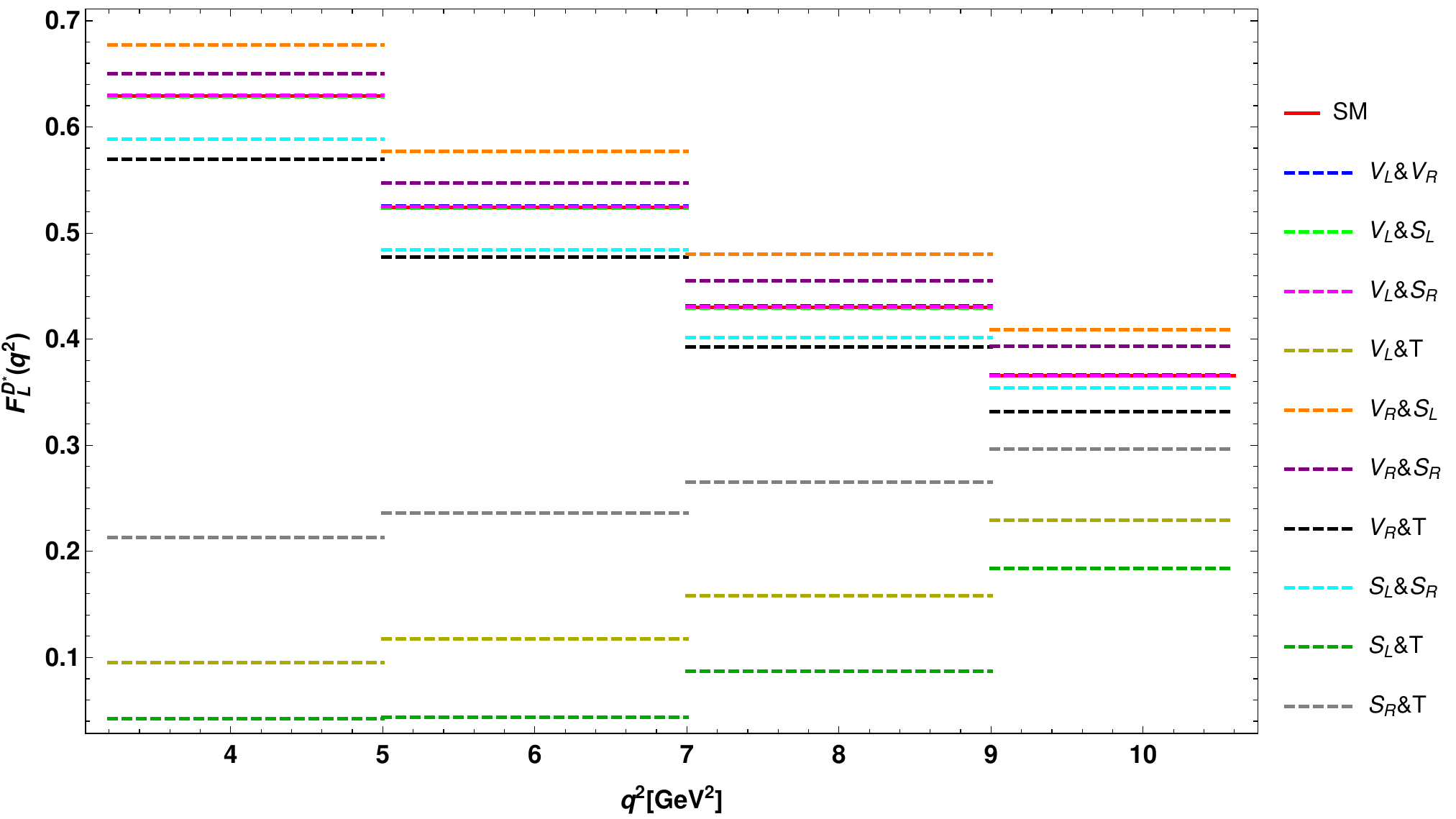}
\quad
\includegraphics[scale=0.38]{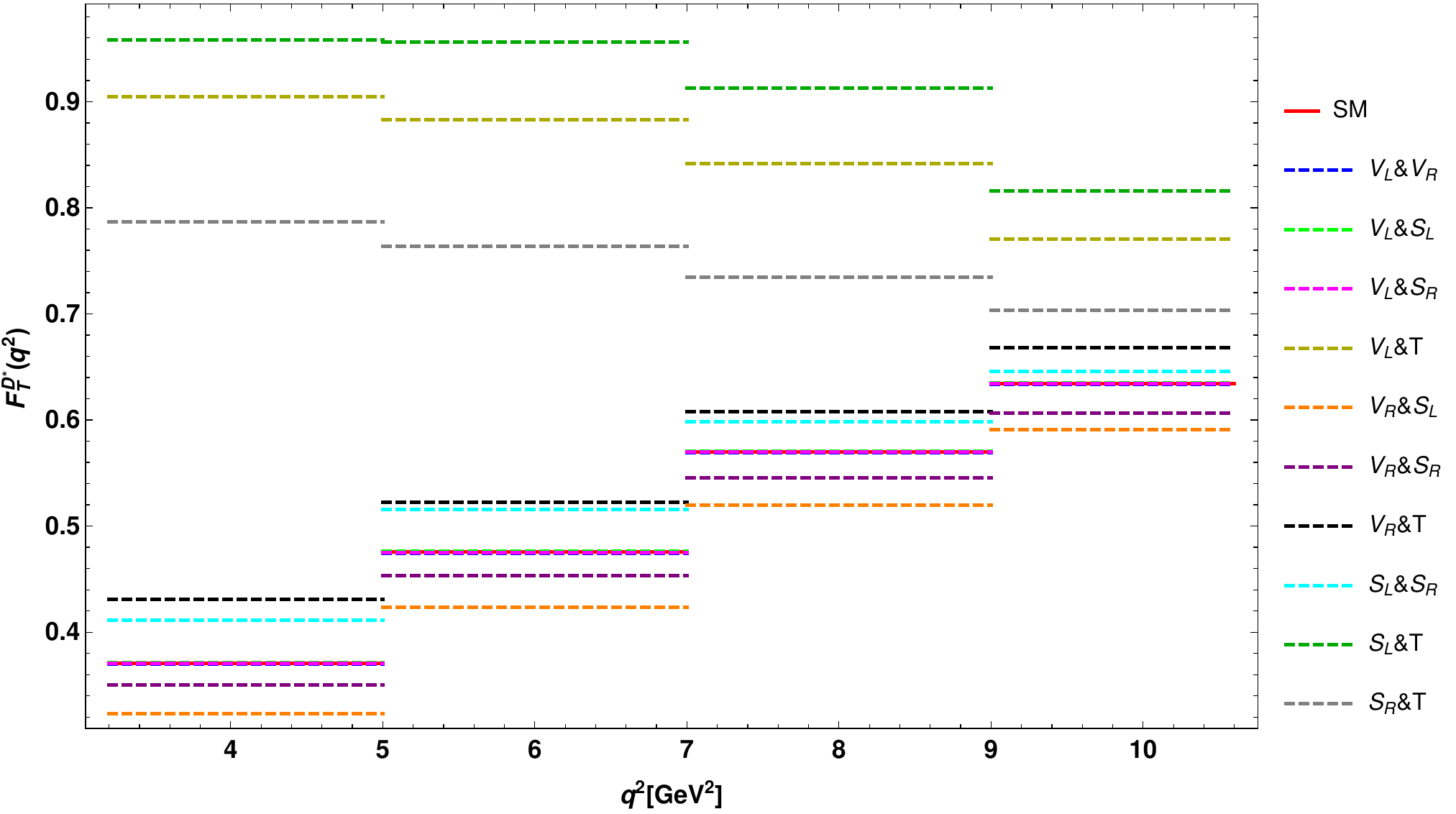}
\quad
\includegraphics[scale=0.38]{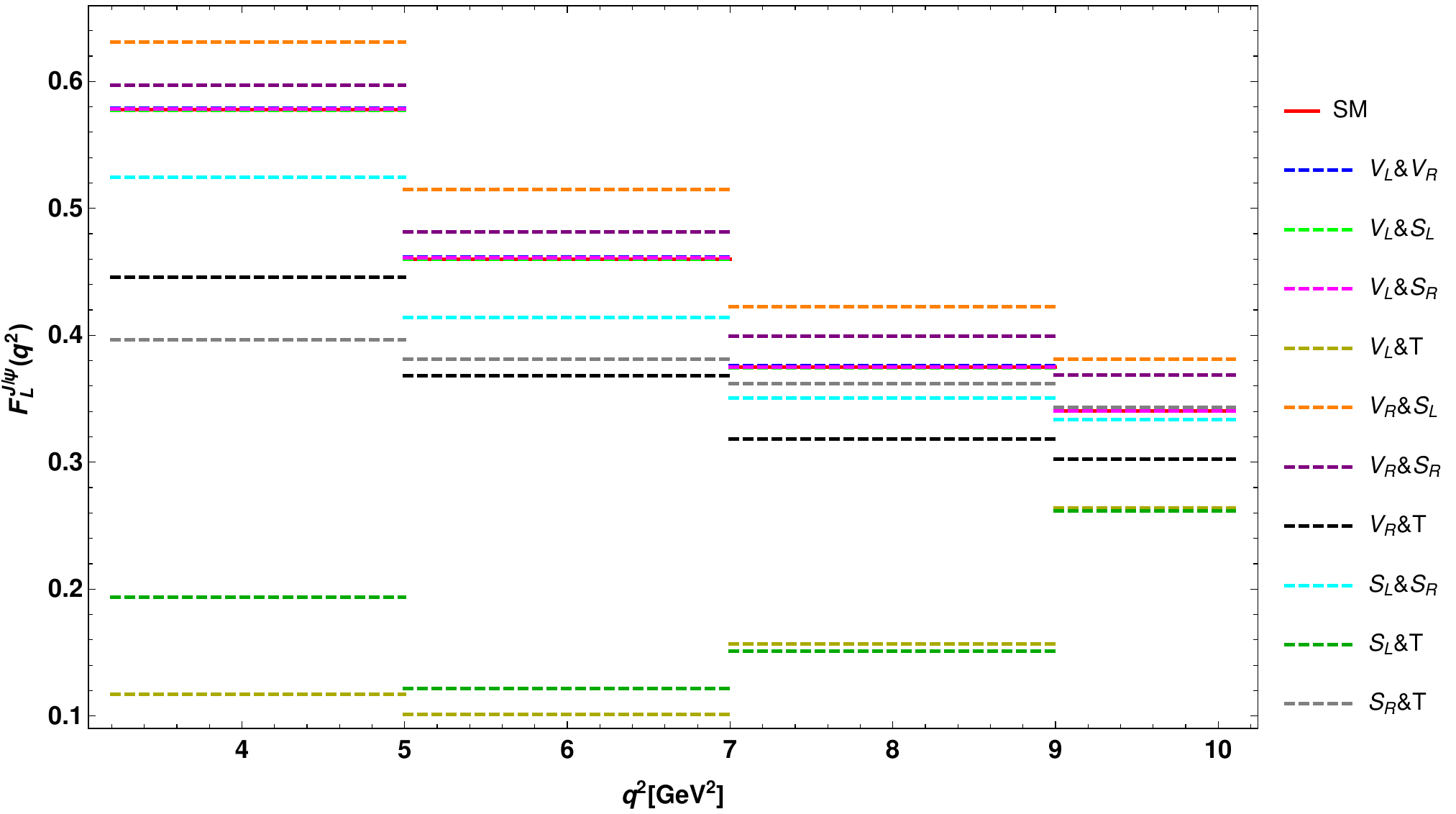}
\quad
\includegraphics[scale=0.38]{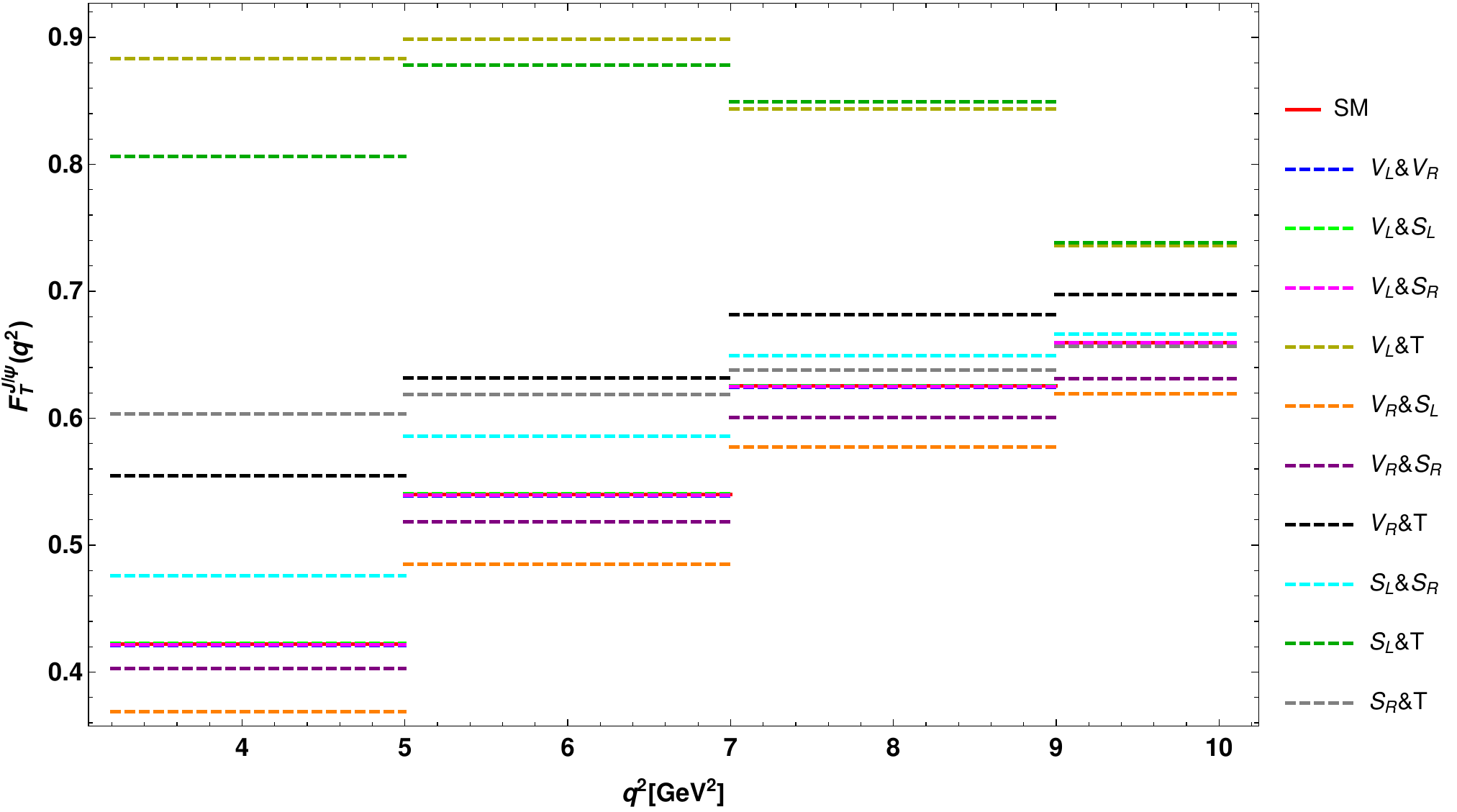}
\quad
\includegraphics[scale=0.38]{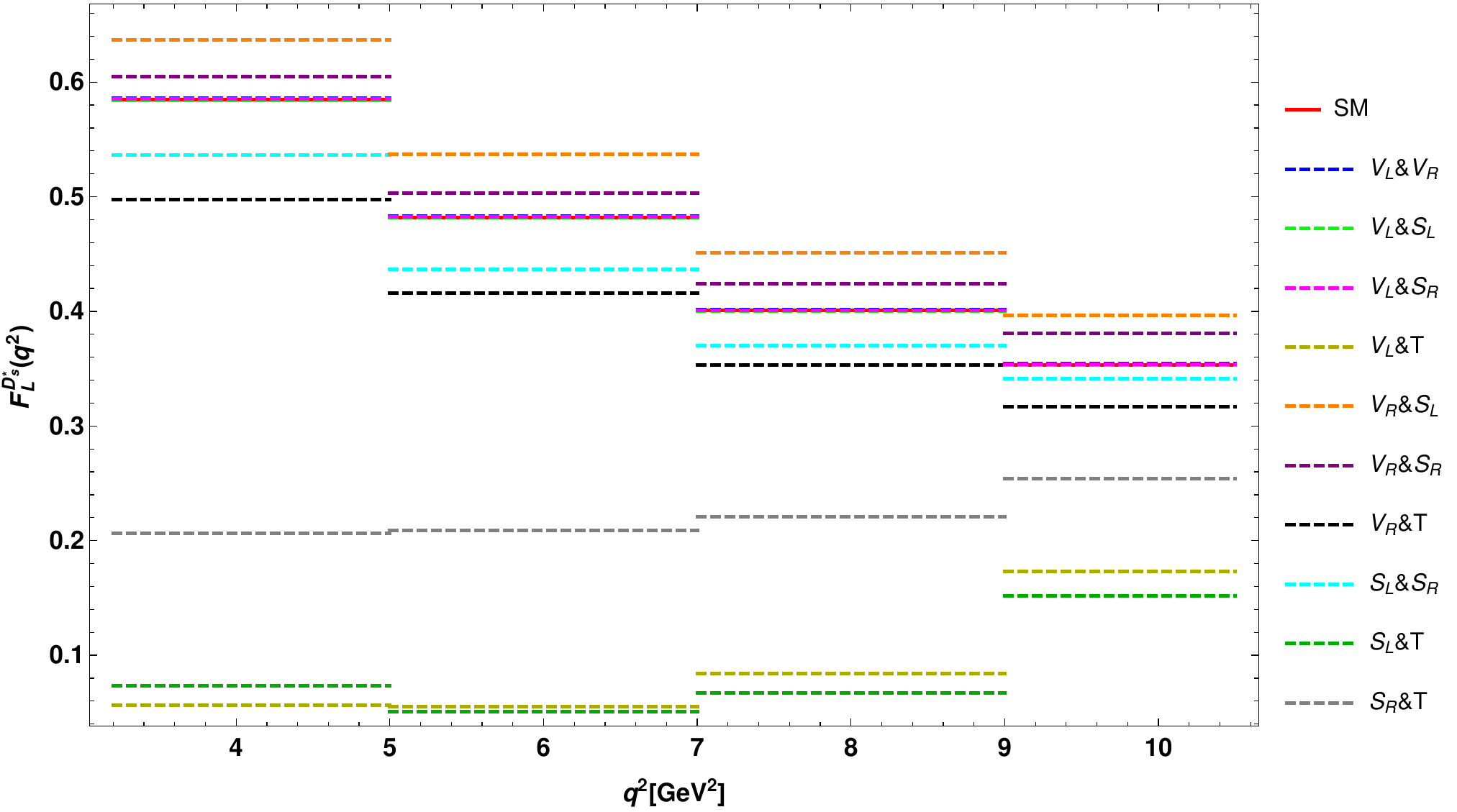}
\quad
\includegraphics[scale=0.38]{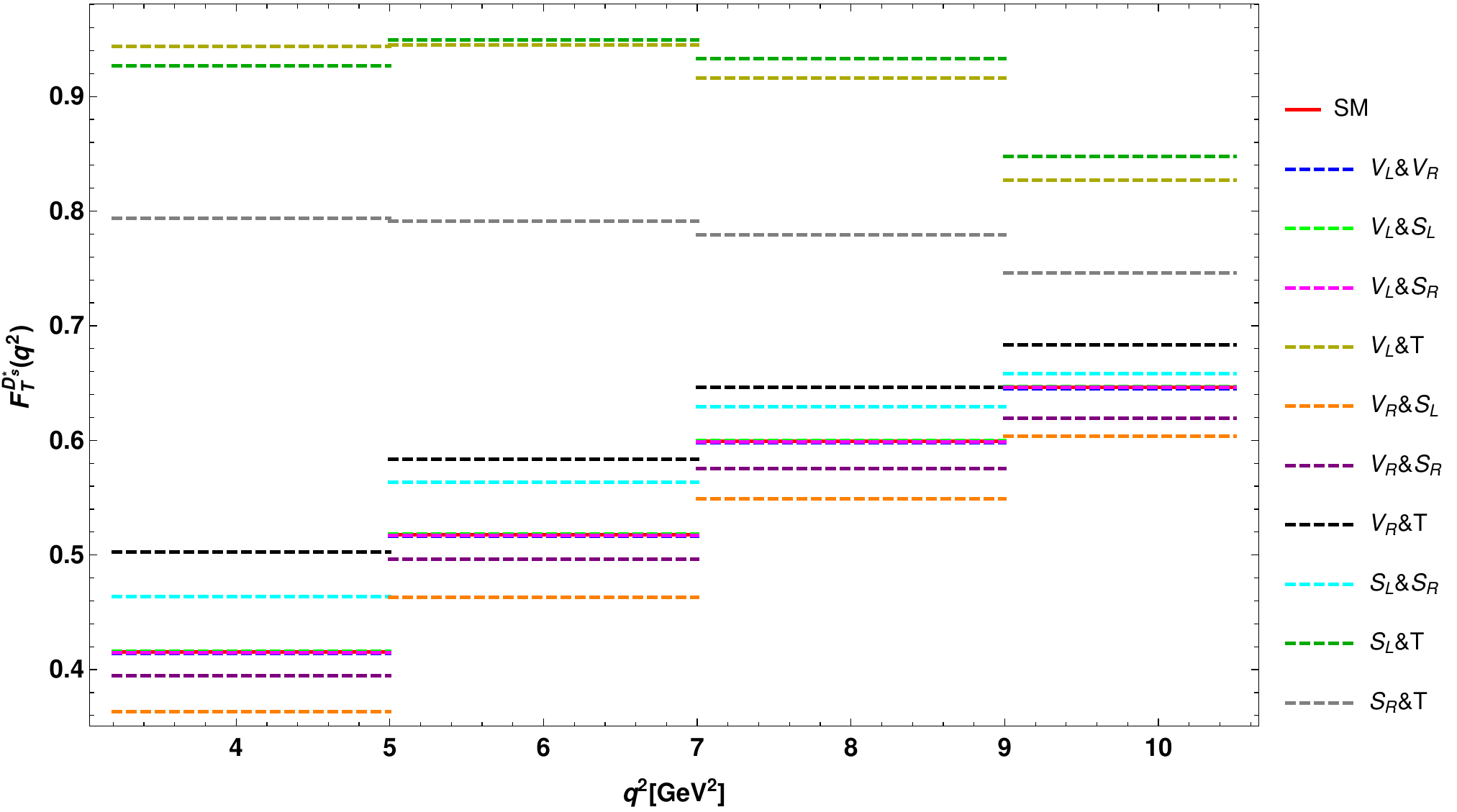}
\caption{ The bin-wise longitudinal (left panel) and transverse (right panel) polarization asymmetry of daughter vector meson of  $\bar B \to D^* \tau \bar \nu_\tau$ (top),  $ B_c^+ \to J/\psi \tau^+  \nu_\tau$ (middle) and $ B_s \to D_s^* \tau \bar \nu_\tau$ (bottom) processes in four $q^2$ bin for case C. }\label{Fig:CC-polarization}
\end{figure}

\section{$\Lambda_b \to \Lambda_c \tau \bar \nu_\tau$}
\begin{figure}[htb]
\includegraphics[scale=0.5]{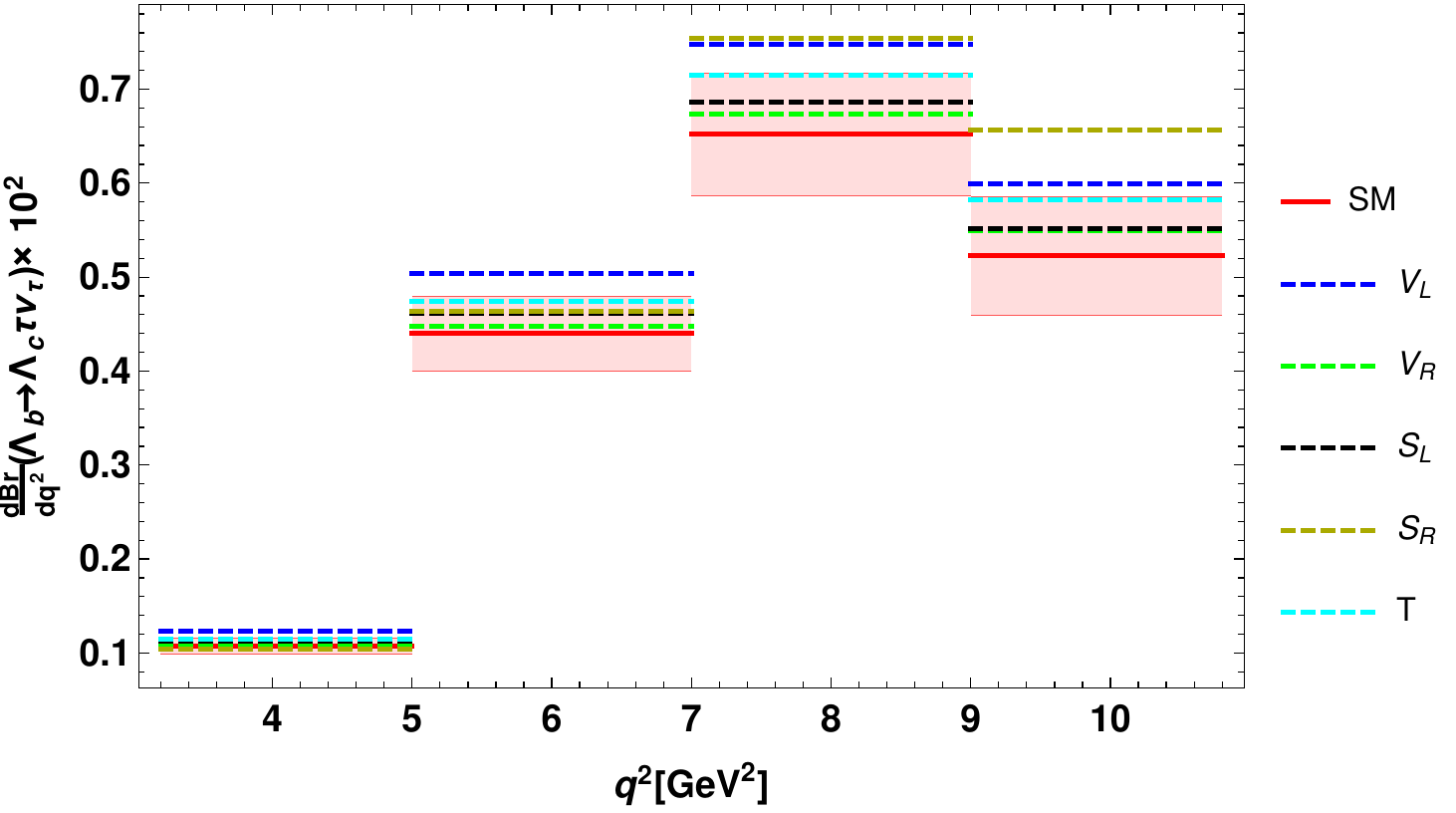}
\quad
\includegraphics[scale=0.5]{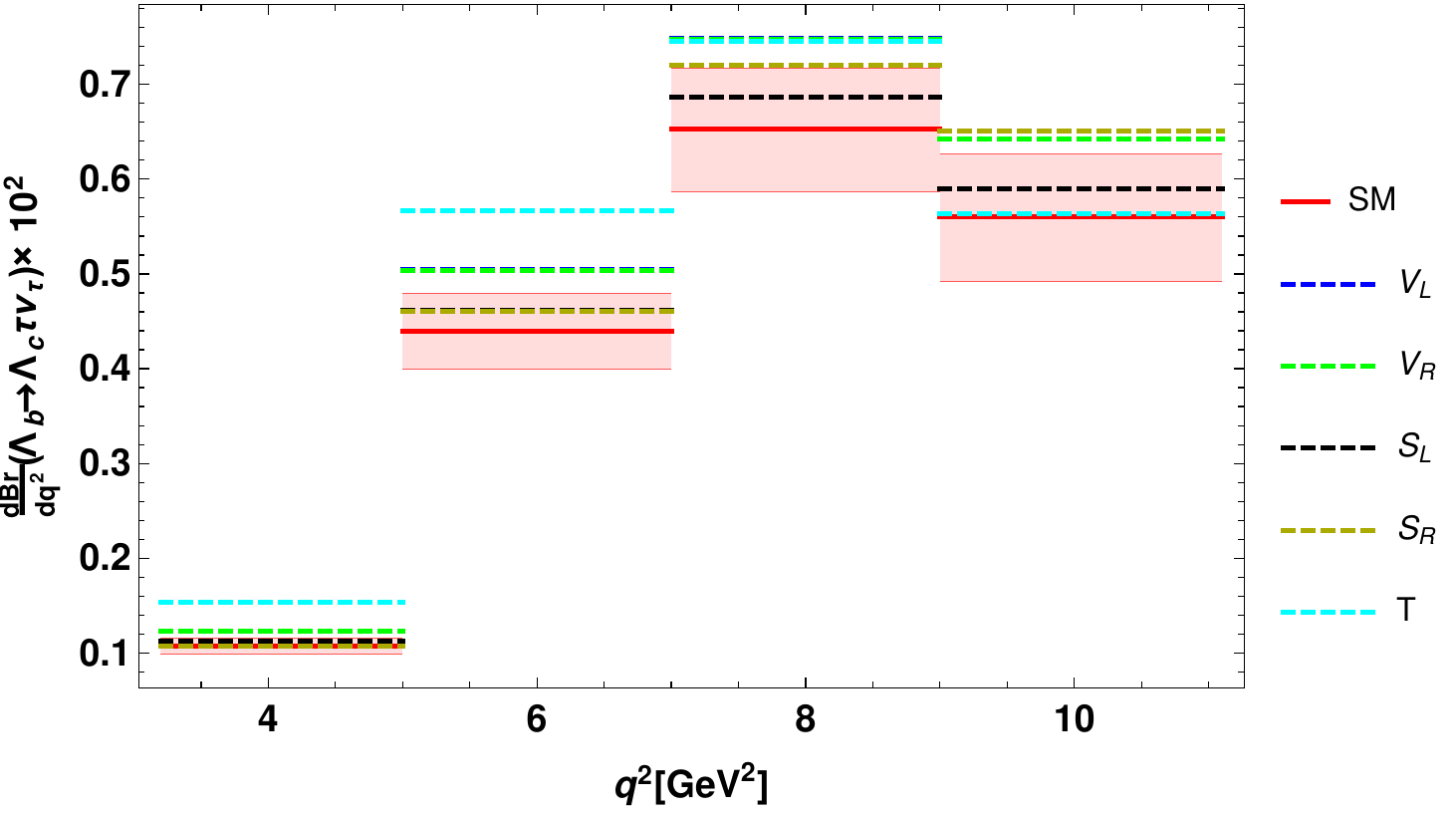}
\quad
\includegraphics[scale=0.5]{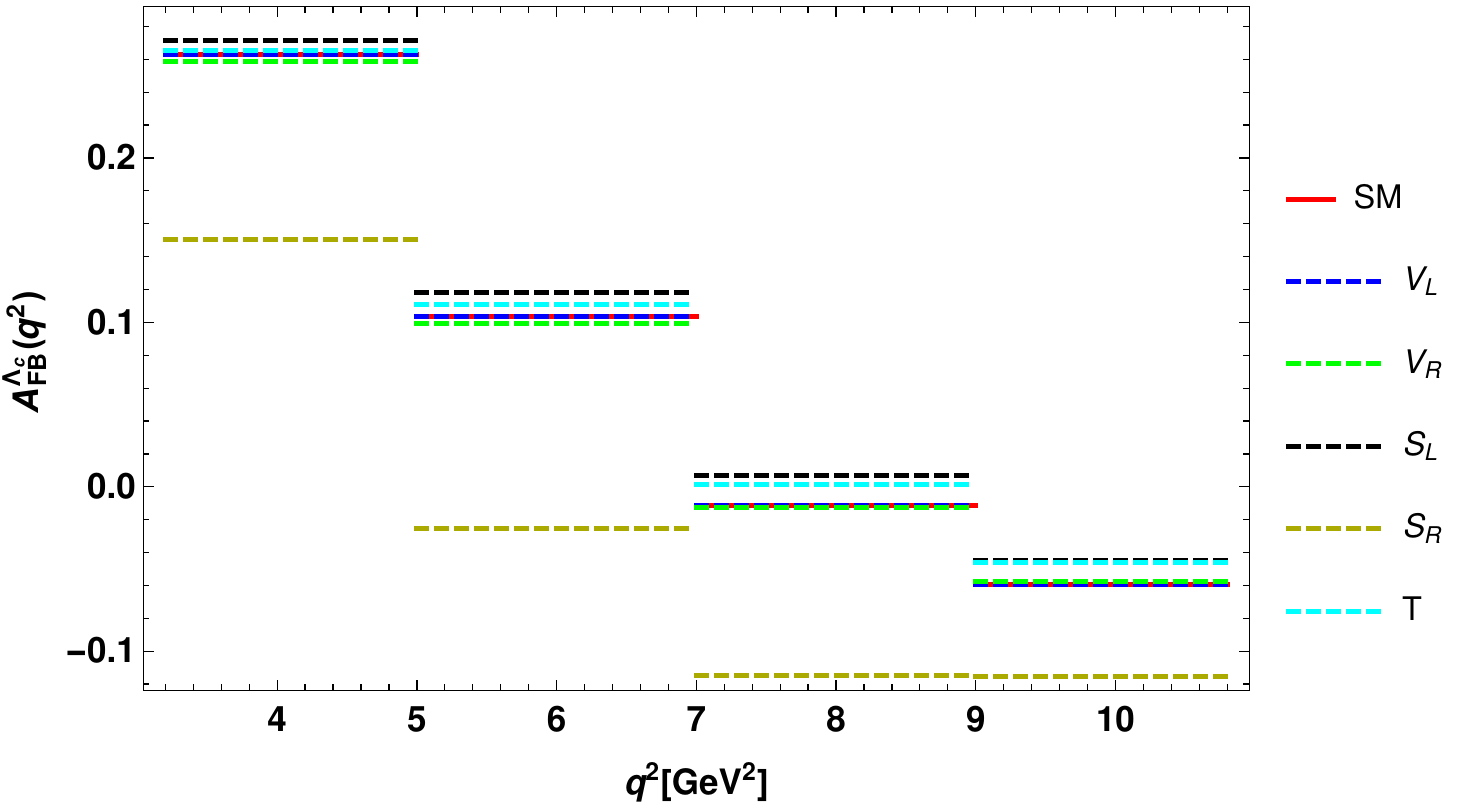}
\quad
\includegraphics[scale=0.5]{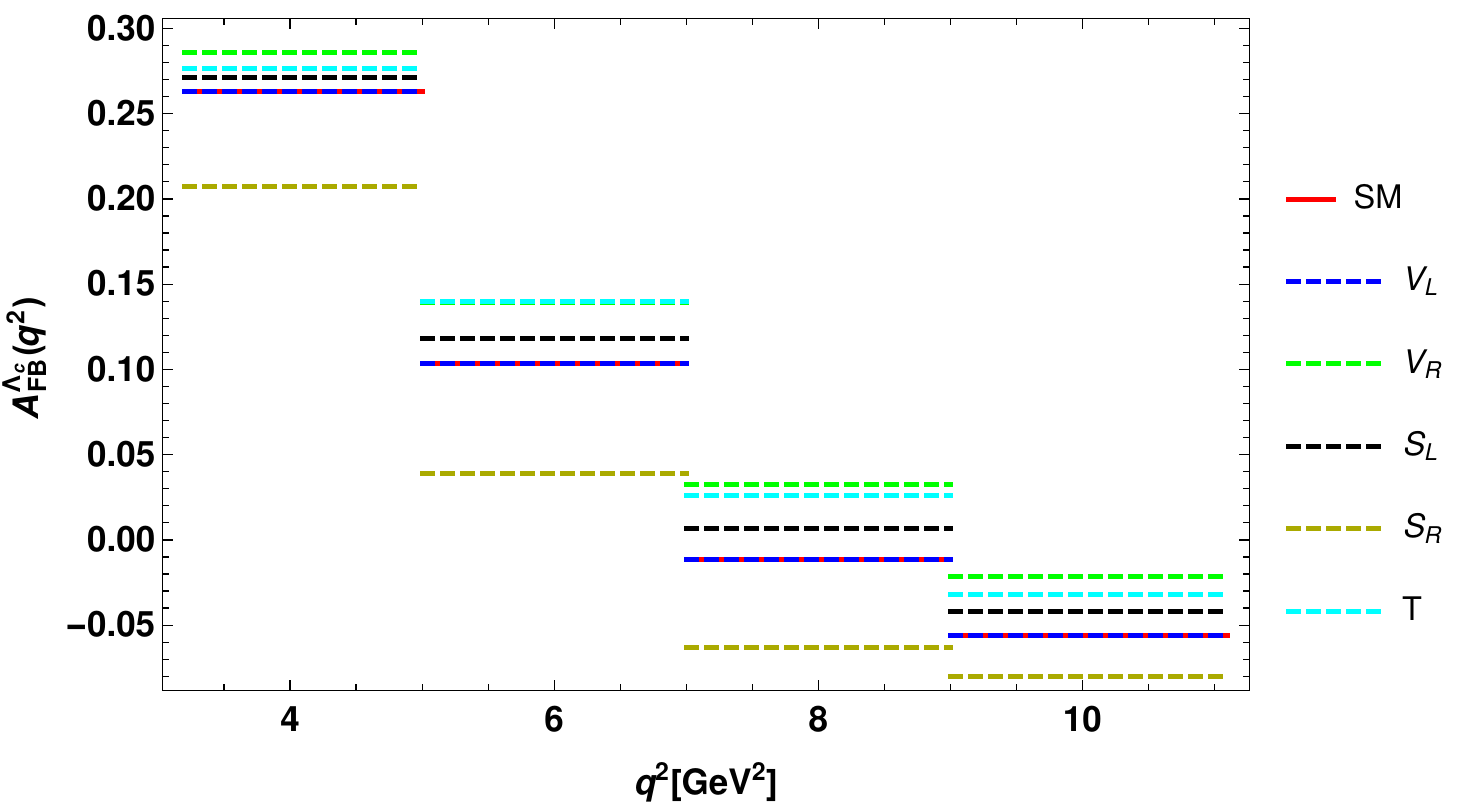}
\quad
\includegraphics[scale=0.5]{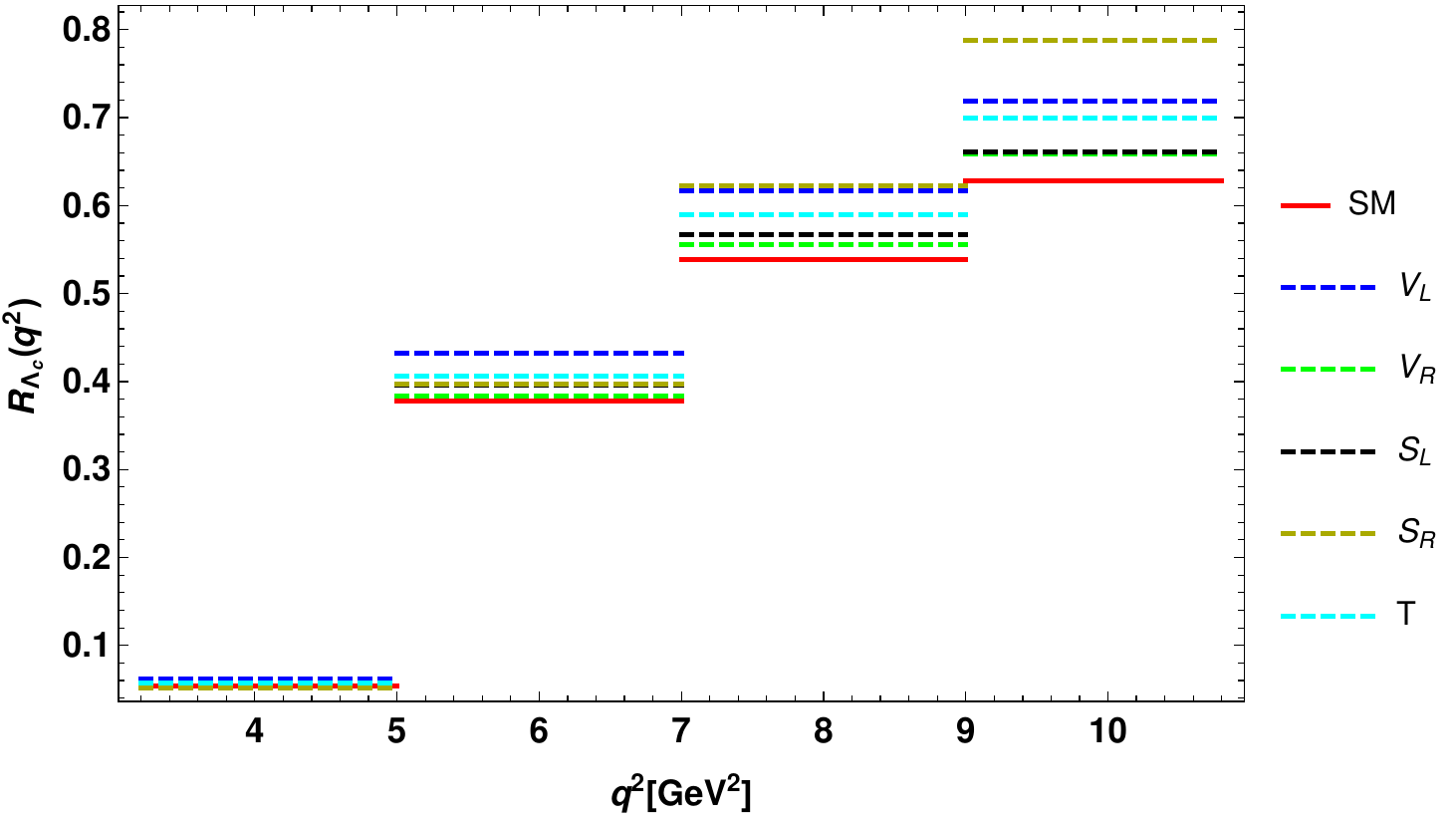}
\quad
\includegraphics[scale=0.5]{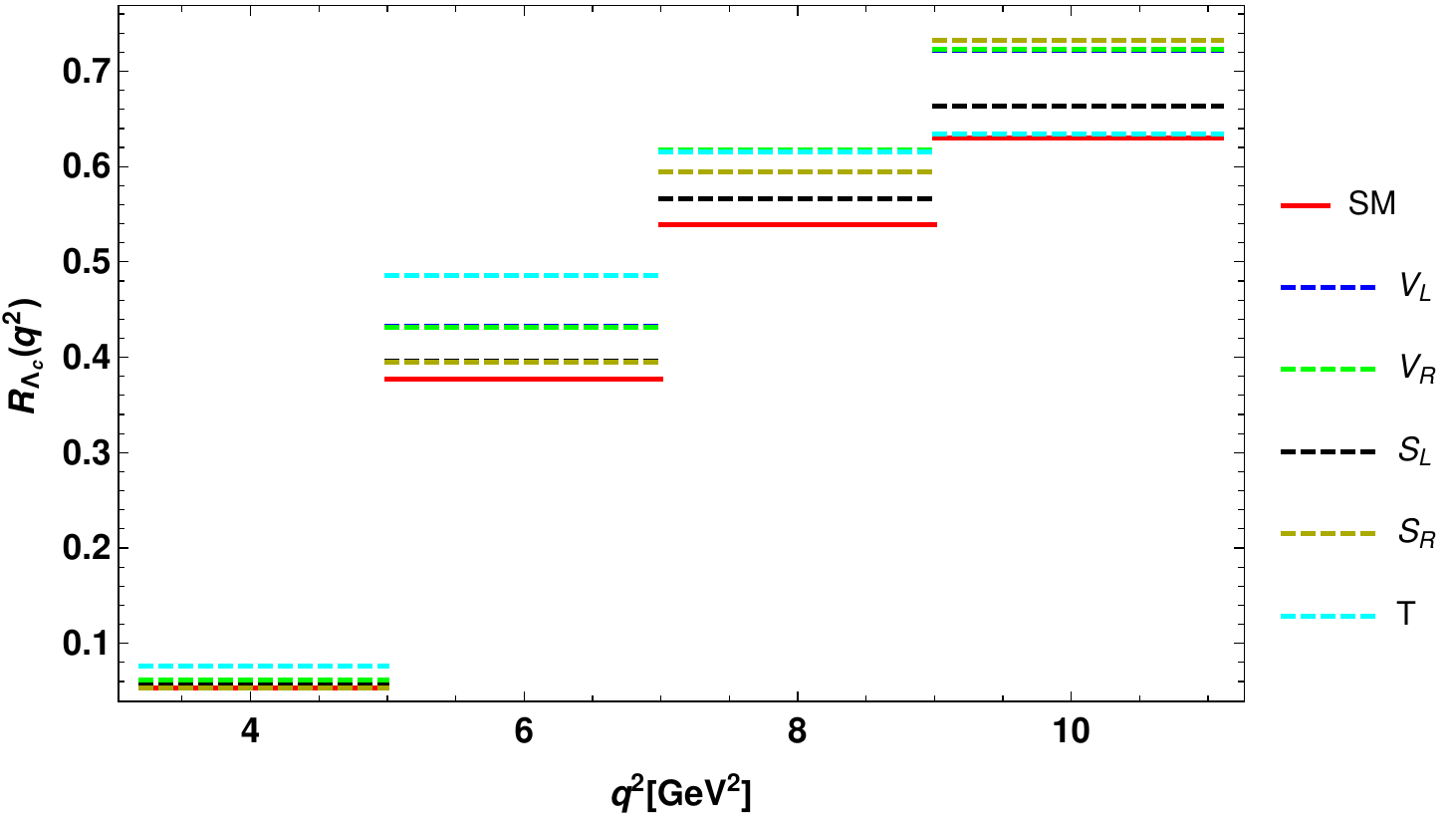}
\quad
\includegraphics[scale=0.5]{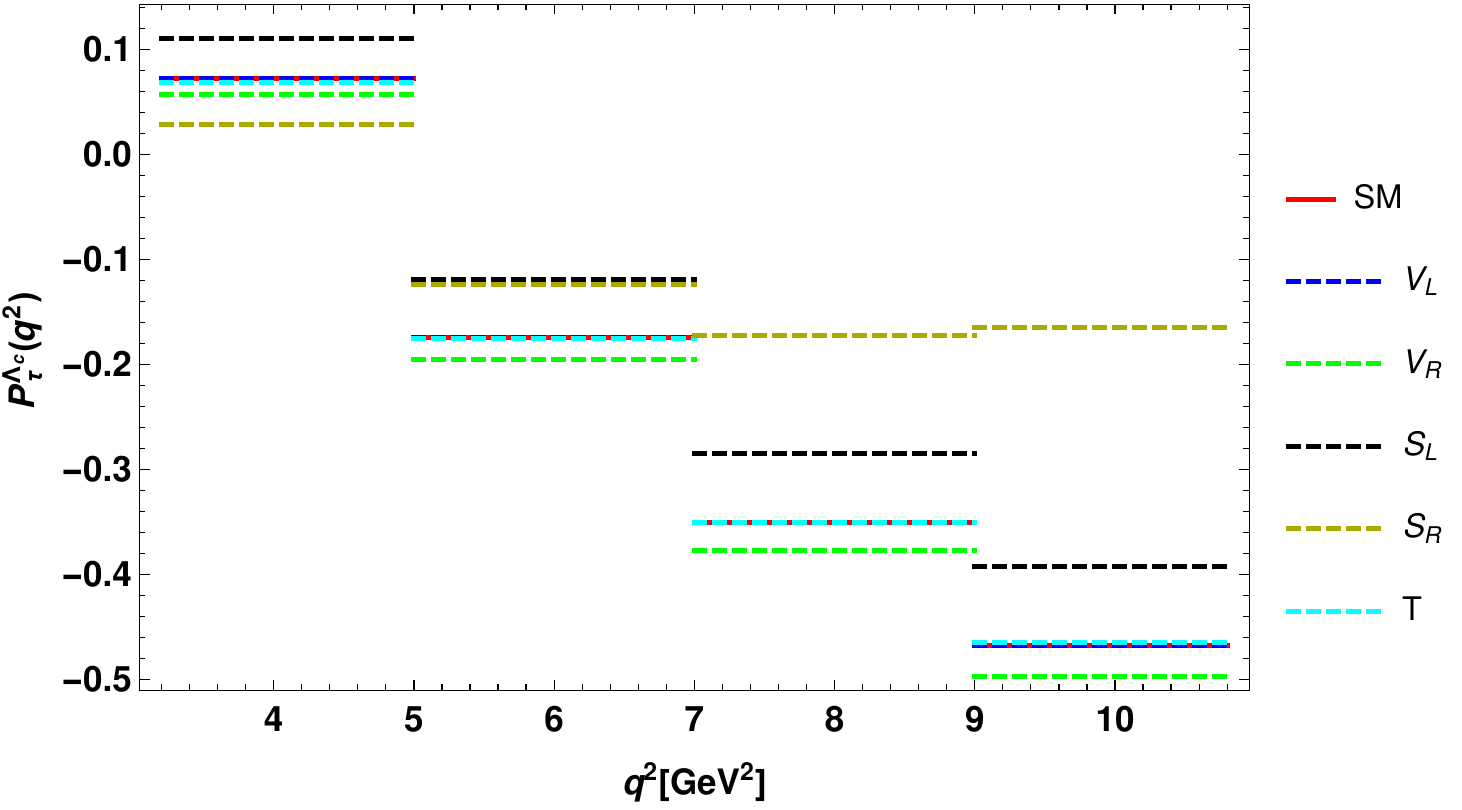}
\quad
\includegraphics[scale=0.5]{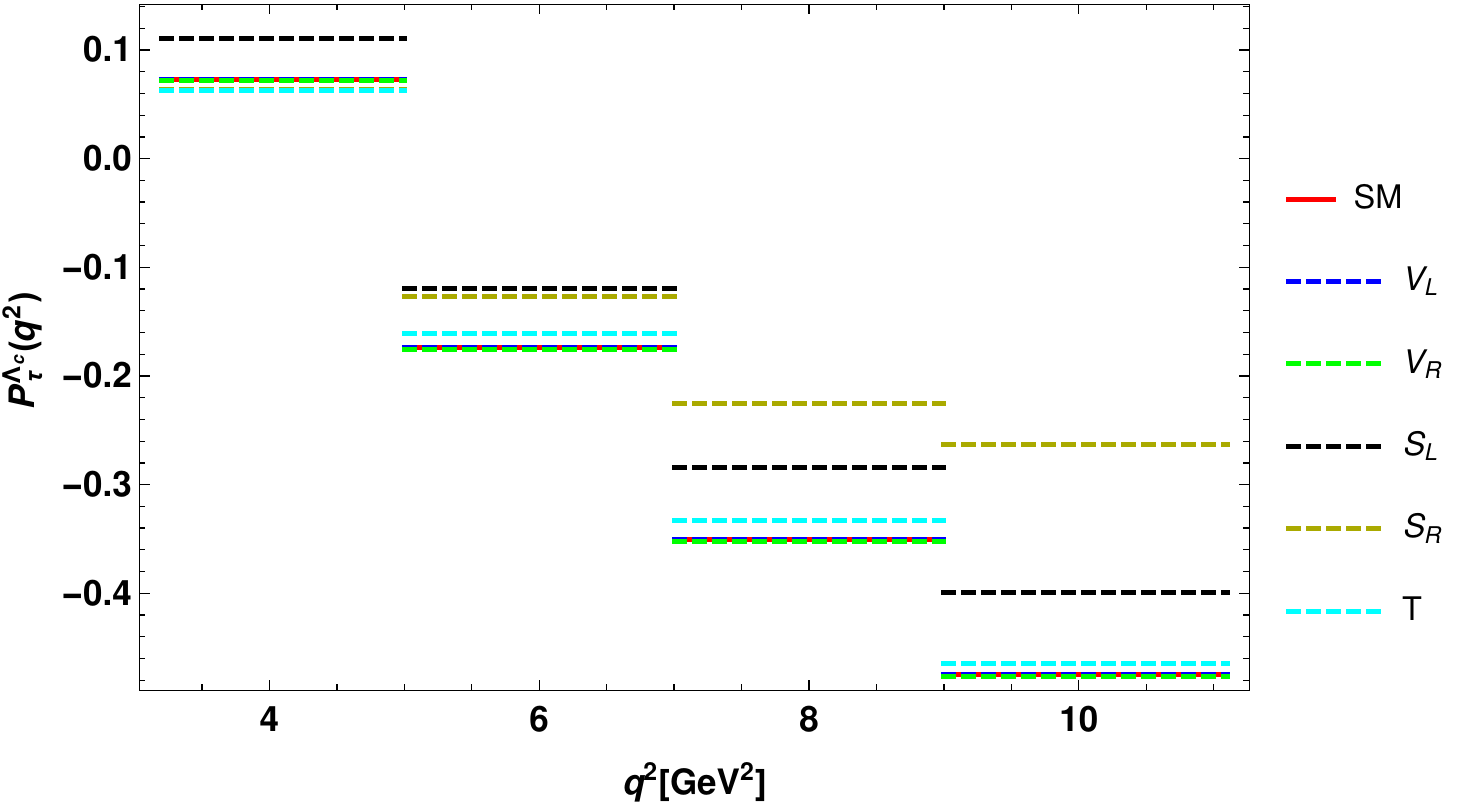}
\quad
\includegraphics[scale=0.5]{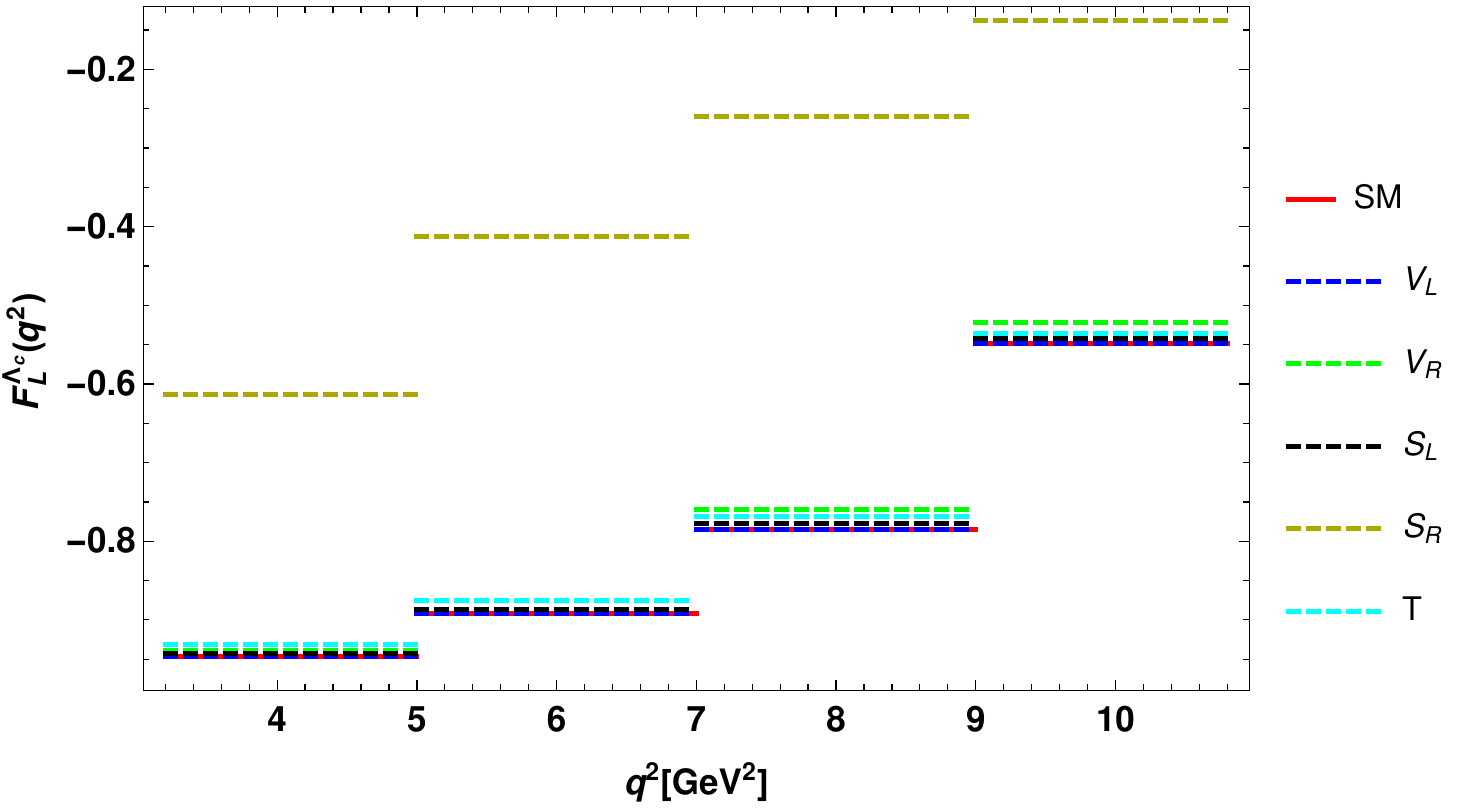}
\quad
\includegraphics[scale=0.5]{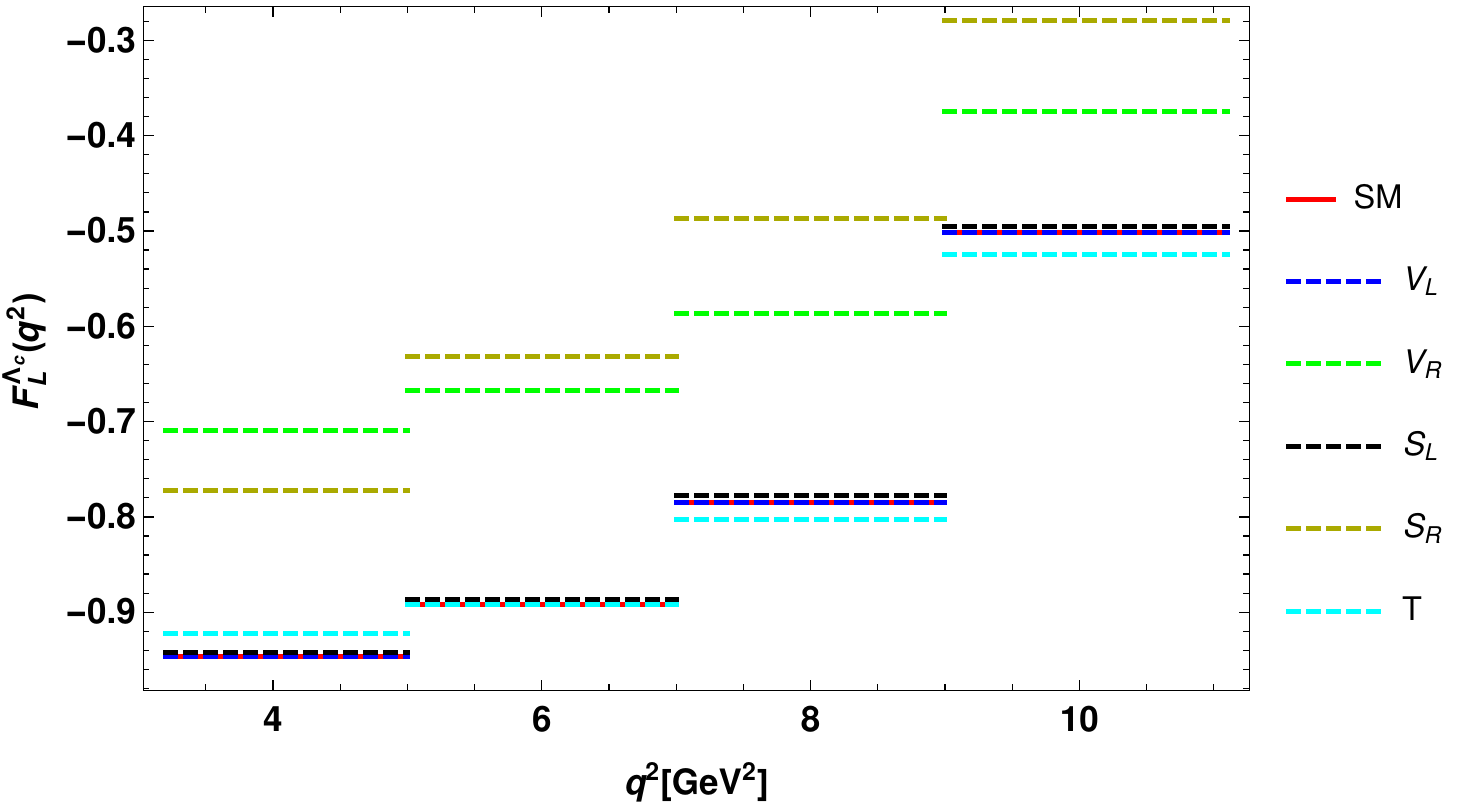}
\caption{ The bin-wise branching ratio (top), forward-backward asymmetry (second from top), $R_{\Lambda_c}$ (third from top), tau (fourth from top) and $\Lambda_c$  (bottom) longitudinal polarization asymmetry of $\Lambda_b \to \Lambda_c  \tau \bar \nu_\tau$  in four $q^2$ bins for case A (left panel) and case B (right panel). }\label{Fig:CA-CB-Lambdab}
\end{figure}

\begin{figure}[htb]
\includegraphics[scale=0.4]{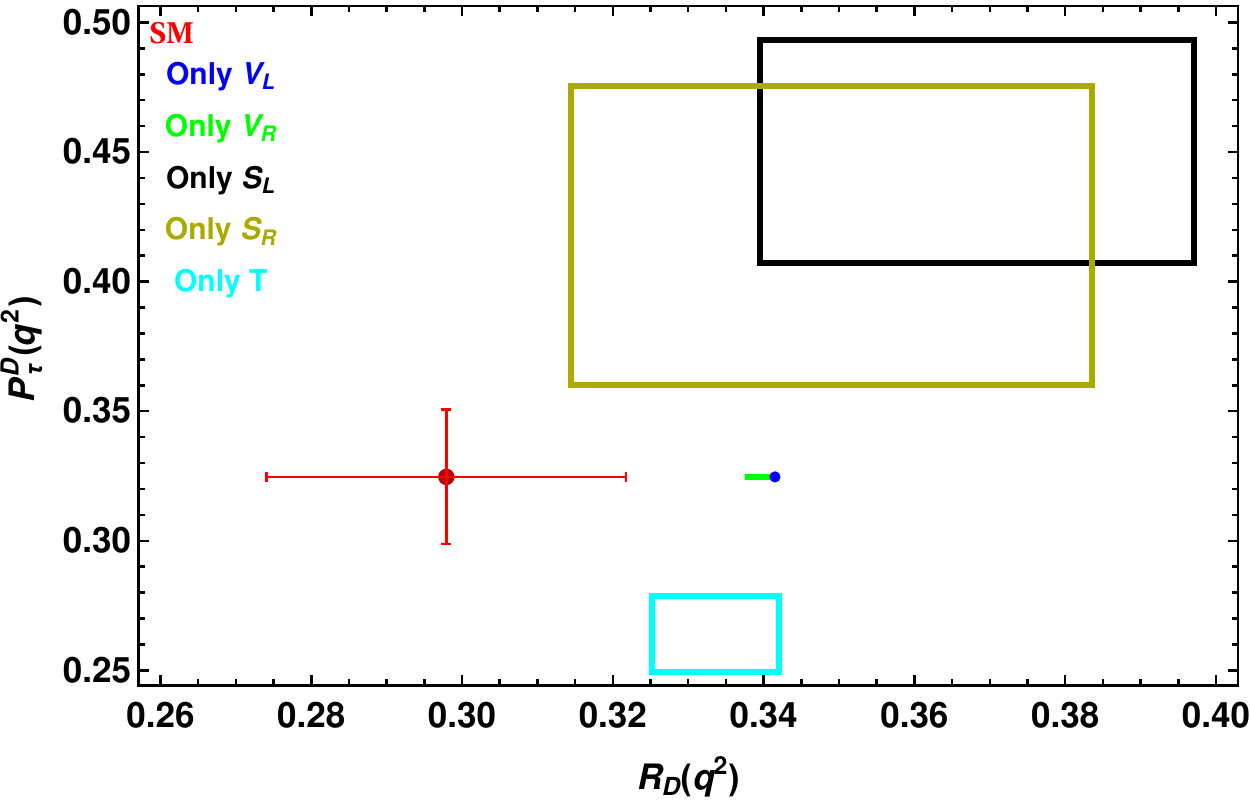}
\quad
\includegraphics[scale=0.4]{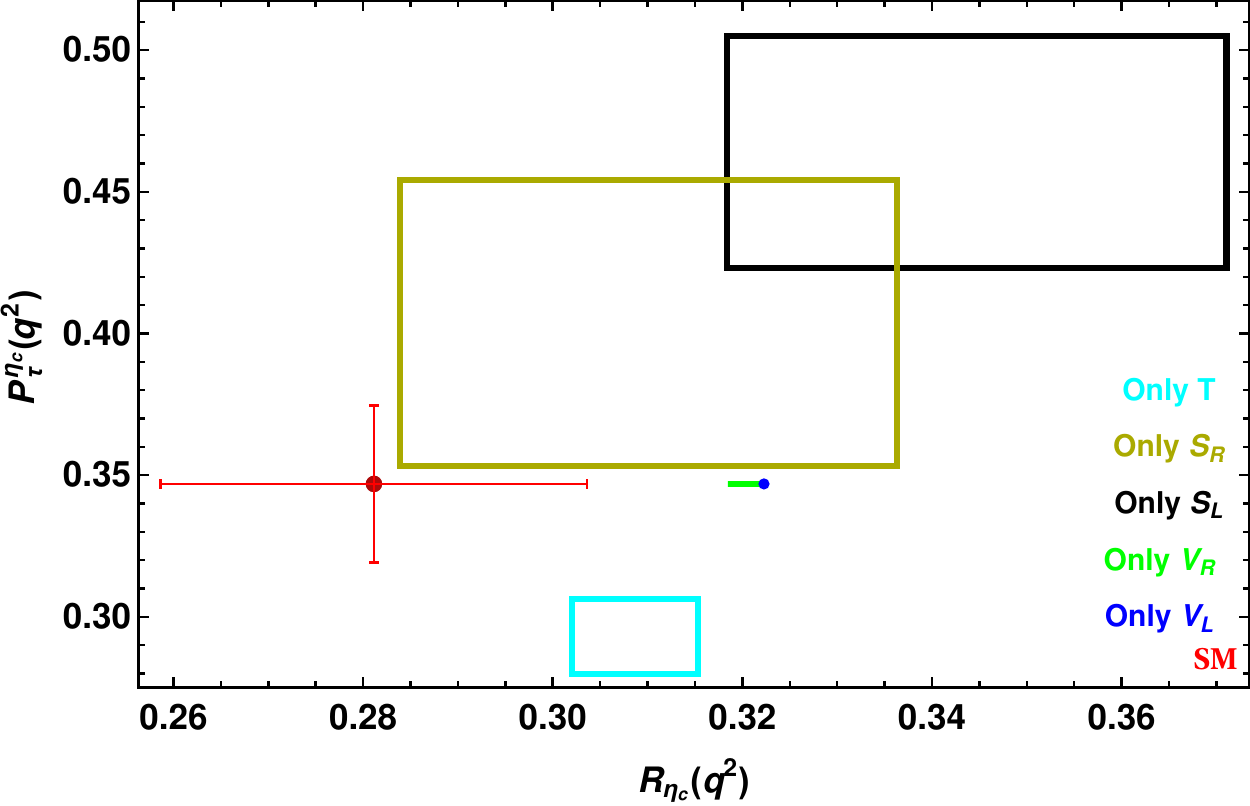}
\quad
\includegraphics[scale=0.4]{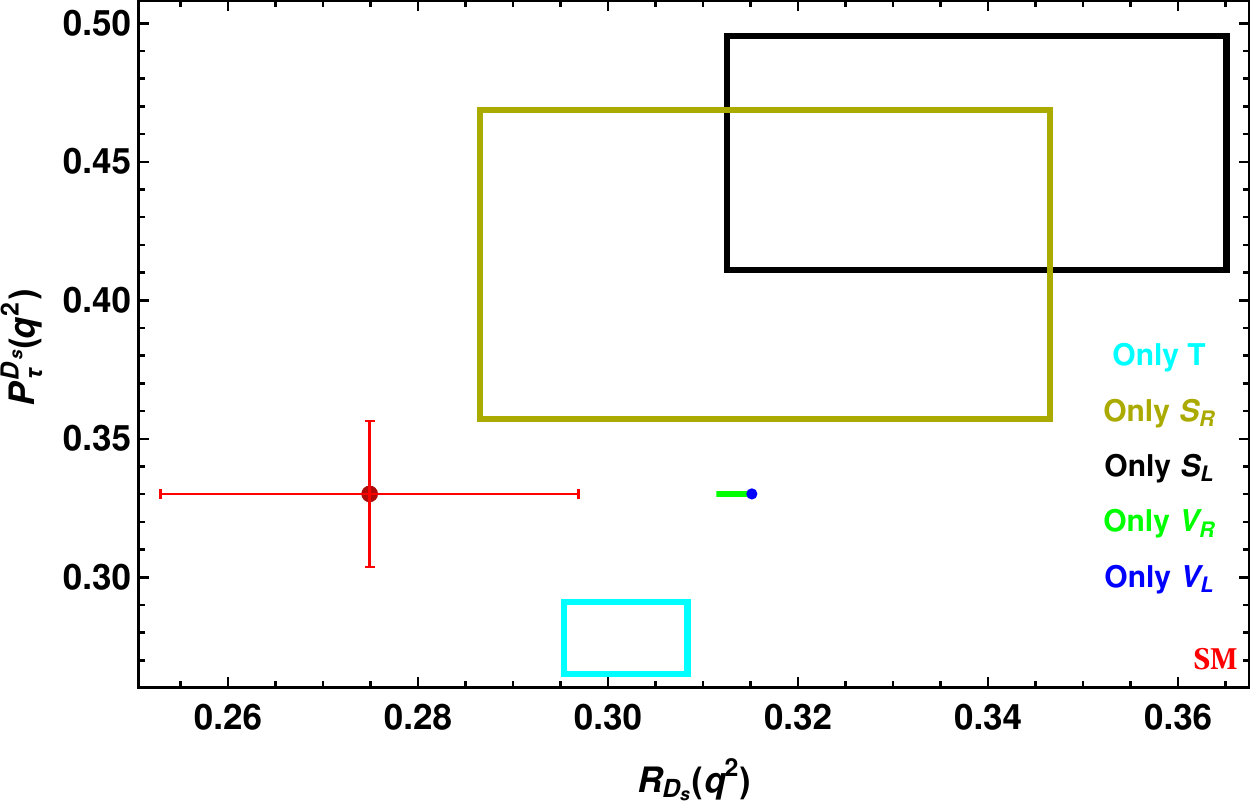}
\quad
\includegraphics[scale=0.4]{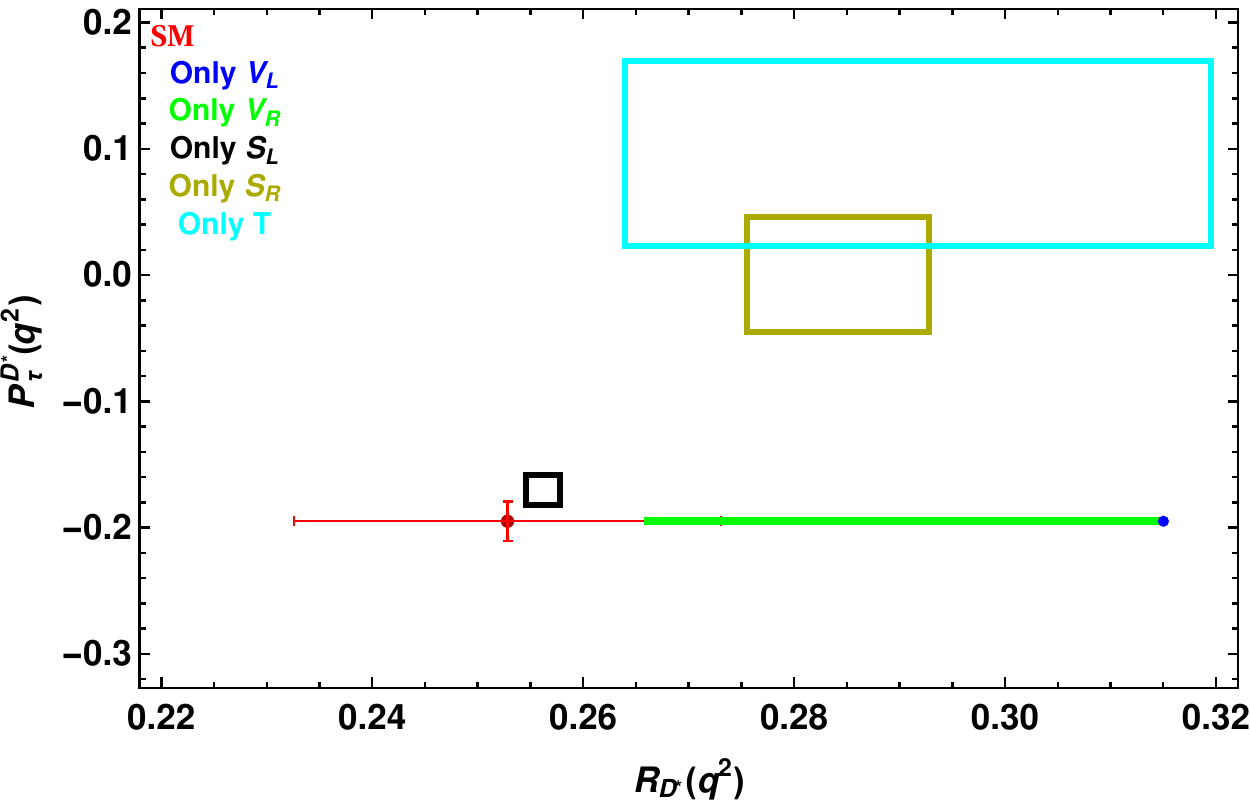}
\quad
\includegraphics[scale=0.4]{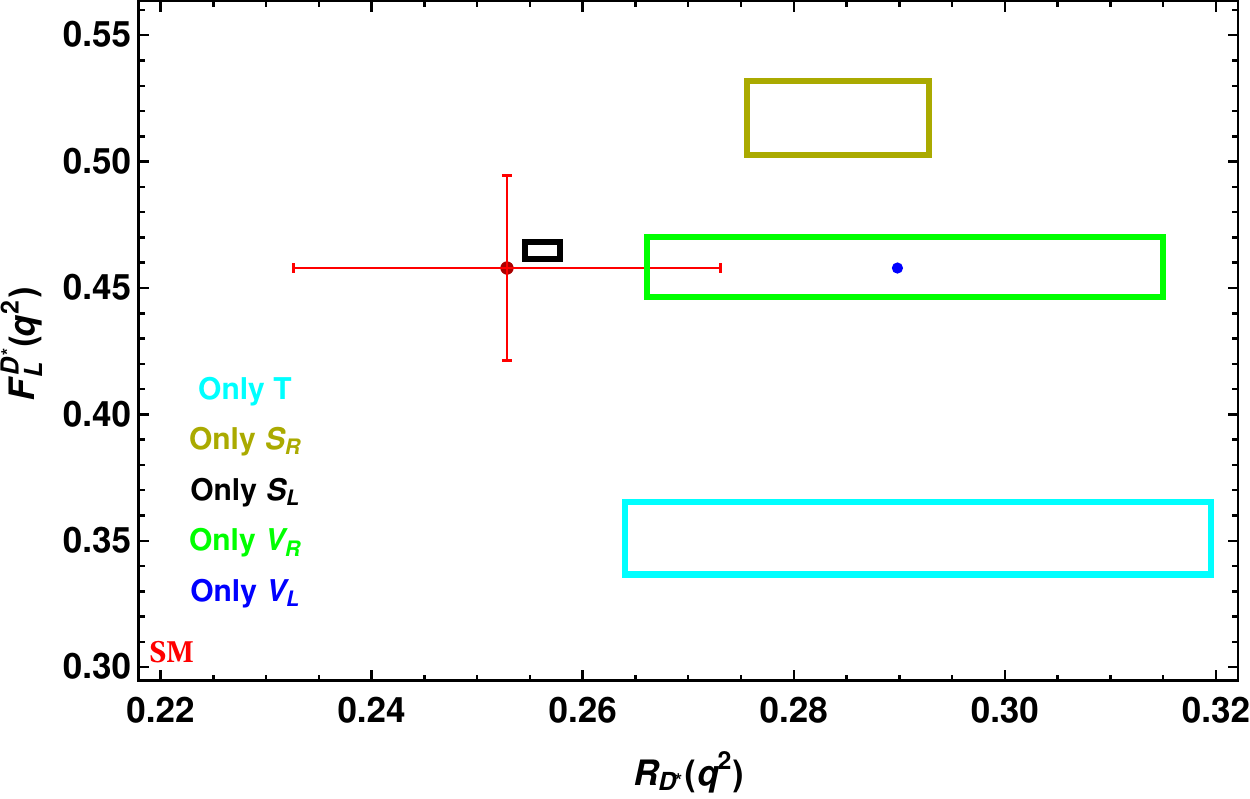}
\quad
\includegraphics[scale=0.4]{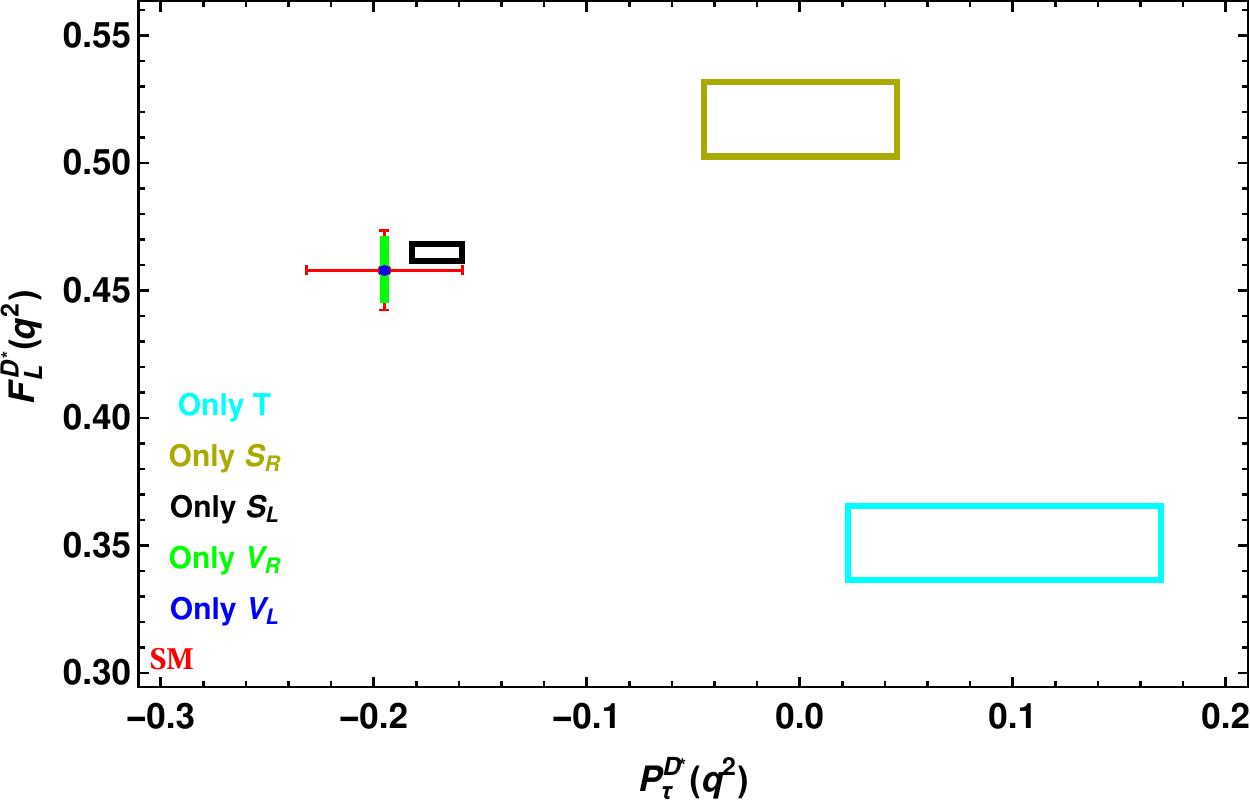}
\quad
\includegraphics[scale=0.4]{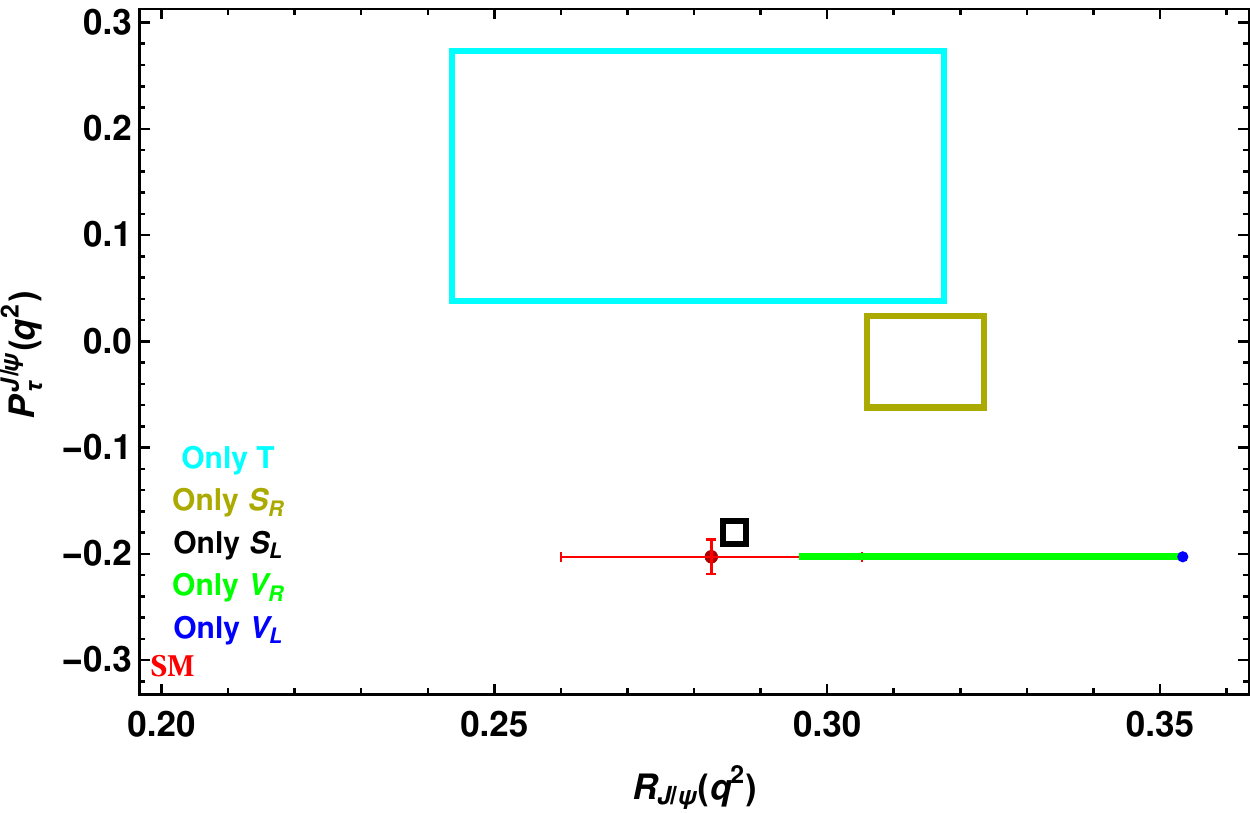}
\quad
\includegraphics[scale=0.4]{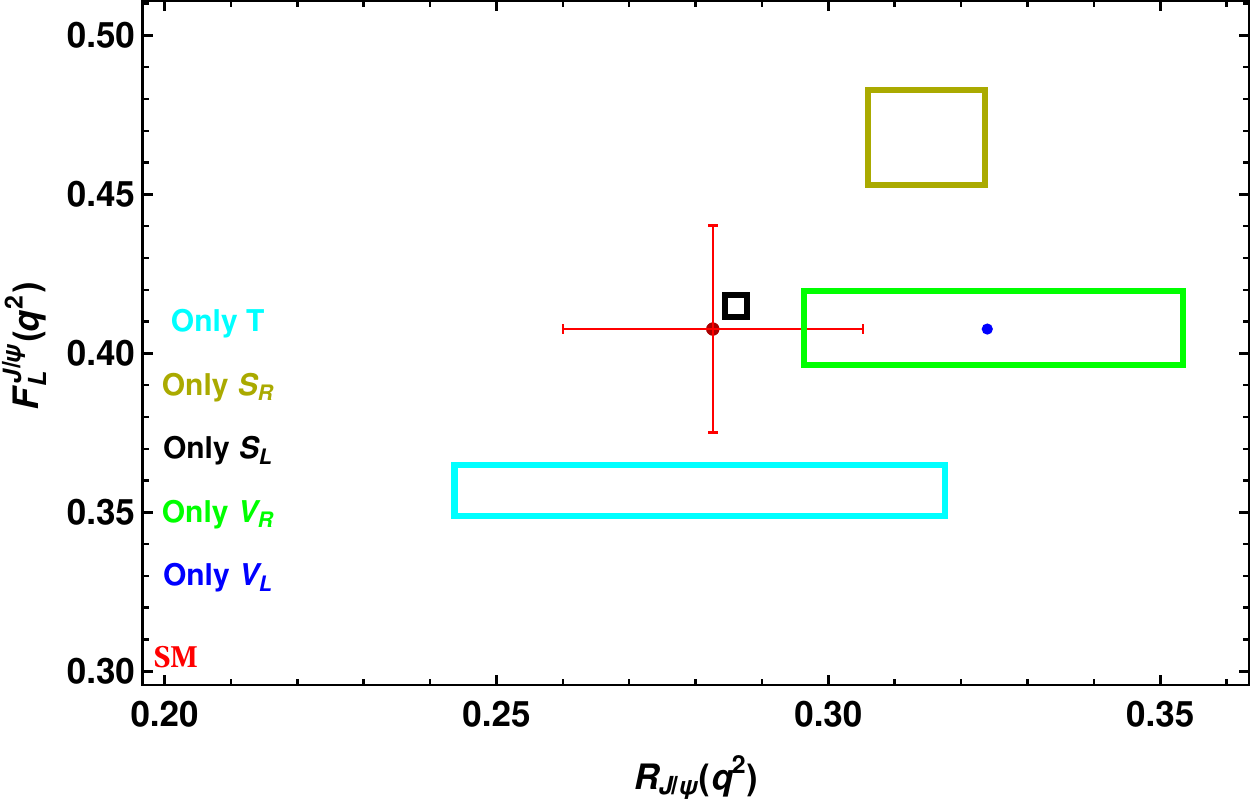}
\quad
\includegraphics[scale=0.4]{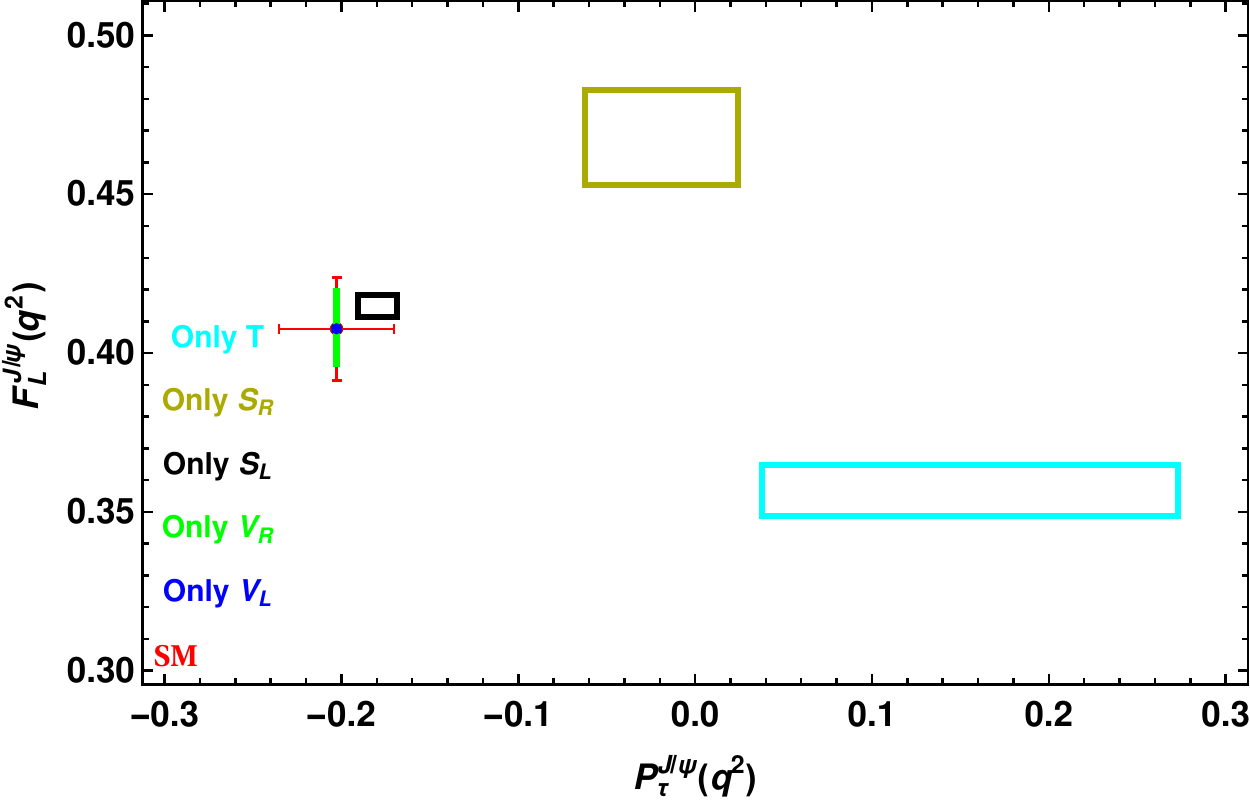}
\quad
\includegraphics[scale=0.4]{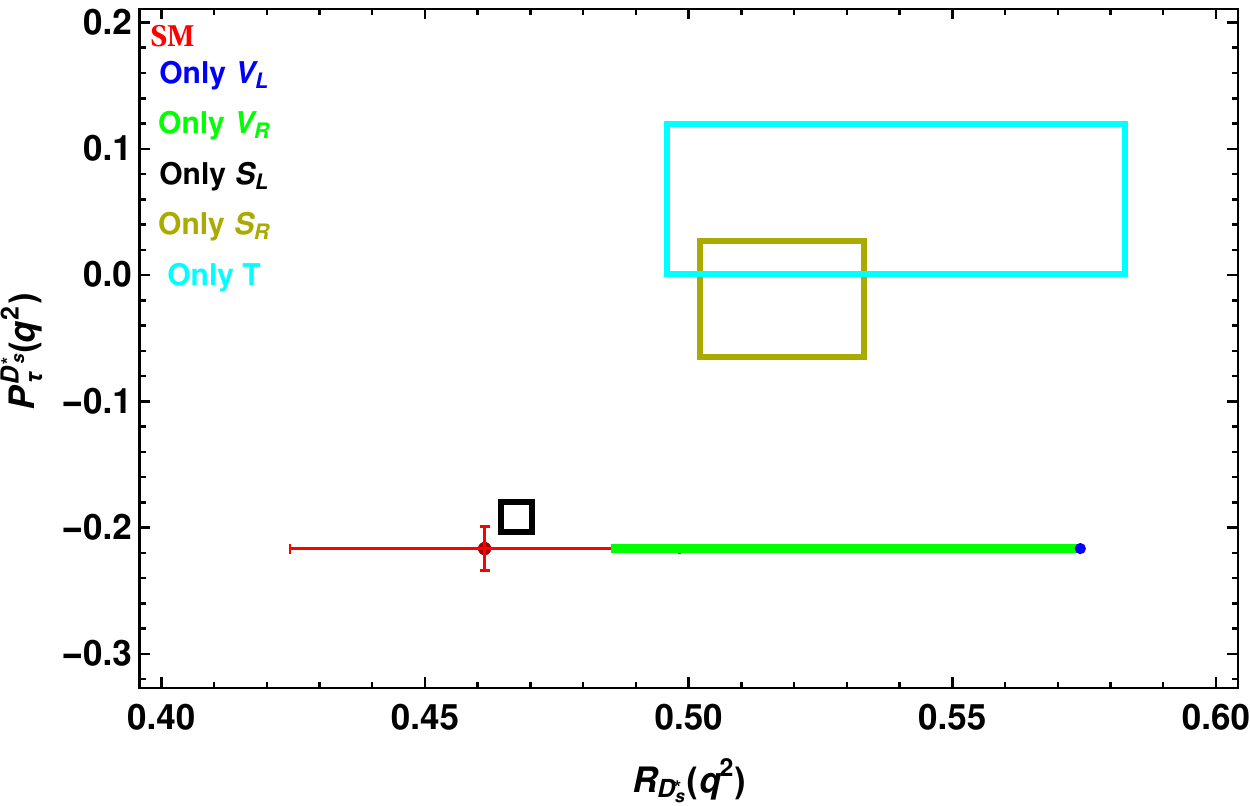}
\quad
\includegraphics[scale=0.4]{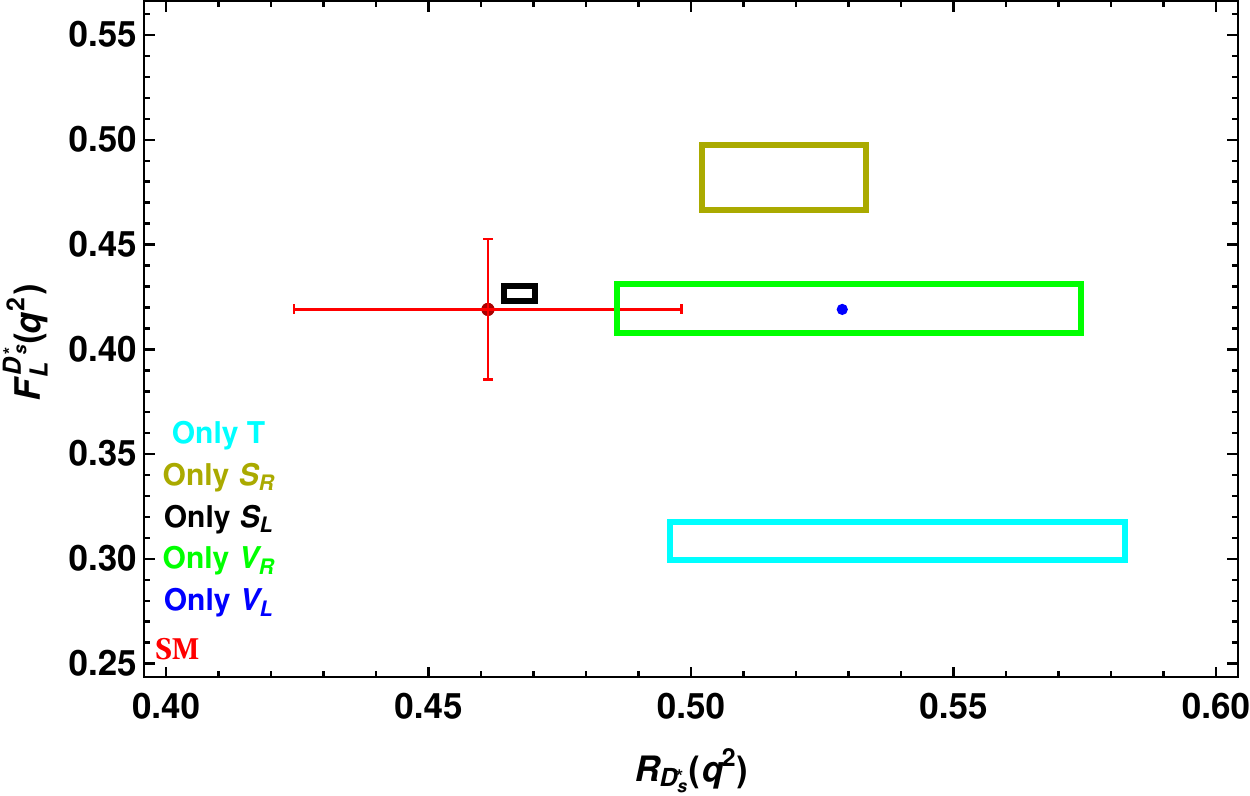}
\quad
\includegraphics[scale=0.4]{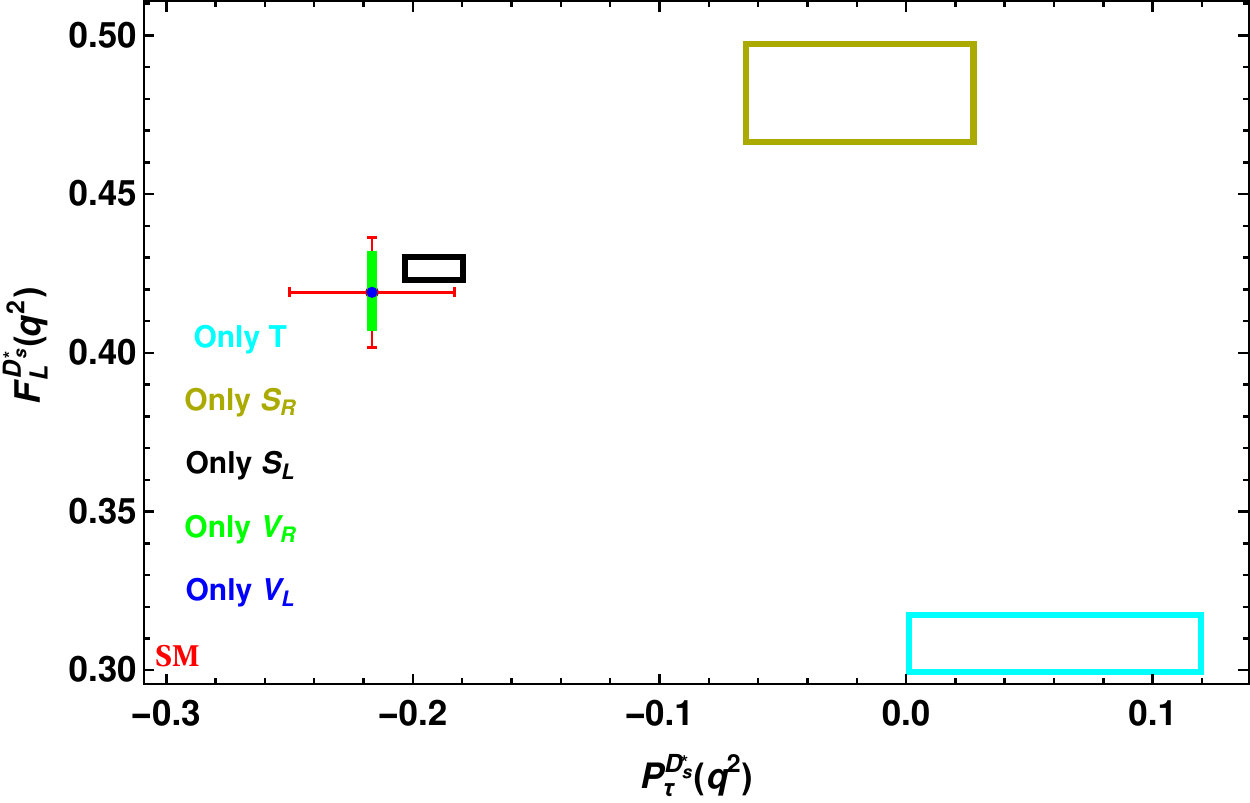}
\quad
\includegraphics[scale=0.4]{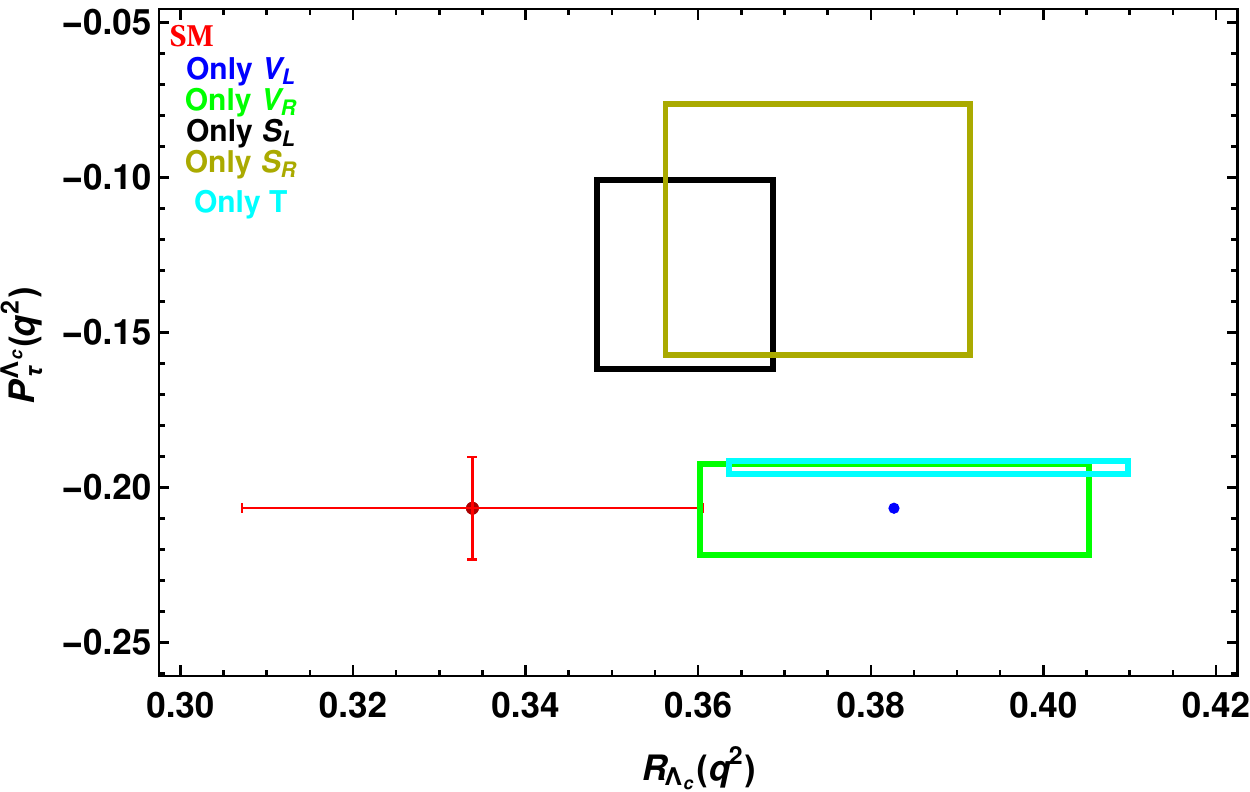}
\quad
\includegraphics[scale=0.4]{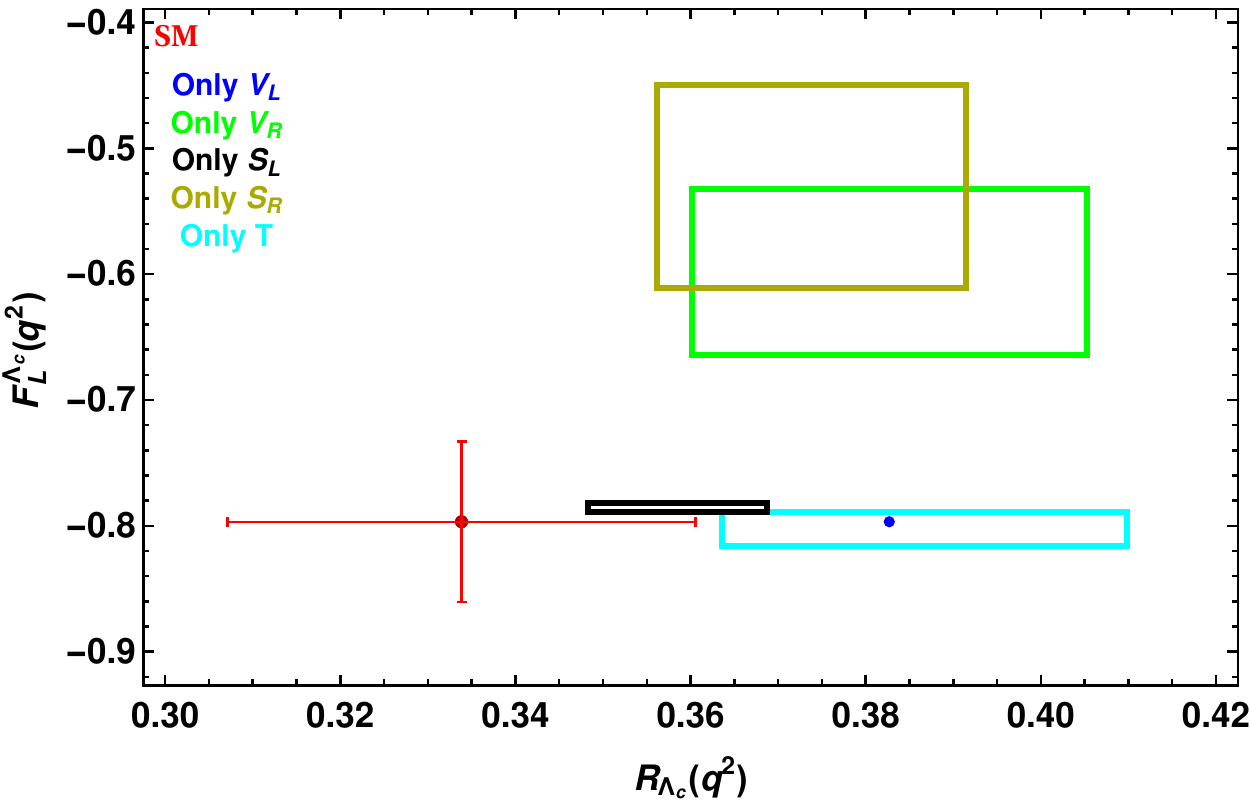}
\quad
\includegraphics[scale=0.4]{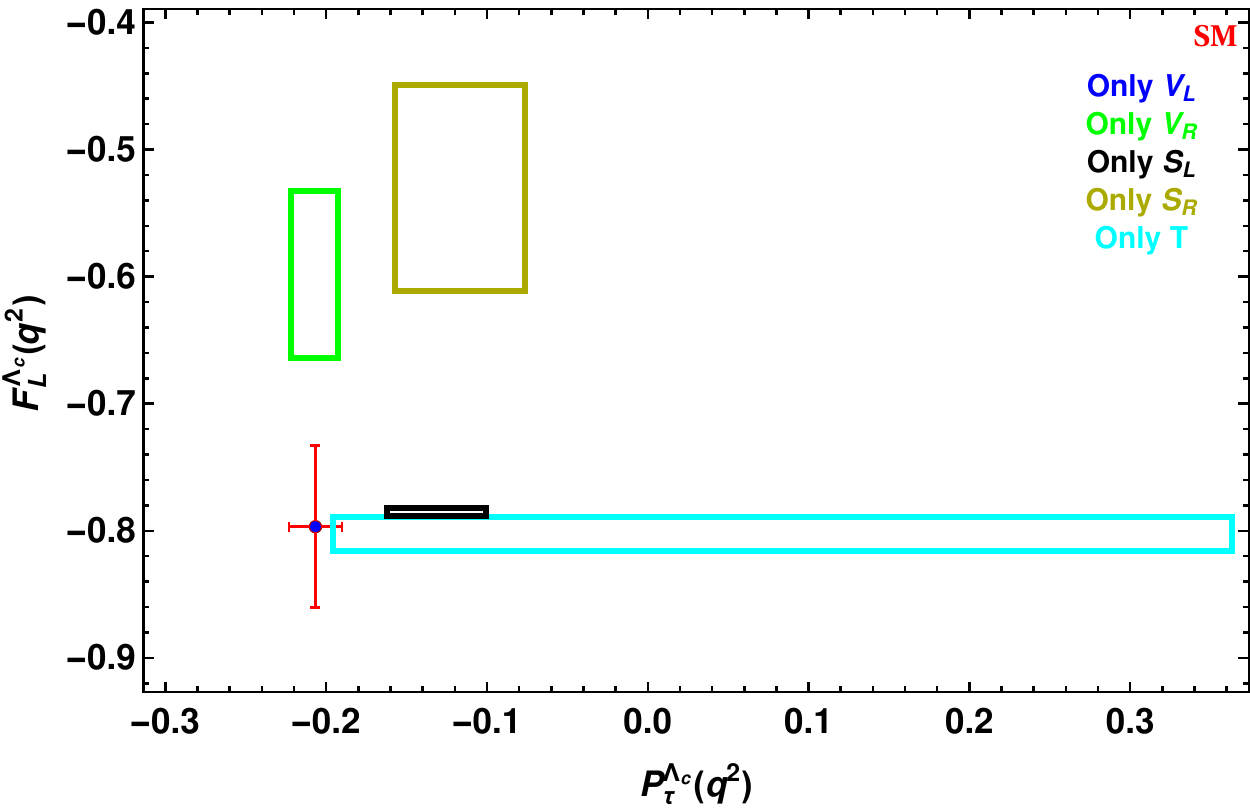}
\caption{Correlation between various angular observables of $b \to c \tau \bar \nu_\tau$ decays for case B.}\label{Fig:correlation}
\end{figure}

\begin{figure}[htb]
\includegraphics[scale=0.4]{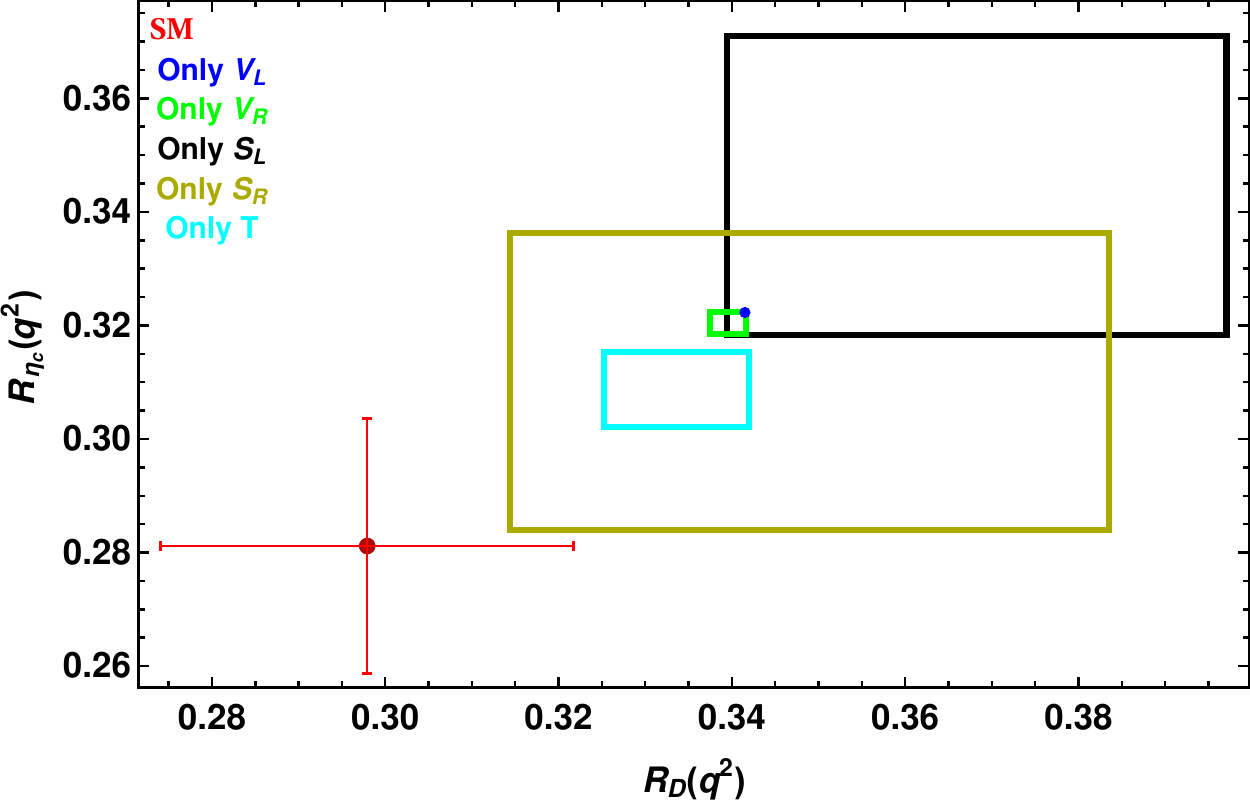}
\quad
\includegraphics[scale=0.4]{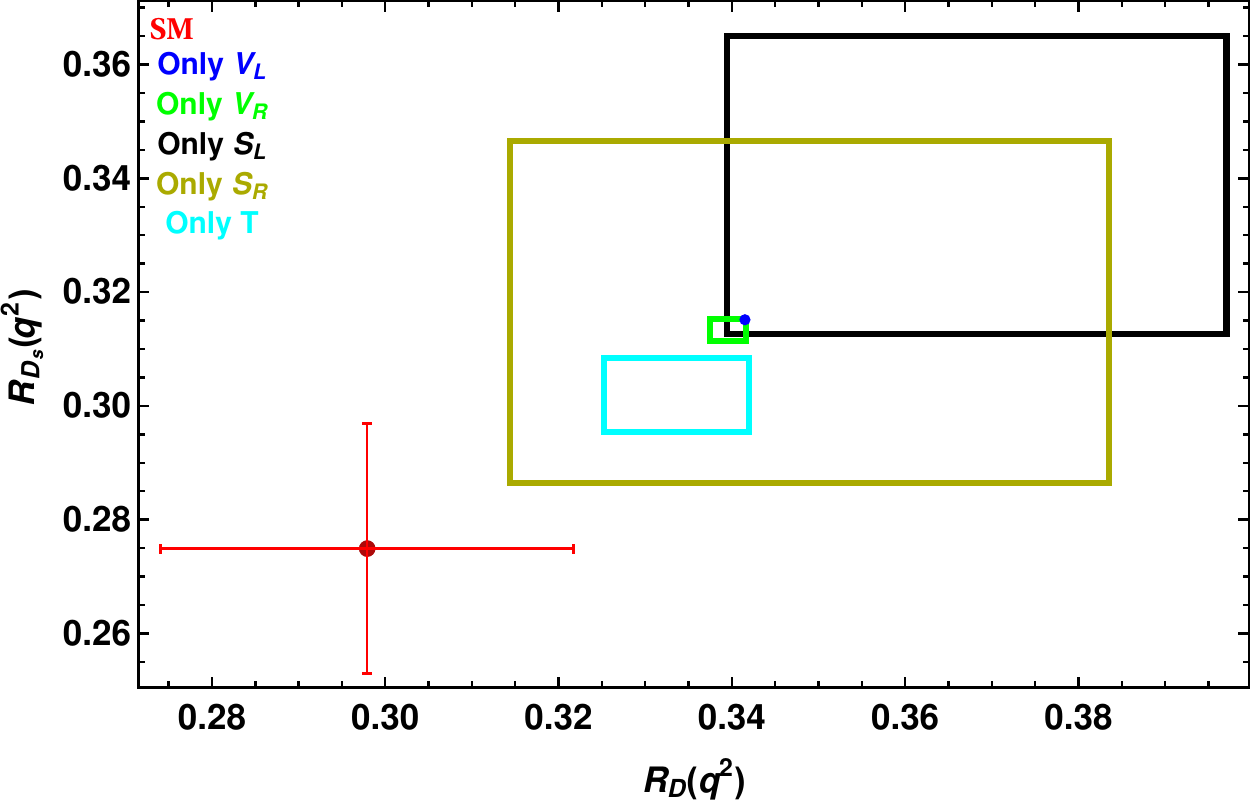}
\quad
\includegraphics[scale=0.4]{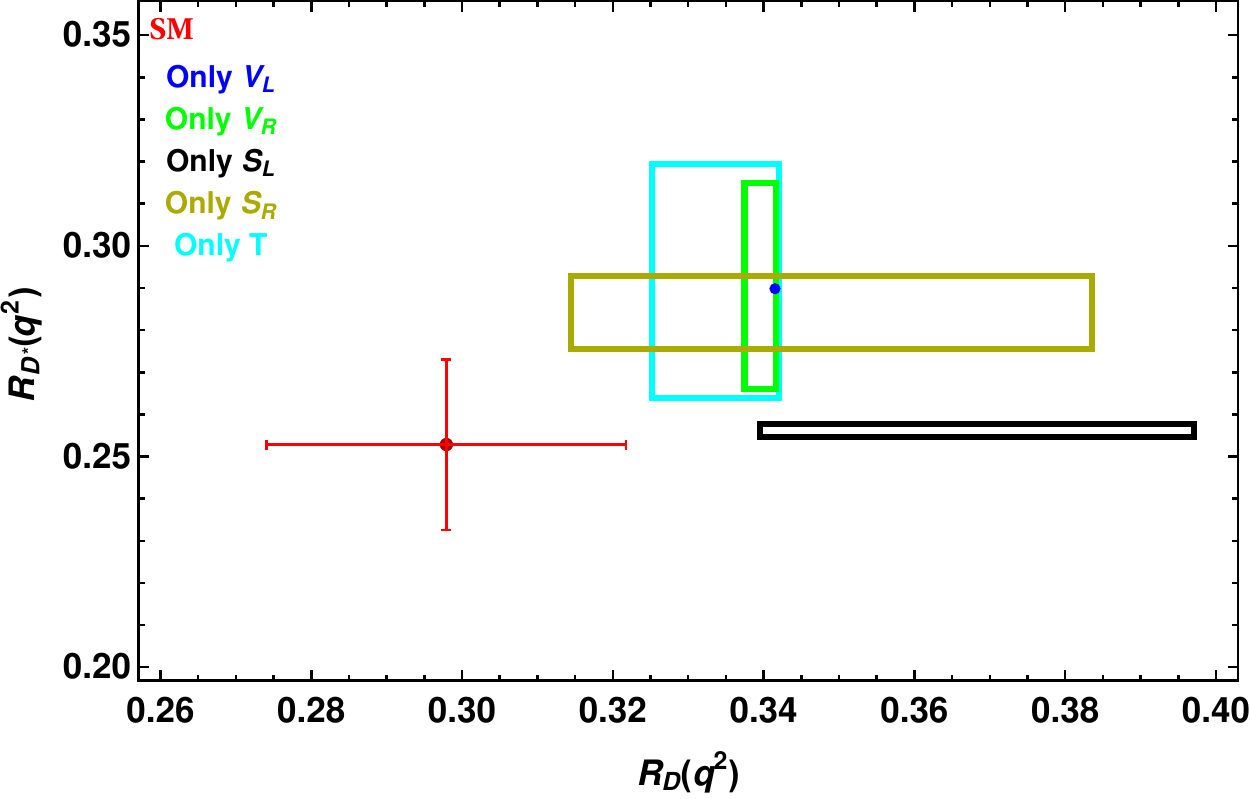}
\quad
\includegraphics[scale=0.4]{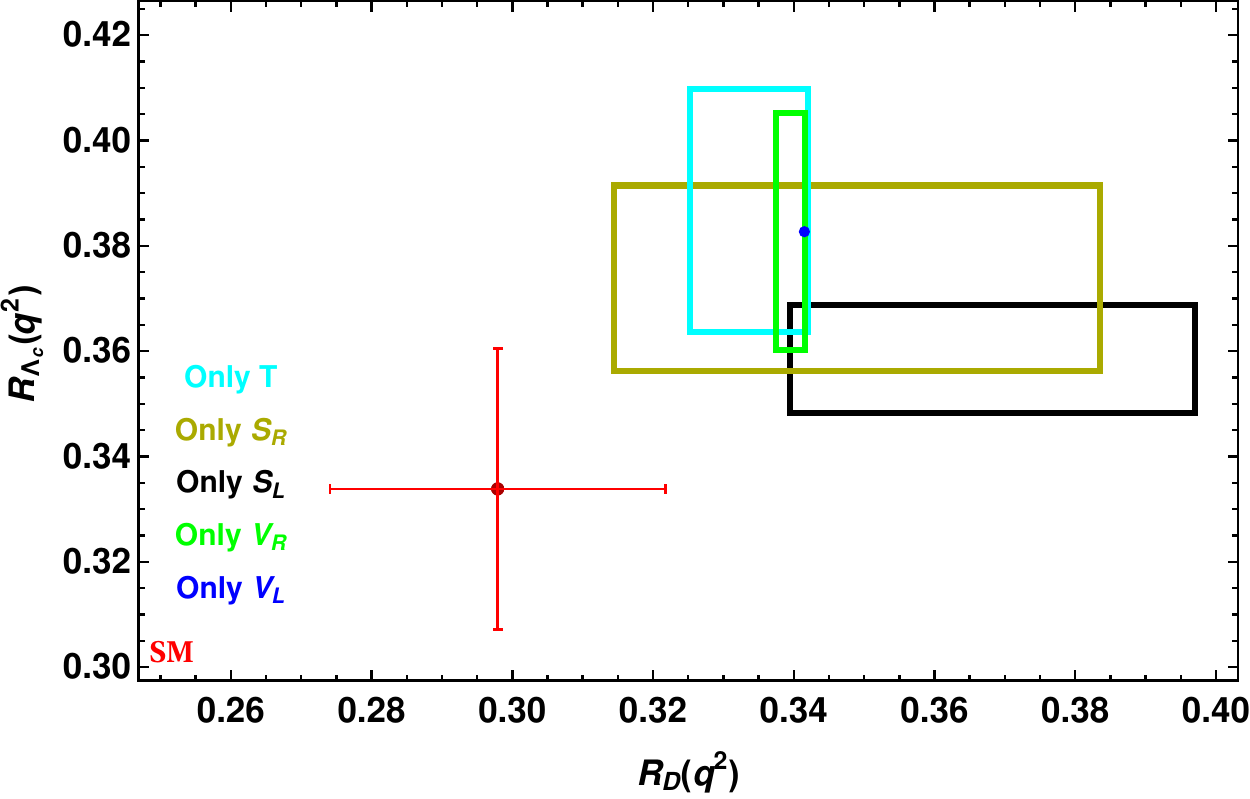}
\quad
\includegraphics[scale=0.4]{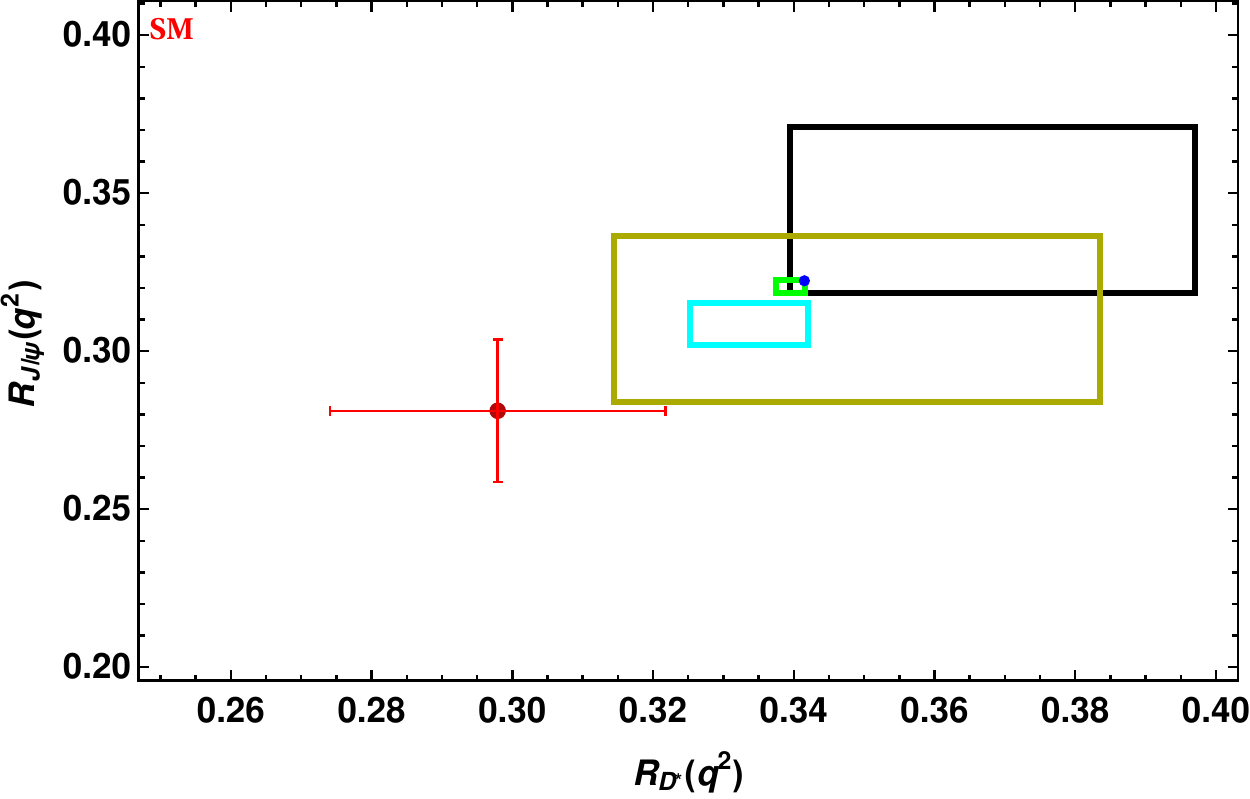}
\quad
\includegraphics[scale=0.4]{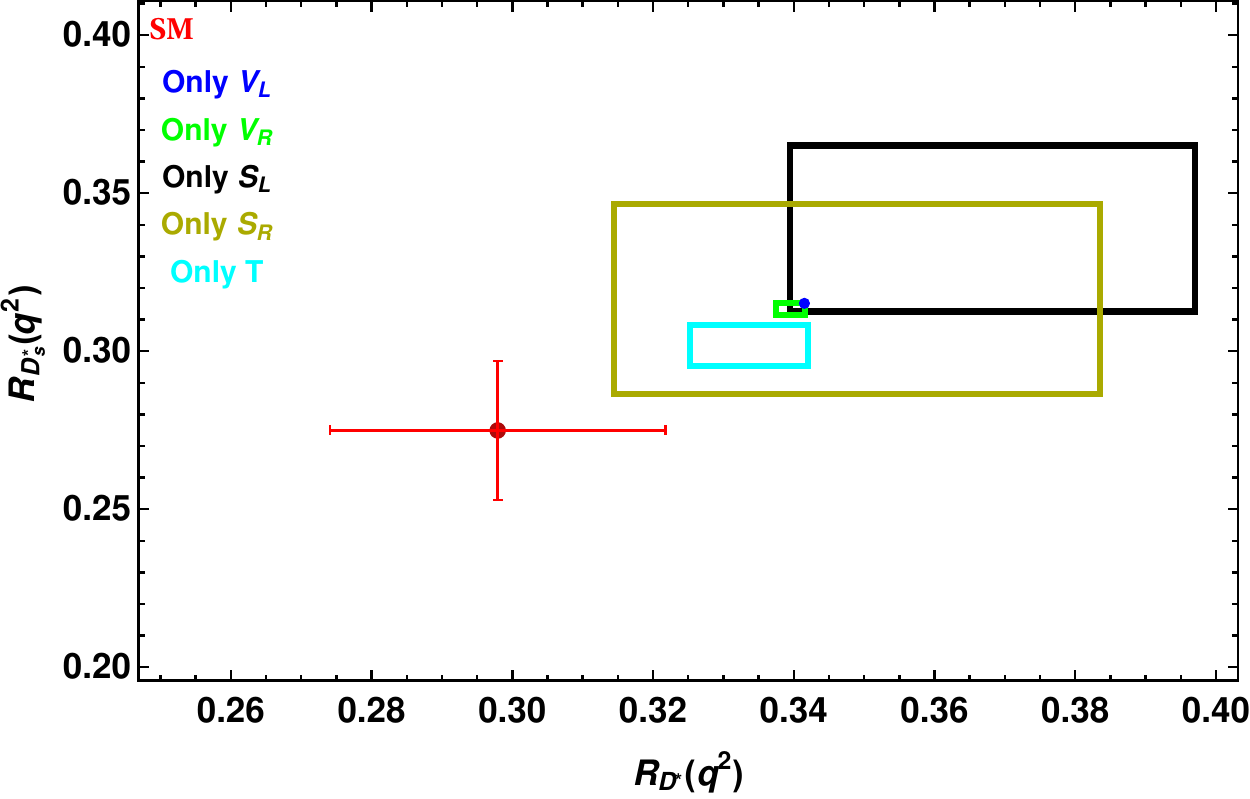}
\quad
\includegraphics[scale=0.4]{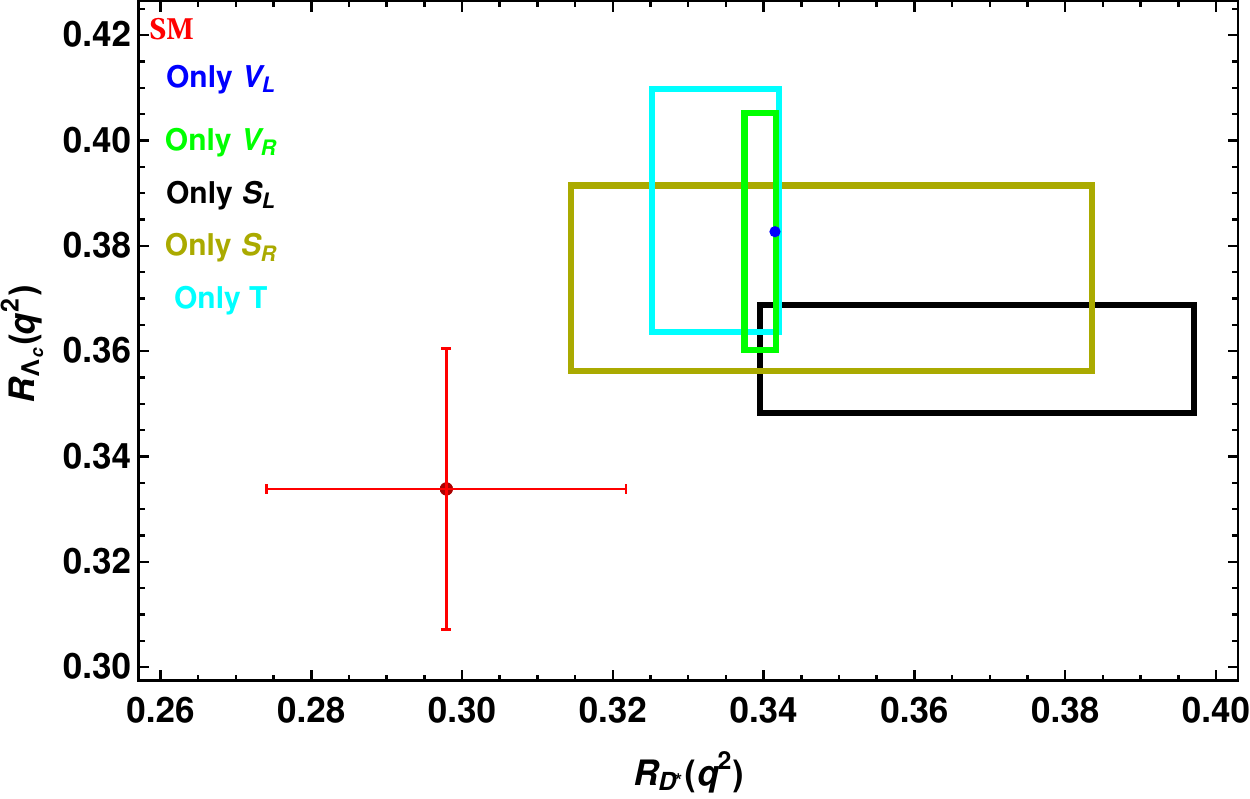}
\quad
\includegraphics[scale=0.4]{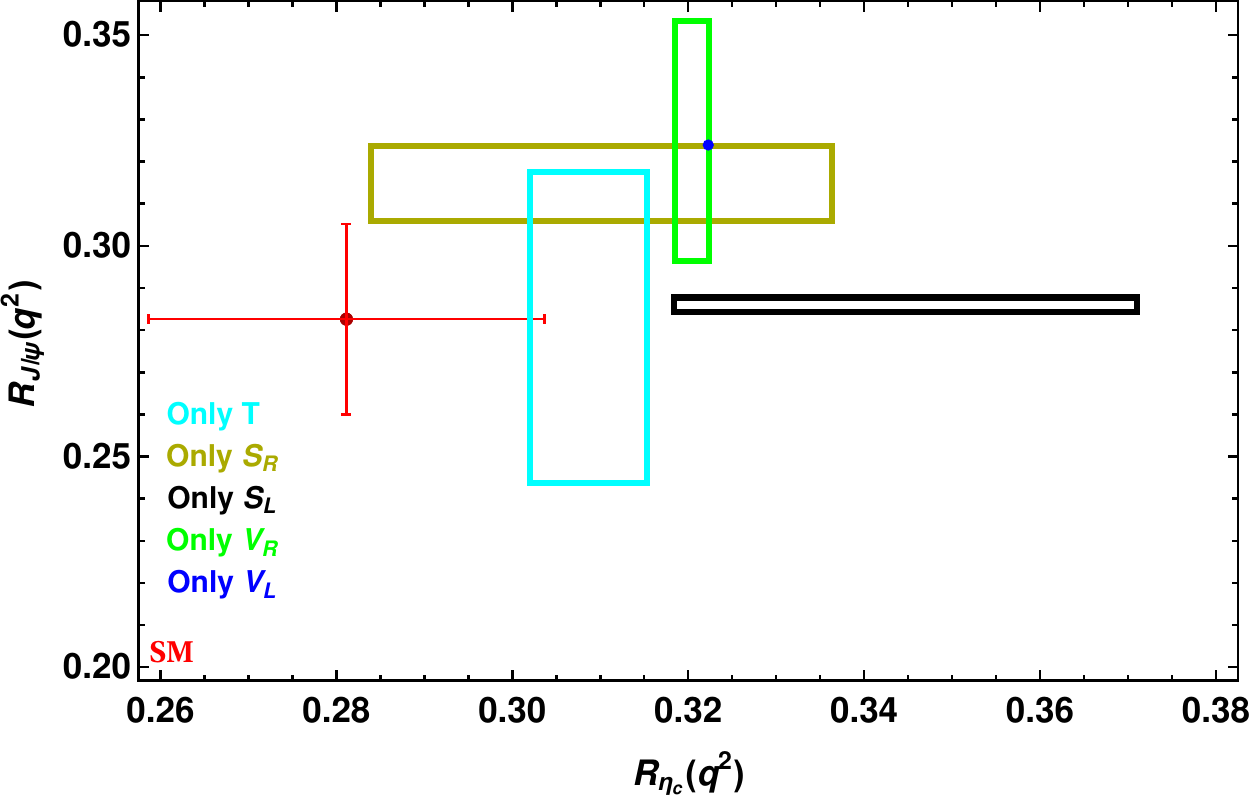}
\quad
\includegraphics[scale=0.4]{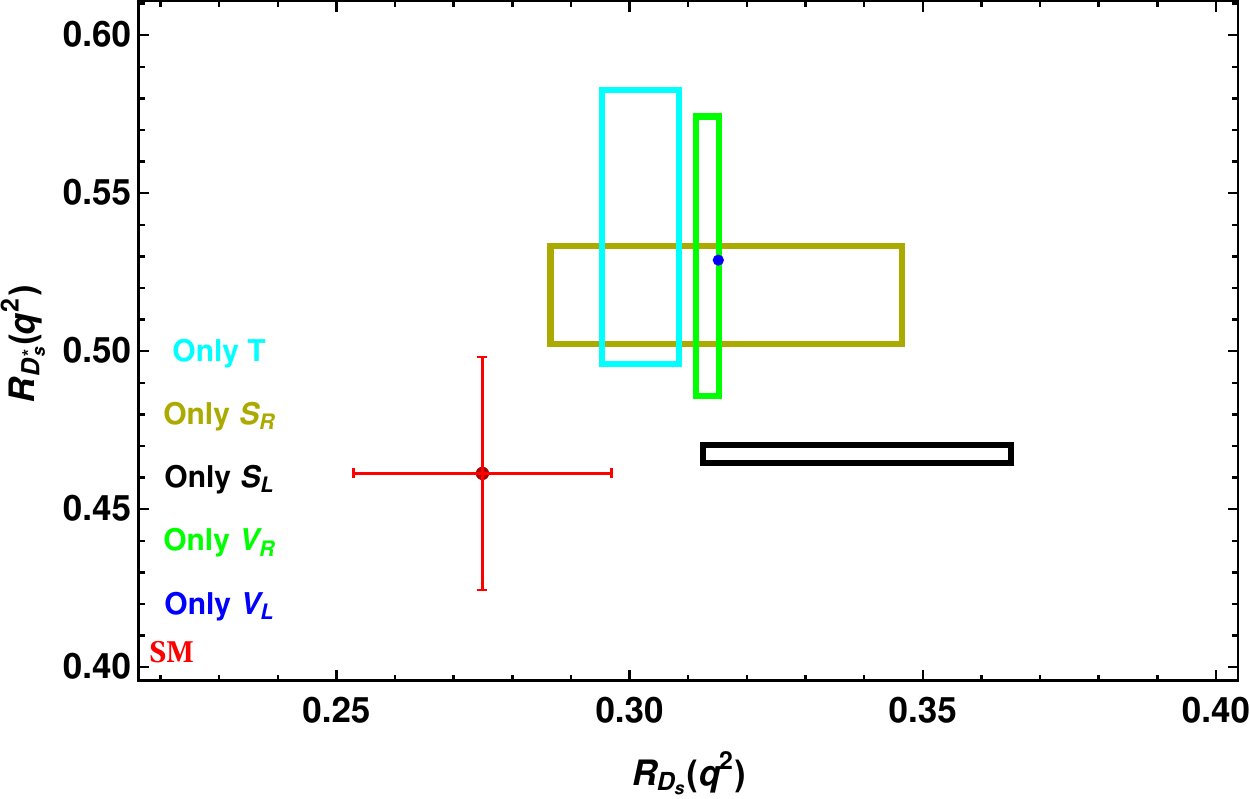}
\caption{Correlation between varrious lepton non-universality parameters for case B.}\label{Fig:LNU-correlation}
\end{figure}

\begin{figure}[htb]
\includegraphics[scale=0.38]{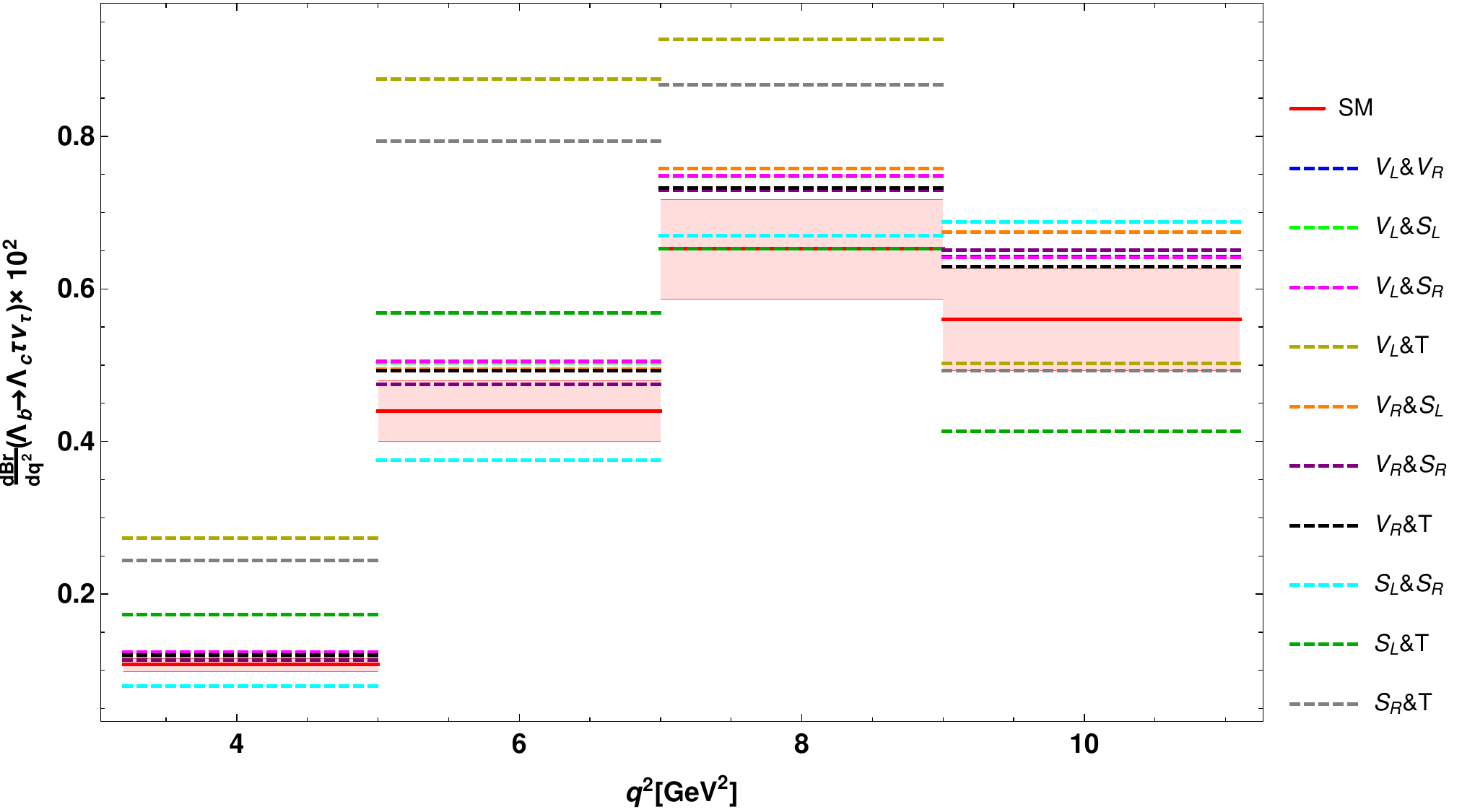}
\quad
\includegraphics[scale=0.38]{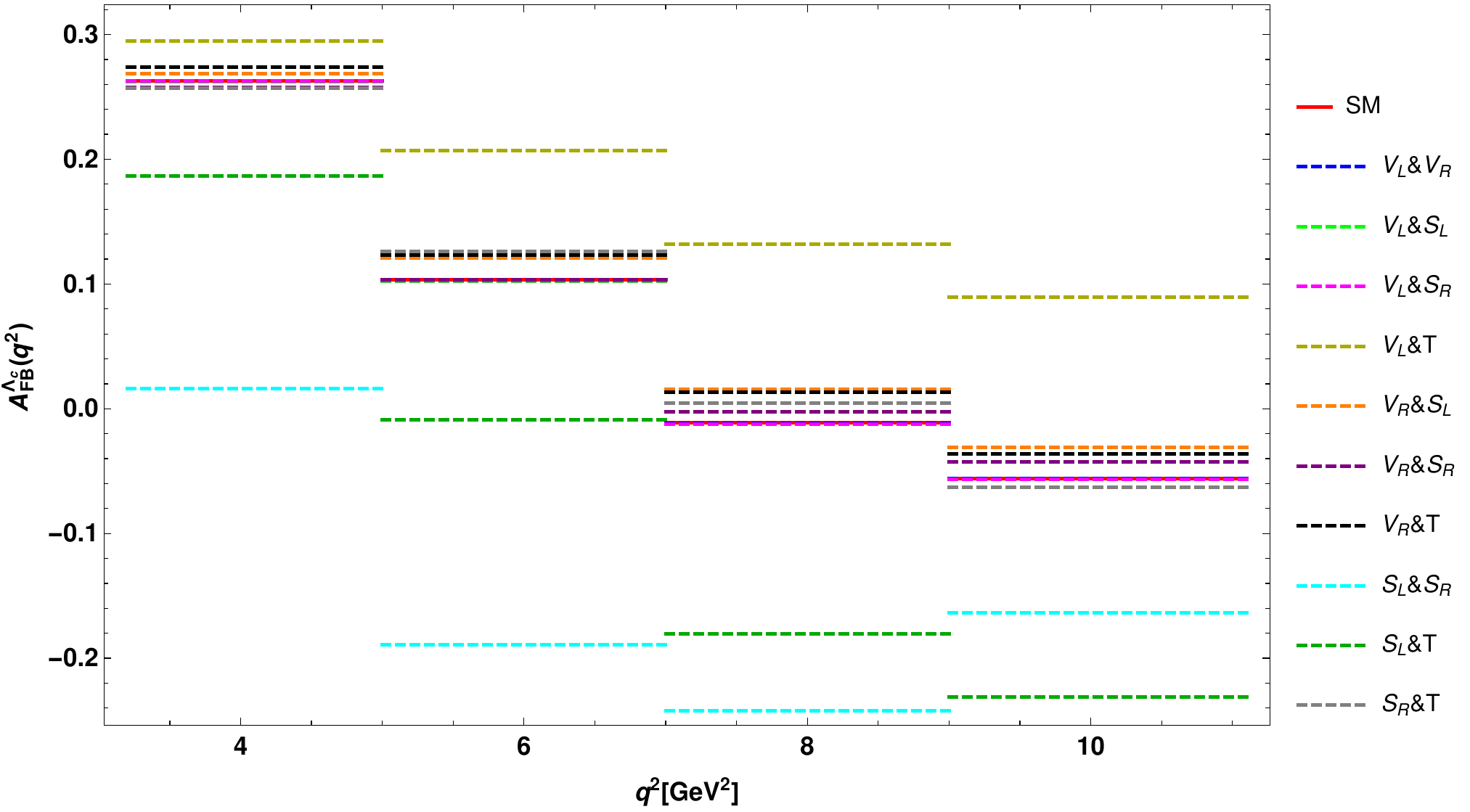}
\quad
\includegraphics[scale=0.38]{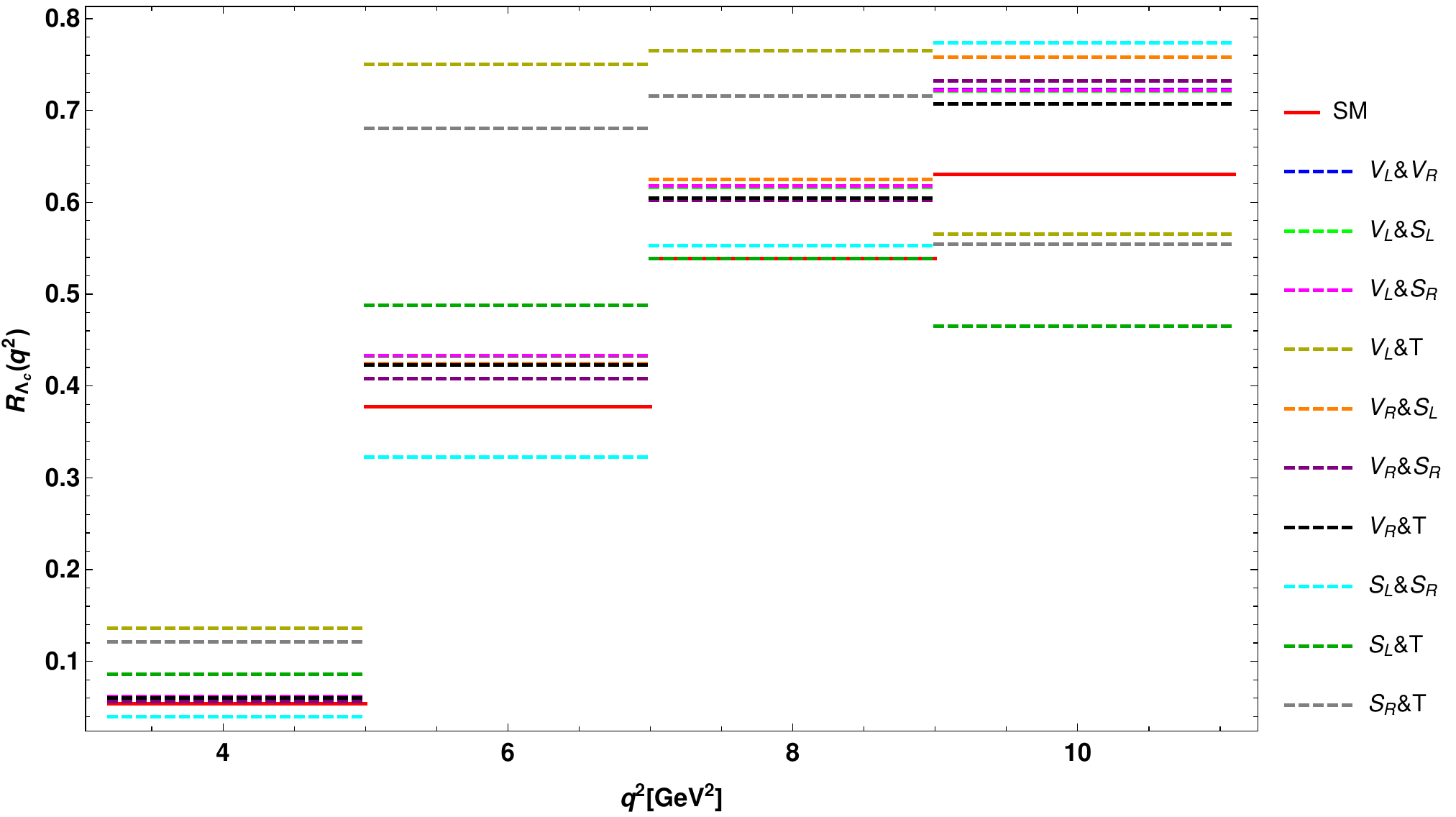}
\quad
\includegraphics[scale=0.38]{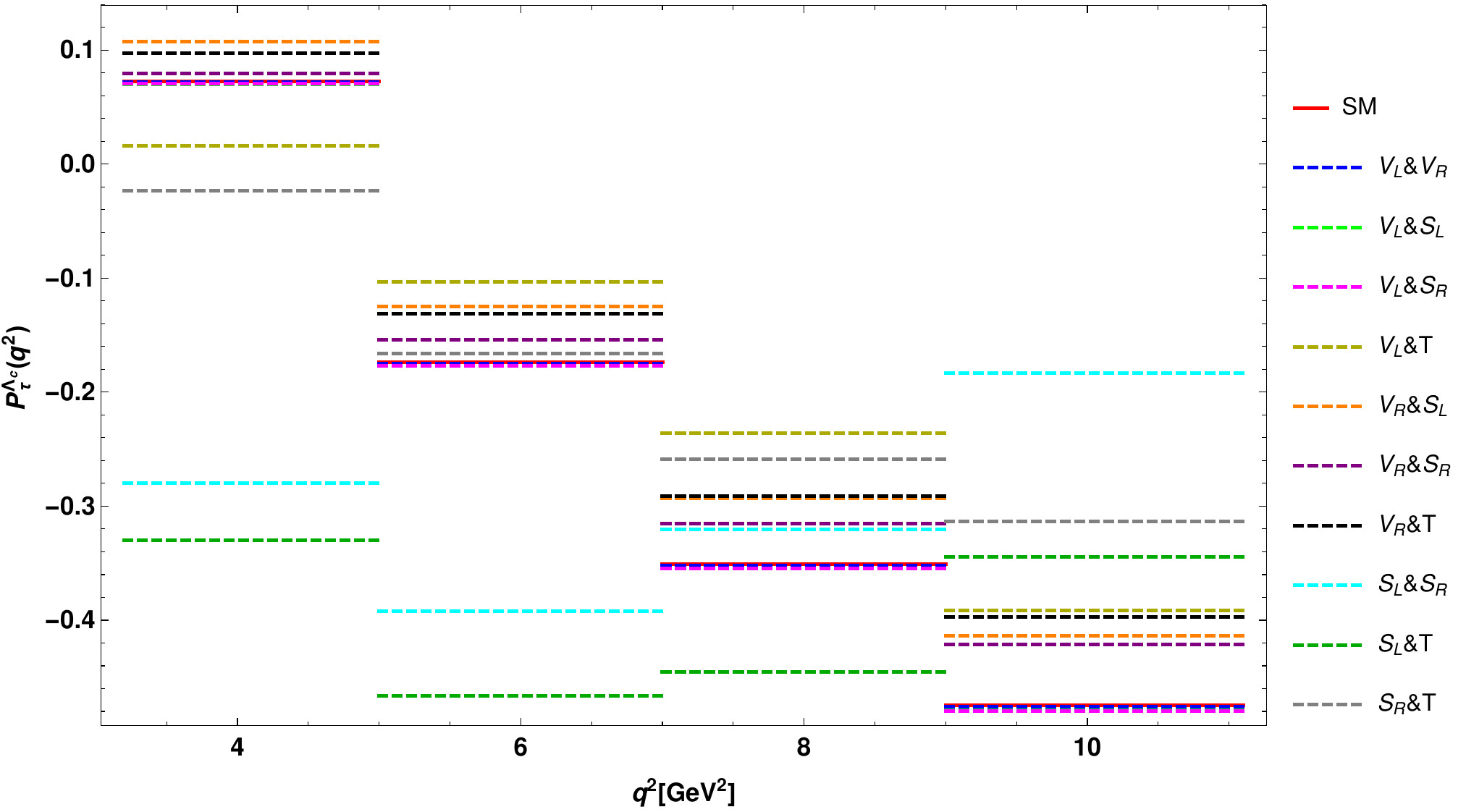}
\quad
\includegraphics[scale=0.38]{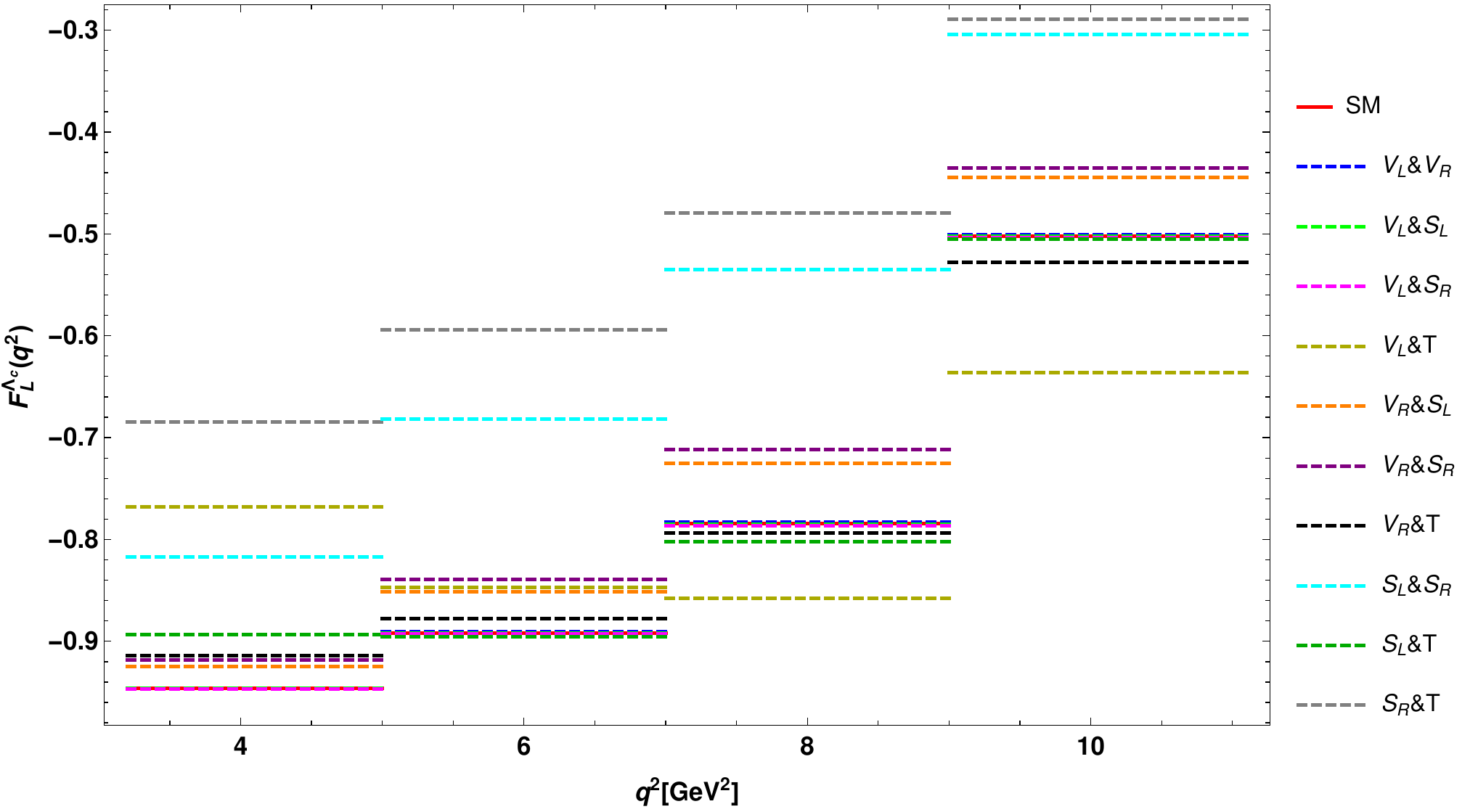}
\caption{ The bin-wise branching ratio (top-left panel), forward-backward asymmetry (top-right panel), $R_{\Lambda_c}$ (middle-left panel), longitudinal $\tau$ (middle-right panel) and $\Lambda_c$  (bottom panel) polarization asymmetry of $\Lambda_b \to \Lambda_c  \tau \bar \nu_\tau$  in four $q^2$ bins for case C. }\label{Fig:CC-Lambdab}
\end{figure}

This section discusses the decay rates and angular observables of $\Lambda_b \to \Lambda_c \tau \bar \nu_\tau$ processes mediated via $b \to c \tau \bar \nu_\tau$ quark level transitions. In our previous work \cite{Ray:2018hrx}\,, we have  presented the complete expressions for all the required observables. The $q^2$ dependence of the helicity form factors of $\Lambda_b \to \Lambda_c$,  computed  in the lattice QCD approach are taken from \cite{Detmold:2015aaa, Datta:2017aue} and the other remaining   input parameters from \cite{Tanabashi:2018oca}. Using the best-fit values of individual new real and complex Wilson coefficients from Table \ref{Tab:Best-fit}\,, we show the bin-wise graphical representation of branching ratios (top), forward-backward-asymmetry (second from top), $R_{\Lambda_c}$  (third from top), $\tau$-polarization (fourth from top) and $\Lambda_c$ polarization (bottom) asymmetries of  $\Lambda_b \to \Lambda_c \tau \bar \nu_\tau$ process in Fig. \ref{Fig:CA-CB-Lambdab}\,. Here the left (right) panel represents the plots for case A (case B). The presence of $V_L~(V_R)$  affects the branching ratio maximally in all the bins (none of the bins)  for case A and in the last three bins (last three) for case B, whereas the  $S_L$ contribution is within SM $1\sigma$ error range.  The $S_R$ has large effects on last two bins for both cases. The new physics  contribution to branching ratio, arising due to an additional $T$ lies within $1\sigma$ uncertainties range of SM for case A and has larger impact in the first three bins  for case B. The inclusion of $S_R$ coefficient provide  profound deviation in $A_{FB}^{\Lambda_c}$ observable from SM predictions  for both cases. The LNU parameters of $\Lambda_b \to \Lambda_c$ deviates significantly from SM in last three bins due to an  additional  coefficients. The real $V_R$ contribution is less in the $2^{\rm nd}$ bin and the complex $T$ affects the first bin of $R_{\Lambda_c}$ profoundly.  The new real and complex scalar contributions has shifted the values of $P_\tau^{\Lambda_c}$ significantly from the SM results and the $F_L^{\Lambda_c}$ is affected by real $S_R$ and complex $V_R/S_R$ coefficients.   The bin-wise numerical values for all observables for complex new Wilson coefficients are shown in Table \ref{Tab:CB-Lambdab}\,. Since, $F_T=1-F_L$, we don't include the $F_T^{\Lambda_c}$ values in  \ref{Tab:CB-Lambdab}\,. The bottom panel of Fig. \ref{Fig:correlation} represents the correlation plots of the $R_{\Lambda_c}$, $P_\tau^{\Lambda_c}$ and $F_L^{\Lambda_c}$ parameters in the full $q^2$ range obtained by using the $1\sigma$ range of new coefficients for case B. The correlation plots of the lepton non-universality parameters of all the discussed $B \to P$, $B \to V$ and $\Lambda_b \to \Lambda_c$ processes are graphically presented in Fig. \ref{Fig:LNU-correlation}\,.  Fig. \ref{Fig:CC-Lambdab}\,. depicts the bin-wise values for the branching ratio (top-left panel), $A_{FB}^{\Lambda_c}$ (top-right panel), $R_{\Lambda_c}$ (middle-left panel), $P_\tau^{\Lambda_c}$ (middle-right panel) and $F_L^{\Lambda_c}$ (bottom panel) observables of $\Lambda_b \to\Lambda_c$ process  graphically for case C. We find that, $V_L \& T$ and $S_R\& T$ has large impact on all observables, expect minor contribution of $S_R\& T$ to forward-backward asymmetry. The  $S_L\& S_R$ ($S_L \& T$) affects  $A_{FB}^{\Lambda_c}$, $P_\tau$ and $F_L^{\Lambda_c}$ ($A_{FB}^{\Lambda_c}$, $R_{\Lambda_c}$ and $P_\tau$) observables. 
\begin{table}[htb]
\scriptsize
\caption{Predicted bin-wise values of branching ratio and angular observables of  $\Lambda_b \to \Lambda_c \tau \bar \nu_\tau$ process in the SM and in the presence of new complex Wilson coefficients (case B).}\label{Tab:CB-Lambdab}
\begin{center}
\begin{tabular}{|c|c|c|c|c|c|c|}
\hline
 ~Observables~&~Values for SM~&~values for $V_L$~&~Values for $V_R$~&~values for $V_L$~&~Values for $V_R$~&~Values for $T$~\\
\hline\hline
~$10^3$ Br$(\Lambda_b \to \Lambda_c)|_{q^2\in [m_\tau^2,5]}$~&~$(1.08\pm 0.09)$~&~$1.233$~&~$1.23$~&~$1.123$~&~$1.08$~&~$1.54$~\\

~$10^3$ Br$(\Lambda_b \to \Lambda_c )|_{q^2\in [5,7]}$~&~$(4.4\pm 0.396)$~&~$5.05$~&~$5.04$~&~$4.62$~&~$4.6$~&~$5.67$\\

~$10^3$ Br$(\Lambda_b \to \Lambda_c)|_{q^2\in [7,9]}$~&~$(6.53\pm 0.653)$~&~$7.481$~&~$7.476$~&~$6.863$~&~$7.2$~&~$7.46$~\\

~$10^3$ Br$(\Lambda_b \to \Lambda_c)|_{q^2\in [9,11.6]}$~&~$(5.6\pm 0.672)$~&~$6.42$~&~$6.424$~&~$5.9$~&~$6.51$~&~$5.635$~\\
\hline

~$\langle A_{FB}^{\Lambda_c} \rangle|_{q^2\in [m_\tau^2,5]}$~&~$(0.263\pm 0.02)$~&~$0.263$~&~$0.286$~&~$0.271$~&~$0.21$~&~$0.277$~\\

~$\langle A_{FB}^{\Lambda_c} \rangle|_{q^2\in [5,7]}$~&~$(0.104\pm 0.009)$~&~$0.103$~&~$0.14$~&~$0.1184$~&~$0.039$~&~$0.14$~\\

~$\langle A_{FB}^{\Lambda_c} \rangle|_{q^2\in [7,9]}$~&~$(-0.0114\pm 0.001)$~&~$-0.0114$~&~$0.0324$~&~$0.0067$~&~$-0.06$~&~$0.0263$~\\

~$\langle A_{FB}^{\Lambda_c} \rangle|_{q^2\in [9,11.6]}$~&~$(-0.056\pm 0.0067)$~&~$-0.056$~&~$-0.0213$~&~$-0.0422$~&~$-0.08$~&~$-0.032$~\\
\hline
~$\langle R_{\Lambda_c} \rangle|_{q^2\in [m_\tau^2,5]}$~&~$0.0536$~&~$0.0615$~&~$0.0613$~&~$0.056$~&~$0.054$~&~$0.0767$~\\

~$\langle  R_{\Lambda_c}\rangle|_{q^2\in [5,7]}$~&~$0.377$~&~$0.433$~&~$0.432$~&~$0.396$~&~$0.395$~&~$0.486$~\\

~$\langle  R_{\Lambda_c} \rangle|_{q^2\in [7,9]}$~&~$0.539$~&~$0.617$~&~$0.617$~&~$0.566$~&~$0.595$~&~$0.616$~\\

~$\langle  R_{\Lambda_c} \rangle|_{q^2\in [9,11.6]}$~&~$0.63$~&~$0.722$~&~$0.723$~&~$0.664$~&~$0.733$~&~$0.34$~\\
\hline

~$\langle P_\tau^{\Lambda_c} \rangle|_{q^2\in [m_\tau^2,5]}$~&~$0.0725$~&~$0.0725$~&~$0.0715$~&~$0.11$~&~$0.0533$~&~$0.063$~\\

~$\langle P_\tau^{\Lambda_c} \rangle|_{q^2\in [5,7]}$~&~$-0.174$~&~$-0.174$~&~$-0.176$~&~$-0.1195$~&~$-0.127$~&~$-0.161$~\\

~$\langle P_\tau^{\Lambda_c} \rangle|_{q^2\in [7,9]}$~&~$-0.351$~&~$-0.351$~&~$-0.353$~&~$-0.2846$~&~$-0.226$~&~$-0.333$~\\

~$\langle P_\tau \rangle|_{q^2\in [9,11.6]}$~&~$-0.474$~&~$-0.474$~&~$-0.4765$~&~$-0.4$~&~$-0.263$~&~$-0.465$~\\
\hline
~$\langle F_L^{\Lambda_c} \rangle|_{q^2\in [m_\tau^2,5]}$~&~$-0.946$~&~$-0.946$~&~$-0.71$~&~$-0.943$~&~$-0.722$~&~$-0.923$~\\

~$\langle F_L^{\Lambda_c} \rangle|_{q^2\in [5,7]}$~&~$-0.892$~&~$-0.892$~&~$-0.67$~&~$-0.8865$~&~$-0.632$~&~$-0.892$~\\

~$\langle F_L^{\Lambda_c} \rangle|_{q^2\in [7,9]}$~&~$-0.7846$~&~$-0.7846$~&~$-0.587$~&~$-0.778$~&~$-0.487$~&~$-0.803$~\\

~$\langle F_L^{\Lambda_c} \rangle|_{q^2\in [9,11.6]}$~&~$-0.5$~&~$-0.5$~&~$-0.375$~&~$-0.496$~&~$-0.279$~&~$-0.525$~\\
\hline
\end{tabular}
\end{center}
\end{table}

\section{$B \to D^{**} \tau \bar \nu_\tau$ processes}

\begin{figure}[htb]
\includegraphics[scale=0.5]{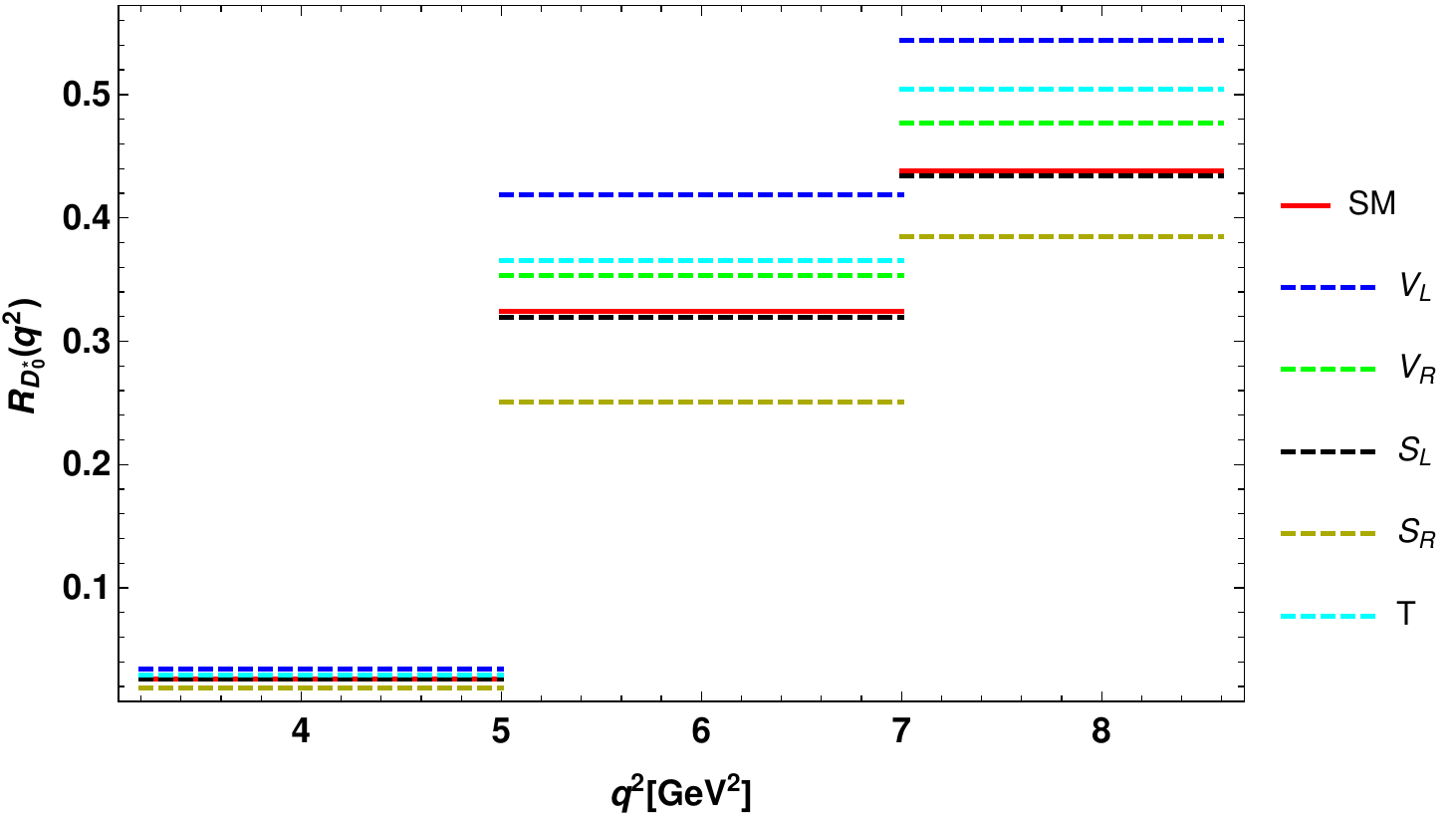}
\quad
\includegraphics[scale=0.5]{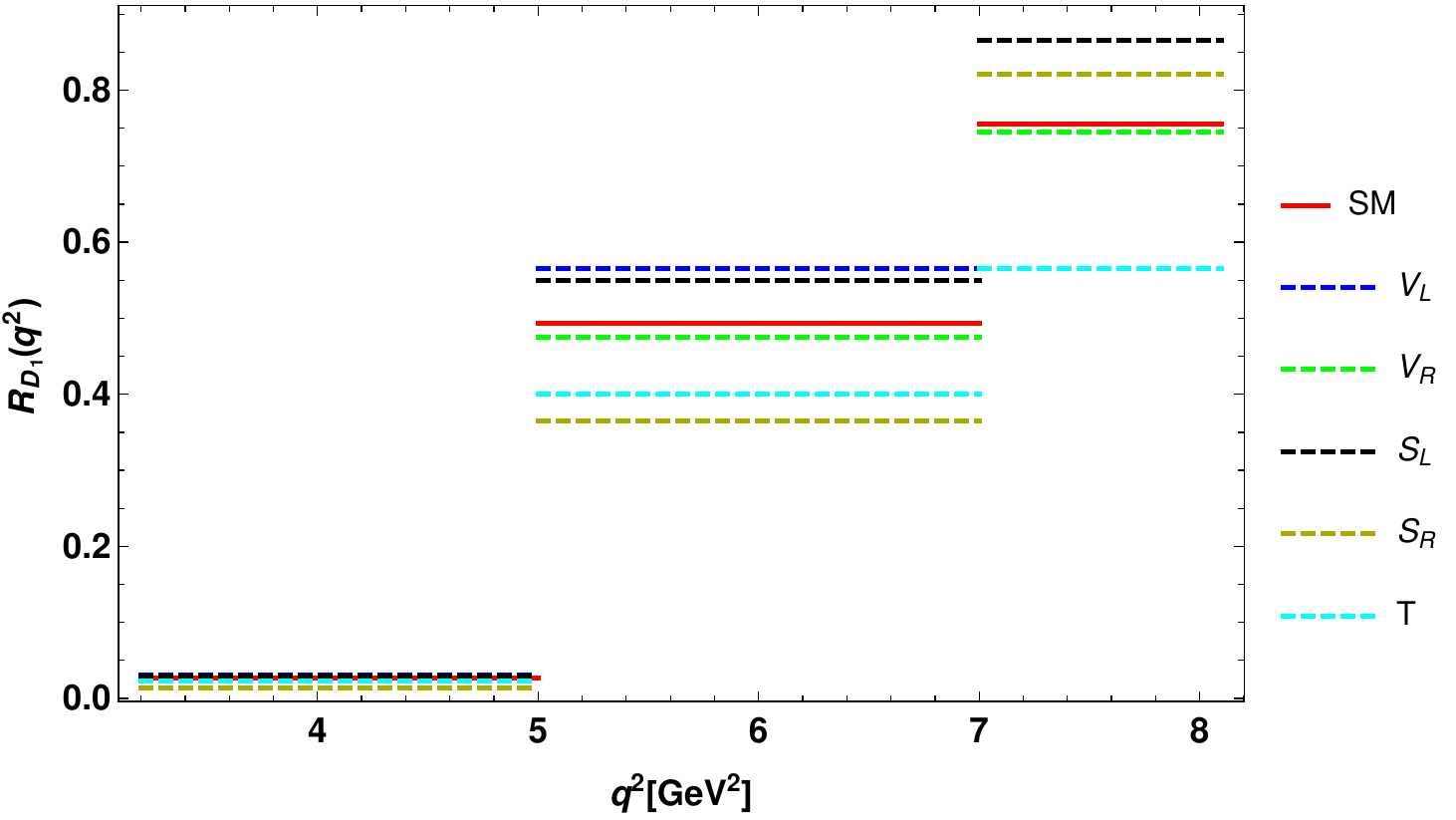}
\quad
\includegraphics[scale=0.5]{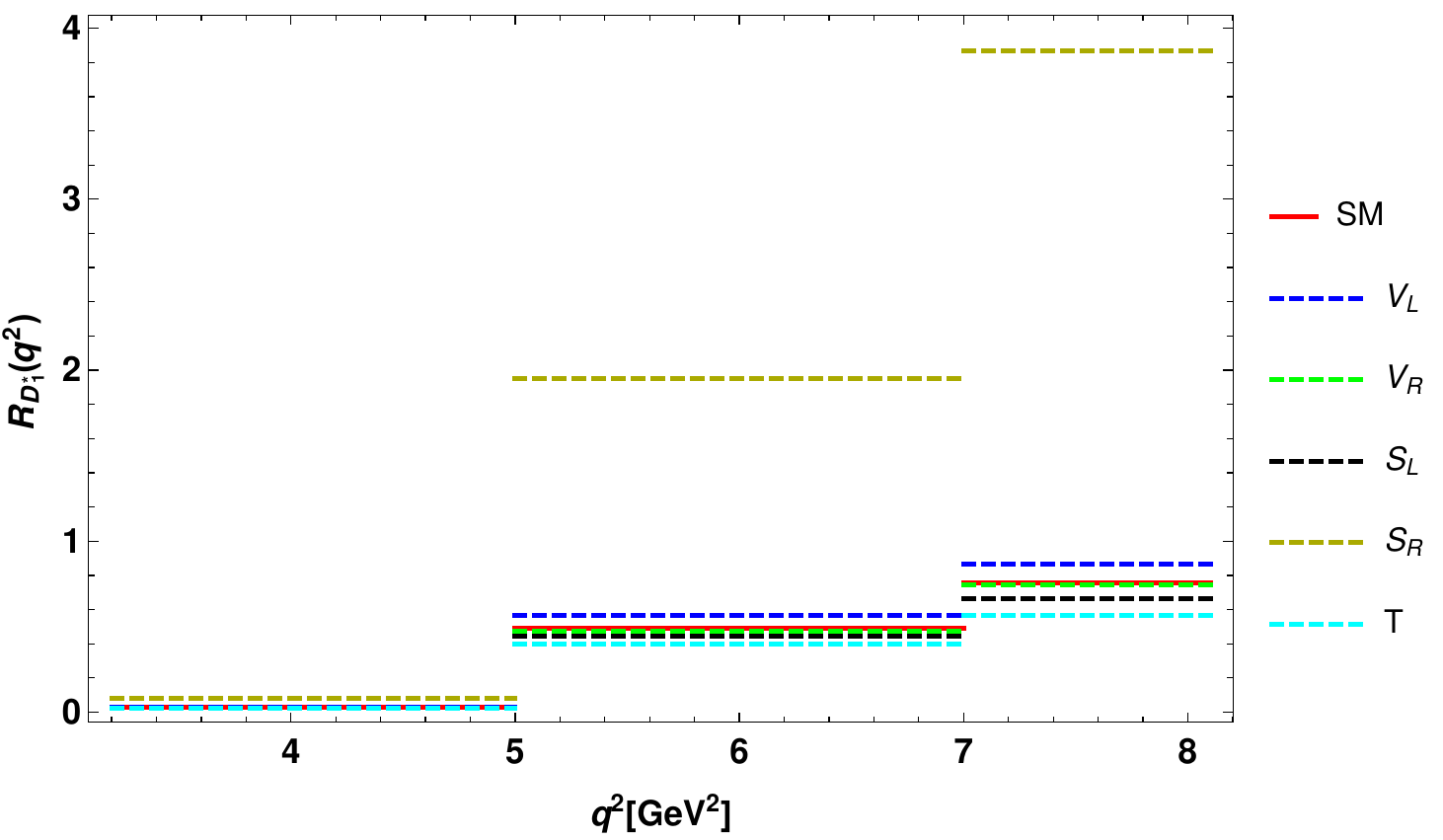}
\quad
\includegraphics[scale=0.5]{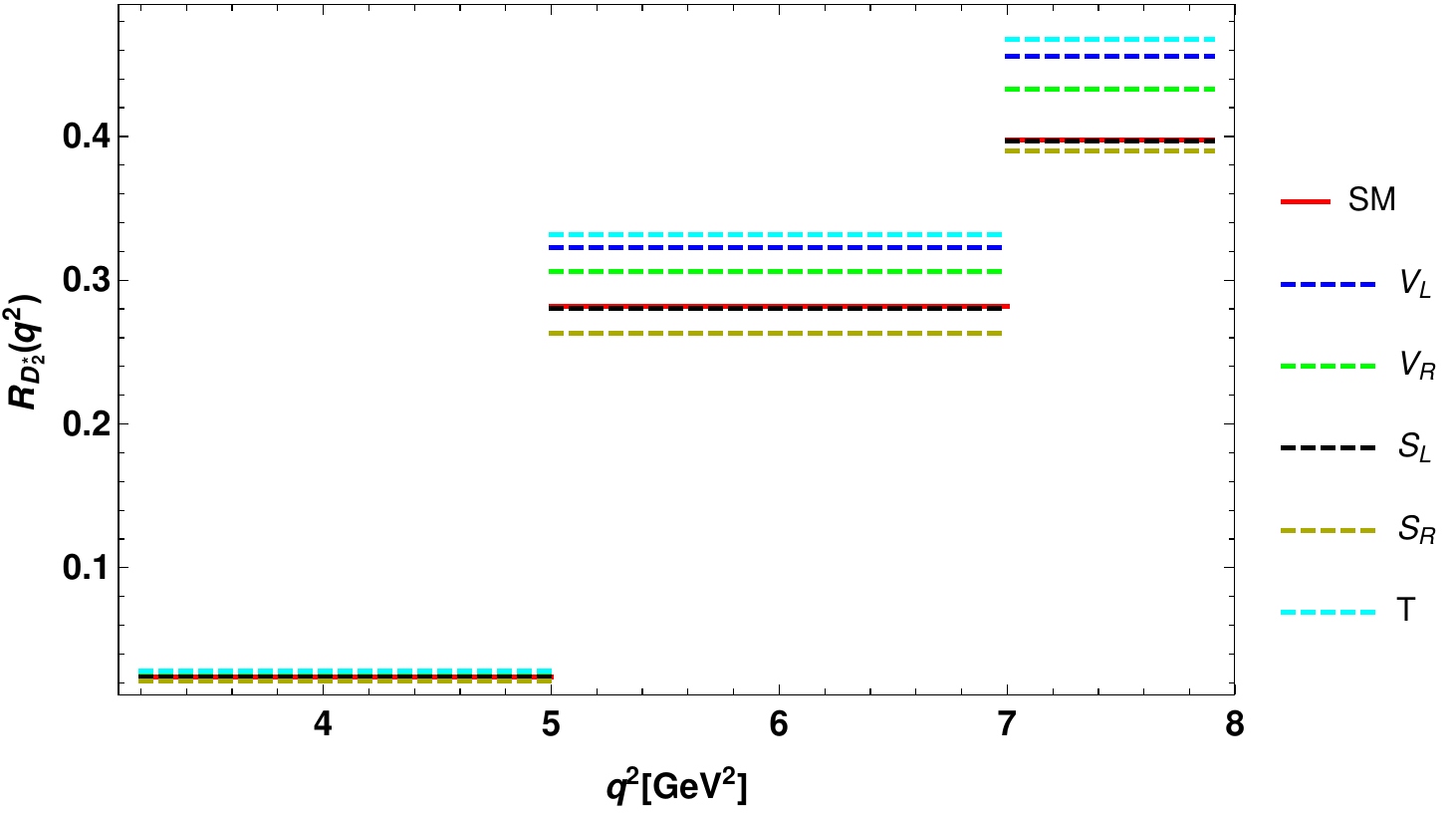}
\caption{ The bin-wise graphical representation of $R_{D^*_0}$ (top-left panel), $R_{D_1}$ (top-right panel), $R_{D^*_1}$ (bottom-left panel) and $R_{D^*_2}$ (bottom-right panel) parameters for case A. }\label{Fig:CA-RD**}
\end{figure}

\begin{figure}[htb]
\includegraphics[scale=0.5]{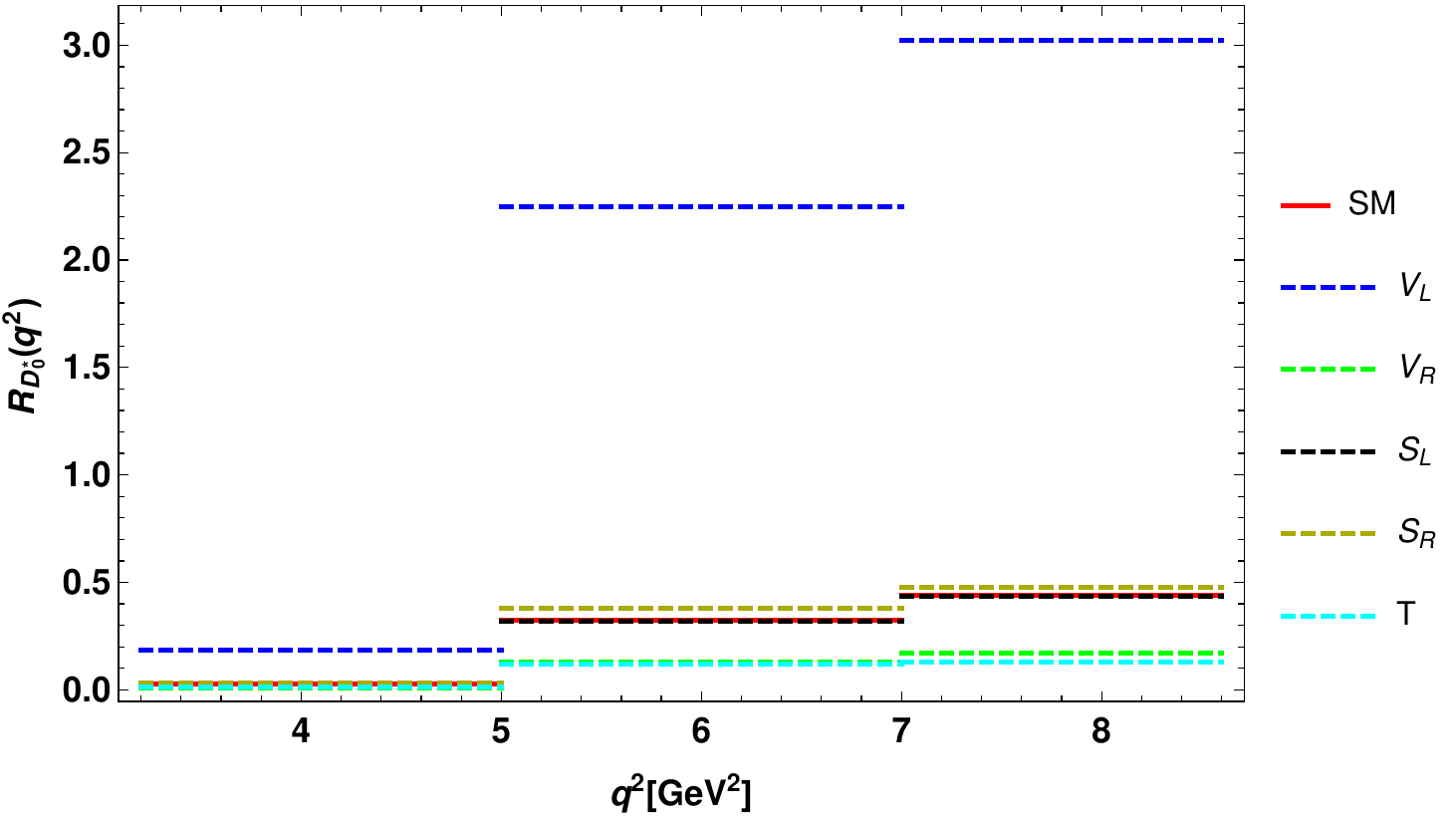}
\quad
\includegraphics[scale=0.5]{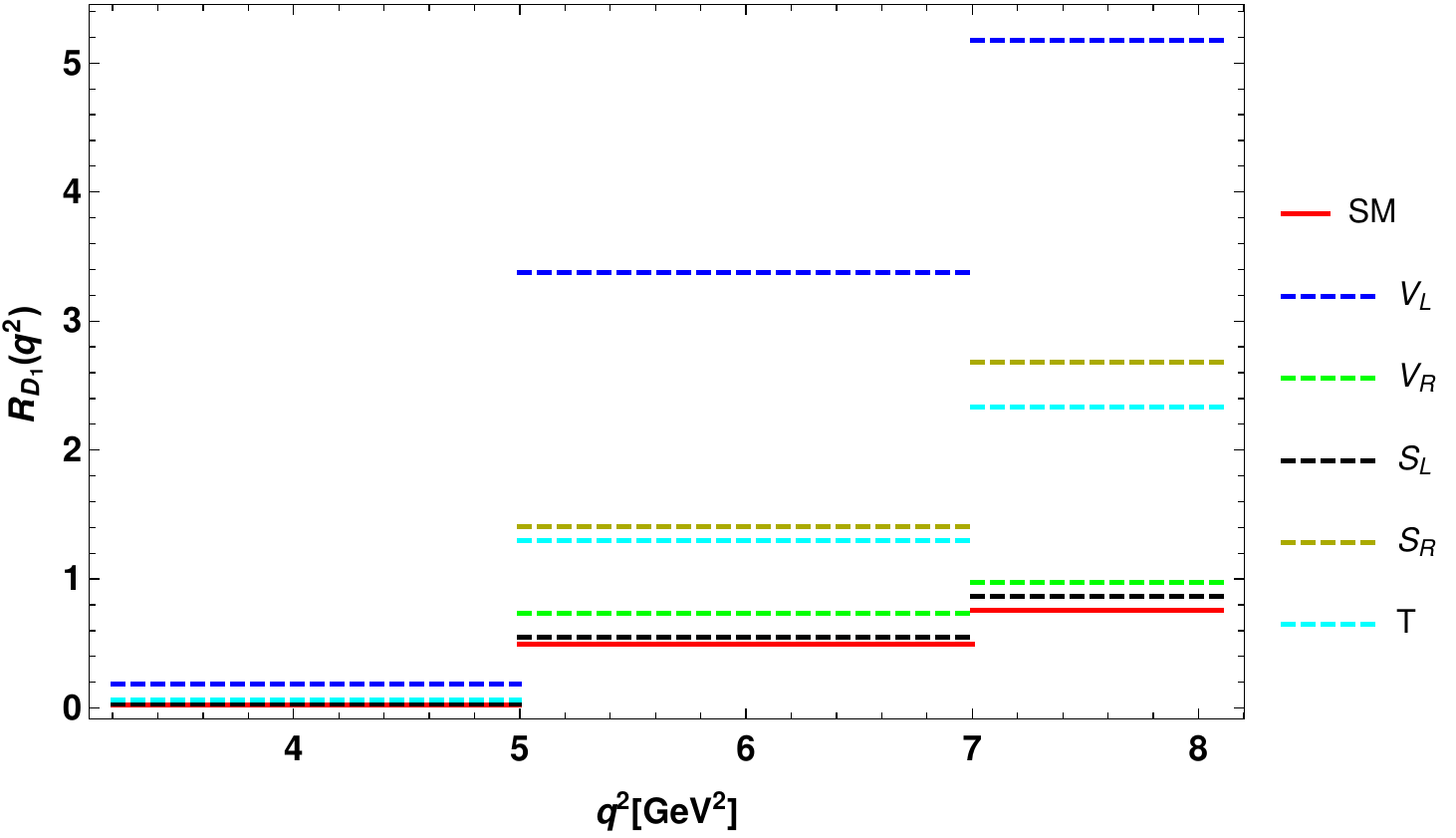}
\quad
\includegraphics[scale=0.5]{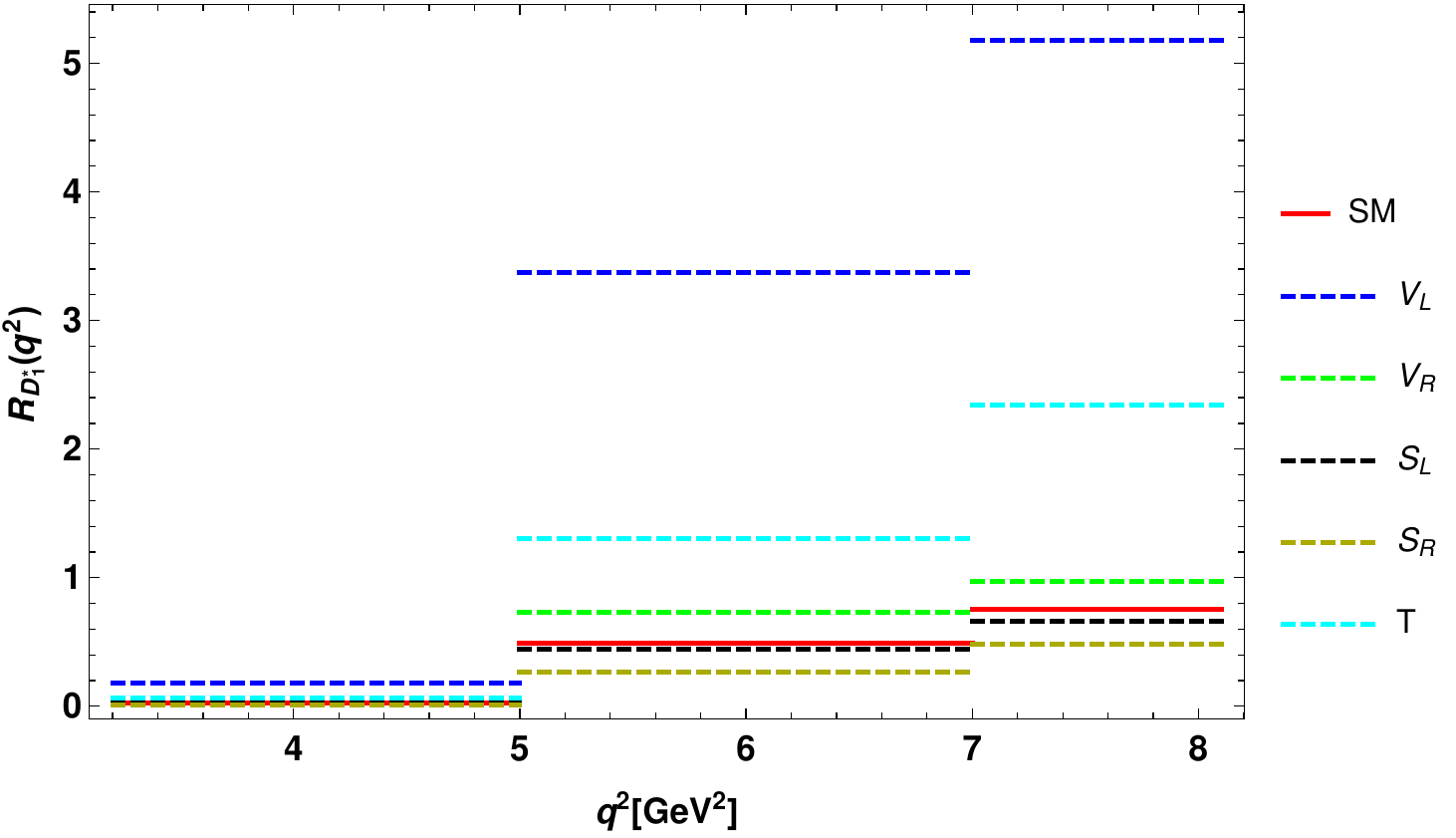}
\quad
\includegraphics[scale=0.5]{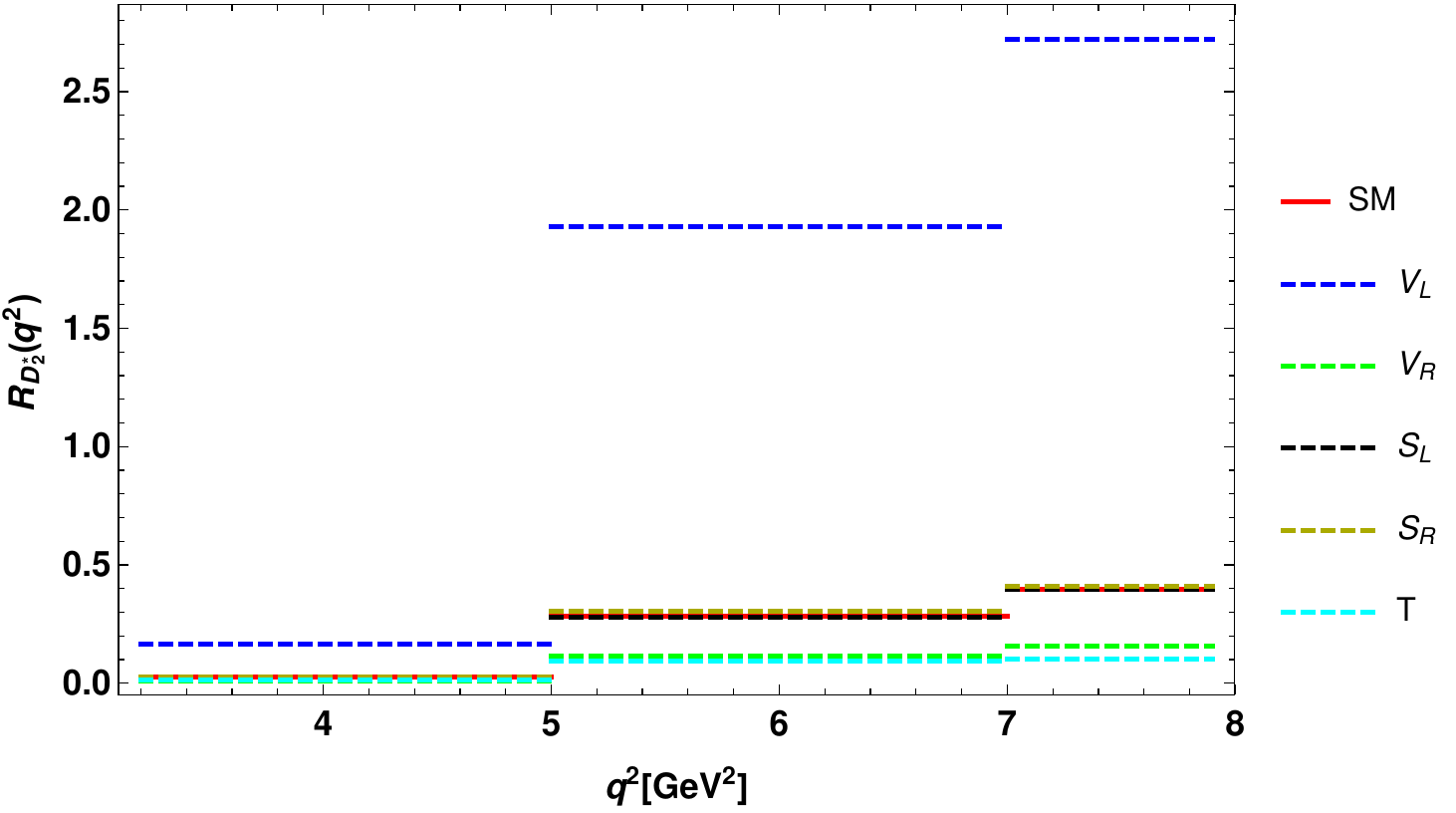}
\caption{ The bin-wise graphical representation of $R_{D^*_0}$ (top-left panel), $R_{D_1}$ (top-right panel), $R_{D^*_1}$ (bottom-left panel) and $R_{D^*_2}$ (bottom-right panel) parameters for case B. }\label{Fig:CB-RD**}
\end{figure}

\begin{figure}[htb]
\includegraphics[scale=0.38]{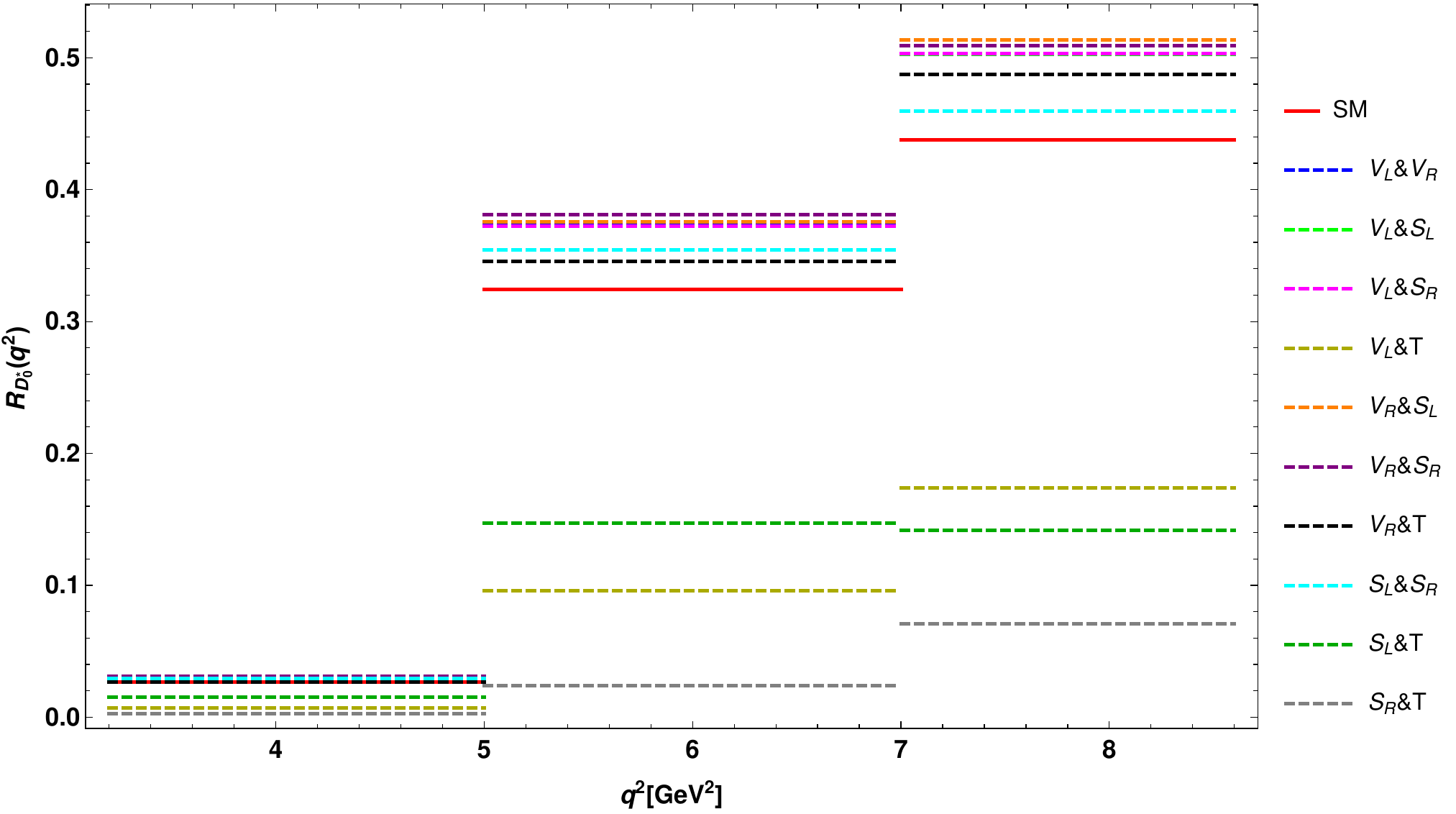}
\quad
\includegraphics[scale=0.38]{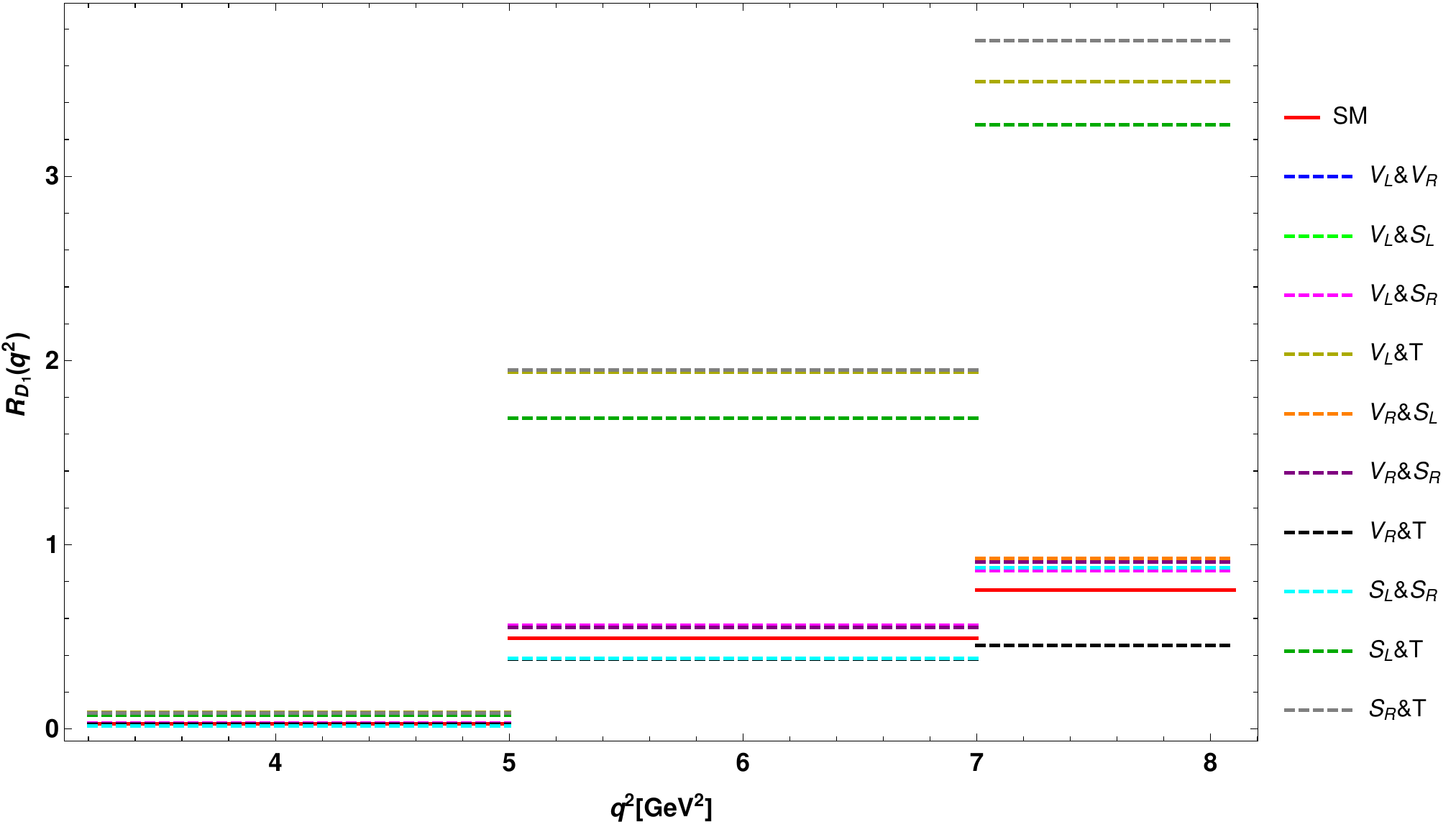}
\quad
\includegraphics[scale=0.38]{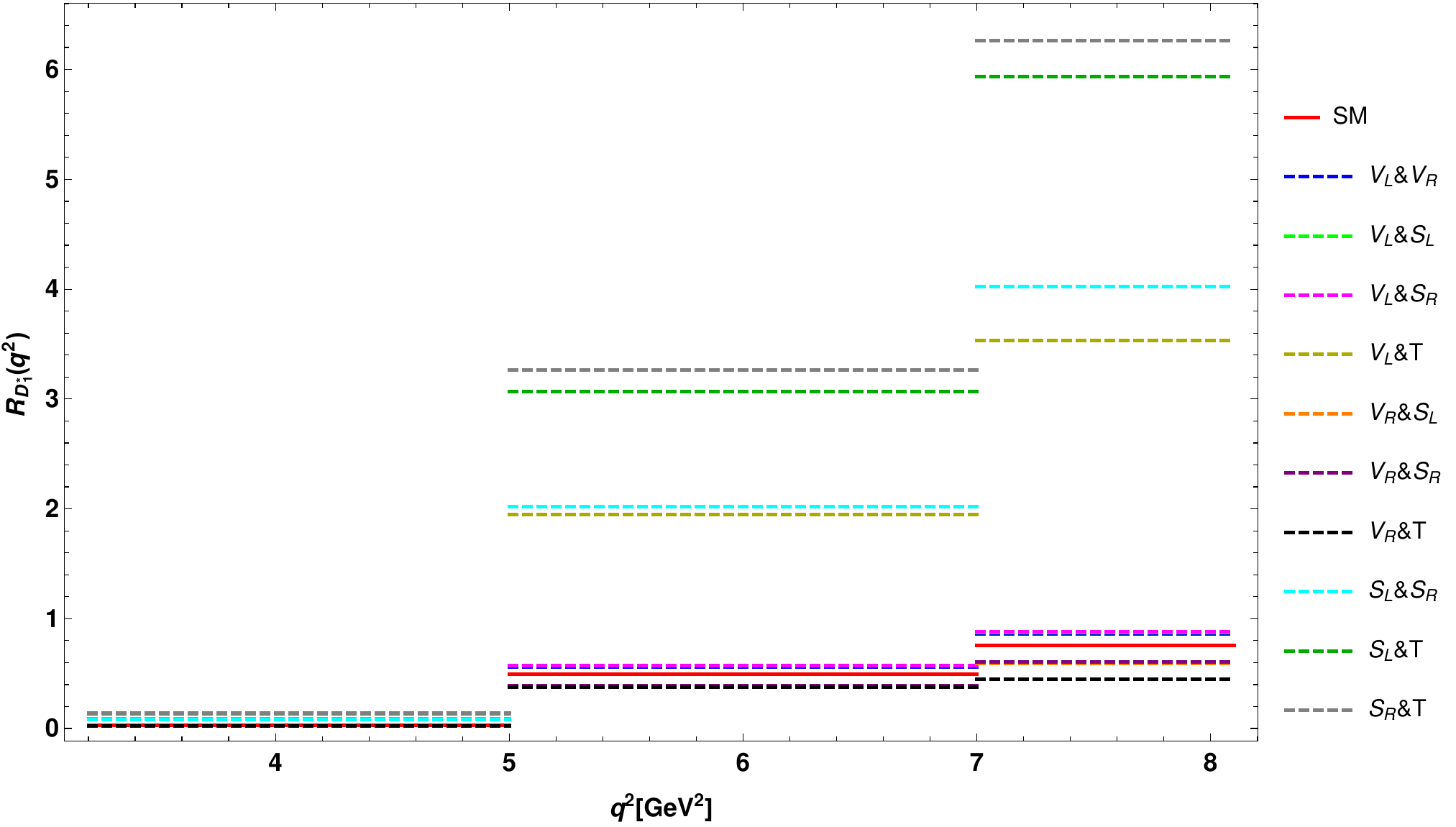}
\quad
\includegraphics[scale=0.38]{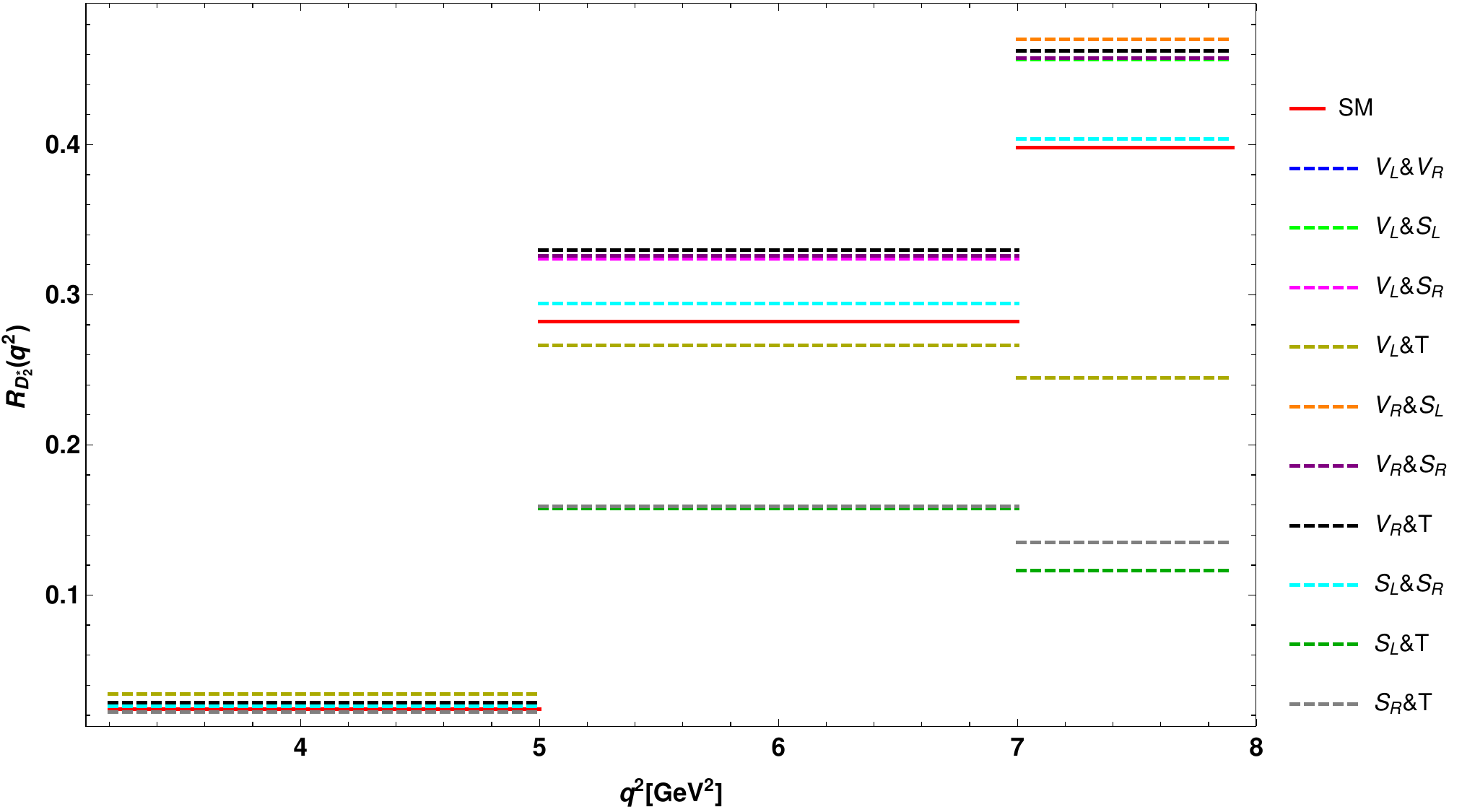}
\caption{ The bin-wise graphical representation of $R_{D^*_0}$ (top-left panel), $R_{D_1}$ (top-right panel), $R_{D^*_1}$ (bottom-left panel) and $R_{D^*_2}$ (bottom-right panel) parameters for case C. }\label{Fig:CC-RD**}
\end{figure}

The implications of additional real and complex coefficients on the  lepton non-universality parameters of $B \to D^{**} \tau \bar \nu_\tau$ processes, where $D^{**} = \{D^*_0, D_1^*, D_1, D_2^*\}$ are the four lightest excited charmed mesons,  will be discussed in this section. The detailed expressions for decay rates in the SM \cite{Bernlochner:2016bci} and in the presence of NP can be found in \cite{Bernlochner:2016bci, Bernlochner:2017jxt}\,. The lepton non-universality ratio $R_{D^{**}}$ is defined as 
\bea
R_{D^{**}} = \frac{{\rm BR}(B \to D^{**} \tau \nu_\tau)}{{\rm BR} (B \to D^{**} l \nu_l)},~~~l=e,\mu\,.
\eea
We use the input parameters from \cite{Tanabashi:2018oca} and the required form factors from \cite{Bernlochner:2016bci, Bernlochner:2017jxt} for numerical analysis. Using the best-fit values from Table \ref{Tab:Best-fit}\,, the bin-wise graphical representation of $R_{D^*_0}$ (top-left panel), $R_{D_1}$ (top-right panel), $R_{D^*_1}$ (bottom-left panel) and $R_{D^*_2}$ (bottom-right panel) parameters for case A, case B and case C are shown in Fig. \ref{Fig:CA-RD**}\,, \ref{Fig:CB-RD**} and \ref{Fig:CC-RD**} respectively. For the case of individual real coefficients, the contribution from $V_L$ provides significant deviation from the SM values of $R_{D^*_{0(2)}}$, $R_{D_1}$ ratios in the last two bins. The $V_R$ coefficient has effectively impact on the   $R_{D^*_{0(2)}}$ parameters in the $2^{\rm nd}$ and $3^{\rm rd}$ bins. The $S_L$ coupling has negligible effects on all LNU  parameters except the   $2^{\rm nd}$ and $3^{\rm rd}$ bins  of $R_{D_1}$. One can notice profound deviation in all the parameters in the last two bins due to an additional contribution form $S_R$ coefficient. Tensor coupling has no effect on $R_{D_1^*}$ and the first $q^2$ bin of remaining three LNU ratios. It should be noted that the $R_i,~i=D_0^*,D_1,D_1^*,D_2^*$ parameters have shifted maximally due to the presence of complex $V_L$ coupling. The deviation due to complex $V_R$ coefficient is minor in the last two bins of all these LNU parameters.   The $S_R$  ($T$) has large impact on  $R_{D_1}$ ($R_{D_1^{(*)}}$) parameters except in the first bin. Among $10$ possible sets of two real coefficients, the $R_i$  ratios deviate significantly from their respective SM results due to the $V_L\& T$, $S_L \& T$ and $S_R \& T$ combinations of coefficients. The last two bins of $R_{D_1^*}$ significantly deviates from SM values due to an additional contribution from $S_L\& S_R$ coefficients. The bin-wise numerical values of the ratio for the sum of the four $D^{**}$ states are given in Table \ref{Tab:CB-RD**}\,.
 
\begin{table}[htb]
\caption{Predicted bin-wise values of $R_{D^{**}}$ parameter in the SM and in the presence of new complex Wilson coefficients.}\label{Tab:CB-RD**}
\begin{center}
\begin{tabular}{|c|c|c|c|}
\hline
 ~Model~&~$R_{D^{**}}~(q^2\in [m_\tau^2,5])$~&~$R_{D^{**}}~(q^2\in [5,7])$~&~$R_{D^{**}}~(q^2\in [7,(m_B-m_i)^2])$~\\
\hline\hline
~SM ~&~$0.026$~&~$0.368$~&~$0.526$~\\
~Only $V_L$ ~&~$0.177$~&~$2.523$~&~$3.61$~\\
~Only $V_R$ ~&~$0.028$~&~$0.353$~&~$0.426$~\\
~Only $S_L$ ~&~$0.026$~&~$0.368$~&~$0.528$~\\
~Only $S_R$ ~&~$0.033$~&~$0.521$~&~$0.825$~\\
~Only $T$ ~&~$0.038$~&~$0.559$~&~$0.84$~\\
\hline
\end{tabular}
\end{center}
\end{table}
\section{Conclusion}
We have scrutinized the relevant semileptonic decays of $B$ mesons and $\Lambda_b$ involving the $b \to c \tau \bar \nu_\tau$ quark level transition  in an effective theory approach. This model independent strategy provides   additional vector, scalar and tensor  contributions to the standard model result. We have considered three cases of new Wilson coefficients: (a) presence of individual real coupling (b) presence of individual complex coupling and (c) presence of two real couplings and we performed a $chi$-square fitting to extract the best-fit values of these new coefficients,  from  the experimental data on $R_{D^{(*)}}$, $R_{J/\psi}$ and Br($B_c^+ \to \tau^+ \nu_\tau$) observables. For case C, we have taken all possible combinations of real coefficients (total $10$).  Using the best-fit values of real/complex Wilson coefficients, we then estimate the branching ratios, forward-backward asymmetry, lepton non-universality, lepton and hadron polarization asymmetries of $\bar B \to D^{(*)} \tau \bar \nu_\tau$, $B_c^+ \to (\eta_c, J/\psi) \tau^+ \nu_\tau$, $B_s \to D_s^{(*)} \tau \bar \nu_\tau$ and $\Lambda_b \to \Lambda_c \tau \bar \nu_\tau$ decay processes in four $q^2$ (in GeV$^2$) bins:  $m_\tau^2\to 5$, $5\to 7$, $7\to 9$ and $9 \to (M_B-M_{P(V)})^2$.   We have also shown the implication of new coefficients on $\bar B \to D^{**} \tau \bar \nu_\tau$ channel,  where $D^{**} = \{D^*_0, D_1^*, D_1, D_2^*\}$ are the four lightest excited charm mesons. As per our main goal, we have checked which type of new couplings are more sensitive to which angular observables of $b \to c \tau \bar \nu_\tau$ processes and specifically to which $q^2$ bins. The forward-backward asymmetry of $B \to P$ channels are independent of $V_L$ coefficients and the dependence of vector couplings drops out in $\tau$-polarization asymmetry parameters. We noticed that the impacts of only real $S_R$ coefficient on almost all the angular observables of $B \to P(V)$ processes are comparatively larger in all four $q^2$ bins (minor in the first bin).  
For the presence of either complex $S_R$ or $T$ coefficient, we observed significant deviation in the angular observables  from their respective SM results. The effects of complex $S_L$ coupling on some of the angular observables are found to be larger. The presence of real/complex vector type coefficients also provide profound deviation in  some angular observables in different $q^2$ bins (complex $V_R$ provides better impacts on observable compared to complex  $V_L$).  Though real/complex $S_R$ coefficient is playing a significant role in all observables, the best-fit values of real/complex $S_R$ doesn't lead to the best $\chi_{\rm min}^2/{\rm d.o.f.}$ value. Out of all possible  combinations of two real coefficients, the branching ratios and angular observables of $b \to c \tau \bar \nu_\tau$ decay modes provide comparatively  significant deviation from their corresponding SM predictions in the presence of $S_L\&S_R$, $S_L\&T$ and $S_R\& T$ sets of coefficients.  Other possible sets also have dominant impact on branching ratios and some angular observables  of $b \to c \tau \bar \nu_\tau$ and can accommodate the experimental limit. We have also shown the correlation between the LNU ratios, tau and hadron polarization asymmetry of $B \to P(V)$ and $\Lambda_b \to \Lambda_c$ processes. To conclude, we have performed a model independent analysis of $b \to c \tau \bar \nu_\tau$ decay processes  and  inspected the branching ratio and angular observables of these channels for both real and complex couplings in four $q^2$ bins. We have shown the bin-wise sensitivity of new coefficients on angular observables which will provide a clear idea on the structure of new physics.

\acknowledgments 
RM   would like to thank Science and Engineering Research Board (SERB), Government of India for financial support through grant No. EMR/2017/001448.

 \bibliography{btoc}

\begin{thebibliography}{75}
\expandafter\ifx\csname natexlab\endcsname\relax\def\natexlab#1{#1}\fi
\expandafter\ifx\csname bibnamefont\endcsname\relax
  \def\bibnamefont#1{#1}\fi
\expandafter\ifx\csname bibfnamefont\endcsname\relax
  \def\bibfnamefont#1{#1}\fi
\expandafter\ifx\csname citenamefont\endcsname\relax
  \def\citenamefont#1{#1}\fi
\expandafter\ifx\csname url\endcsname\relax
  \def\url#1{\texttt{#1}}\fi
\expandafter\ifx\csname urlprefix\endcsname\relax\def\urlprefix{URL }\fi
\providecommand{\bibinfo}[2]{#2}
\providecommand{\eprint}[2][]{\url{#2}}

\bibitem[{\citenamefont{Aaij et~al.}(2014)}]{Aaij:2014ora}
\bibinfo{author}{\bibfnamefont{R.}~\bibnamefont{Aaij}} \bibnamefont{et~al.}
  (\bibinfo{collaboration}{LHCb}), \bibinfo{journal}{Phys. Rev. Lett.}
  \textbf{\bibinfo{volume}{113}}, \bibinfo{pages}{151601}
  (\bibinfo{year}{2014}), \eprint{1406.6482}.

\bibitem[{\citenamefont{Aaij et~al.}(2019)}]{Aaij:2019wad}
\bibinfo{author}{\bibfnamefont{R.}~\bibnamefont{Aaij}} \bibnamefont{et~al.}
  (\bibinfo{collaboration}{LHCb}) (\bibinfo{year}{2019}), \eprint{1903.09252}.

\bibitem[{\citenamefont{Aaij et~al.}(2017)}]{Aaij:2017vbb}
\bibinfo{author}{\bibfnamefont{R.}~\bibnamefont{Aaij}} \bibnamefont{et~al.}
  (\bibinfo{collaboration}{LHCb}), \bibinfo{journal}{JHEP}
  \textbf{\bibinfo{volume}{08}}, \bibinfo{pages}{055} (\bibinfo{year}{2017}),
  \eprint{1705.05802}.

\bibitem[{\citenamefont{Abdesselam
  et~al.}(2019{\natexlab{a}})}]{Abdesselam:2019wac}
\bibinfo{author}{\bibfnamefont{A.}~\bibnamefont{Abdesselam}}
  \bibnamefont{et~al.} (\bibinfo{collaboration}{Belle})
  (\bibinfo{year}{2019}{\natexlab{a}}), \eprint{1904.02440}.

\bibitem[{\citenamefont{Huschle et~al.}(2015)}]{Huschle:2015rga}
\bibinfo{author}{\bibfnamefont{M.}~\bibnamefont{Huschle}} \bibnamefont{et~al.}
  (\bibinfo{collaboration}{Belle}), \bibinfo{journal}{Phys. Rev.}
  \textbf{\bibinfo{volume}{D92}}, \bibinfo{pages}{072014}
  (\bibinfo{year}{2015}), \eprint{1507.03233}.

\bibitem[{\citenamefont{Abdesselam
  et~al.}(2016{\natexlab{a}})}]{Abdesselam:2016cgx}
\bibinfo{author}{\bibfnamefont{A.}~\bibnamefont{Abdesselam}}
  \bibnamefont{et~al.} (\bibinfo{collaboration}{Belle}), in
  \emph{\bibinfo{booktitle}{{Proceedings, 51st Rencontres de Moriond on
  Electroweak Interactions and Unified Theories: La Thuile, Italy, March 12-19,
  2016}}} (\bibinfo{year}{2016}{\natexlab{a}}), \eprint{1603.06711},
  \urlprefix\url{http://inspirehep.net/record/1431982/files/arXiv:1603.06711.pdf}.

\bibitem[{\citenamefont{Abdesselam
  et~al.}(2016{\natexlab{b}})}]{Abdesselam:2016xqt}
\bibinfo{author}{\bibfnamefont{A.}~\bibnamefont{Abdesselam}}
  \bibnamefont{et~al.} (\bibinfo{year}{2016}{\natexlab{b}}),
  \eprint{1608.06391}.

\bibitem[{\citenamefont{Hirose et~al.}(2018)}]{Hirose:2017dxl}
\bibinfo{author}{\bibfnamefont{S.}~\bibnamefont{Hirose}} \bibnamefont{et~al.}
  (\bibinfo{collaboration}{Belle}), \bibinfo{journal}{Phys. Rev.}
  \textbf{\bibinfo{volume}{D97}}, \bibinfo{pages}{012004}
  (\bibinfo{year}{2018}), \eprint{1709.00129}.

\bibitem[{\citenamefont{Hirose et~al.}(2017)}]{Hirose:2016wfn}
\bibinfo{author}{\bibfnamefont{S.}~\bibnamefont{Hirose}} \bibnamefont{et~al.}
  (\bibinfo{collaboration}{Belle}), \bibinfo{journal}{Phys. Rev. Lett.}
  \textbf{\bibinfo{volume}{118}}, \bibinfo{pages}{211801}
  (\bibinfo{year}{2017}), \eprint{1612.00529}.

\bibitem[{\citenamefont{Aaij et~al.}(2018{\natexlab{a}})}]{Aaij:2017tyk}
\bibinfo{author}{\bibfnamefont{R.}~\bibnamefont{Aaij}} \bibnamefont{et~al.}
  (\bibinfo{collaboration}{LHCb}), \bibinfo{journal}{Phys. Rev. Lett.}
  \textbf{\bibinfo{volume}{120}}, \bibinfo{pages}{121801}
  (\bibinfo{year}{2018}{\natexlab{a}}), \eprint{1711.05623}.

\bibitem[{\citenamefont{Aaij et~al.}(2015{\natexlab{a}})}]{Aaij:2015yra}
\bibinfo{author}{\bibfnamefont{R.}~\bibnamefont{Aaij}} \bibnamefont{et~al.}
  (\bibinfo{collaboration}{LHCb}), \bibinfo{journal}{Phys. Rev. Lett.}
  \textbf{\bibinfo{volume}{115}}, \bibinfo{pages}{111803}
  (\bibinfo{year}{2015}{\natexlab{a}}), \bibinfo{note}{[Erratum: Phys. Rev.
  Lett.115,no.15,159901(2015)]}, \eprint{1506.08614}.

\bibitem[{\citenamefont{Aaij et~al.}(2018{\natexlab{b}})}]{Aaij:2017deq}
\bibinfo{author}{\bibfnamefont{R.}~\bibnamefont{Aaij}} \bibnamefont{et~al.}
  (\bibinfo{collaboration}{LHCb}), \bibinfo{journal}{Phys. Rev.}
  \textbf{\bibinfo{volume}{D97}}, \bibinfo{pages}{072013}
  (\bibinfo{year}{2018}{\natexlab{b}}), \eprint{1711.02505}.

\bibitem[{\citenamefont{Aaij et~al.}(2018{\natexlab{c}})}]{Aaij:2017uff}
\bibinfo{author}{\bibfnamefont{R.}~\bibnamefont{Aaij}} \bibnamefont{et~al.}
  (\bibinfo{collaboration}{LHCb}), \bibinfo{journal}{Phys. Rev. Lett.}
  \textbf{\bibinfo{volume}{120}}, \bibinfo{pages}{171802}
  (\bibinfo{year}{2018}{\natexlab{c}}), \eprint{1708.08856}.

\bibitem[{\citenamefont{Lees et~al.}(2012)}]{Lees:2012xj}
\bibinfo{author}{\bibfnamefont{J.~P.} \bibnamefont{Lees}} \bibnamefont{et~al.}
  (\bibinfo{collaboration}{BaBar}), \bibinfo{journal}{Phys. Rev. Lett.}
  \textbf{\bibinfo{volume}{109}}, \bibinfo{pages}{101802}
  (\bibinfo{year}{2012}), \eprint{1205.5442}.

\bibitem[{\citenamefont{Lees et~al.}(2013)}]{Lees:2013uzd}
\bibinfo{author}{\bibfnamefont{J.~P.} \bibnamefont{Lees}} \bibnamefont{et~al.}
  (\bibinfo{collaboration}{BaBar}), \bibinfo{journal}{Phys. Rev.}
  \textbf{\bibinfo{volume}{D88}}, \bibinfo{pages}{072012}
  (\bibinfo{year}{2013}), \eprint{1303.0571}.

\bibitem[{\citenamefont{{Heavy Flavor Averaging Group}}(2019)}]{HFLAV}
\bibinfo{author}{\bibnamefont{{Heavy Flavor Averaging Group}}}
  (\bibinfo{year}{2019}),
  \urlprefix\url{https://hflav-eos.web.cern.ch/hflav-eos/semi/spring19/html/RDsDsstar/RDRDs.html}.

\bibitem[{\citenamefont{Bobeth et~al.}(2007)\citenamefont{Bobeth, Hiller, and
  Piranishvili}}]{Bobeth:2007dw}
\bibinfo{author}{\bibfnamefont{C.}~\bibnamefont{Bobeth}},
  \bibinfo{author}{\bibfnamefont{G.}~\bibnamefont{Hiller}}, \bibnamefont{and}
  \bibinfo{author}{\bibfnamefont{G.}~\bibnamefont{Piranishvili}},
  \bibinfo{journal}{JHEP} \textbf{\bibinfo{volume}{12}}, \bibinfo{pages}{040}
  (\bibinfo{year}{2007}), \eprint{0709.4174}.

\bibitem[{\citenamefont{Capdevila et~al.}(2018)\citenamefont{Capdevila,
  Crivellin, Descotes-Genon, Matias, and Virto}}]{Capdevila:2017bsm}
\bibinfo{author}{\bibfnamefont{B.}~\bibnamefont{Capdevila}},
  \bibinfo{author}{\bibfnamefont{A.}~\bibnamefont{Crivellin}},
  \bibinfo{author}{\bibfnamefont{S.}~\bibnamefont{Descotes-Genon}},
  \bibinfo{author}{\bibfnamefont{J.}~\bibnamefont{Matias}}, \bibnamefont{and}
  \bibinfo{author}{\bibfnamefont{J.}~\bibnamefont{Virto}},
  \bibinfo{journal}{JHEP} \textbf{\bibinfo{volume}{01}}, \bibinfo{pages}{093}
  (\bibinfo{year}{2018}), \eprint{1704.05340}.

\bibitem[{\citenamefont{Na et~al.}(2015)\citenamefont{Na, Bouchard, Lepage,
  Monahan, and Shigemitsu}}]{Na:2015kha}
\bibinfo{author}{\bibfnamefont{H.}~\bibnamefont{Na}},
  \bibinfo{author}{\bibfnamefont{C.~M.} \bibnamefont{Bouchard}},
  \bibinfo{author}{\bibfnamefont{G.~P.} \bibnamefont{Lepage}},
  \bibinfo{author}{\bibfnamefont{C.}~\bibnamefont{Monahan}}, \bibnamefont{and}
  \bibinfo{author}{\bibfnamefont{J.}~\bibnamefont{Shigemitsu}}
  (\bibinfo{collaboration}{HPQCD}), \bibinfo{journal}{Phys. Rev.}
  \textbf{\bibinfo{volume}{D92}}, \bibinfo{pages}{054510}
  (\bibinfo{year}{2015}), \bibinfo{note}{[Erratum: Phys.
  Rev.D93,no.11,119906(2016)]}, \eprint{1505.03925}.

\bibitem[{\citenamefont{Fajfer et~al.}(2012{\natexlab{a}})\citenamefont{Fajfer,
  Kamenik, and Nisandzic}}]{Fajfer:2012vx}
\bibinfo{author}{\bibfnamefont{S.}~\bibnamefont{Fajfer}},
  \bibinfo{author}{\bibfnamefont{J.~F.} \bibnamefont{Kamenik}},
  \bibnamefont{and}
  \bibinfo{author}{\bibfnamefont{I.}~\bibnamefont{Nisandzic}},
  \bibinfo{journal}{Phys. Rev.} \textbf{\bibinfo{volume}{D85}},
  \bibinfo{pages}{094025} (\bibinfo{year}{2012}{\natexlab{a}}),
  \eprint{1203.2654}.

\bibitem[{\citenamefont{Fajfer et~al.}(2012{\natexlab{b}})\citenamefont{Fajfer,
  Kamenik, Nisandzic, and Zupan}}]{Fajfer:2012jt}
\bibinfo{author}{\bibfnamefont{S.}~\bibnamefont{Fajfer}},
  \bibinfo{author}{\bibfnamefont{J.~F.} \bibnamefont{Kamenik}},
  \bibinfo{author}{\bibfnamefont{I.}~\bibnamefont{Nisandzic}},
  \bibnamefont{and} \bibinfo{author}{\bibfnamefont{J.}~\bibnamefont{Zupan}},
  \bibinfo{journal}{Phys. Rev. Lett.} \textbf{\bibinfo{volume}{109}},
  \bibinfo{pages}{161801} (\bibinfo{year}{2012}{\natexlab{b}}),
  \eprint{1206.1872}.

\bibitem[{\citenamefont{Wang et~al.}(2013)\citenamefont{Wang, Fan, and
  Xiao}}]{Wen-Fei:2013uea}
\bibinfo{author}{\bibfnamefont{W.-F.} \bibnamefont{Wang}},
  \bibinfo{author}{\bibfnamefont{Y.-Y.} \bibnamefont{Fan}}, \bibnamefont{and}
  \bibinfo{author}{\bibfnamefont{Z.-J.} \bibnamefont{Xiao}},
  \bibinfo{journal}{Chin. Phys.} \textbf{\bibinfo{volume}{C37}},
  \bibinfo{pages}{093102} (\bibinfo{year}{2013}), \eprint{1212.5903}.

\bibitem[{\citenamefont{Ivanov et~al.}(2005)\citenamefont{Ivanov, Korner, and
  Santorelli}}]{Ivanov:2005fd}
\bibinfo{author}{\bibfnamefont{M.~A.} \bibnamefont{Ivanov}},
  \bibinfo{author}{\bibfnamefont{J.~G.} \bibnamefont{Korner}},
  \bibnamefont{and}
  \bibinfo{author}{\bibfnamefont{P.}~\bibnamefont{Santorelli}},
  \bibinfo{journal}{Phys. Rev.} \textbf{\bibinfo{volume}{D71}},
  \bibinfo{pages}{094006} (\bibinfo{year}{2005}), \bibinfo{note}{[Erratum:
  Phys. Rev.D75,019901(2007)]}, \eprint{hep-ph/0501051}.

\bibitem[{\citenamefont{Dutta and Bhol}(2017)}]{Dutta:2017xmj}
\bibinfo{author}{\bibfnamefont{R.}~\bibnamefont{Dutta}} \bibnamefont{and}
  \bibinfo{author}{\bibfnamefont{A.}~\bibnamefont{Bhol}},
  \bibinfo{journal}{Phys. Rev.} \textbf{\bibinfo{volume}{D96}},
  \bibinfo{pages}{076001} (\bibinfo{year}{2017}), \eprint{1701.08598}.

\bibitem[{\citenamefont{Tanaka and Watanabe}(2013)}]{Tanaka:2012nw}
\bibinfo{author}{\bibfnamefont{M.}~\bibnamefont{Tanaka}} \bibnamefont{and}
  \bibinfo{author}{\bibfnamefont{R.}~\bibnamefont{Watanabe}},
  \bibinfo{journal}{Phys. Rev.} \textbf{\bibinfo{volume}{D87}},
  \bibinfo{pages}{034028} (\bibinfo{year}{2013}), \eprint{1212.1878}.

\bibitem[{\citenamefont{Adamczyk}(2019)}]{Adamczyk:2019wyt}
\bibinfo{author}{\bibfnamefont{K.}~\bibnamefont{Adamczyk}}
  (\bibinfo{collaboration}{Belle, Belle-II}), in
  \emph{\bibinfo{booktitle}{{10th International Workshop on the CKM Unitarity
  Triangle (CKM 2018) Heidelberg, Germany, September 17-21, 2018}}}
  (\bibinfo{year}{2019}), \eprint{1901.06380}.

\bibitem[{\citenamefont{Abdesselam
  et~al.}(2019{\natexlab{b}})}]{Abdesselam:2019wbt}
\bibinfo{author}{\bibfnamefont{A.}~\bibnamefont{Abdesselam}}
  \bibnamefont{et~al.} (\bibinfo{collaboration}{Belle}), in
  \emph{\bibinfo{booktitle}{{10th International Workshop on the CKM Unitarity
  Triangle (CKM 2018) Heidelberg, Germany, September 17-21, 2018}}}
  (\bibinfo{year}{2019}{\natexlab{b}}), \eprint{1903.03102}.

\bibitem[{\citenamefont{Alok et~al.}(2017)\citenamefont{Alok, Kumar, Kumbhakar,
  and Sankar}}]{Alok:2016qyh}
\bibinfo{author}{\bibfnamefont{A.~K.} \bibnamefont{Alok}},
  \bibinfo{author}{\bibfnamefont{D.}~\bibnamefont{Kumar}},
  \bibinfo{author}{\bibfnamefont{S.}~\bibnamefont{Kumbhakar}},
  \bibnamefont{and} \bibinfo{author}{\bibfnamefont{S.~U.}
  \bibnamefont{Sankar}}, \bibinfo{journal}{Phys. Rev.}
  \textbf{\bibinfo{volume}{D95}}, \bibinfo{pages}{115038}
  (\bibinfo{year}{2017}), \eprint{1606.03164}.

\bibitem[{\citenamefont{Bhol}(2014)}]{Bhol:2014jta}
\bibinfo{author}{\bibfnamefont{A.}~\bibnamefont{Bhol}}, \bibinfo{journal}{EPL}
  \textbf{\bibinfo{volume}{106}}, \bibinfo{pages}{31001}
  (\bibinfo{year}{2014}).

\bibitem[{\citenamefont{Li et~al.}(2009)\citenamefont{Li, Lu, and
  Wang}}]{Li:2009wq}
\bibinfo{author}{\bibfnamefont{R.-H.} \bibnamefont{Li}},
  \bibinfo{author}{\bibfnamefont{C.-D.} \bibnamefont{Lu}}, \bibnamefont{and}
  \bibinfo{author}{\bibfnamefont{Y.-M.} \bibnamefont{Wang}},
  \bibinfo{journal}{Phys. Rev.} \textbf{\bibinfo{volume}{D80}},
  \bibinfo{pages}{014005} (\bibinfo{year}{2009}), \eprint{0905.3259}.

\bibitem[{\citenamefont{Li et~al.}(2010)\citenamefont{Li, Shao, and
  Wang}}]{Li:2010bb}
\bibinfo{author}{\bibfnamefont{G.}~\bibnamefont{Li}},
  \bibinfo{author}{\bibfnamefont{F.-l.} \bibnamefont{Shao}}, \bibnamefont{and}
  \bibinfo{author}{\bibfnamefont{W.}~\bibnamefont{Wang}},
  \bibinfo{journal}{Phys. Rev.} \textbf{\bibinfo{volume}{D82}},
  \bibinfo{pages}{094031} (\bibinfo{year}{2010}), \eprint{1008.3696}.

\bibitem[{\citenamefont{Atoui et~al.}(2014{\natexlab{a}})\citenamefont{Atoui,
  Becirevic, Morénas, and Sanfilippo}}]{Atoui:2013mqa}
\bibinfo{author}{\bibfnamefont{M.}~\bibnamefont{Atoui}},
  \bibinfo{author}{\bibfnamefont{D.}~\bibnamefont{Becirevic}},
  \bibinfo{author}{\bibfnamefont{V.}~\bibnamefont{Morénas}}, \bibnamefont{and}
  \bibinfo{author}{\bibfnamefont{F.}~\bibnamefont{Sanfilippo}},
  \bibinfo{journal}{PoS} \textbf{\bibinfo{volume}{LATTICE2013}},
  \bibinfo{pages}{384} (\bibinfo{year}{2014}{\natexlab{a}}),
  \eprint{1311.5071}.

\bibitem[{\citenamefont{Atoui et~al.}(2014{\natexlab{b}})\citenamefont{Atoui,
  Morénas, Be?irevic, and Sanfilippo}}]{Atoui:2013zza}
\bibinfo{author}{\bibfnamefont{M.}~\bibnamefont{Atoui}},
  \bibinfo{author}{\bibfnamefont{V.}~\bibnamefont{Morénas}},
  \bibinfo{author}{\bibfnamefont{D.}~\bibnamefont{Be?irevic}},
  \bibnamefont{and}
  \bibinfo{author}{\bibfnamefont{F.}~\bibnamefont{Sanfilippo}},
  \bibinfo{journal}{Eur. Phys. J.} \textbf{\bibinfo{volume}{C74}},
  \bibinfo{pages}{2861} (\bibinfo{year}{2014}{\natexlab{b}}),
  \eprint{1310.5238}.

\bibitem[{\citenamefont{Bailey et~al.}(2012)}]{Bailey:2012rr}
\bibinfo{author}{\bibfnamefont{J.~A.} \bibnamefont{Bailey}}
  \bibnamefont{et~al.}, \bibinfo{journal}{Phys. Rev.}
  \textbf{\bibinfo{volume}{D85}}, \bibinfo{pages}{114502}
  (\bibinfo{year}{2012}), \bibinfo{note}{[Erratum: Phys.
  Rev.D86,039904(2012)]}, \eprint{1202.6346}.

\bibitem[{\citenamefont{Monahan et~al.}(2016)\citenamefont{Monahan, Na,
  Bouchard, Lepage, and Shigemitsu}}]{Monahan:2016qxu}
\bibinfo{author}{\bibfnamefont{C.~J.} \bibnamefont{Monahan}},
  \bibinfo{author}{\bibfnamefont{H.}~\bibnamefont{Na}},
  \bibinfo{author}{\bibfnamefont{C.~M.} \bibnamefont{Bouchard}},
  \bibinfo{author}{\bibfnamefont{G.~P.} \bibnamefont{Lepage}},
  \bibnamefont{and}
  \bibinfo{author}{\bibfnamefont{J.}~\bibnamefont{Shigemitsu}},
  \bibinfo{journal}{PoS} \textbf{\bibinfo{volume}{LATTICE2016}},
  \bibinfo{pages}{298} (\bibinfo{year}{2016}), \eprint{1611.09667}.

\bibitem[{\citenamefont{Na et~al.}(2012)\citenamefont{Na, Monahan, Davies,
  Horgan, Lepage, and Shigemitsu}}]{Na:2012kp}
\bibinfo{author}{\bibfnamefont{H.}~\bibnamefont{Na}},
  \bibinfo{author}{\bibfnamefont{C.~J.} \bibnamefont{Monahan}},
  \bibinfo{author}{\bibfnamefont{C.~T.~H.} \bibnamefont{Davies}},
  \bibinfo{author}{\bibfnamefont{R.}~\bibnamefont{Horgan}},
  \bibinfo{author}{\bibfnamefont{G.~P.} \bibnamefont{Lepage}},
  \bibnamefont{and}
  \bibinfo{author}{\bibfnamefont{J.}~\bibnamefont{Shigemitsu}},
  \bibinfo{journal}{Phys. Rev.} \textbf{\bibinfo{volume}{D86}},
  \bibinfo{pages}{034506} (\bibinfo{year}{2012}), \eprint{1202.4914}.

\bibitem[{\citenamefont{Monahan et~al.}(2018)\citenamefont{Monahan, Bouchard,
  Lepage, Na, and Shigemitsu}}]{Monahan:2018lzv}
\bibinfo{author}{\bibfnamefont{C.~J.} \bibnamefont{Monahan}},
  \bibinfo{author}{\bibfnamefont{C.~M.} \bibnamefont{Bouchard}},
  \bibinfo{author}{\bibfnamefont{G.~P.} \bibnamefont{Lepage}},
  \bibinfo{author}{\bibfnamefont{H.}~\bibnamefont{Na}}, \bibnamefont{and}
  \bibinfo{author}{\bibfnamefont{J.}~\bibnamefont{Shigemitsu}},
  \bibinfo{journal}{Phys. Rev.} \textbf{\bibinfo{volume}{D98}},
  \bibinfo{pages}{114509} (\bibinfo{year}{2018}), \eprint{1808.09285}.

\bibitem[{\citenamefont{Chen et~al.}(2012)\citenamefont{Chen, Fu, Kim, and
  Wang}}]{Chen:2011ut}
\bibinfo{author}{\bibfnamefont{X.~J.} \bibnamefont{Chen}},
  \bibinfo{author}{\bibfnamefont{H.~F.} \bibnamefont{Fu}},
  \bibinfo{author}{\bibfnamefont{C.~S.} \bibnamefont{Kim}}, \bibnamefont{and}
  \bibinfo{author}{\bibfnamefont{G.~L.} \bibnamefont{Wang}},
  \bibinfo{journal}{J. Phys.} \textbf{\bibinfo{volume}{G39}},
  \bibinfo{pages}{045002} (\bibinfo{year}{2012}), \eprint{1106.3003}.

\bibitem[{\citenamefont{Fan et~al.}(2014)\citenamefont{Fan, Wang, and
  Xiao}}]{Fan:2013kqa}
\bibinfo{author}{\bibfnamefont{Y.-Y.} \bibnamefont{Fan}},
  \bibinfo{author}{\bibfnamefont{W.-F.} \bibnamefont{Wang}}, \bibnamefont{and}
  \bibinfo{author}{\bibfnamefont{Z.-J.} \bibnamefont{Xiao}},
  \bibinfo{journal}{Phys. Rev.} \textbf{\bibinfo{volume}{D89}},
  \bibinfo{pages}{014030} (\bibinfo{year}{2014}), \eprint{1311.4965}.

\bibitem[{\citenamefont{Monahan et~al.}(2017)\citenamefont{Monahan, Na,
  Bouchard, Lepage, and Shigemitsu}}]{Monahan:2017uby}
\bibinfo{author}{\bibfnamefont{C.~J.} \bibnamefont{Monahan}},
  \bibinfo{author}{\bibfnamefont{H.}~\bibnamefont{Na}},
  \bibinfo{author}{\bibfnamefont{C.~M.} \bibnamefont{Bouchard}},
  \bibinfo{author}{\bibfnamefont{G.~P.} \bibnamefont{Lepage}},
  \bibnamefont{and}
  \bibinfo{author}{\bibfnamefont{J.}~\bibnamefont{Shigemitsu}},
  \bibinfo{journal}{Phys. Rev.} \textbf{\bibinfo{volume}{D95}},
  \bibinfo{pages}{114506} (\bibinfo{year}{2017}), \eprint{1703.09728}.

\bibitem[{\citenamefont{Dutta and Rajeev}(2018)}]{Dutta:2018jxz}
\bibinfo{author}{\bibfnamefont{R.}~\bibnamefont{Dutta}} \bibnamefont{and}
  \bibinfo{author}{\bibfnamefont{N.}~\bibnamefont{Rajeev}},
  \bibinfo{journal}{Phys. Rev.} \textbf{\bibinfo{volume}{D97}},
  \bibinfo{pages}{095045} (\bibinfo{year}{2018}), \eprint{1803.03038}.

\bibitem[{\citenamefont{Aaij et~al.}(2015{\natexlab{b}})}]{Aaij:2015bfa}
\bibinfo{author}{\bibfnamefont{R.}~\bibnamefont{Aaij}} \bibnamefont{et~al.}
  (\bibinfo{collaboration}{LHCb}), \bibinfo{journal}{Nature Phys.}
  \textbf{\bibinfo{volume}{11}}, \bibinfo{pages}{743}
  (\bibinfo{year}{2015}{\natexlab{b}}), \eprint{1504.01568}.

\bibitem[{\citenamefont{Fiore}(2015)}]{Fiore:2015cmx}
\bibinfo{author}{\bibfnamefont{M.}~\bibnamefont{Fiore}}, in
  \emph{\bibinfo{booktitle}{{Proceedings, Meeting of the APS Division of
  Particles and Fields (DPF 2015): Ann Arbor, Michigan, USA, 4-8 Aug 2015}}}
  (\bibinfo{year}{2015}), \eprint{1511.00105}.

\bibitem[{\citenamefont{Patrignani et~al.}(2016)}]{Patrignani:2016xqp}
\bibinfo{author}{\bibfnamefont{C.}~\bibnamefont{Patrignani}}
  \bibnamefont{et~al.} (\bibinfo{collaboration}{Particle Data Group}),
  \bibinfo{journal}{Chin. Phys.} \textbf{\bibinfo{volume}{C40}},
  \bibinfo{pages}{100001} (\bibinfo{year}{2016}).

\bibitem[{\citenamefont{Hsiao and Geng}(2017)}]{Hsiao:2017umx}
\bibinfo{author}{\bibfnamefont{Y.~K.} \bibnamefont{Hsiao}} \bibnamefont{and}
  \bibinfo{author}{\bibfnamefont{C.~Q.} \bibnamefont{Geng}},
  \bibinfo{journal}{Eur. Phys. J.} \textbf{\bibinfo{volume}{C77}},
  \bibinfo{pages}{714} (\bibinfo{year}{2017}), \eprint{1705.00948}.

\bibitem[{\citenamefont{Woloshyn}(2013)}]{Woloshyn:2014hka}
\bibinfo{author}{\bibfnamefont{R.~M.} \bibnamefont{Woloshyn}},
  \bibinfo{journal}{PoS} \textbf{\bibinfo{volume}{Hadron2013}},
  \bibinfo{pages}{203} (\bibinfo{year}{2013}).

\bibitem[{\citenamefont{Wu}(2015)}]{Wu:2015yqa}
\bibinfo{author}{\bibfnamefont{W.}~\bibnamefont{Wu}}, Master's thesis,
  \bibinfo{school}{Mississippi U.} (\bibinfo{year}{2015}), \eprint{1505.03418},
  \urlprefix\url{http://search.proquest.com/docview/1697862095}.

\bibitem[{\citenamefont{Shivashankara et~al.}(2015)\citenamefont{Shivashankara,
  Wu, and Datta}}]{Shivashankara:2015cta}
\bibinfo{author}{\bibfnamefont{S.}~\bibnamefont{Shivashankara}},
  \bibinfo{author}{\bibfnamefont{W.}~\bibnamefont{Wu}}, \bibnamefont{and}
  \bibinfo{author}{\bibfnamefont{A.}~\bibnamefont{Datta}},
  \bibinfo{journal}{Phys. Rev.} \textbf{\bibinfo{volume}{D91}},
  \bibinfo{pages}{115003} (\bibinfo{year}{2015}), \eprint{1502.07230}.

\bibitem[{\citenamefont{Gutsche et~al.}(2016)\citenamefont{Gutsche, Ivanov,
  Korner, Lyubovitskij, and Santorelli}}]{Gutsche:2015rrt}
\bibinfo{author}{\bibfnamefont{T.}~\bibnamefont{Gutsche}},
  \bibinfo{author}{\bibfnamefont{M.~A.} \bibnamefont{Ivanov}},
  \bibinfo{author}{\bibfnamefont{J.~G.} \bibnamefont{Korner}},
  \bibinfo{author}{\bibfnamefont{V.~E.} \bibnamefont{Lyubovitskij}},
  \bibnamefont{and}
  \bibinfo{author}{\bibfnamefont{P.}~\bibnamefont{Santorelli}},
  \bibinfo{journal}{Phys. Rev.} \textbf{\bibinfo{volume}{D93}},
  \bibinfo{pages}{034008} (\bibinfo{year}{2016}), \eprint{1512.02168}.

\bibitem[{\citenamefont{Gutsche et~al.}(2015)\citenamefont{Gutsche, Ivanov,
  Körner, Lyubovitskij, Santorelli, and Habyl}}]{Gutsche:2015mxa}
\bibinfo{author}{\bibfnamefont{T.}~\bibnamefont{Gutsche}},
  \bibinfo{author}{\bibfnamefont{M.~A.} \bibnamefont{Ivanov}},
  \bibinfo{author}{\bibfnamefont{J.~G.} \bibnamefont{Körner}},
  \bibinfo{author}{\bibfnamefont{V.~E.} \bibnamefont{Lyubovitskij}},
  \bibinfo{author}{\bibfnamefont{P.}~\bibnamefont{Santorelli}},
  \bibnamefont{and} \bibinfo{author}{\bibfnamefont{N.}~\bibnamefont{Habyl}},
  \bibinfo{journal}{Phys. Rev.} \textbf{\bibinfo{volume}{D91}},
  \bibinfo{pages}{074001} (\bibinfo{year}{2015}), \bibinfo{note}{[Erratum:
  Phys. Rev.D91,no.11,119907(2015)]}, \eprint{1502.04864}.

\bibitem[{\citenamefont{Detmold et~al.}(2015)\citenamefont{Detmold, Lehner, and
  Meinel}}]{Detmold:2015aaa}
\bibinfo{author}{\bibfnamefont{W.}~\bibnamefont{Detmold}},
  \bibinfo{author}{\bibfnamefont{C.}~\bibnamefont{Lehner}}, \bibnamefont{and}
  \bibinfo{author}{\bibfnamefont{S.}~\bibnamefont{Meinel}},
  \bibinfo{journal}{Phys. Rev.} \textbf{\bibinfo{volume}{D92}},
  \bibinfo{pages}{034503} (\bibinfo{year}{2015}), \eprint{1503.01421}.

\bibitem[{\citenamefont{Dutta}(2016)}]{Dutta:2015ueb}
\bibinfo{author}{\bibfnamefont{R.}~\bibnamefont{Dutta}},
  \bibinfo{journal}{Phys. Rev.} \textbf{\bibinfo{volume}{D93}},
  \bibinfo{pages}{054003} (\bibinfo{year}{2016}), \eprint{1512.04034}.

\bibitem[{\citenamefont{Pervin et~al.}(2005)\citenamefont{Pervin, Roberts, and
  Capstick}}]{Pervin:2005ve}
\bibinfo{author}{\bibfnamefont{M.}~\bibnamefont{Pervin}},
  \bibinfo{author}{\bibfnamefont{W.}~\bibnamefont{Roberts}}, \bibnamefont{and}
  \bibinfo{author}{\bibfnamefont{S.}~\bibnamefont{Capstick}},
  \bibinfo{journal}{Phys. Rev.} \textbf{\bibinfo{volume}{C72}},
  \bibinfo{pages}{035201} (\bibinfo{year}{2005}), \eprint{nucl-th/0503030}.

\bibitem[{\citenamefont{Faustov and Galkin}(2016)}]{Faustov:2016yza}
\bibinfo{author}{\bibfnamefont{R.~N.} \bibnamefont{Faustov}} \bibnamefont{and}
  \bibinfo{author}{\bibfnamefont{V.~O.} \bibnamefont{Galkin}},
  \bibinfo{journal}{Eur. Phys. J.} \textbf{\bibinfo{volume}{C76}},
  \bibinfo{pages}{628} (\bibinfo{year}{2016}), \eprint{1610.00957}.

\bibitem[{\citenamefont{Datta et~al.}(2017)\citenamefont{Datta, Kamali, Meinel,
  and Rashed}}]{Datta:2017aue}
\bibinfo{author}{\bibfnamefont{A.}~\bibnamefont{Datta}},
  \bibinfo{author}{\bibfnamefont{S.}~\bibnamefont{Kamali}},
  \bibinfo{author}{\bibfnamefont{S.}~\bibnamefont{Meinel}}, \bibnamefont{and}
  \bibinfo{author}{\bibfnamefont{A.}~\bibnamefont{Rashed}},
  \bibinfo{journal}{JHEP} \textbf{\bibinfo{volume}{08}}, \bibinfo{pages}{131}
  (\bibinfo{year}{2017}), \eprint{1702.02243}.

\bibitem[{\citenamefont{Li et~al.}(2017)\citenamefont{Li, Yang, and
  Zhang}}]{Li:2016pdv}
\bibinfo{author}{\bibfnamefont{X.-Q.} \bibnamefont{Li}},
  \bibinfo{author}{\bibfnamefont{Y.-D.} \bibnamefont{Yang}}, \bibnamefont{and}
  \bibinfo{author}{\bibfnamefont{X.}~\bibnamefont{Zhang}},
  \bibinfo{journal}{JHEP} \textbf{\bibinfo{volume}{02}}, \bibinfo{pages}{068}
  (\bibinfo{year}{2017}), \eprint{1611.01635}.

\bibitem[{\citenamefont{Di~Salvo et~al.}(2018)\citenamefont{Di~Salvo,
  Fontanelli, and Ajaltouni}}]{DiSalvo:2018ngq}
\bibinfo{author}{\bibfnamefont{E.}~\bibnamefont{Di~Salvo}},
  \bibinfo{author}{\bibfnamefont{F.}~\bibnamefont{Fontanelli}},
  \bibnamefont{and} \bibinfo{author}{\bibfnamefont{Z.~J.}
  \bibnamefont{Ajaltouni}} (\bibinfo{year}{2018}), \eprint{1804.05592}.

\bibitem[{\citenamefont{Bernlochner
  et~al.}(2018{\natexlab{a}})\citenamefont{Bernlochner, Ligeti, Robinson, and
  Sutcliffe}}]{Bernlochner:2018kxh}
\bibinfo{author}{\bibfnamefont{F.~U.} \bibnamefont{Bernlochner}},
  \bibinfo{author}{\bibfnamefont{Z.}~\bibnamefont{Ligeti}},
  \bibinfo{author}{\bibfnamefont{D.~J.} \bibnamefont{Robinson}},
  \bibnamefont{and} \bibinfo{author}{\bibfnamefont{W.~L.}
  \bibnamefont{Sutcliffe}} (\bibinfo{year}{2018}{\natexlab{a}}),
  \eprint{1808.09464}.

\bibitem[{\citenamefont{Ray et~al.}(2019{\natexlab{a}})\citenamefont{Ray,
  Sahoo, and Mohanta}}]{Ray:2018hrx}
\bibinfo{author}{\bibfnamefont{A.}~\bibnamefont{Ray}},
  \bibinfo{author}{\bibfnamefont{S.}~\bibnamefont{Sahoo}}, \bibnamefont{and}
  \bibinfo{author}{\bibfnamefont{R.}~\bibnamefont{Mohanta}},
  \bibinfo{journal}{Phys. Rev.} \textbf{\bibinfo{volume}{D99}},
  \bibinfo{pages}{015015} (\bibinfo{year}{2019}{\natexlab{a}}),
  \eprint{1812.08314}.

\bibitem[{\citenamefont{Bernlochner and Ligeti}(2017)}]{Bernlochner:2016bci}
\bibinfo{author}{\bibfnamefont{F.~U.} \bibnamefont{Bernlochner}}
  \bibnamefont{and} \bibinfo{author}{\bibfnamefont{Z.}~\bibnamefont{Ligeti}},
  \bibinfo{journal}{Phys. Rev.} \textbf{\bibinfo{volume}{D95}},
  \bibinfo{pages}{014022} (\bibinfo{year}{2017}), \eprint{1606.09300}.

\bibitem[{\citenamefont{Bernlochner
  et~al.}(2018{\natexlab{b}})\citenamefont{Bernlochner, Ligeti, and
  Robinson}}]{Bernlochner:2017jxt}
\bibinfo{author}{\bibfnamefont{F.~U.} \bibnamefont{Bernlochner}},
  \bibinfo{author}{\bibfnamefont{Z.}~\bibnamefont{Ligeti}}, \bibnamefont{and}
  \bibinfo{author}{\bibfnamefont{D.~J.} \bibnamefont{Robinson}},
  \bibinfo{journal}{Phys. Rev.} \textbf{\bibinfo{volume}{D97}},
  \bibinfo{pages}{075011} (\bibinfo{year}{2018}{\natexlab{b}}),
  \eprint{1711.03110}.

\bibitem[{\citenamefont{Sakaki et~al.}(2013)\citenamefont{Sakaki, Tanaka,
  Tayduganov, and Watanabe}}]{Sakaki:2013bfa}
\bibinfo{author}{\bibfnamefont{Y.}~\bibnamefont{Sakaki}},
  \bibinfo{author}{\bibfnamefont{M.}~\bibnamefont{Tanaka}},
  \bibinfo{author}{\bibfnamefont{A.}~\bibnamefont{Tayduganov}},
  \bibnamefont{and} \bibinfo{author}{\bibfnamefont{R.}~\bibnamefont{Watanabe}},
  \bibinfo{journal}{Phys. Rev.} \textbf{\bibinfo{volume}{D88}},
  \bibinfo{pages}{094012} (\bibinfo{year}{2013}), \eprint{1309.0301}.

\bibitem[{\citenamefont{Biancofiore et~al.}(2013)\citenamefont{Biancofiore,
  Colangelo, and De~Fazio}}]{Biancofiore:2013ki}
\bibinfo{author}{\bibfnamefont{P.}~\bibnamefont{Biancofiore}},
  \bibinfo{author}{\bibfnamefont{P.}~\bibnamefont{Colangelo}},
  \bibnamefont{and} \bibinfo{author}{\bibfnamefont{F.}~\bibnamefont{De~Fazio}},
  \bibinfo{journal}{Phys. Rev.} \textbf{\bibinfo{volume}{D87}},
  \bibinfo{pages}{074010} (\bibinfo{year}{2013}), \eprint{1302.1042}.

\bibitem[{\citenamefont{Bailey et~al.}(2015)}]{Lattice:2015rga}
\bibinfo{author}{\bibfnamefont{J.~A.} \bibnamefont{Bailey}}
  \bibnamefont{et~al.} (\bibinfo{collaboration}{MILC}), \bibinfo{journal}{Phys.
  Rev.} \textbf{\bibinfo{volume}{D92}}, \bibinfo{pages}{034506}
  (\bibinfo{year}{2015}), \eprint{1503.07237}.

\bibitem[{\citenamefont{Caprini et~al.}(1998)\citenamefont{Caprini, Lellouch,
  and Neubert}}]{Caprini:1997mu}
\bibinfo{author}{\bibfnamefont{I.}~\bibnamefont{Caprini}},
  \bibinfo{author}{\bibfnamefont{L.}~\bibnamefont{Lellouch}}, \bibnamefont{and}
  \bibinfo{author}{\bibfnamefont{M.}~\bibnamefont{Neubert}},
  \bibinfo{journal}{Nucl. Phys.} \textbf{\bibinfo{volume}{B530}},
  \bibinfo{pages}{153} (\bibinfo{year}{1998}), \eprint{hep-ph/9712417}.

\bibitem[{\citenamefont{Bailey et~al.}(2014)}]{Bailey:2014tva}
\bibinfo{author}{\bibfnamefont{J.~A.} \bibnamefont{Bailey}}
  \bibnamefont{et~al.} (\bibinfo{collaboration}{Fermilab Lattice, MILC}),
  \bibinfo{journal}{Phys. Rev.} \textbf{\bibinfo{volume}{D89}},
  \bibinfo{pages}{114504} (\bibinfo{year}{2014}), \eprint{1403.0635}.

\bibitem[{\citenamefont{Amhis et~al.}(2014)}]{Amhis:2014hma}
\bibinfo{author}{\bibfnamefont{Y.}~\bibnamefont{Amhis}} \bibnamefont{et~al.}
  (\bibinfo{collaboration}{Heavy Flavor Averaging Group (HFAG)})
  (\bibinfo{year}{2014}), \eprint{1412.7515}.

\bibitem[{\citenamefont{Kurimoto et~al.}(2003)\citenamefont{Kurimoto, Li, and
  Sanda}}]{Kurimoto:2002sb}
\bibinfo{author}{\bibfnamefont{T.}~\bibnamefont{Kurimoto}},
  \bibinfo{author}{\bibfnamefont{H.-n.} \bibnamefont{Li}}, \bibnamefont{and}
  \bibinfo{author}{\bibfnamefont{A.~I.} \bibnamefont{Sanda}},
  \bibinfo{journal}{Phys. Rev.} \textbf{\bibinfo{volume}{D67}},
  \bibinfo{pages}{054028} (\bibinfo{year}{2003}), \eprint{hep-ph/0210289}.

\bibitem[{\citenamefont{Watanabe}(2018)}]{Watanabe:2017mip}
\bibinfo{author}{\bibfnamefont{R.}~\bibnamefont{Watanabe}},
  \bibinfo{journal}{Phys. Lett.} \textbf{\bibinfo{volume}{B776}},
  \bibinfo{pages}{5} (\bibinfo{year}{2018}), \eprint{1709.08644}.

\bibitem[{\citenamefont{Aoki et~al.}(2014)}]{Aoki:2013ldr}
\bibinfo{author}{\bibfnamefont{S.}~\bibnamefont{Aoki}} \bibnamefont{et~al.},
  \bibinfo{journal}{Eur. Phys. J.} \textbf{\bibinfo{volume}{C74}},
  \bibinfo{pages}{2890} (\bibinfo{year}{2014}), \eprint{1310.8555}.

\bibitem[{\citenamefont{Chiu et~al.}(2007)\citenamefont{Chiu, Hsieh, Huang, and
  Ogawa}}]{Chiu:2007km}
\bibinfo{author}{\bibfnamefont{T.-W.} \bibnamefont{Chiu}},
  \bibinfo{author}{\bibfnamefont{T.-H.} \bibnamefont{Hsieh}},
  \bibinfo{author}{\bibfnamefont{C.-H.} \bibnamefont{Huang}}, \bibnamefont{and}
  \bibinfo{author}{\bibfnamefont{K.}~\bibnamefont{Ogawa}}
  (\bibinfo{collaboration}{TWQCD}), \bibinfo{journal}{Phys. Lett.}
  \textbf{\bibinfo{volume}{B651}}, \bibinfo{pages}{171} (\bibinfo{year}{2007}),
  \eprint{0705.2797}.

\bibitem[{\citenamefont{Tanabashi et~al.}(2018)}]{Tanabashi:2018oca}
\bibinfo{author}{\bibfnamefont{M.}~\bibnamefont{Tanabashi}}
  \bibnamefont{et~al.} (\bibinfo{collaboration}{Particle Data Group}),
  \bibinfo{journal}{Phys. Rev.} \textbf{\bibinfo{volume}{D98}},
  \bibinfo{pages}{030001} (\bibinfo{year}{2018}).

\bibitem[{\citenamefont{Akeroyd and Chen}(2017)}]{Akeroyd:2017mhr}
\bibinfo{author}{\bibfnamefont{A.~G.} \bibnamefont{Akeroyd}} \bibnamefont{and}
  \bibinfo{author}{\bibfnamefont{C.-H.} \bibnamefont{Chen}},
  \bibinfo{journal}{Phys. Rev.} \textbf{\bibinfo{volume}{D96}},
  \bibinfo{pages}{075011} (\bibinfo{year}{2017}), \eprint{1708.04072}.

\bibitem[{\citenamefont{Huang et~al.}(2018)\citenamefont{Huang, Li, Lu,
  Paracha, and Wang}}]{Huang:2018nnq}
\bibinfo{author}{\bibfnamefont{Z.-R.} \bibnamefont{Huang}},
  \bibinfo{author}{\bibfnamefont{Y.}~\bibnamefont{Li}},
  \bibinfo{author}{\bibfnamefont{C.-D.} \bibnamefont{Lu}},
  \bibinfo{author}{\bibfnamefont{M.~A.} \bibnamefont{Paracha}},
  \bibnamefont{and} \bibinfo{author}{\bibfnamefont{C.}~\bibnamefont{Wang}},
  \bibinfo{journal}{Phys. Rev.} \textbf{\bibinfo{volume}{D98}},
  \bibinfo{pages}{095018} (\bibinfo{year}{2018}), \eprint{1808.03565}.

\bibitem[{\citenamefont{Ray et~al.}(2019{\natexlab{b}})\citenamefont{Ray,
  Sahoo, and Mohanta}}]{Ray:2019gkv}
\bibinfo{author}{\bibfnamefont{A.}~\bibnamefont{Ray}},
  \bibinfo{author}{\bibfnamefont{S.}~\bibnamefont{Sahoo}}, \bibnamefont{and}
  \bibinfo{author}{\bibfnamefont{R.}~\bibnamefont{Mohanta}},
  \bibinfo{journal}{Eur. Phys. J.} \textbf{\bibinfo{volume}{C79}},
  \bibinfo{pages}{670} (\bibinfo{year}{2019}{\natexlab{b}}),
  \eprint{1907.13586}.

\end{thebibliography}

\end{document}